\documentclass{article}
\usepackage{titling}
\usepackage[margin=1in]{geometry}
\usepackage{authblk}
\usepackage{minitoc}
\usepackage[toc,page,header]{appendix}
\usepackage{lineno}
\usepackage[url=false,doi=false,isbn=false,style=numeric-comp,sorting=none]{biblatex}
\usepackage{acronym}
\usepackage{siunitx}
\usepackage{graphicx}
\usepackage{multirow}
\usepackage{subcaption}
\usepackage{hyperref}
\usepackage{amsmath}
\usepackage{amssymb}
\usepackage{algorithm,algpseudocode}
\usepackage{textcomp, gensymb}   
\usepackage{booktabs}
\usepackage{placeins} 
\usepackage{longtable} 
\usepackage{tabularx}
\usepackage{amsthm}
\theoremstyle{plain}%
\newtheorem{theorem}{Theorem} 

\usepackage{chemformula}
\let\ce\ch

\usepackage{cleveref}   

\crefformat{equation}{eq.~#2#1#3}
\Crefformat{equation}{Eq.~#2#1#3}
\crefmultiformat{equation}
  {eqs.~#2#1#3}
  { and~#2#1#3}
  {, #2#1#3}
  { and~#2#1#3}
\Crefmultiformat{equation}
  {Eqs.~#2#1#3}
  { and~#2#1#3}
  {, #2#1#3}
  { and~#2#1#3}

\usepackage{makecell}
\usepackage{adjustbox}
\usepackage{array}
\newcolumntype{T}{>{\collectcell\makecell[tl]}l<{\endcollectcell}}

\AtEveryBibitem{
    \clearfield{urlyear}
    \clearfield{urlmonth}
}

\newcommand{\labelphantom}[1]{%
  \parbox{0pt}{\phantomsubcaption\label{#1}}%
}

\newcommand{\sn}[2]{\ensuremath{#1\times 10^{#2}}}

\newcommand{\smiles}[1]{\texttt{#1}}

\newcommand{\ci}[3]{#1 [\ensuremath{#2}, \ensuremath{#3}]}
\newcommand{\cistacked}[3]{\makecell{#1 \\ \relax [#2, #3]}}

\newcommand{\datadrop}[1][doi:10.5281/zenodo.17527149]{\href{https://doi.org/10.5281/zenodo.17527149}{#1}}

\newcommand{\ghcommit}[2]{\href{https://github.com/#1/tree/#2}{#1}}

\newcommand{\hfrepo}[1]{\href{https://huggingface.co/#1}{#1}}

\let\oldcite\cite
\renewcommand{\cite}[1]{\unskip~\oldcite{#1}}

\ExplSyntaxOn
\NewDocumentCommand{\tok}{m}
 {
    \seq_set_from_clist:Nn \l_tmpa_seq { #1 }
    \seq_indexed_map_inline:Nn \l_tmpa_seq {
        \int_compare:nF { ##1 = 1 } {\  }
        {
          \fboxsep=0.1em
          \fbox{\texttt{##2}}
        }
    }
 }
\ExplSyntaxOff

\usepackage{xcolor}

\definecolor{maize}{cmyk}{0.0, 0.18, 1.0, 0.0}
\definecolor{blue}{cmyk}{1.0, 0.60, 0.0, 0.6}
\definecolor{cat:blue}{RGB}{31, 119, 180}
\definecolor{cat:green}{RGB}{44, 160, 44}

\acrodef{MIST}{Molecular Insight SMILES Transformers}

\acrodef{NLP}{Natural Language Processing}
\acrodef{CV}{Computer Vision}
\acrodef{MLM}{Masked Language Modeling}
\acrodef{MLP}{Multi-Layer Perceptron}
\acrodef{LLM}{Large Language Model}
\acrodef{PCA}{Principle Component Analysis}
\acrodef{t-SNE}{t-distributed Stochastic Neighbor Embedding}
\acrodef{UMAP}{Uniform Manifold Approximation and Projection}
\acrodef{SVM}{Support Vector Machines}
\acrodef{LAMB}{Layerwise Adaptive Large Batch}
\acrodef{GNN}{Graph Neural Network}
\acrodef{DDP}{Distributed Data Parallel}
\acrodef{MR}{Multi-Replica}
\acrodef{GELU}{Gaussian Error Linear Units}
\acrodef{SciFM}{Scientific Foundation Model}
\acrodef{FM}{Foundation Model}
\acrodef{GAS}{Gradient Accumulation Steps}
\acrodef{SciML}{Scientific Machine Learning}

\acrodef{MAE}{Mean Absolute Error}
\acrodef{RMSE}{Root-Mean-Squared Error}
\acrodef{RMSD}{Root-Mean-Squared Deviation}
\acrodef{MSE}{Mean-Squared Error}
\acrodef{MCMC}{Markov Chain Monte Carlo}
\acrodef{NUTS}{No-U-Turn Sampler}
\acrodef{WAIC}{Watanabe–Akaike Information Criterion}
\acrodef{AIC}{Akaike Information Criteria}
\acrodef{BIC}{Bayesian Information Criteria}
\acrodef{DIC}{Deviance Information Criteria}
\acrodef{MAPE}{Mean Absolute Percentage Error}
\acrodef{MLE}{Maximum Likelihood Estimate}
\acrodef{ESS}{Effective Sample Size}
\acrodef{AUROC}{Area Under the Receiver Operating Characteristic}
\acrodef{RMT}{Random Matrix Theory}
\acrodef{HT-SR}{Heavy-Tailed Self-Regularization}
\acrodef{ESD}{Empirical Spectral Density}
\acrodef{UMAP}{Uniform Manifold Approximation and Projection}
\acrodef{HYDRA}{Hyperbolic Distance Recovery and Approximation}
\acrodef{TDA}{Topological Data Analysis}
\acrodef{LHS}{Latin Hypercube Sampling}

\acrodef{HOMO}{Highest Occupied Molecular Orbital}
\acrodef{LUMO}{Lowest Unoccupied Molecular Orbital}
\acrodef{HOMO-LUMO}{Highest Occupied/ Lowest Unoccupied Molecular Orbitals}
\acrodef{PAH}{Polycyclic Aromatic Hydrocarbon}
\acrodef{DFT}{Density Functional Theory}
\acrodef{QSAR}{Quantitative Structure-Activity Relationships}
\acrodef{KT}{Kamlet-Taft}
\acrodef{NMR}{Nuclear Magnetic Resonance}
\acrodef{DNN}{Deep Neural Network}
\acrodef{VFT}{Vogel-Fulcher-Tammann}
\acrodef{AEM}{Advanced Electrolyte Model}
\acrodef{SMILES}{Simplified Molecular Input Line Entry System}
\acrodef{LHCE}{Localized High-Concentration Electrolytes}
\acrodef{SEI}{Solid Electrolyte Interphase}
\acrodef{CT}{Characteristic Temperature}
\acrodef{SOAP}{Smooth Overlap of Atomic Positions}
\acrodef{DN}{Donor Number}
\acrodef{DMSO}{Dimethyl Sulfoxide}

\acrodef{ACN}{Acetonitrile}
\acrodef{BC}{Butylene Carbonate}
\acrodef{EGDA}{Ethylene glycol diacetate}
\acrodef{EC}{Ethylene Carbonate}
\acrodef{PC}{Propylene Carbonate}
\acrodef{DMC}{Dimethyl Carbonate}
\acrodef{EMC}{Ethyl Methyl Carbonate}
\acrodef{FEC}{Fluoroethylene Carbonate}
\acrodef{DEC}{Diethyl Carbonate}
\acrodef{DME}{Dimethyl Ether}
\acrodef{NPA}{Natural Population Analysis}

\acrodef{OR}{Olfactory-Receptor}
\acrodef{QSAR}{Quantitative Structure-Activity Relationship}
\acrodef{RO5}{Rule of Five}
\acrodef{FLOP}{Floating Point Operation}
\acrodef{LDPC}{Low-Density Parity-Check}
\acrodef{BEC}{Binary Erasure Channel}
\acrodef{BEQ}{Binary Erasure Quantization}
\acrodef{SEMF}{Semi-empirical Mass Formula}
\acrodef{GSD}{Generalized Subset Design}
\acrodef{SI}{Supporting Information}
\acrodef{DREAM}{Dialogue for Reverse Engineering Assessment and Methods}
\acrodef{POM}{Perceptual Odour Map}

\usepackage{ifthen}
\usepackage{xcolor}
\usepackage{currfile} 
\newboolean{hidecomments}
\setboolean{hidecomments}{false}   

\makeatletter
\newwrite\comments@out
\AtBeginDocument{\immediate\openout\comments@out=\jobname.comment}
\AtEndDocument{\immediate\closeout\comments@out}

\newcommand{\comments@write}[3]{
  \begingroup
    \edef\comments@page{\thepage}%
    \edef\comments@file{%
      \ifdefined\currfiledir
        \currfiledir\currfilename
      \else
        \ifdefined\currfileabspath
          \currfileabspath
        \else
          \currfilename
        \fi
      \fi
    }
    \edef\comments@line{\ifdefined\inputlineno\the\inputlineno\else?\fi}%
    \immediate\write\comments@out{#2 [#1] @ p.\comments@page\space (\comments@file:\comments@line): \detokenize{#3}}%
  \endgroup
}

\newcommand{\commenter}[3][red]{
  \expandafter\DeclareRobustCommand\csname #2\endcsname[1]{%
    \comments@write{#2}{#3}{##1}%
    \ifthenelse{\boolean{hidecomments}}{}{%
      \textcolor{#1}{\sffamily[\relax #3: ##1]}%
    }%
  }%
}
\makeatother

\commenter[red]{michael}{Michael}
\commenter[brown]{kareem}{Kareem}
\commenter[green]{alw}{Alex}
\commenter[cyan]{ab}{Anoushka}
\commenter[blue]{austin}{Austin}

\setboolean{hidecomments}{false}   

\begin{document}

\title{Foundation Models for Discovery and Exploration \\ in Chemical Space}

\newcommand{\fnm}[1]{#1}
\newcommand{\sur}[1]{#1}
\newcommand{\orgdiv}[1]{#1}
\newcommand{\orgname}[1]{#1}
\newcommand{\orgaddress}[1]{#1}
\newcommand{\city}[1]{#1}
\newcommand{\state}[1]{#1}
\newcommand{\country}[1]{#1}
\newcommand{\email}[1]{}
\newcommand{\equalcont}{}
\newcommand{\equalcontxt}{}

\author[1]{\fnm{Alexius} \sur{Wadell}$^*$}\equalcont
\author[1]{\fnm{Anoushka} \sur{Bhutani}$^*$}\equalcont
\author[2]{\fnm{Victor} \sur{Azumah}} 
\author[3]{\fnm{Austin R.} \sur{Ellis-Mohr}} 
\author[4]{\fnm{Andrew J.} \sur{Stier}} 
\author[5,6]{\fnm{Kareem} \sur{Hegazy}} 
\author[7,8]{\fnm{Alexander} \sur{Brace}} 
\author[1]{\fnm{Hancheng} \sur{Zhao}} 
\author[1]{\fnm{Celia} \sur{Kelly}} 
\author[3]{\fnm{Anuj K.} \sur{Nayak}} 
\author[1]{\fnm{Yuhan} \sur{Chen}} 
\author[1]{\fnm{Dimitrios} \sur{Simatos}} 
\author[1]{\fnm{Hongyi} \sur{Lin}} 
\author[8]{\fnm{Murali} \sur{Emani}} 
\author[8]{\fnm{Venkatram} \sur{Vishwanath}} 
\author[9]{\fnm{Kevin} \sur{Gering}} 
\author[10]{\fnm{Melisa} \sur{Alkan}} 
\author[10]{\fnm{Tom} \sur{Gibbs}} 
\author[10]{\fnm{Jack} \sur{Wells}} 
\author[11]{\fnm{Wesley W.} \sur{Qian}} 
\author[11]{\fnm{Richard C.} \sur{Gerkin}} 
\author[11]{\fnm{Benjamin} \sur{Amorelli}} 
\author[11]{\fnm{Alexander B.} \sur{Wiltschko}} 
\author[3, 12, 13]{\fnm{Lav R.} \sur{Varshney}} 
\author[14]{\fnm{Bharath} \sur{Ramsundar}} 
\author[1,15]{\fnm{Karthik} \sur{Duraisamy}} 
\author[5,6,16]{\fnm{Michael W.} \sur{Mahoney}} 
\author[7,8]{\fnm{Arvind} \sur{Ramanathan}} 
\author[1,15]{\fnm{Venkatasubramanian} \sur{Viswanathan}}\email{venkvis@umich.edu} 

\affil[*]{Equal contribution}
\affil[1]{\orgdiv{Department of Mechanical Engineering}, \orgname{University of Michigan}, \orgaddress{\city{Ann Arbor}, \state{MI}, \country{USA}}}
\affil[2]{\orgdiv{Department of Chemical Engineering}, \orgname{University of Michigan}, \orgaddress{\city{Ann Arbor}, \state{MI}, \country{USA}}}
\affil[3]{\orgdiv{Department of Electrical and Computer Engineering}, \orgname{University of Illinois at Urbana-Champaign}, \orgaddress{\city{Urbana}, \state{IL}, \country{USA}}}
\affil[4]{\orgname{The Santa Fe Institute}, \orgaddress{\city{Santa Fe},  \state{NM}, \country{USA}}}
\affil[5]{\orgname{International Computer Science Institute}, \orgaddress{\city{Berkeley}, \state{CA}, \country{USA}}}
\affil[6]{\orgdiv{Department of Statistics}, \orgname{University of California}, \orgaddress{\city{Berkeley}, \state{CA}, \country{USA}}}
\affil[7]{\orgdiv{Department of Computer Science}, \orgname{University of Chicago}, \orgaddress{\city{Chicago}, \state{IL}, \country{USA}}}
\affil[8]{\orgname{Argonne National Laboratory}, \orgaddress{\city{Lemont}, \state{IL}, \country{USA}}}
\affil[9]{\orgname{Idaho National Laboratory}, \orgaddress{\city{Idaho Falls}, \state{ID}, \country{USA}}}
\affil[10]{\orgname{NVIDIA Corporation}, \orgaddress{\city{Santa Clara}, \state{CA}, \country{USA}}}
\affil[11]{\orgname{Osmo Labs, PBC}, \orgaddress{\city{New York}, \state{NY}, \country{USA}}}
\affil[12]{\orgname{AI Innovation Institute, Stony Brook University}, \orgaddress{\city{Stony Brook}, \state{NY}, \country{USA}}}
\affil[13]{\orgname{Brookhaven National Laboratory}, \orgaddress{\city{Upton}, \state{NY}, \country{USA}}}
\affil[14]{\orgname{Deep Forest Sciences}, \orgaddress{\city{Palo Alto}, \state{CA}, \country{USA}}}
\affil[15]{\orgdiv{Department of Aerospace Engineering}, \orgname{University of Michigan}, \orgaddress{\city{Ann Arbor}, \state{MI}, \country{USA}}}
\affil[16]{\orgname{Lawrence Berkeley National Laboratory}, \orgaddress{\city{Berkeley}, \state{CA}, \country{USA}}}

\begin{refsegment}
\maketitle

\clearpage
\doparttoc 
\faketableofcontents 
\part{} 
    
\begin{abstract}
  Accurate prediction of atomistic, thermodynamic, and kinetic properties from molecular structures underpins materials innovation.
Existing computational and experimental approaches lack the scalability required to navigate chemical space efficiently~\cite{DobChemicalSpaceBiology2004,MFC+HowMachineLearning2021}.
Scientific foundation models trained on large unlabelled datasets offer a path towards navigating chemical space across application domains.
Here, we develop MIST, a family of molecular foundation models with up to an order of magnitude more parameters and data than prior works~\cite{RBC+LargescaleChemicalLanguage2022,SSB+LargeEncoderDecoderFamily2024}. 
Trained using a novel tokenizer, Smirk, which comprehensively captures nuclear, electronic, and geometric information, \acs{MIST} learns a diverse range of molecules.
MIST models have been fine-tuned to predict more than 400 structure-property relationships and have been shown to match or exceed state-of-the-art performance across diverse benchmarks, from physiology to electrochemistry and quantum chemistry. 
We demonstrate the ability of these models to solve real-world problems across chemical space --- multiobjective electrolyte solvent screening, isotope half-life prediction, stereochemical reasoning for chiral organometallic compounds and mixture property prediction.  
The clearest demonstration of a foundation model is its ability to solve problems that were neither explicit targets of training nor central to the intentions of its developers.  
We identify olfactory perception mapping as that domain, and show that \acs{MIST} accurately predicted scent profiles and learned a hierarchical representation of olfactory space consistent with hyperbolic geometry.  
To experimentally validate these capabilities, we worked with the authors of POM\cite{leePrincipalOdorMap2023} to analyze discordant triplets, a challenging chemistry problem deliberately designed so that structural similarity fails as a guide to perception, and showed that MIST correctly classified 64\% of the triplets while the prior state-of-the-art achieved 50\%.
We formulated hyperparameter aware Bayesian neural scaling laws which eliminate the need for hyperparameter sweeps at every scale, making training large compute-optimal models feasible on a limited compute budget.
The methods and findings presented here represent a significant step towards accelerating materials discovery, design, and optimization using foundation models; they provide valuable guidance on training compute-optimal scientific foundation models.
All code, model weights, and training recipes have been open-sourced to accelerate further exploration of chemical space.
\end{abstract}

Chemical space is extraordinarily vast, with estimates placing the number of small molecules on the order of \(10^{60}\)~\cite{DobChemicalSpaceBiology2004}.
Moreover, it is heterogeneous, encompassing everything from simple organic molecules to inorganic salts to complex organometallics and mixtures of them~\cite{kirkpatrickChemicalSpace2004,kimApproachExpandingChemical2024}.
Navigating this space to find molecules with desirable chemical, thermal, biological, or perceptual properties has traditionally relied on capital- and time-intensive wet lab experimentation and/or computationally expensive first-principles simulations~\cite{MFC+HowMachineLearning2021}.
The cost of these methods becomes intractable at scale, limiting current exploration to only a minuscule fraction of chemical space~\cite{kirkpatrickChemicalSpace2004,lipinskiNavigatingChemicalSpace2004,gaoGenerativeAINavigating2025}.

Over the past decade, molecular machine learning has accelerated material development cycles, improved efficiency, and reduced costs~\cite{MFC+HowMachineLearning2021}.
However, most state-of-the-art molecular property prediction models still rely on labelled training data, which is sparse, scarce, and expensive to generate~\cite{MFC+HowMachineLearning2021,FDHLWhyBigData2022,ZRA+DifferentiableModelingOptimization2024}.
Consequently, these models are plagued by poor generalization, hampering their utility in real-world applications where novel molecules of interest differ substantially from those contained in curated training datasets~\cite{FDHLWhyBigData2022}.

The recent success of transfer learning in \ac{NLP}~\cite{DCLTBERTPretrainingDeep2019} and \ac{CV}~\cite{radfordLearningTransferableVisual2021} has led to interest in the development of large-scale \acp{SciFM}~\cite{CGRChemBERTaLargeScaleSelfSupervised2020,RBC+LargescaleChemicalLanguage2022,SSB+LargeEncoderDecoderFamily2024,xiangVisionLanguageFoundation2025}.
Rather than train on scarce labelled datasets, \acp{SciFM} leverage large unlabelled scientific corpora to uncover representations broadly applicable to downstream tasks~\cite{SSB+LargeEncoderDecoderFamily2024,OYIWInterpretableScientificFoundation2024}, thereby alleviating the need for ever-larger labelled datasets.
Molecular chemistry is especially well-positioned to benefit from \ac{NLP} \ac{FM} breakthroughs, as large libraries of text-encoded molecules are readily available~\cite{EnaREALSpace2024}.
Despite this, existing chemical \acp{FM} remain specialized to particular subdomains, thus failing to provide a general foundation on which to build~\cite{BHA+OpportunitiesRisksFoundation2022}, and with limited ability to generalize across chemical space~\cite{CGRChemBERTaLargeScaleSelfSupervised2020,RBC+LargescaleChemicalLanguage2022,SSB+LargeEncoderDecoderFamily2024}.
There is a need for a single model capable of learning from billions of diverse molecules, while resolving isotopic, stereochemical, structural, and electronic nuances.
Such a model can then discover unifying scientific principles and accelerate chemical design across application domains, from energy to olfaction.

Here, we introduce \acf{MIST}, a family of molecular \ac{FM}s with expanded coverage of chemical space trained using Smirk~\cite{WBVTokenizationMolecularFoundation2026}, a novel tokenization scheme which captures a comprehensive representation of molecular structure including nuclear, electronic, and geometric features.
Motivated by scaling trends in \ac{NLP}~\cite{HBM+TrainingComputeOptimalLarge2022,KMH+ScalingLawsNeural2020,BCC+DeepSeekLLMScaling2024},
the largest \ac{MIST} models were trained with an order of magnitude more parameters and data than prior work~\cite{SSB+LargeEncoderDecoderFamily2024,RBC+LargescaleChemicalLanguage2022,CGRChemBERTaLargeScaleSelfSupervised2020},
matching or exceeding the state-of-the-art across diverse chemical benchmarks.
We demonstrate \ac{MIST}'s efficacy in solving problems across chemical space: multiobjective electrolyte solvent screening, organometallic stereochemical reasoning, isotope half-life prediction, mixture property prediction and olfactory perception mapping.
Moreover, we show that generalizable chemical concepts are learned by the model during training.
We examine every layer of the model for encoded chemical knowledge and find identifiable patterns, such as H\"{u}ckel's aromaticity rule and Lipinski's \ac{RO5}, not explicitly labelled in the training data.
These findings position \ac{MIST} as a powerful tool for discovery and exploration in chemical space.

To train \ac{MIST} compute-optimally, we augment the existing formulation of neural scaling laws~\cite{HBM+TrainingComputeOptimalLarge2022} to derive new insights about the importance of data quality from the scaling exponents, which we expect to inform the data curation for \acp{FM} and \acp{SciFM} across scientific disciplines.
Neural scaling laws introduce the notion of a compute-optimal frontier with respect to dataset size, model size, and compute budget~\cite{HBM+TrainingComputeOptimalLarge2022}.
However, current formulations assume all other hyperparameters (e.g., learning rate) have already been optimally tuned~\cite{KMH+ScalingLawsNeural2020,HBM+TrainingComputeOptimalLarge2022}.
Here, we introduce penalty terms to explicitly model the impact of off-optimal hyperparameter selection on data, model, and compute scaling.
We parameterize these penalized neural scaling laws using Bayesian parameter estimation, providing a statistically rigorous framework for identifying scaling exponents.
We show how the fitted scaling of data versus model, $D \propto N^{\frac{\alpha}{\beta}}$, can provide a rich signal of the model’s operating regime and potential limitations of the pretraining corpus.

\ac{MIST} establishes a paradigm for \ac{FM}-guided, accelerated design of molecular materials and mixture formulations.
Spanning chemical space from organics to organometallics and encompassing nuclear, electronic, and geometric diversity, the \ac{MIST} models are poised to deepen our understanding of molecules and materials.
Simultaneously, this work provides a path towards economizing and optimizing the training of future \acp{SciFM} by augmenting the existing compute-optimal neural scaling laws with hyperparameter penalties and linking data quality to scaling efficiency.

\begin{figure}
    \centering
    \labelphantom{fig:mist_overview}
    \labelphantom{fig:mist_arch_diagram}
    \includegraphics[width=\linewidth]{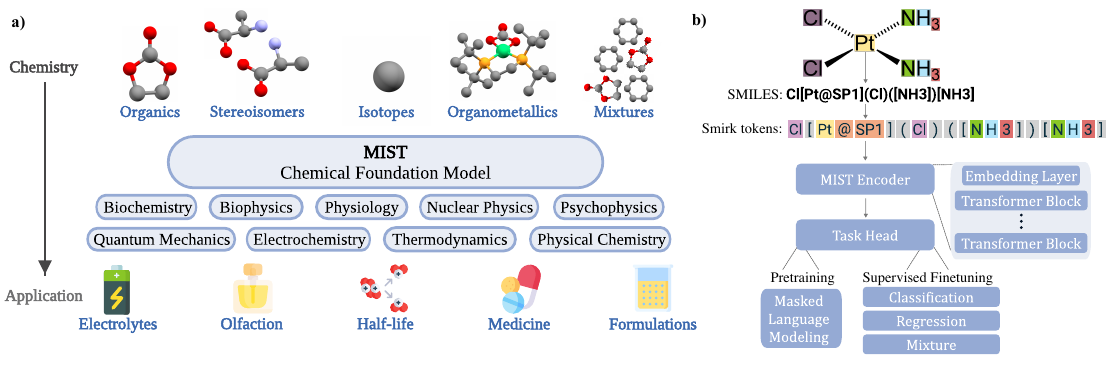}
    \caption{
        \textbf{\acs{MIST} models molecules across chemical space, accelerating a broad range of materials design tasks.}
        (\subref*{fig:mist_overview})~\acf{MIST} is a family of molecular \ac{FM}s which match or exceed state-of-the-art performance across diverse benchmarks, from physiology to quantum chemistry.
        The models solve real-world problems across chemical space --- ranging from multiobjective electrolyte solvent screening and olfactory perception mapping to stereochemical reasoning for organometallic compounds and mixture property prediction.
        (\subref*{fig:mist_arch_diagram})~The \ac{MIST} models are encoder-only transformers~\cite{VSP+AttentionAllYou2017,LOG+RoBERTaRobustlyOptimized2019} pretrained using \acl{MLM} and a novel tokenization scheme, Smirk~\cite{WBVTokenizationMolecularFoundation2026}, to learn informative vector embeddings of molecules from their \acs{SMILES} representations.
        During fine-tuning, the molecular embeddings are processed by small task networks, to predict properties of interest.
    }
\end{figure}

\section{Results}
Molecular \acp{FM} in the \ac{MIST} family range in size from a few million up to 1.8 billion parameters, and are pretrained on the \ac{SMILES} representations of up to 6B molecules.
The two primary pretrained encoders (Methods~\cref{tab:arch_table} and Supplementary~\cref{sec:si:pretraining}) discussed in this paper are MIST-28M (28 million parameters trained on 245 million molecules) and MIST-1.8B (1.8 billion parameters trained on 2 billion molecules).
\ac{MIST} models use an encoder-only transformer~\cite{VSP+AttentionAllYou2017,LOG+RoBERTaRobustlyOptimized2019} architecture and were pretrained using the \acf{MLM} objective~\cite{DCLTBERTPretrainingDeep2019} (Methods~\cref{sec:methods:pretraining}) on the Enamine REALSpace Dataset~\cite{EnaREALSpace2024} (Supplementary~\cref{sec:si:pretraining}), which consists of synthetically accessible organic molecules.

 \ac{MIST} models have been fine-tuned (Methods~\cref{sec:methods:fine-tuning}) on over 400 molecular and formulation property prediction tasks (Supplementary~\cref{sec:si:fine-tuning}), and demonstrate state-of-the-art performance across many chemical machine learning benchmarks (Extended Data~\cref{tab:molnet_class_bench,tab:molnet_reg_bench,tab:ionic_cond}).
Fine-tuned \ac{MIST} models consist of the pretrained \ac{MIST} encoder followed by a task network.
The task networks (except for the specialized mixture task networks described in \cref{sec:mixtures}) are two-layer \acp{MLP}.

To train \ac{MIST} models on \ac{SMILES} sequences for a diverse range of molecules, the Smirk~\cite{WBVTokenizationMolecularFoundation2026} tokenization algorithm was developed.
In contrast to existing tokenizers for chemistry, Smirk captures nuclear, electronic, and geometric features enabling \acp{SciFM} to learn a rich representation of molecular structure~\cite{WBVTokenizationMolecularFoundation2026}.
This capacity enables fine-tuning on chemistries including radioactive isotopes and organometallic complexes.
Interestingly, we find that despite the limited diversity of the pretraining dataset (discussed further in~\cref{sec:scaling} and Supplementary~\cref{sec:si:data_quality}), the downstream performance of \ac{MIST} models benefits from the pretraining stage, even on tasks involving far more diverse chemical species (\cref{sec:organometallics} and Supplementary~\cref{sec:si:pretraining_adv}).

We developed and leveraged hyperparameter-penalized neural scaling laws (\cref{sec:scaling}) to remain on the compute-optimal frontier while pretraining \ac{MIST} models.
We reduced the cost of fitting scaling laws, using Bayesian parameter estimation and design of experiments~\cite{SVH+GeneralizedSubsetDesigns2017}.
This reduced our compute cost by an estimated 10 petaflop-days, an order of magnitude reduction, relative to the grid search widely used to fit neural scaling laws~\cite{li2025misfittingsurveyscalinglaws,HBK+ScalingLawsComputeOptimal2024,KMH+ScalingLawsNeural2020}.

\subsection{MIST has Applications Across Chemical Space}
\begin{figure}
    \centering
    \labelphantom{fig:lio2_screening_pareto}
    \labelphantom{fig:isotope_clusters}
    \labelphantom{fig:tmQM}
     \labelphantom{fig:mist_mix_excess_arch}
    \labelphantom{fig:mist_mix_ionic_arch}
    \labelphantom{fig:soap_outliers}
    \labelphantom{fig:skew_parity}
    \labelphantom{fig:ionic_prediction}
    \includegraphics[width=\linewidth]{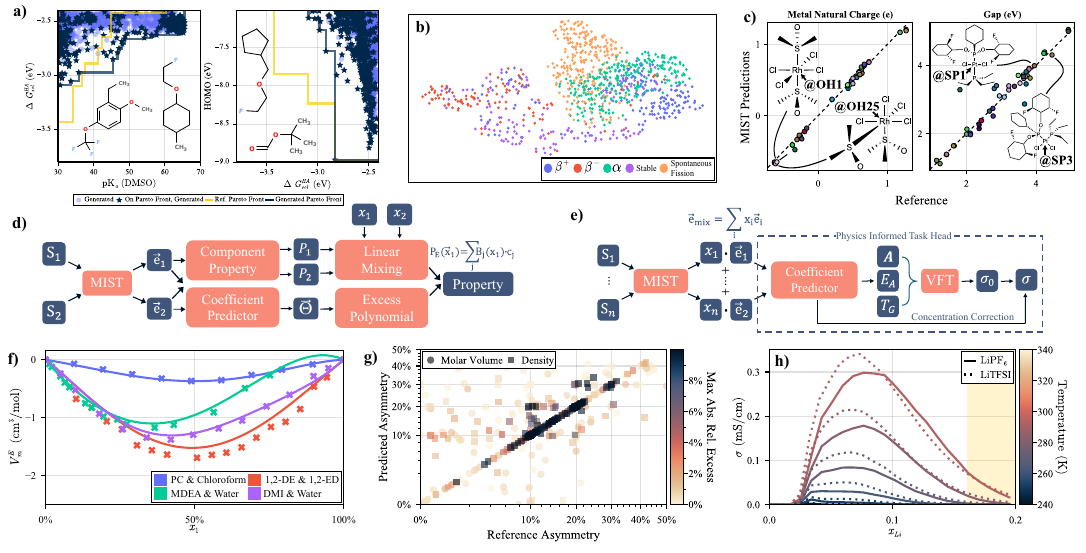}
    \caption{
        \textbf{\acs{MIST} rapidly explores chemical space.}
        (\subref*{fig:lio2_screening_pareto})~\ac{MIST}-28M accurately predicts quantum, chemical, and thermodynamic descriptors for electrolyte design --- including orbital energy levels and solvatochromic parameters --- when fine-tuned on labelled datasets as small as $\sim$180 samples.
        Using fine-tuned MIST-28M models, we identified 390 solvent molecules for lithium-air batteries on the Pareto front, improving stability against hydrogen abstraction and encouraging solution-mediated peroxide growth relative to our reference set.
        (\subref*{fig:isotope_clusters})~A \ac{UMAP} projection of embedding vectors from a \ac{MIST}-28M variant fine-tuned on the half-lives of radioisotopes reveals nuclide clusters organized by dominant decay channels.
        (\subref*{fig:tmQM})~Parity plots for \ac{MIST}-28M models, fine-tuned on transition metal organometallic complexes, with annotated examples of stereoisomer pairs.
        Stereochemical nuances are accurately captured by the Smirk tokenizer, enabling \ac{MIST} to differentiate between isomers.
        (\subref*{fig:mist_mix_excess_arch})~To fine-tune \ac{MIST} models on binary mixture properties, we use a specialized permutation-invariant task network.
        This task network uses \ac{MIST} embeddings $\vec{e_i}$ to compute the linear mixing $P_L$ and excess $P_E$ contributions to the mixture property $P_{\mathrm{mix}}$.
        (\subref*{fig:mist_mix_ionic_arch})~Electrolyte ionic conductivity is predicted using a physics-informed task network from a concentration-weighted sum of \ac{MIST} embeddings.
        (\subref*{fig:soap_outliers})~
        \ac{MIST} mixture property models accurately predict excess properties, including for challenging mixtures not captured by similarity-based descriptors~\cite{KADVExcessDensityDescriptor2025}.
        (\subref*{fig:skew_parity})~Identifying additives with high relative excess at small concentrations is critical during electrolyte formulation~\cite{SVArtificialIntelligenceElectrolyte2025}.
        \ac{MIST} models accurately estimate the composition at the excess extrema.
        (\subref*{fig:ionic_prediction})~\ac{MIST} models learn physically consistent trends in ionic conductivity with salt mole fraction ($x_{Li}$), correctly capturing the expected decay outside the training data range ($x_{Li} > 0.15$, highlighted in yellow).
    }
\end{figure}

The core advantage of a foundation model is that it can be adapted to a wide range of downstream tasks given a small number of labelled examples~\cite{BHA+OpportunitiesRisksFoundation2022}.
We demonstrate the \ac{MIST} models' efficacy as \acp{SciFM} by fine-tuning variants of \ac{MIST} to predict over 400 properties --- including quantum mechanical, thermodynamic, biochemical, and psychophysical properties --- from a molecule’s \ac{SMILES} representation (\cref{fig:mist_overview} and Supplementary~\cref{sec:si:datasets}).
The \ac{MIST} encoders are fine-tuned on labelled datasets as small as $\sim$180 examples.
The encoders are fine-tuned on single molecule property prediction (classification and regression) tasks by attaching a small two-layer \ac{MLP} (Methods~\cref{sec:methods:fine-tuning}).
They are fine-tuned on mixture property prediction tasks using a physics-informed task network architecture (Methods~\cref{sec:methods:mixtures}).
As we show in the following sections, these fine-tuned models unlock problem-solving capabilities across vast regions of chemical space at multiple scales, from single-molecule electrolyte solvents to large chiral organometallic structures and complex multi-component mixtures.
Notably, in olfaction—a stringent test of transfer beyond the intended scope of pretraining—we experimentally validated \ac{MIST}'s ability to resolve structure–odour discordances, demonstrating that its learned representations capture perceptual relationships not evident from structure alone.
\subsubsection{High-Throughput Screening for Electrolytes}
\label{sec:screening}

Energy storage and conversion devices require electrolyte materials that enable functionality for various applications such as electric mobility, grid storage and chemicals production~\cite{XuNonaqueousLiquidElectrolytes2004,XuElectrolytesInterphasesLiIon2014}.
Finding new molecules that meet the full spectrum of physicochemical property criteria for use in energy storage devices --- electrochemical stability, a wide operating liquid range, and good transport properties --- is a challenging multiobjective optimization problem~\cite{XuElectrolytesInterphasesLiIon2014}.
High-throughput computational screening has emerged as an efficient and cost-effective strategy to accelerate electrolyte discovery, reducing the number of wet-lab experiments required~\cite{Cheng2015}.
Fine-tuned \ac{MIST} models provide fast and accurate predictions for a broad range of electrolyte properties, making them particularly effective in these pipelines.

We built a high-throughput screening pipeline for electrolyte design (Methods~\cref{sec:methods:screen}) using fine-tuned \ac{MIST}-28M models.
The screening pipeline begins with a seed inventory from which novel molecules are generated using FASMIFRA~\cite{BTMolecularGenerationFast2021}.
The generated molecules are screened against the specified criteria using \ac{MIST} models fine-tuned to predict the relevant properties.
We demonstrate the efficacy of this pipeline in two settings: first, a broad search for solvents with improved electrochemical and thermophysical stability (Supplementary~\cref{fig:si:screening_electrolytes}) and second, a more tightly constrained problem of solvent discovery for lithium--air batteries (Supplementary~\cref{sec:si:lithium_air_screening}).

Electrochemical stability was evaluated using a MIST-28M model fine-tuned on the QM9 dataset~\cite{RDRVQuantumChemistryStructures2014} and
thermal stability with models fine-tuned on a \ac{CT} dataset to predict melting and boiling points.
For lithium--air screening, additional models were fine-tuned to predict $pK_a$ in \ac{DMSO} (Supplementary~\cref{sec:si:pKa_dmso}), \acf{KT} solvent parameters (Supplementary~\cref{sec:si:kt}) and normalized Dimroth--Reichardt solvent polarity ($E_T^{N}$,~\cref{sec:si:dimroth_eth}).
Across a curated set of 92 electrolyte solvents, these models recovered expected property relationships, including the established linear correlation between flash point and boiling point provided by~Hess et al.\cite{HWWFlammabilityLiIonBattery2015} (\cref{fig:si:screening_electrolytes}).
Selected predictions on generated molecules were further checked using \ac{DFT} as discussed in Supplementary~\cref{sec:si:qm9_replication}.

Applied to the joint electrochemical and thermophysical screen, this workflow evaluated 90 million molecules in 8 hours on 8 NVIDIA A100 GPUs and identified 63 Pareto-optimal solvent candidates.
Critically, the generated candidates improved upon both the electronic and thermal properties of commonly used electrolytes (\cref{fig:si:screening_pareto}).

Analyzing molecules on the Pareto front, we found that generating and evaluating novel molecules during a screening campaign is important.
We evaluated the novelty of a molecule using \cref{eq:molecular_surprise}, which we call the \emph{molecular surprise}:
\begin{equation}\label{eq:molecular_surprise}
    - \sum_{i=0}^N \ln \frac{\exp Q(x_i)}{\sum_{x \in V} \exp Q(x)} ,
\end{equation}
where \(Q(x_i)\) gives the model's logit score for token \(i\)  and the denominator marginalizes over all other tokens in the vocabulary \(V\).
Access to more surprising molecules yielded several Pareto front candidates with improved electrochemical stability (\cref{fig:si:screening_creativity}), highlighting the importance of generating and evaluating novel molecules during a screening campaign.

Pareto front molecules (Supplementary~\cref{fig:pareto_front_p1,fig:pareto_front_p2,fig:pareto_front_p3,fig:pareto_front_p4,fig:pareto_front_p5}) were consistent with established electrolyte heuristics --- 57\% were fluorinated and esters were the most common functional group~\cite{XuElectrolytesInterphasesLiIon2014}.
Additionally, several Pareto-optimal candidates also exhibit mixed halogenation, including fluorine/bromine substitution in 1,8-dibromoperfluorooctane (Molecule 4, Supplementary~\cref{fig:pareto_front_p1}) and fluorine/chlorine substitution in (3S,5R)-pentachloro-nonafluorohexane (Molecule 29, Supplementary~\cref{fig:pareto_front_p2}). 
Their emergence suggests that the workflow recovers not only canonical fluorination-based design rules~\cite{zhangNotAllFluorination2020}, but also an underexplored and chemically interpretable space in which partial replacement of fluorine with heavier halogens tunes the trade-off between electrochemical robustness and volatility. 
Follow-up analysis  (Supplementary~\cref{sec:si:mixed_halogenation_validation}) showed that fluorine-dominant mixed halogenation can preserve electrochemical stability, whereas heavier halogens tend to raise boiling and flash points relative to fully fluorinated analogues.

We next applied the framework to the more challenging problem of solvent discovery for lithium--air batteries, where viable candidates must satisfy a narrow and mechanistically constrained design window. 
Imposing additional bounds associated with solution-mediated peroxide growth and resistance to hydrogen abstraction (Supplementary~\cref{sec:si:lithium_air_screening})  yielded 390 Pareto-front candidate solvents. 
These molecules were enriched in chemically plausible Li--O$_2$ electrolyte motifs, including haloalkanes, ethers, arenes, amines and fluorinated scaffolds.
This indicates that the model identifies structurally interpretable solvent families rather than arbitrary admissible structures (Supplementary~\cref{sec:si:lio2_design_rules}). 
This result provides a stringent demonstration that fine-tuned \ac{MIST} models can translate literature-derived electrolyte design rules into a scalable molecular discovery workflow in a chemically meaningful regime.
\subsubsection{Predicting Isotope Half-Life}
Predicting the half-life of a radioactive isotope is a longstanding challenge in nuclear science.
Accurate half-life predictions are crucial for understanding nuclear stability, the synthesis of heavy elements, and practical applications like nuclear medicine.
Theoretical calculations of the half-life remain difficult because the decay constants depend nonlinearly on small errors in the binding energy and nuclear transition matrix elements.
Competing decay channels ($\alpha$, $\beta^\pm$, fission) further compound these uncertainties.
Hence, the field has recently turned to machine learning based approaches~\cite{jalilialphaDecayHalflife2024}.

The Smirk tokenizer's ability to represent all isotopic variants for all elements enables us to fine-tune a \ac{MIST} model for half-life prediction~\cite{WBVTokenizationMolecularFoundation2026}.
Projecting embeddings from the fine-tuned model using \ac{UMAP} reveals structure in the embedding space (\cref{fig:isotope_clusters}) explained by nuclide stability and dominant decay channels (Supplementary~\cref{sec:si:isotope_stability}).
We use this as a demonstration of the Smirk tokenizer's ability to parse isotopic information in \ac{SMILES} (for example \smiles{[2H]} and \smiles{[13C]}).
Although a \ac{SMILES}-based model is not necessary for this particular task, this shows \ac{MIST}'s ability to tackle molecular property prediction problems with isotopic variation.

\subsubsection{Modelling Organometallic Complexes\label{sec:organometallics}}
Transition metal complexes are fundamental for a wide range of applications, from biochemistry~\cite{JulCatalysisExcitedState2025} to photochemistry~\cite{MRWMolecularDesignPrinciples2025}.
Understanding the electronic structure and properties of transition metal organometallics is key to their functionality~\cite{MRWMolecularDesignPrinciples2025}.
After fine-tuning \ac{MIST} on a dataset of transition metal complexes\cite{BSTmQMDatasetQuantum2020}, the model accurately predicts quantum mechanical properties, with \ac{MIST}-28M and \ac{MIST}-1.8B achieving validation $R^2$ scores of 0.76 and 0.80, respectively.
Notably, the Smirk tokenizer~\cite{WBVTokenizationMolecularFoundation2026} tokenizes the relevant chiral tags (e.g., \smiles{@SP1} versus \smiles{@SP3}), enabling \ac{MIST} to differentiate between non-tetrahedral isomers (\cref{fig:tmQM}).
By contrast, prior molecular \ac{FM}s used tokenizers that obscured stereochemical features~\cite{WBVTokenizationMolecularFoundation2026}.
We also show that the \ac{MIST} pretraining stage, while limited to organic molecules, improves performance on organometallic tasks (Supplementary~\cref{sec:si:pretraining_adv}).

\subsubsection{Mixture Property Prediction}
\label{sec:mixtures}

Mapping molecular properties with macroscale mixture properties, such as density, viscosity, and ionic conductivity, is challenging due to the combinatorial nature of the problem, complex non-ideal interactions between mixture components, and the scarcity of data~\cite{ZRA+DifferentiableModelingOptimization2024}.
Physics-based models, such as molecular dynamics simulations, can be used to study component interactions inside complex mixture systems.
However, their computational cost can be prohibitive.
Empirical relationships, such as the Arrhenius equation or \ac{QSAR}, which simplify intermolecular interactions, are widely used to inform formulation design.
However, the accuracy of these models typically depends on adequate parameterization with experimental data, and their ability to generalize across chemical space is limited~\cite{ZRA+DifferentiableModelingOptimization2024}.

Machine learning based approaches have emerged as a feasible and scalable solution for mixture property prediction.
Building on prior work~\cite{ZRA+DifferentiableModelingOptimization2024,zohairChemicalFoundationModel2025a}, we extend \ac{MIST}’s molecular property prediction capabilities to mixture property prediction using downstream task networks informed by chemical thermodynamics.
Fine-tuned \ac{MIST} models achieve state-of-the-art performance on existing benchmark datasets for excess molar volume, excess molar enthalpy, and ionic conductivity (Extended Data~\cref{tab:ionic_cond}),
outperforming purpose built \acp{GNN} including those which require 3D geometrical information~\cite{ZRA+DifferentiableModelingOptimization2024}.

\paragraph{Excess Property Prediction.}
Excess properties measure the deviation of mixture properties from ideal mixing behaviour.
These deviations arise from molecular interactions between components~\cite{engelbrecht2022md}.
Understanding these deviations is particularly relevant when identifying additives that can meaningfully impact performance metrics at low concentrations~\cite{SVArtificialIntelligenceElectrolyte2025}.
Mixture properties can be decomposed into a linear mixing and an excess term.
For a binary mixture of substances $S_i \;|\;i \in \{ 1, 2 \}$, a mixture property \(P_{\mathrm{mix}}\) can be expressed as:
\begin{align}
    P_{\mathrm{mix}} &= \underbrace{\sum_{i=1}^2 x_iP_i}_{P_L\; (\text{linear mixing})}
    + \underbrace{\sum_{j = 1}^n \Theta_j \cdot f_j (x_1)}_{P_E\;(\text{excess term})}  ,
\label{eq:mix_thermo}
\end{align}
where $x_1$ and $x_2$ are the mole fractions of the two components in the mixture,
$P_i$ is the property of the pure substance $S_i$,
$f_j$ is the $j^{th}$ basis of a polynomial and $\Theta_j$ is the corresponding empirically fit coefficient.
The excess term is commonly modelled using a Redlich–Kister polynomial basis to enforce \(P_{\mathrm{E}} = 0\) when \(x_1 = 0\) or \( x_2 = 0 \)~\cite{RKAlgebraicRepresentationThermodynamic1948,ZRA+DifferentiableModelingOptimization2024}.

Our binary mixture property prediction task network is informed by the thermodynamic statement of excess properties (\cref{eq:mix_thermo,fig:mist_mix_excess_arch}), and preserves key physical principles:
\(P_{\mathrm{E}} = 0\) in the degenerate case of a pure compound, and excess properties are permutation invariant.
We fine-tune \ac{MIST} to predict \(P_i\) and \(\Theta_j\) of \cref{eq:mix_thermo} using a permutation-invariant task network (\cref{fig:mist_mix_excess_arch}) that produces smooth, physically consistent excess-property curves (Methods~\cref{sec:methods:mixtures}).

We use the \ac{MIST} model to accurately estimate the ``excess asymmetry'' (\cref{fig:skew_parity}) in density and molar volume.
We define the ``excess asymmetry'' as the deviation of the composition of the excess extremum from that of a regular solution model, that is 0.5.
This prediction enables identification of additives with high relative excess at low concentrations, which is critical for additive design in electrolytes and liquid formulation problems~\cite{SVArtificialIntelligenceElectrolyte2025}.
The \ac{MIST} model accurately identifies the emergence of excess properties in mixtures that are not well modelled by existing similarity-based descriptors.
We demonstrate this by identifying a set of mixtures that disagree with the negative correlation between structural similarity and excess magnitude observed in prior work~\cite{KADVExcessDensityDescriptor2025} (Supplementary~\cref{fig:soap_corr}) and show that the model accurately predicts the excess properties of these mixtures (\cref{fig:soap_outliers}).

\paragraph{Multi-Component Electrolyte Property Prediction.}
\ac{MIST} models were fine-tuned to predict the ionic conductivity of electrolytes (salts in ternary solvent systems)~\cite{ZRA+DifferentiableModelingOptimization2024}.
A specialized multi-component property prediction task head is used to enable this.
The task head constructs a mixture embedding, $\vec{e}_{mix}$, as the sum of the component embeddings $\vec{e}_i$ weighted by their mole ratios in the mixture $x_i$:
\begin{equation}
    \vec{e}_{mix} = \sum_{i=1}^{n} x_i \vec{e}_i
    \label{eq:e_mix}
\end{equation}
This mixture embedding is used to predict two sets of coefficients.
The first set of coefficients parameterize the \ac{VFT} relation~\cite{ngaiIntroductionProblemsRelaxation2011}, which the model uses to learn the dependence of ionic conductivity on temperature.
We show that the \ac{VFT} relation accurately describes the temperature dependence seen in electrolytes in Supplementary~\cref{sec:si:why_vft}.
The second set of coefficients parameterize an empirical correction term used by the model to learn the dependence of ionic conductivity on concentration (Methods~\cref{sec:methods:ionic}).

These physics-informed relationships allow the model to generalize to higher salt mole fractions and concentrations than those seen during training; as shown in \cref{fig:ionic_prediction}, the model correctly captures the expected decay outside the training data range ($x_{Li} > 0.15$).
Using the learnt \ac{VFT} pseudo-activation energy $E_a$ and Vogel temperature $T_0$, we compute the associated excess quantities (deviation from linear mixing across solvent fractions, see Supplementary~\cref{sec:si:conductivity_excess_ea,sec:si:conductivity_t0}).
Sweeping along the salt and solvent concentration axes, we disentangle solvent–solvent and salt–solvent effects on ion transport across the ternary composition space.
The model recovers the expected trend that linear carbonate-rich solvents exhibit smaller \(E_a\) and \(T_0\) (Supplementary~\cref{fig:ea_dec_pf6,fig:ea_dec_tfsi,fig:ea_emc_pf6,fig:ea_emc_tfsi,fig:t0_dec_pf6,fig:t0_dec_tfsi,fig:t0_emc_pf6,fig:t0_emc_tfsi}), whereas viscous cyclic carbonates (\ac{EC}, \ac{PC}, and \ac{FEC}) show the opposite behaviour (Supplementary~\cref{fig:ea_fec_pf6,fig:ea_fec_tfsi,fig:t0_fec_pf6,fig:t0_fec_tfsi}).
The ternary excess surfaces of \(E_a\) and \(T_0\) reveal nonideal mixing, with a positive excess near linear carbonate rich corners (Supplementary~\cref{fig:excess_dec_pf6,fig:excess_dec_tfsi,fig:excess_t0_dec_pf6,fig:excess_t0_dec_tfsi}) and a negative excess in moderate cyclic carbonate compositions, where viscosity decreases without a corresponding increase in association.
The model captures a subtle yet important design principle: adding small amounts of linear carbonates to cyclic-carbonate mixtures reduces the viscosity while maintaining high ionic dissociation.
The \ac{MIST} model's predictions suggest an interesting effect of the salt in the electrolyte, LiPF$_6$ exhibits higher pseudo-activation steepness ($E_a/T$) than LiTFSI, indicating stronger temperature-activated barriers for LiPF$_6$ (Supplementary~\cref{sec:si:conductivity_comparing}).
Together, the learnt \ac{VFT} coefficients and their excesses provide mechanistic composition-resolved guidance for selecting solvent ratios and salts that reduce fragility and improve cold-temperature performance.
Supplementary~\cref{sec:si:conductivity} provides additional details and results for this analysis.

\subsubsection{Mapping Olfactory Perception}
\label{sec:olfaction}

\begin{figure}
    \centering
    \labelphantom{fig:discordance_triads}
    \labelphantom{fig:olfaction_clusters}
    \labelphantom{fig:olfaction_hyperbolic}
    \labelphantom{fig:olfaction_mixtures}
    \includegraphics[width=\linewidth]{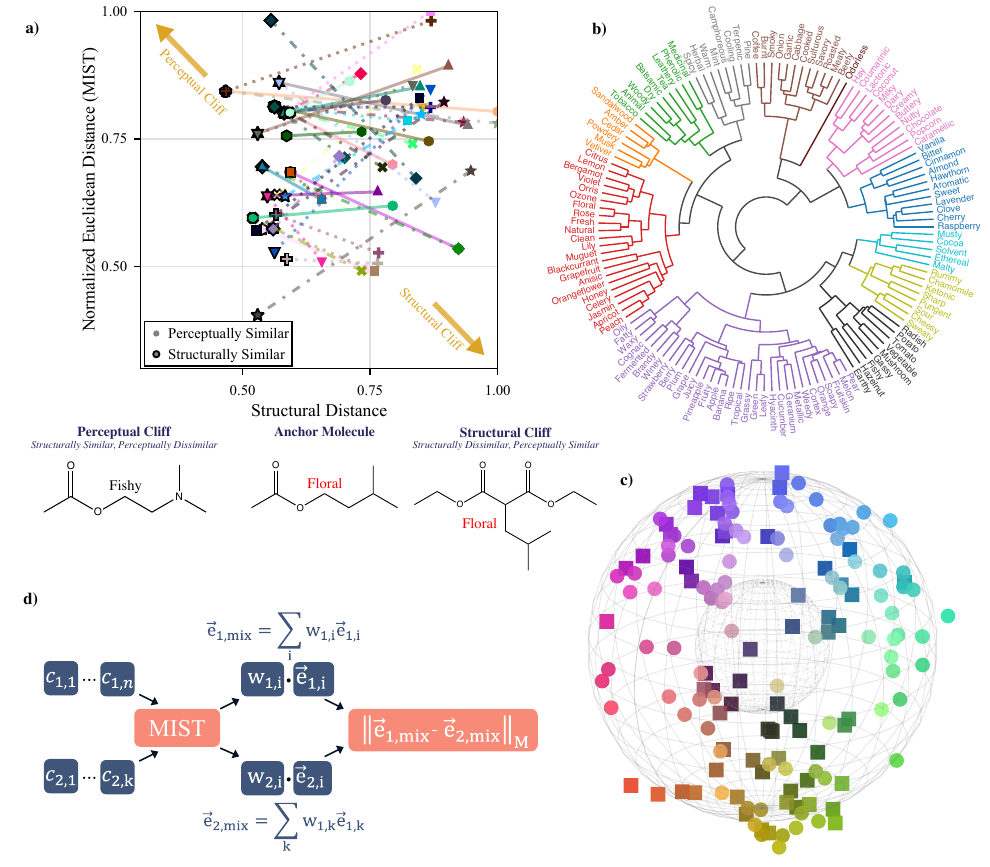}
    \caption{
        \label{fig:olfaction_panel}
        \textbf{\ac{MIST} captures discontinuities in structure--odour mapping and the hierarchical geometry of olfactory space.}
        (\subref*{fig:discordance_triads}) Structure alone often fails to explain odour similarity: across 41 odour-discordant triplets, the experimentally closer-smelling molecule is frequently not the structurally closest one. Using Euclidean distance between perceptual profiles predicted by \ac{MIST}, we are able to correctly identify 64.3\% of these discordant triplets, outperforming the prior state of the art (50.0\%).
        (\subref*{fig:olfaction_clusters}) Hierarchical clustering of \ac{MIST} scent logits recovers human-interpretable scent relationships, including grouped savoury and fruity descriptors and a distinct ``odourless'' cluster.
        (\subref*{fig:olfaction_hyperbolic}) Shell-constrained \ac{HYDRA} embedding visualizes the learned odour representation in hyperbolic space. 
        Circles/rectangles fall on the near/far side of the sphere. 
        The RGB color scales were proportional to the XYZ coordinates of points.
        (\subref*{fig:olfaction_mixtures}) The \ac{MIST} model was extended to predict perceptual distances between multi-component odour mixtures.
    }
\end{figure}

The clearest demonstration of a foundation model is its ability to solve problems that were neither explicit targets of training nor central to the intentions of its developers~\cite{BHA+OpportunitiesRisksFoundation2022}. 
Predicting the odour of a molecule remains a longstanding challenge in chemistry, psychophysics and neuroscience because the mapping from molecular structure to perception is highly nonlinear, discontinuous and variable across individuals~\cite{kellerPredictingHumanOlfactory2017,malnicCombinatorialReceptorCodes1999,acheboucheApplicationArtificialIntelligence2022}. Minimal chemical modifications can produce large perceptual changes, whereas structurally dissimilar molecules can evoke near-identical odours~\cite{leePrincipalOdorMap2023}.
We therefore identified olfaction as a particularly difficult test of \ac{MIST}: it lies outside the applications motivating model development, yet success in this domain would suggest that \ac{MIST} has learned molecular representations that transfer beyond the tasks targeted during pretraining~\cite{U.S2024INCITEFact2023,jollyBuildingAIFoundation2025}.

After fine-tuning \ac{MIST}-28M on a dataset of \(\sim5,000\) molecules curated by~Lee et al.\cite{leePrincipalOdorMap2023}, we find the \ac{MIST} model exceeds the accuracy of the median rater on an expert panel and existing state-of-the-art models~\cite{leePrincipalOdorMap2023}.
The fine-tuned model has an \ac{AUROC} of 0.915, improving on the 0.894 \ac{AUROC} reported by Lee et al.\cite{leePrincipalOdorMap2023} for their \ac{GNN}.

The \ac{MIST} model's performance scales with the chemical specificity of scent descriptors.
Scents associated with particular functional groups are highly separable; for example, fruity scents caused by the presence of an ester group (``banana'', ``apple'' and ``cognac'') all have \acp{AUROC} above 0.9.
The model accurately captures dramatic changes in odour caused by small changes in molecular structure;
for example, it accurately labels acetone \smiles{CC(=O)C} as ``ethereal'' and that a sulfur substitution (thioacetone \smiles{CC(=S)C}) leads to a ``sulfurous'' and ``meaty'' odour (Supplementary~\cref{fig:discontinuous_odour}).
We validate the \ac{MIST} model's robustness to discontinuities in structure-odour mapping by using it to discover previously uncharacterized ``discordant'' triplets (\cref{fig:discordance_triads}, Supplementary~\cref{sec:si:discordance}).
Discordant triplets are triplets of molecular odourants in which the structurally similar pair is not the perceptually similar pair~\cite{leePrincipalOdorMap2023}.
We evaluated structural similarity using Tanimoto similarity and perceptual similarity using the Euclidean distance between predicted \ac{MIST} odour profiles. 
The experimentally determined labels for these triplets were not publicly available at the time of prediction; we subsequently obtained them through direct correspondence with Lee et al. providing a prospective, blinded evaluation. 
The \ac{MIST} model correctly resolved 64.3\% of the discordant triplets (Supplementary~\cref{fig:si:mist_euclidean_discord}), outperforming the \ac{GNN} of~Lee et al.\cite{leePrincipalOdorMap2023}, which resolved 50\% (Supplementary~\cref{fig:si:gnn_discord}).

We observe that the model learns a hierarchical clustering of scents (\cref{fig:olfaction_clusters}, Methods~\cref{sec:methods:olfaction}) consistent with human perceptual organization;
for example, ``beefy'', ``meaty'', ``roasted'', ``savory'' all form a cluster, fruity scents (such as ``strawberry'', ``berry'', ``plum'' and ``grape'') form another and notably ``odourless'' molecules are clustered alone.
The existence of this hierarchical structure is consistent with the model having learned the hyperbolic geometry of the olfactory space~\cite{zhouHyperbolicGeometryOlfactory2018}.
To validate this, we modeled the olfactory profiles in hyperbolic and Euclidean space.
We compared the empirical Betti curves of the \ac{MIST} logit similarity structure with those generated from best-fit Euclidean and hyperbolic model spaces using ALBATROSS~\cite{albatross,stierALBATROSSCheapFiltration2025} (Methods~\cref{sec:methods:olfaction}).
The \ac{MIST} logit geometry was well fit by a hyperbolic shell model (\(p=1.0\), \(\chi^2=0\)) (Supplementary~\cref{sec:si:tda_olfaction}). 
Together, these results indicate that the learned olfactory representation is statistically consistent with a low-dimensional hyperbolic geometry.
This finding is consistent with prior work showing that both natural odour co-occurrence statistics and human olfactory perceptual spaces are well described by low-dimensional hyperbolic geometry~\cite{zhouHyperbolicGeometryOlfactory2018}.
We visualized the learned perceptual representation as a hyperbolic shell (\cref{fig:olfaction_hyperbolic}) using shell-constrained \ac{HYDRA} embeddings~\cite{kellerressel2019hydramethodstrainminimizinghyperbolic,albatross} (Methods~\cref{sec:methods:olfaction}).

Finally, we extend the model's capabilities beyond predicting the odour of a single molecule to comparing the odour profiles of multi-component mixtures.
We fine-tuned \ac{MIST} to predict experimentally measured perceptual distances ranging from 0 (indistinguishable odour) to 1 (maximally distinct odours) for pairs of mixtures with variable numbers of components (see Supplementary~\cref{sec:si:olf_mix_dataset} for dataset details).
We use a weighted sum of single molecule embeddings (with learned weights) to obtain a mixture embedding (\cref{fig:olfaction_mixtures}), 
The fine-tuned \ac{MIST} achieved a Pearson correlation of $0.62 \pm 0.05$ on a held-out validation split, demonstrating comparable performance to an ensemble of top-performing models submitted to the \ac{DREAM} Olfaction Challenge 2024, which reported a Pearson correlation of 0.57~\cite{satarifardHighFidelityTuningOlfactory2025}.

\subsection{Learning a Chemically Meaningful Representation}
\begin{figure}
    \centering
    \labelphantom{fig:interp_schematic}
    \labelphantom{fig:token_emb_tsne}
    \labelphantom{fig:interp_emb_rings}
    \labelphantom{fig:interp_emb_pah}
    \labelphantom{fig:hc_trends_size}

    \includegraphics[width=\linewidth]{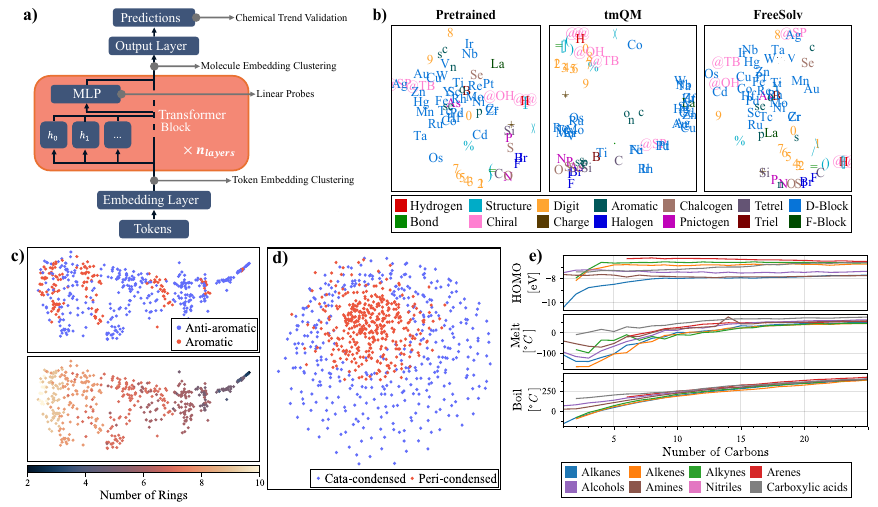}
    \caption{
        \label{fig:mech_interp}
        \textbf{Interpretability analysis reveals \ac{MIST}'s potential as a robust tool for exploration and discovery.}
        (\subref*{fig:interp_schematic})~Interpretable chemical features were extracted from every stage of \acs{MIST} models --- token embeddings to downstream predictions.
        (\subref*{fig:token_emb_tsne})~A t-SNE projection of token embeddings shows dataset-specific structure.
        All three datasets shown cluster the halogens and digits.
        The transition metal Quantum Mechanics dataset (tmQM) forms clusters for the left and right transition metal block.
        FreeSolv orders the frequently seen p-block elements (\tok{Si, P, N, O, S, I, Br, F} and  \tok{C}) near the bottom of the figure.
        (\subref*{fig:interp_emb_pah}) \& (\subref*{fig:interp_emb_rings}) \ac{UMAP} projections reveal that the pretrained embeddings encode ring aromaticity and \aclp{PAH} classes, despite the model never training on labelled examples of these classes, suggesting the existence of chemically coherent structures in \ac{MIST}'s embedding space.
        (\subref*{fig:hc_trends_size}) Using fine-tuned \ac{MIST}-28M variants, we recover expected trends (increasing and converging) in \acs{HOMO} energies and phase transition temperatures (Melt \& Boil) with carbon chain length.
    }
\end{figure}

The potential of \acp{SciFM}, such as \ac{MIST}, for scientific discovery extends beyond their raw predictive accuracy.
For exploration, scientists need to be able to extract interpretable features aligned with scientific concepts and relationships which can be understood and steered to generate hypotheses or diagnose failure modes~\cite{murdochDefinitionsMethodsApplications2019,OYIWInterpretableScientificFoundation2024}.
In \ac{NLP} and \ac{CV}, mechanistic interpretability has revealed feature vectors within \acp{FM} correlated with semantically meaningful concepts across modalities and languages~\cite{TCM+ScalingMonosemanticityExtracting2024,joseph2025prismaopensourcetoolkit}.
Yet comparable efforts for \acp{SciFM} remain limited~\cite{OYIWInterpretableScientificFoundation2024}.
Prior \ac{SciFM} work has examined attention maps~\cite{CGRChemBERTaLargeScaleSelfSupervised2020,RBC+LargescaleChemicalLanguage2022} and low-dimensional projections of embedding space~\cite{ZBH+GenSLMsGenomescaleLanguage2023,SSB+LargeEncoderDecoderFamily2024,RBC+LargescaleChemicalLanguage2022} to probe internal behaviour.

We systematically search for interpretable, concept-aligned features across all layers, and we demonstrate that \ac{MIST} models consistently encode chemical concepts that generalize beyond their training objectives (\cref{fig:mech_interp}).
Such concept-aligned features can be used to enable the discovery of new material design rules and robust generalization to novel tasks, thereby positioning \ac{MIST} as a practical tool for hypothesis generation and systematic discovery across chemical space~\cite{OYIWInterpretableScientificFoundation2024}.

\subsubsection{Interpreting Token Embeddings}
Input molecules are tokenized and each token index is mapped to a trainable embedding vector before being processed by the transformer architecture.
During pretraining, these token embeddings are updated to reflect the semantic meaning of each token~\cite{VSP+AttentionAllYou2017,DCLTBERTPretrainingDeep2019,LOG+RoBERTaRobustlyOptimized2019}.
We analyse the token embeddings of pretrained and fine-tuned \ac{MIST} models.
Due to the pretraining dataset's limited chemical diversity~\cite{WBVTokenizationMolecularFoundation2026}, many tokens in \ac{MIST}’s vocabulary are rarely or never seen during pretraining, and the token embeddings decay towards zero due to weight decay.
Fine-tuning provides a second opportunity for the model to learn representations for these tokens.
As shown in \cref{fig:token_emb_tsne}, the most substantial embedding updates align with differences between the pretraining and fine-tuning datasets.
For example, when fine-tuning on tmQM (an organometallic dataset) d-block token embeddings are updated forming two distinctive clusters.
Our benchmark results suggest that learning token embeddings during fine-tuning is sufficient to achieve state-of-the-art performance (Extended Data \cref{tab:molnet_class_bench,tab:molnet_reg_bench}).

\subsubsection{Interpreting Hidden States}
\label{sec:hidden_states}
The \ac{RO5}, introduced by Lipinski in 1997~\cite{LLDFExperimentalComputationalApproaches1997}, defines four simple thresholds (each a multiple of five) associated with drug-likeness.
Aside from lipophilicity, no variant of MIST was trained to predict these features, making \ac{RO5} a natural testbed for probing learned representations.
To test whether \ac{RO5} signals are linearly decodable from \ac{MIST}’s internal representations of chemistry, we trained linear classifier probes~\cite{ABUnderstandingIntermediateLayers2018} on frozen activations from every layer of pretrained and fine-tuned \ac{MIST}-28M models.
Linear classifier probes are logistic regressors of the form $y_i = \sigma(\vec{f}_i  \cdot \vec{x} + b_i)$ which map the model's hidden states at a layer, $\vec{x}$, to a binary \ac{RO5} criterion, $y_i$, where \(i\) indexes over criterion and $f_i$ is the learned feature vector.
Further details on probe training appear in Supplementary~\cref{sec:si:probes}.
Across both pretrained and fine-tuned models, linear probes recover chemically interpretable features \(\vec{f}_i\) for the \ac{RO5} criteria (Supplementary~\cref{fig:lipinski_linear_probes}), indicating that \ac{RO5}-relevant structure is learned during pretraining and retained under fine-tuning.

Additionally, we used low-dimensional projections of molecule embeddings from the pretrained \ac{MIST}-1.8B encoder to further analyse chemical knowledge encoded by the model (Methods~\cref{sec:methods:embeddings}).
We found features correlated with \(\pi\)-bonding orbitals, ring counts, and \acp{PAH} subclasses.
We observe a banding pattern (\cref{fig:interp_emb_rings}) that separates anti-aromatic and aromatic compounds, consistent with H\"uckel's rule for \(\pi\)-orbital stability~\cite{kikuchiHistoryStructuralTheory1997}.
We also find a clear separation between the embeddings of \emph{cata-} and \emph{peri-} condensed \acp{PAH} (\cref{fig:interp_emb_pah}).
Critically, this structure is seen in unsupervised projections of embeddings from models that were not trained on the highlighted features, making the emergence of a clear structure surprising.
The separability of these chemical classes from the projected embeddings was validated using a \ac{SVM} (Supplementary~\cref{sec:si:embeddings}).

Together, the linear probe and unsupervised projection analyses suggest that the representation of chemical space learned by the \ac{MIST} models organizes chemical concepts along physically meaningful directions.
The combination of linearly decodable reasoning and interpretable embeddings makes \ac{MIST} a practical instrument for hypothesis generation and systematic exploration across chemical space.

\subsubsection{Consistency of MIST Predictions with Chemical Intuition}\label{sec:trends}

Using fine-tuned variants of MIST-28M, we evaluated trends in quantum, chemical, and thermodynamic properties across simple hydrocarbons, fatty acids, and electrolyte solvents (\cref{fig:hc_trends_size} and Supplementary~\cref{fig:fatty_acids}).
The model successfully recovers expected structure–property relationships, including decreases in Gibbs free energy and increases in polarizability with chain length, as well as the distinction between melting and boiling points for common organics (\cref{fig:hc_trends_size}).
For fatty acids, the \ac{MIST} model captures the effects of chain length and unsaturation on phase change temperatures and stability, correctly predicting that saturation improves packing and elevates melting points, while unsaturation introduces \(\pi\)-bonding and disrupts cohesion (Supplementary~\cref{fig:fatty_acids}).
These findings (Supplementary~\cref{sec:si:chemical_consistency}) provide qualitative evidence that fine-tuned \ac{MIST} models learn chemically meaningful trends across diverse chemical functional groups.

\begin{figure}
    \centering
    \labelphantom{fig:bayes_schematic}
    \labelphantom{fig:bayes_covariance}
    \labelphantom{fig:lr_scaling_residual}
    \labelphantom{fig:loss_compute_scaling}
    \labelphantom{fig:compute_optimal_frontier}
    \labelphantom{fig:compute_scaling_parameters}
    \includegraphics{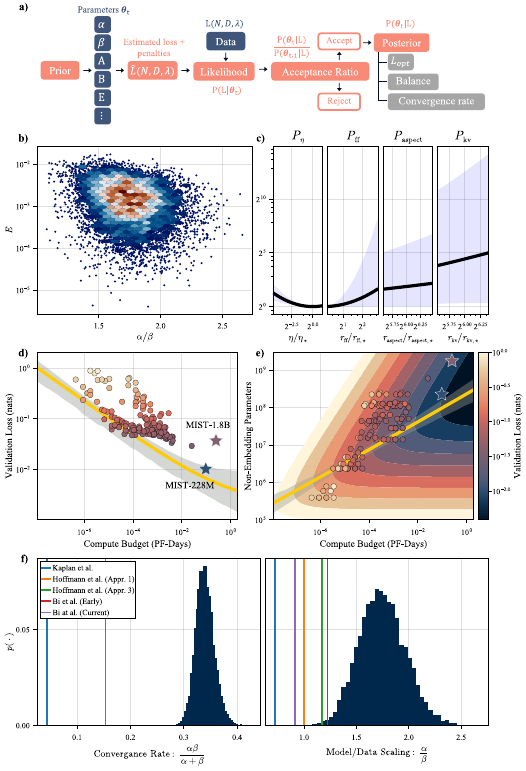}
    \caption{
        \label{fig:neural_scaling_laws}
        \textbf{Compute-optimal training was critical to efficiently scaling \ac{MIST}.}
        (\subref*{fig:bayes_schematic})~We adopted \acs{MCMC} sampling to parameterize neural scaling laws, with penalty terms accounting for off-optimal hyperparameter selection, enabling robust predictions of the compute-optimal frontier.
        (\subref*{fig:bayes_covariance})~Covariance plot of posterior samples for \(E\) and \(\alpha/\beta\) after fitting penalized neural scaling laws shows low variance in \(\alpha/\beta\) but higher variance in \(E\).
        (\subref*{fig:lr_scaling_residual})~Fitted penalty terms (with 95\% credible interval) for learning rate (\(P_{\eta}\)) and encoder shape (\(P_{ff}\), \(P_{aspect}\) \& \(P_{kv}\)), showing estimated impact of off-optimal hyperparameters on \(L\).
        (\subref*{fig:loss_compute_scaling})~95\% credible interval for the compute-optimal frontier (\(L_{\mathrm{opt}}\), \cref{eq:compute_optimal_loss}).
        (\subref*{fig:compute_optimal_frontier})~Posterior mean estimate of the neural scaling law and 95\% credible interval for \(L_{\mathrm{opt}}\).
        Overlaid scatter plots in \subref*{fig:loss_compute_scaling}\& \subref*{fig:compute_optimal_frontier} show the minimum validation loss of the model used to fit the neural scaling law.
        Largest models trained are marked with \(\bigstar\).
        Overall, the hyperparameter-penalized neural scaling law (\cref{eq:penalized_neural_scaling}) provides strong predictive accuracy, achieving a Spearman's rank correlation ($\uparrow$) of \(0.812\) and \acs{MAPE} ($\downarrow$) of 29.1\%.
        This is a marked improvement over the fit without penalty terms, which achieved 0.776 and 38.5\%, respectively.
        (\subref*{fig:compute_scaling_parameters})~Histograms of convergence rate and model/data balance for \ac{MIST} models.
        \ac{MIST}'s scaling exponents deviate notably from \acs{NLP} models.
    }
\end{figure}

\subsection{Hyperparameter Penalized Bayesian Neural Scaling Laws}
\label{sec:scaling}

Scaling \ac{MIST} to 1.8 billion parameters was an extremely computationally expensive endeavour, in large part because the dataset and model size needed to achieve the desired accuracy was unknown {\em a priori}.
Neural scaling laws, first developed for \ac{NLP} models~\cite{KMH+ScalingLawsNeural2020,HBM+TrainingComputeOptimalLarge2022}, provide a framework for predicting how a model's loss decreases with increasing model or dataset size.
Following~Hoffmann et al.\cite{HBM+TrainingComputeOptimalLarge2022}, the cross-entropy loss \(L\) can be modelled as a function of the number of non-embedding parameters \(N\) and the dataset size \(D\):
\begin{equation}
\label{eq:hoffmann_scaling}
    L(N, D) = \frac{A}{N^{\alpha}} + \frac{B}{D^{\beta}} + E .
\end{equation}

Assuming training compute, \(C \approx 6ND\), the compute-optimal model size for a budget \(C\) is the one that minimizes \(L\); this is detailed in Supplementary~\cref{sec:si:compute_optimal_scaling} and derived by~Hoffmann et al.\cite{HBM+TrainingComputeOptimalLarge2022}.
Standard formulations implicitly assume that all hyperparameters are tuned optimally~\cite{KMH+ScalingLawsNeural2020,HBM+TrainingComputeOptimalLarge2022} and yield only point-estimates of the neural scaling coefficients (\(A, \alpha , B, \beta\) \& \(E\)).
However, optimal hyperparameters often vary with model size~\cite{SKNeuralScalingLaw2020,VSP+AttentionAllYou2017}.
The shape of the transformer layers can also weakly impact \(L\)~\cite{KMH+ScalingLawsNeural2020,HBM+TrainingComputeOptimalLarge2022}.
Kaplan et al.\cite{KMH+ScalingLawsNeural2020} reported model loss to be most sensitive to the feedforward ratio (the ratio of the feedforward network width and model width, see Supplementary~\cref{sec:si:bayesian_neural_scaling}).
The importance of model shape for \acp{SciFM} is largely unexplored in prior work~\cite{RBC+LargescaleChemicalLanguage2022,SSB+LargeEncoderDecoderFamily2024,NPD+SequenceModelingDesign2024}.
Instead, these hyperparameters \(\lambda\) are often selected based on heuristics~\cite{SKNeuralScalingLaw2020,VSP+AttentionAllYou2017}, grid searches~\cite{NPD+SequenceModelingDesign2024,BCC+DeepSeekLLMScaling2024}, or are otherwise tuned~\cite{KMH+ScalingLawsNeural2020,HBM+TrainingComputeOptimalLarge2022}, incurring significant costs at scale.

To address these limitations within the context of \acp{SciFM}, we extend neural scaling laws (\cref{eq:hoffmann_scaling}) in two key ways.
First, we explicitly model deviations from optimal hyperparameter selection using quadratic penalty terms.
Second, we fit neural scaling coefficients using Bayesian parameter estimation, rather than non-linear optimization.
Together, this allows us to (i) model the impact of off-optimal hyperparameter selection, (ii) propagate parameter uncertainty through the neural scaling law, and (iii) encode prior knowledge from the literature into weakly informative priors.
We refer to this approach as \emph{hyperparameter penalized Bayesian neural scaling laws} and fit models of the form:
\begin{align}
L &\sim \mbox{LogNormal}(\hat{L}, \sigma^2) \\
\hat{L}(N, D, \lambda) &= \left( \frac{A}{N^{\alpha}} + \frac{B}{D^{\beta}} + E \right) \times \prod_i \exp(P_i(\lambda_i)) \label{eq:penalized_neural_scaling} \\
P_i(\lambda_i) &= c_i \left( \ln \lambda_i - \ln \lambda_{\star,i} \right)^2 ,
\label{eq:penalty_term}
\end{align}

\noindent
where \(i\) indexes hyperparameters.
By construction, the penalty terms have a single global optimum located at \(\lambda_{\star,i}\) such that \(P_i(\lambda_{\star,i}) = 0\).
Additionally, a quadratic penalty for off-optimal parameter selection is qualitatively similar to the behaviour observed by~Kaplan et al. for the feed-forward and aspect ratio~\cite{KMH+ScalingLawsNeural2020}.
We incorporated penalty terms for the learning rate \(\eta\) and shape of the transformer layers (feed-forward ratio, aspect ratio, and Key-Query-Value size, Supplementary~\cref{sec:si:bayesian_neural_scaling}).

Our procedure for fitting the hyperparameter penalized neural scaling laws is illustrated in \cref{fig:bayes_schematic} and described in Supplementary~\cref{sec:si:bayesian_neural_scaling}.
We use \ac{MCMC} sampling (Methods~\cref{sec:methods:bayesian}) to estimate the posterior distribution of the parameters in \cref{eq:penalized_neural_scaling}.
The additional computational overhead of Bayesian parameter estimation is negligible compared to a full hyperparameter sweep:
\ac{MCMC} sampling to fit \cref{eq:penalty_term} required less than one CPU-hour, trivial compared to even initializing the training of a \ac{FM} across hundreds of GPUs.

Bayesian estimation quantifies both parameter and prediction uncertainty, which helped guide \ac{MIST} training.
For example, while some parameters remained uncertain (for example, \(E\)), low-variance estimates for others (for example, \(\alpha/\beta\)) provided reliable and actionable guidance (\cref{fig:bayes_covariance}).
These insights were critical in selecting hyperparameters for our frontier model, MIST-1.8B.

Penalized neural scaling laws (\cref{eq:penalized_neural_scaling}) had strong predictive accuracy, achieving a Spearman’s rank correlation of 0.812 and \ac{MAPE} of 29.1\%,
while non-penalized neural scaling laws (\cref{eq:hoffmann_scaling}) scored 0.776 and 38.5\%, respectively.
Posterior means and 95\% credible intervals for \(A, \alpha, B, \beta\) \& \(E\) for both variants are provided in Extended Data~\cref{tab:scaling_law_parameters}.
Alternative formulations for modeling off-optimal hyperparameter selection are discussed in Supplementary~\cref{sec:si:alt_scaling_laws}.

Encoding prior knowledge into weakly informative priors reduced the amount of data (model training runs) needed to parameterize our neural scaling laws.
Inspired by power-law heuristics in prior work~\cite{VSP+AttentionAllYou2017,KMH+ScalingLawsNeural2020,YLR+LargeBatchOptimization2020}, we modeled the optimal learning rate as scaling with both batch size (\(\mathcal{B}\)) and model size (\(N\)):
\(\eta_{\star} = \eta_0 N^{\gamma}\mathcal{B}^{\delta}\).
We encoded both MoLFormer's baseline learning rate (\sn{1.6}{-4}~\cite{RBC+LargescaleChemicalLanguage2022}) and the square-root scaling with batch size~\cite{KriOneWeirdTrick2014} using the following priors:
\begin{align}
\eta_0 &\sim \mathrm{LogNormal}(-5.3, 0.5) \\
\delta &\sim \mathcal{N}(0.5, 0.05) \\
\eta_{\star} &= \eta_0 N^\gamma \mathcal{B}^{\delta} .
\label{eq:optimal_lr}
\end{align}
These priors were guided by our initial training efforts;
all other priors are discussed in Supplementary~\cref{sec:si:bayesian_priors}.

A few hyperparameter sweeps remain necessary,
but they can be integrated into sweeps over model and data size within the overall experimental design for neural scaling laws.
This reduces the total number of model runs needed to develop a neural scaling law.
Using design of experiments~\cite{SVH+GeneralizedSubsetDesigns2017}, we selected hyperparameters for our development sweeps (Supplementary~\cref{sec:si:scaling_campaign}).
To fit our neural scaling laws, we trained 138 \ac{MIST} models (Supplementary~\cref{tab:scaling_campaign}), spanning 393k--603M parameters at the cost of 4,760 GPU-hours.
We estimate an exhaustive grid search, widely used to fit neural scaling laws\cite{li2025misfittingsurveyscalinglaws,KMH+ScalingLawsNeural2020,HBK+ScalingLawsComputeOptimal2024}, would consume at least an order of magnitude more compute (Supplementary~\cref{tab:scaling_campaign_doe}).

Once fit, the central question for neural scaling laws is the behaviour of \(L_{\text{opt}}(C)\), the minimal achievable loss given a compute budget (\cref{fig:loss_compute_scaling,fig:compute_optimal_frontier}) as defined in~\cref{eq:compute_optimal_loss}:
\begin{equation}
\label{eq:compute_optimal_loss}
L_{\text{opt}}(C) = E
+ \left(A G^{-\alpha} + B G^{\beta} \right)
\left( \frac{C}{6} \right)^{-\frac{\alpha\beta}{\alpha + \beta}} 
\end{equation}

\noindent
where \(G=\left( \frac{\alpha A}{\beta B} \right)^{\frac{1}{\alpha+\beta}}\).
In the limit of infinite compute \(C \to \infty\), \(L_{\text{opt}}\) converges to \(E\), the entropy of the true data distribution~\cite{HBM+TrainingComputeOptimalLarge2022}.
The convergence rate, governed by \(\frac{\alpha \beta}{\alpha + \beta}\), is maximized when \(\alpha/\beta = 1\) for \(\alpha, \beta > 0\) (as detailed in Supplementary~\cref{sec:si:alpha_beta_optim}).
This aligns with prior results that parameters and data should be scaled at equal rates, \(D_{\mathrm{opt}} \propto N_{\mathrm{opt}}^{\alpha/\beta}\) with \(\alpha/\beta \approx 1\), for compute-optimal training~\cite{HBM+TrainingComputeOptimalLarge2022,BCC+DeepSeekLLMScaling2024} (Extended Data~\cref{tab:scaling_coeffs}).
However, for \ac{MIST} models we compute the ratio \(\alpha / \beta\) as \(\ci{1.75}{1.34}{2.30}\) (posterior mean and 95\% credible interval), indicating that dataset size should scale substantially faster than model size for compute-optimal training (\cref{fig:compute_scaling_parameters}).
We note that \(\alpha/\beta \neq 1\) results from the nature of the model and pretraining dataset, rather than the Bayesian parameter estimation procedure.

Our estimate for \(\alpha \approx 1\) (Extended Data~\cref{tab:scaling_coeffs}), suggests that \ac{MIST} operates in the variance-limited regime of parameter scaling~\cite{BDK+ExplainingNeuralScaling2024}.
Our estimate for \(\beta\) suggests resolution-limited scaling with respect to the data, where the model effectively interpolates between points on a \(d\)-dimensional data manifold~\cite{sharmaScalingLawsData2022}.
In particular, \(\beta \approx 0.54\) implies that our pretraining dataset (REALSpace) lies on a low-dimensional manifold with effective dimension \(d \gtrsim  7.4\)~\cite{BDK+ExplainingNeuralScaling2024}.
Together, the scaling exponents indicate that \ac{MIST} scales efficiently with model size, shifting the primary bottleneck to data.
Given the low inferred manifold dimension, this constraint likely arises from insufficient quality or coverage of the data manifold.

To explain the regime we observed ($\alpha/\beta > 1$), we can characterize learning as an iterative decoding process over a bipartite graph~\cite{NVInformationTheoryComputeOptimal2024,AALTheoryInferenceCompute2025}.
Specifically, we demonstrate (\cref{sec:si:compute_optimal_scaling}) that when concept exposure frequencies follow a Zipf distribution \(p(r) \propto r^{-\alpha/\beta}\), this framework predicts super-linear scaling of data with model size along the compute-optimal frontier, \(D_{\mathrm{opt}} \propto N_{\mathrm{opt}}^{\alpha/\beta}\) for \(\alpha/\beta > 1\).
This reflects the increased learning difficulty induced by long-tailed concept distributions, aligning with empirical findings that repeated exposures yield diminishing returns~\cite{MRB+ScalingDataConstrainedLanguage2023}.
Taken together with~Bi et al.'s observation linking data quality to the compute-optimal allocation of model and data~\cite{BCC+DeepSeekLLMScaling2024} and prior work showing the benefits of data pruning~\cite{sorscherNeuralScalingLaws2023}, our results underscore the importance of diverse, high-quality pre-training datasets to support continued scaling of molecular foundation models.

\section{Discussion}
\label{sec:discussion}
We have presented \ac{MIST}, a family of molecular \acp{FM} trained on billions of tokens across chemical space.
Fine-tuned \ac{MIST} models enable property prediction at the atomistic, molecular, and mixture scales.
These fine-tuned models accelerate the search for novel materials by surrogatizing expensive computational or wet-lab experimentation.
Building on the Smirk tokenizer, \ac{MIST} is capable of reasoning over the geometric, nuclear, electronic, and chemical features encoded with \ac{SMILES}.
\ac{MIST}'s internal molecular representations contain interpretable and chemically meaningful features, indicating that \ac{MIST} has credibly learned chemistry beyond that on which it was trained.
We demonstrate this through applications across numerous domains.
Notably, in olfaction—a stringent test of transfer beyond the intended scope of pretraining—\ac{MIST} achieved strong predictive performance, learned a hierarchical representation consistent with hyperbolic geometry, and prospectively resolved experimentally validated structure–perception discontinuities.

Training \acp{FM} or domain-specific \acp{SciFM} remains an expensive endeavour.
To help mitigate this, we developed hyperparameter penalized Bayesian neural scaling laws.
These account for the impact of off-optimal hyperparameter selection, they propagate uncertainties, and they partially alleviate the need for extensive hyperparameter tuning.
This method is applicable for developing neural scaling laws for other \acp{SciFM} or \acp{FM}.

While \ac{MIST} is a significant step towards \ac{FM}-guided predictive chemical modeling, a few aspects need further improvement.
The Smirk tokenizer enables comprehensive coverage of \ac{SMILES} encodings.
However, \ac{SMILES} are themselves a lossy representation of molecular structure.
They cannot capture information about 3D geometry, conformer ensembles, molecular interactions, or environmental effects.
These effects have a significant impact on the properties of organometallics, charged species, and mixtures in many real-world applications.
Additionally, \ac{MIST} does not have a robust mechanism for quantifying prediction uncertainty.
Uncertainty quantification is important for the model to be used in an exploratory setting, for example, in an active learning loop.

We expect that molecular \acp{FM} will continue to benefit from scaling up model size, dataset size, and compute budgets based on trends seen in \ac{NLP} and other \acp{SciFM}~\cite{BMR+LanguageModelsAre2020,NPD+SequenceModelingDesign2024,BCC+DeepSeekLLMScaling2024}.
However, we found that the quality of the pretraining dataset limited the scalability of \ac{MIST} models.
This prompts the need for higher quality and more diverse chemical datasets.
\Ac{NLP} works have previously emphasized the importance of dataset selection and curation~\cite{BCC+DeepSeekLLMScaling2024}, but investigations of how dataset quality influences model scalability and strategies for curating optimal training datasets remain limited.
Identifying quality metrics predictive of scaling efficiency would help inform dataset curation and pruning for future models, allowing researchers to trade data quantity for quality, thereby potentially lowering the cost of training \acp{SciFM}.
Ultimately, \ac{MIST} presents a path towards \acp{SciFM} that function as reliable scientific instruments: readable, steerable, and extensible tools for systematic exploration and discovery across chemical space.

\FloatBarrier

\section{Methods}
\subsection{Pretraining}
\label{sec:methods:pretraining}

\begin{table}[ht!]
\centering
\begin{tabular}{lccc}
\toprule
                             & \textbf{MIST-28M} & \textbf{MIST-1.8B} \\
\midrule
\textbf{Parameters}          & 28M                              & 1.8B               \\
\textbf{Layers}              & 8                                 & 28                 \\
\textbf{Hidden Size}         & 512                             & 2304               \\
\textbf{Intermediate Size}   & 2048                             & 9216               \\
\textbf{Attention Heads}     & 8                                & 18                 \\
\textbf{Max Sequence Length} & 2048                          & 2048               \\
\midrule
\textbf{Training Steps}     & 30,000    & 500,000 \\
\textbf{Learning Rate}      & \sn{1.6}{-4} & \sn{3}{-4}         \\
\textbf{Effective Batch Size} & 8,192 & 4,096 \\
\midrule
\textbf{Total Tokens}        & 12B                             & 116B               \\
\textbf{Masked Tokens}       & 2B                              & 17B                \\
\textbf{Molecules}           & 246M                              & 2B                 \\
\bottomrule
\end{tabular}
\caption{
\label{tab:arch_table}
Model architecture and training configurations for \ac{MIST} base models. 
Effective Batch size is the number of examples per optimization step.
In a \acs{DDP} setting, this is the product of the per-device batch size, the number of devices (world size) and \acl{GAS}.
}
\end{table}

For our production models (\cref{tab:arch_table}), we trained HuggingFace's \texttt{RoBERTa-PreLayerNorm}~\cite{HuggingfaceTransformersTransformers2024} model on molecules from Enamine REALSpace~\cite{EnaREALSpace2024} with the Smirk tokenizer~\cite{WBVTokenizationMolecularFoundation2026} and absolute position embeddings.
MIST-28M was trained as part of an earlier work~\cite{WBVTokenizationMolecularFoundation2026}.
Architectural hyperparameters for the two base model variants (\ac{MIST}-28M and \ac{MIST}-1.8B) are summarized in \cref{tab:arch_table}.

We trained \ac{MIST} using \acf{MLM}~\cite{DCLTBERTPretrainingDeep2019}, whereby approximately 15\% of the tokens within each tokenized \ac{SMILES}
sequence are randomly selected and replaced with a special \texttt{[MASK]} token.
The model is trained to predict each masked token using the surrounding unmasked context --- i.e., to estimate the conditional distribution \(p(x_i \mid X)\), where \(x_i\) is a masked token and \(X\) is the mutated input sequence.
During training the model weights are updated to minimize the cross-entropy loss between the predicted and true tokens.

Training was parallelized across up to 40 GPUs using DeepSpeed's  \ac{DDP}~\cite{MicrosoftDeepSpeed2024} and the PyTorch Lightning framework~\cite{FtPyTorchLightning2023,MicrosoftDeepSpeed2024}.
The training pipeline was computationally efficient for higher GPU counts (\cref{fig:gas_scaling}).
However, since we use \ac{DDP}, a larger GPU count resulted in a larger effective batch size, which in turn led to poor data efficiency.
We used DeepSpeed's FusedLAMB~\cite{YLR+LargeBatchOptimization2020} optimizer with \(\beta_1 = 0.87\) and \(\beta_2 = 0.997\).
The $\beta_2$ value was key to stabilizing training as explained in Supplementary~\cref{sec:si:spikes}.
Learning rates were linearly warmed up to the maximum learning rate over \(2( 1 - \beta_2)^{-1} = 667\) steps~\cite{MYAdequacyUntunedWarmup2021}, before decaying 10$\times$ following a cosine decay schedule.
The learning rate and duration of MIST-1.8B were selected based on our fitted hyperparameter-penalized neural scaling laws.
All hyperparameter settings, loss curves, system utilization and throughput measurements for our production models are tabulated in human-readable JSON files provided in our data release.

\subsection{Fine-tuning}
\label{sec:methods:fine-tuning}
Fine-tuned \ac{MIST} models consist of the pretrained \ac{MIST} model (the encoder), followed by a task network.
The task networks (except for the specialized mixture task networks described in \cref{sec:mixtures}) consist of a two-layer \ac{MLP} with \ac{GELU} activations and dropout.
The final hidden state vectors for all tokens in the sequence are pooled to produce a single embedding vector.
Consistent with prior works, the encoder hidden states were pooled by taking the hidden state of the first token~\cite{RBC+LargescaleChemicalLanguage2022}.
Models were fine-tuned using the AdamW~\cite{LHDecoupledWeightDecay2019} optimizer, with the same linear warmup~\cite{MYAdequacyUntunedWarmup2021} and cosine-decay schedule used during pretraining; the maximum learning rates were tuned as needed.
Models listed as ``frozen'' in Extended Data \cref{tab:molnet_reg_bench,tab:molnet_class_bench} were trained by only updating the task network weights;
the weights of the encoder remained fixed.
As with the pretrained models, all hyperparameters are documented in the log files provided in our data release.
Datasets used for fine-tuning are tabulated in Supplementary~\cref{sec:si:datasets}.
Metrics used to evaluate fine-tuned models are defined in Supplementary~\cref{sec:si:metrics}.
The mean and standard deviation for tabulated performance metrics were computed from 100 bootstrapped replicates as implemented by \texttt{torchmetrics}~\cite{LightningAITorchmetrics2024}.
Unless otherwise specified, fine-tuning datasets were split randomly into training, validation, and test sets (80\%, 10\%, and 10\% respectively).

\subsection{High-Throughput Screening Pipeline}
\label{sec:methods:screen}
The high-throughput screening pipeline consisted of a fragmentation and recombination based generator, followed by fine-tuned \ac{MIST} model evaluators.
Molecules for screening were generated online using FASMIFRA~\cite{BTMolecularGenerationFast2021}, a fast molecular generator for recombining molecular fragments using Markov sampling, similar to BRICS molecular generation~\cite{DWZRArtCompilingUsing2008}.
Molecular fragments were created from a set of <100 curated electrolyte molecules (\cref{fig:si:screening_electrolytes}) and molecules from the ChEMBL~\cite{ZFH+ChEMBLDatabase20232024} and ZINC20~\cite{ITY+ZINC20AFreeUltralargeScale2020} datasets.

To minimize communication overhead, each GPU processed molecules generated by a separate instance of FASMIFRA.
As constructed, the pipeline reaches a peak evaluation throughput of 4,544 molecules per GPU per second.
For the general electrolyte solvent screening case, evaluation using \ac{MIST} models was parallelized over 8 NVIDIA A100 GPUs and the workflow identified 1,936 candidate molecules in 8 hours (wall-time) after evaluating 90M molecules.
For the lithium-air electrolyte solvent screening case, evaluation using \ac{MIST} models was parallelized over 4 NVIDIA A100 GPUs and the workflow identified 33,524 candidate molecules in 8 hours (wall-time) after evaluating 40M molecules.

\subsection{Mixture Property Prediction}
\label{sec:methods:mixtures}
The binary property prediction model (\cref{sec:mixtures}, Supplementary~\cref{fig:mixture_dataset}) was fine-tuned on two excess property data sets, described in~\cite{ZRA+DifferentiableModelingOptimization2024}, and an expanded data set, described in Supplementary~\cref{sec:si:datasets}.
The multi-component ionic conductivity model (\cref{sec:mixtures}) was fine-tuned using a dataset curated by~Zhu et al.\cite{ZRA+DifferentiableModelingOptimization2024}.
This dataset contains 24,822 mixtures of single-salt, ternary-solvent electrolyte solutions generated by the \ac{AEM} developed by~Gering\cite{gering2017prediction} at Idaho National Laboratory.
The architecture for both task networks is described below.
All model weights were updated during fine-tuning, including those of the pretrained encoder.
The task network contained approximately 3M learnable parameters and was trained using the AdamW~\cite{LHDecoupledWeightDecay2019} optimizer.

\subsubsection{Excess Property Prediction Architecture}
The architecture of our mixture excess property prediction task network was informed by the chemical thermodynamic framing of excess properties, as shown in \cref{fig:mist_mix_excess_arch}.
To learn smooth, physically consistent excess curves, we propose fine-tuning the model to predict \(P_{\mathrm{E}}\) at multiple fixed mole ratios \(\vec{x}' = [x'_1, \dots, x'_n]\).
We preserve permutation invariance by taking the sum of \(g_{\mathrm{E}}(\vec{e}_{12}, \vec{x}')\) and \(\mathbb{J} \cdot g_{\mathrm{E}}(\vec{e}_{21}, 1 - \vec{x}')\), where \(g_{\mathrm{E}}\) is a \ac{MLP}, \(\mathbb{J}\) is the exchange matrix, and \(\vec{e}_{12}\) and \(\vec{e}_{21}\) are embedding vectors computed using a permutation equivariant fusion operation.
This fusion operation \(\mathrm{fusion}(\vec{e}_1, \vec{e}_2)\) combines single molecule embedding vectors from MIST (\(\vec{e}_{1}\) and \(\vec{e}_{2}\)).
We use the predicted excess property values at the control points \(P_E(\vec{x}')\) to construct an interpolating polynomial and evaluate the excess property at the desired composition \(x_1\).
The proposed architecture can be summarized as follows:
\begin{align*}
    \vec{e}_{12},\ \vec{e}_{21} &= \mathrm{fusion}(\vec{e}_1, \vec{e}_2) \\
    P_{\mathrm{E}}(\vec{x}') &= g_{\mathrm{E}}(\vec{e}_{12}, \vec{x}') + \mathbb{J} \cdot g_{\mathrm{E}}(\vec{e}_{21}, 1- \vec{x}')\\
    \vec{c} &= \mathbf{B^{-1}}  P_{\mathrm{E}}(\vec{x}') \\
    P_{\mathrm{E}}(x_1) &= \sum_{j=0}^N B_j(x_1) \cdot c_j ,
\end{align*}
where \(\mathbf{B}\) is the basis matrix for an $N$ degree polynomial and $c_i$ are the evaluated coefficients.
Similar to the Redlich-Kister polynomial~\cite{RKAlgebraicRepresentationThermodynamic1948}, we explicitly enforce \(P_{\mathrm{E}} = 0\) when \(x_1 = 0\) or \(x_2 = 0\) when computing our interpolating polynomial.
We use a Legendre polynomial basis evaluated at rescaled Chebyshev nodes for numerical stability.
A second \ac{MLP} is used to predict the pure compound properties \(P_i\) which are then used to compute the linear mixing component of the mixture property.

Permutation equivariant fusion operations are used to obtain the mixture embeddings $\vec{e}_{12}$ and $\vec{e}_{21}$ for the component embeddings $\vec{e}_1$ and  $\vec{e}_2$.
Permutation equivariance is needed to maintain the consistency of \(P_{\mathrm{E}}(\vec{x}')\) under permutation.
We evaluated task networks using the following fusion operations.

\paragraph{Permutation Invariant Embedding Fusion Strategies.}
We employed the following fusion approaches to obtain $\vec{e}_{1,2}$ and $\vec{e}_{2,1}$.
\begin{description}
    \item[Cross Attention.]
        We encode each molecule as a sequence of token embeddings $\{ \vec{e}_{1,k} \}$ and $ \{\vec{e}_{2,k}\}$,
        where $k$ indexes over tokens.
        Two multi-head cross-attention passes (one for each compound) capture fine-grained, token-level interactions in both directions.
        Each attended sequence is pooled to yield two fixed-length vectors.
        This bidirectional attention plus symmetric pooling ensures that swapping the inputs simply swaps the two pooled outputs, preserving equivariance under component exchange while still allowing asymmetric mixtures to be represented.
    \item[Cross Product.]
        Molecule embeddings from the MIST encoder are first split into $H$ heads of dimension $d$.
        Each head is down-projected to a 3-dimensional subspace, and a 3D cross-product is computed between corresponding head embeddings of $A$ and $B$.
        The result is then up-projected back to the original head dimension.
        By anti-symmetry of the cross product ($AB = -BA$), this fusion inherently encodes exchange information;
        a simple mean of the two directions yields a permutation-invariant representation.
    \item[Difference.]
        It has been shown that excess properties are correlated with the structural similarity between solvent molecules~\cite{KADVExcessDensityDescriptor2025}.
        Based on this we test the use of the element-wise difference between molecule embeddings from MIST as the mixture embedding.
    \item[Concatenation.]
        This approaches concatenates the embeddings for components 1 and 2 in both orderings.
        Since order is preserved this embedding is inherently permutation equivariant.
\end{description}

\noindent
Ultimately, we selected concatenation for our fusion operator for its simplicity and empirical effectiveness.
Training logs for our ablation sweeps are provided in our data release.

\subsubsection{Ionic Conductivity Prediction Architecture \label{sec:methods:ionic}}
The \ac{MIST} ionic conductivity task network uses a sum of the component embeddings $\vec{e}_i$ weighted by their mole ratio $x_i$ to construct a mixture embedding  \(\vec{e}_{mix}\).
A two-layer \ac{MLP} uses the mixture embedding \(\vec{e}_{mix}\) to predict coefficients for physics-based temperature and concentration dependence expressions.
The task network learns temperature dependence through the \ac{VFT} relation~\cite{ngaiIntroductionProblemsRelaxation2011} and a concentration correction dependent on the salt mole ratio ($x_{Li}$):
\begin{align}
\ln\;\sigma_0 &=  \ln \; A  - \frac{E_a}{ T - T_0} \label{eq:vft} \\
\gamma &= (1 - \beta) \cdot \exp\left({\frac{x_{Li} - \alpha }{\lambda}}\right) + \beta \\
\ln\;\sigma &=
\begin{cases}
\gamma \cdot \ln\;\sigma_0 & \text{if } x_{Li} > \alpha \\
\ln\;\sigma_0 & \text{otherwise} ,
\end{cases}
\end{align}
where the pseudo-activation energy $E_a$ and parameters $T_0$,  $A$ , $\alpha$ and $\beta$ are predicted by the model, which in turn, are used to compute the log of the ionic conductivity \(\ln \sigma\).
Liquid electrolytes commonly used in Li-ion batteries exhibit non-Arrhenius temperature dependence in their transport properties~\cite{XuNonaqueousLiquidElectrolytes2004}.
Experiments consistently show that electrolyte ionic conductivity does not follow a straight-line Arrhenius plot; instead, it is well described by the \ac{VFT} equation (\cref{eq:vft})~\cite{porionComparativeStudyTransport2013}.

\paragraph{Concentration Correction.}
The concentration correction ($\gamma$) is informed by the ion interaction behaviour in different concentration regimes~\cite{ZBVPredictingElectrolyteConductivity2022,XuNonaqueousLiquidElectrolytes2004}.
At low salt concentrations (\(x_{Li} \ll \alpha\)), ionic conductivity generally tends to increase linearly with salt concentration.
This low concentration region is characterized by singly solvated cations whose solvation shells are absent of other charged species.
The model learns the dynamics in this region in terms of only the parameters in the \ac{VFT} relation since  $\ln\;\sigma = \ln \;\sigma_0$~\cite{hwang2018ionic}.
As salt concentration increases, the number of singly solvated cations reduce;
the ability of the solvent to fully solvate the salt ions reduces and coulombic interactions increase.
This first results in the formation of solvent-separated ion pairs --- oppositely charged ions whose direct coordination is prevented by the presence of a solvent;
these reduce the conductivity of the solution as the effective charge held in the solvation shell reduces.
As the salt concentration continues to increase, larger aggregates form in the electrolyte solution, with the emergence of triple ion pairs.
At high concentrations (\(x_{Li} \gg \alpha\)), diffusivity effects begin to dominate, and the formation of larger solvated structures further increases the solution's viscosity~\cite{hwang2018ionic}.
The concentration correction (\(\gamma \ln \sigma_0\)) enables the model to capture the effects of solvent-separated ion pairs and aggregates, which inhibit conductivity.

\subsection{Linear Probes}
We used linear classifier probes to find feature vectors associated with \acf{RO5} criteria in \cref{sec:hidden_states}.
Linear classifier probes~\cite{ABUnderstandingIntermediateLayers2018} of the form \(y_i = \sigma(\vec{f}_i\cdot\vec{x} + b_i)\) were trained to predict molecule-level features from token-level activations \(\vec{x}\) within pretrained or fine-tuned \ac{MIST} models;
where $i$ indexes over \ac{RO5} criteria.
We used PyTorch's gradient hook interface to instrument \ac{MIST} models and capture their internal activation while evaluating molecules from the positive and negative class (that is, molecules that do or do not satisfy a particular \ac{RO5} criterion).
To generate a suitable dataset for training, we curated a dataset of molecules from MoleculeNet~\cite{WRF+MoleculeNetBenchmarkMolecular2018} and used RDKit~\cite{GreRDKitOpensourceCheminformatics2024} to annotate the \ac{RO5} criterion for each molecule.
Using iterative proportional refitting, we downsampled the initial dataset to provide a balanced mix of positive and negative examples across all four individual and the overall \ac{RO5} criterion;
additional details are provided in Supplementary~\cref{sec:si:probes}.
Probes were trained for 10,000 steps using the AdamW~\cite{LHDecoupledWeightDecay2019} optimizer with a constant learning rate of \sn{1}{-3}.
The probe with the lowest binary cross-entropy loss was selected as the trained probe.
We trained \ac{RO5} probes for MIST-28M (Pretrained, MoleculeNet and tmQM) and MIST-1.8B (Pretrained).
Notably, by construction training linear probes will converge to the global optimum, as identifying the parameters of a linear classifier trained to minimize the cross-entropy loss is a convex problem~\cite{ABUnderstandingIntermediateLayers2018}.

\subsection{Embeddings}
\label{sec:methods:embeddings}
To visualize the embedding space of the model in \cref{fig:interp_emb_pah} and \cref{fig:interp_emb_rings}, molecular embedding vectors (hidden states from the last layer of the encoder for the first token) were projected to 2D using \ac{UMAP}.
In order to investigate the separation between \emph{cata-} and \emph{peri-} condensed structures (\cref{fig:interp_emb_pah}) as well as aromatic and anti-aromatic \ac{PAH} (\cref{fig:interp_emb_rings}) we used a set of \acp{PAH} sampled randomly from the COMPAS collection~\cite{wahabCOMPASProjectComputational2022,mayoyanesCOMPAS2DatasetCatacondensed2024,wahabCOMPAS3DataSet2024}.
Cata-condensed \acp{PAH} have rings that share at most two carbon atoms, forming linear or angular chains.
Peri-condensed \acp{PAH} have rings that share more than two carbon atoms, creating compact clustered structures.
Labels for cata and peri-condensed are used as specified in the COMPAS datasets.
Aromaticity labels are assigned using RDKit's \texttt{GetIsAromatic} method which is based on H\"{u}ckel's rule; a ring, or fused ring system, is considered to be aromatic if it obeys the 4N+2 $\pi$ electrons rule~\cite{huckel,LTK+RdkitRdkit2024_09_42024}.
We labelled molecules where all bonds are aromatic as aromatic in \cref{fig:interp_emb_rings}.

\subsection{Hierarchical Clustering and Hyperbolic Projection of Odours}
\label{sec:methods:olfaction}
The dendrogram in \cref{fig:olfaction_clusters} was plotted by clustering logits from the \ac{MIST}-28M variant fine-tuned on olfaction.
The fine-tuned model acts as a multi-label classifier predicting a logit vector for an input molecule, where each element in the vector provides the log-odds for the presence of particular scent descriptors.
To group descriptors by shared co-activation across molecules, we computed a descriptor–descriptor similarity matrix using the Pearson correlation across molecules and converted it to a dissimilarity distance:
\begin{equation}
d_{ij} \;=\; \sqrt{\,2\bigl(1-\rho_{ij}\bigr)}\,
\end{equation}
where $\rho_{ij}$ is the Pearson correlation between descriptors $i$ and $j$.

Hierarchical agglomerative clustering was performed on this distance matrix using the average linkage criterion (qualitatively similar results were obtained with complete linkage).
The full merge tree is visualized as a circular dendrogram, placing leaves in the order returned by the clustering and mapping internal node radii to normalized merge heights.
All analyses were implemented in Julia~\cite{BEKSJuliaFreshApproach2017} using \texttt{Clustering.jl} for hierarchical clustering.

To test if the logit geometry of the \ac{MIST} model was well fit by a hyperbolic shell, we followed a framework similar to the one described in Ref.~\cite{zhouHyperbolicGeometryOlfactory2018}.
Using ALBATROSS~\cite{albatross,stierALBATROSSCheapFiltration2025}, a statistical topological data analysis framework that uses stochastic sub-sampling, we constructed a pairwise similarity structure from the \ac{MIST} olfactory logits, generated filtered clique complexes across similarity thresholds, and summarized their topology using the first three Betti curves, whose shapes and integrated areas are sensitive to global geometry while being insensitive to monotone rescalings of the underlying similarities. 
ALBATROSS fits Euclidean and hyperbolic generative models by matching the integrated Betti values and the L1 and Wasserstein distances between empirical and model-derived Betti curves. 
It then tests whether the observed topology is statistically consistent with each candidate geometry~\cite{albatross,stierALBATROSSCheapFiltration2025}.
After confirming the hyperbolic geometry, for visualization of the \ac{MIST}-predicted perceptual profiles on a hyperbolic shell, we used a shell-constrained version of \ac{HYDRA}~\cite{kellerressel2019hydramethodstrainminimizinghyperbolic}, a method for strain-minimizing hyperbolic embedding, implemented in ALBATROSS, where the empirical dissimilarities were first quantile-mapped onto the geodesic distance scale of the best-fit ALBATROSS hyperbolic shell, after which hyperbolic multidimensional scaling was performed in the Poincaré ball with the embedded points softly constrained to lie between $r_{min}$ and $r_{max}$ (the inner/outer radii of the best-fit ALBATROSS shell).
Hyperparameters and full logs for the geometry identification and embedding projection are provided in our \datadrop[data release].

\subsection{Bayesian Parameter Estimation using \acs{MCMC} Sampling}
\label{sec:methods:bayesian}
Neural scaling laws were fit using a No-U-Turn sampler~\cite{TDD+TpappDynamicHMCjlV3502025,HGNouturnSamplerAdaptively2014} to sample model parameters $\theta = \{A, B, \alpha, \beta, \allowbreak E, \dots\}$ from the posterior after a burn-in of 2,000 steps.
Chains were checked for convergence using the potential scale reduction factor \(\hat{R}\), with convergence defined as \(\hat{R} \leq 1.01\)~\cite{VGS+RankNormalizationFoldingLocalization2021,GCS+BayesianDataAnalysis2014}.
We also evaluated the \ac{ESS} and adjusted sampling duration to ensure an \ac{ESS} of at least 400~\cite{VGS+RankNormalizationFoldingLocalization2021,GCS+BayesianDataAnalysis2014}.
Rather than fitting a closed-form posterior, we generate predictions by evaluating the model over samples in the posterior distribution.
Detailed fit statistics and parameter covariances are reported in Supplementary~\cref{sec:si:bayesian_neural_scaling}.
Source code for this analysis and \ac{MCMC} samples for all fitted models are provided in our data release.
Scaling coefficients (\(A, B, \alpha, \beta\) \& \(E\)) and quality metrics for our scaling laws are additionally provided in Extended Data \cref{tab:scaling_law_parameters}.

To fit our neural scaling laws, we performed a sweep over model size (0.4M to 603M), learning rate, feed-forward ratio (1, 4), batch size (\(2^{11}\) to \(2^{15}\)), and dataset size (8M to 2B).
For this sweep, we assumed square root scaling of learning rate with effective batch size assuming \(\eta \propto \sqrt{\mathcal{B}}\) and sweeping the proportionality constant.
We estimate that a full factorial sweep over this space would require over 6,000 runs and 12 petaflop-days of compute.
To reduce this, we used generalized subset design~\cite{SVH+GeneralizedSubsetDesigns2017} to prune the sweep to 138 runs, reducing our compute budget to 1.2 petaflop-days.
Training logs for these runs, in addition to the over 20,000 training runs conducted in this work, are provided as JSON files in our data release.

Our model for the optimal learning rate (\cref{eq:optimal_lr}) was motivated by existing heuristics for estimating learning rates~\cite{KMH+ScalingLawsNeural2020,VSP+AttentionAllYou2017,KriOneWeirdTrick2014}.
In particular, our prior beliefs that \(\delta \approx 1/2\) and \(\gamma \approx 0\) were motivated by early success using square-root scaling to stabilize training (Supplementary~\cref{sec:si:spikes}).
Fitted parameters confirmed our priors: \(\delta = \ci{0.55}{0.46}{0.63}\), \(\gamma = \ci{0.04}{-0.05}{0.15}\).
As a retrospective, we refit our model replacing \(N\) with hidden size \(d_{\text{model}}\), following~Vaswani et al.\cite{VSP+AttentionAllYou2017}:
\(\eta_{\star} = \eta_0 d_{\text{model}}^{\gamma} \mathcal{B}^{\delta}\),
where \(\mathcal{B}\) is the effective batch size and \(\eta_0, \gamma, \delta\) are fitted coefficients.
This alternative model (Penalized, \(\eta_{\star} = f(d_{\text{model}})\) in Extended Data~\cref{tab:scaling_law_parameters}) outperformed 
our previous fits (Penalized, Baseline) with a \ac{MAPE} of 20\% and \ac{WAIC} of -580; lower is better for both metrics.
It also reaffirmed square-root scaling \(\delta = \ci{0.47}{0.39}{0.56}\), while also identifying non-degenerate scaling with \(d_{\text{model}}\): \(\gamma = \ci{-0.74}{-0.91}{-0.56}\).

\section{Code and Data Availability}\label{sec:code_availability}
The source code, model weights, fine-tuning datasets, training logs, and all other data assets generated in this study will be available on Zenodo.
Additionally, our source code will be published on GitHub (\url{https://github.com/BattModels/mist}), including our full commit history throughout the development process.

\FloatBarrier

\clearpage
\section{Extended Data}
\begin{table}[ht!]
\resizebox{\textwidth}{!}{%
\begin{tabular}{lcccccc}
\toprule
Model                    & BBBP($\uparrow$)        & HIV($\uparrow$)         & BACE($\uparrow$)        & Tox21($\uparrow$)       & ClinTox($\uparrow$)     & SIDER($\uparrow$)       \\ \midrule
MAT\cite{MDM+MoleculeAttentionTransformer2020}                      & 0.728          &                &                &                &                &                \\
Mol-BERT\cite{LJMolBERTEffectiveMolecular2021}                 & 0.875          &                &                & 0.839          & 0.923          & 0.695          \\
SELFormer\cite{YUUDSELFormerMolecularRepresentation2023}       & 0.902          & 0.681          & 0.832          & 0.653        &                & 0.745          \\
FP-BERT-10M\cite{WLZ+FingerprintsBasedMolecular2022}           & 0.714          & 0.776          &                &                &                &                \\
ChemBERTa-10M\cite{CGRChemBERTaLargeScaleSelfSupervised2020}            & 0.643          & 0.622          &                &                &                &                \\
ChemBERTa-2-77M\cite{ASC+ChemBERTa2ChemicalFoundation2022}    & 0.728          &                & 0.799          &                & 0.563          &                \\
MolFormer-XL\cite{RBC+LargescaleChemicalLanguage2022}             & 0.937          & 0.822          & 0.863          & 0.805          & 0.948          & 0.655          \\
SMI-TED\cite{SSB+LargeEncoderDecoderFamily2024}                  & 0.923          & 0.769          & \textbf{0.882}          & 0.819          & 0.943          & 0.657          \\ \hline
MIST-28M (unfrozen)              &   0.967 $\pm$ 0.014 & \textbf{0.838 $\pm$  0.001} &   \textbf{0.881 $\pm$  0.034}      &    0.844 $\pm$ 0.013    &      \textbf{0.999   $\pm$ 0.001} &  0.827 $\pm$ 0.012  \\
MIST-28M (frozen)                &   0.956 $\pm$ 0.015 & 0.809 $\pm$ 0.021 &       0.839 $\pm$  0.035    &     \textbf{0.855  $\pm$ 0.011}      &       0.987 $\pm$ 0.003 & 0.823 $\pm$ 0.008 \\
MIST-1.8B (unfrozen)               &        \textbf{0.971  $\pm$ 0.013}   &   \textbf{0.838 $\pm$  0.002}     &    0.850 $\pm$     0.036       &   0.840 $\pm$  0.013         &     0.998    $\pm$  0.009      &     \textbf{0.828 $\pm$ 0.011}  \\
MIST-1.8B (frozen)          &      0.970  $\pm$  0.020 &    0.728 $\pm$ 0.021    &   0.837   $\pm$   0.034        &      0.851 $\pm$ 0.012          &           0.998  $\pm$   0.011 &   0.784 $\pm$ 0.013     \\
\bottomrule
\end{tabular}
}
\caption{
    \label{tab:molnet_class_bench}
    Performance (\ac{AUROC}) of text based molecular property prediction models on a subset of MoleculeNet\cite{WRF+MoleculeNetBenchmarkMolecular2018} classification tasks.
    Scores for SMI-TED and MolFormer-XL are as reported in \cite{SSB+LargeEncoderDecoderFamily2024}.
    Scores for all other models are as reported in \cite{SSMVTransformersMolecularProperty2024a}.
    Errors reported for MIST variants were evaluated on the validation set with confidence intervals evaluated using bootstrapping on 100 samples.
}
\end{table}

\begin{table}[ht!]
\resizebox{\textwidth}{!}{%
\begin{tabular}{lccccc}
\toprule
Model                    & ESOL($\downarrow$)        & FreeSolv($\downarrow$)    & Lipo ($\downarrow$) & QM8($\downarrow$)          & QM9($\downarrow$) \\ \midrule
ST\cite{HSUSMILESTransformerPretrained2019} + (MLP/L2/LightGBM)   & 0.720          & 1.650          & 0.921            &                 &        \\
MAT\cite{MDM+MoleculeAttentionTransformer2020}                      & \textbf{0.285} & 0.263          &                  &                 &        \\
RT\cite{BMRegressionTransformerEnables2023} + fine-tuning         & 0.690          & 1.030          & 0.740            &                 &        \\
MolBERT\cite{LJMolBERTEffectiveMolecular2021} + NN (fine-tune) & 0.531          & 0.948          & 0.561            &                 &        \\
SELFormer\cite{YUUDSELFormerMolecularRepresentation2023}                & 0.386          & 1.005          & 0.674            &                 &        \\
FP-BERT-10M\cite{WLZ+FingerprintsBasedMolecular2022}           & 0.670          & 1.070          & 0.670            &                 &        \\
ChemFormer\cite{IDHBChemformerPretrainedTransformer2022} (Combined)    & 0.633          & 1.230          & 0.598            &                 &        \\
MolFormer-XL\cite{RBC+LargescaleChemicalLanguage2022}             & 0.880          & 2.342          & 0.529           & 0.0102          & 1.5894 \\
X-Mol\cite{XZC+XMOLLargescalePretraining2022}                    & 0.578          & 1.108          & 0.596            &                 &        \\
ChemBERTa-2\cite{ASC+ChemBERTa2ChemicalFoundation2022}              & 0.89           &                & 0.798            &                 &        \\
SMI-TED\cite{SSB+LargeEncoderDecoderFamily2024}                  & 0.611          & 1.223          & 0.552            & 0.0095          & \textbf{1.325} \\ \hline
MIST-28M (unfrozen)            &     0.632 $\pm$ 0.062   &  \textbf{0.249 $\pm$ 0.032} &       0.699 $\pm$ 0.031     &  \textbf{0.0089 $\pm$ 0.0003} &   3.782 $\pm$ 0.057   \\
MIST-28M (frozen)              &    0.859 $\pm$  0.057 &  0.350 $\pm$ 0.045 &        0.945 $\pm$ 0.032      &   0.0206  $\pm$ 0.0004   &  15.610 $\pm$ 0.190    \\
MIST-1.8B (unfrozen)        &         0.664 $\pm$ 0.066  &    0.275 $\pm$ 0.035    &  \textbf{0.507 $\pm$ 0.031} &     0.0103 $\pm$ 0.0003   & 1.766 $\pm$ 0.324       \\
MIST-1.8B (frozen)        &     0.781 $\pm$ 0.082  &     0.277   $\pm$  0.034 &  0.863 $\pm$ 0.035  &     0.0190   $\pm$ 0.0003  &    5.499 $\pm$ 0.199   \\
\bottomrule
\end{tabular}
}
\caption{
    \label{tab:molnet_reg_bench}
    Performance (\acs{MAE}) of text based molecular property prediction models on a subset of MoleculeNet\cite{WRF+MoleculeNetBenchmarkMolecular2018} regression tasks.
    Scores for SMI-TED and MolFormer-XL are as reported in \cite{SSB+LargeEncoderDecoderFamily2024}.
    Scores for all other models are as reported in \cite{SSMVTransformersMolecularProperty2024a}.
}
\end{table}

\begin{table}[ht!]
\resizebox{\textwidth}{!}{%
\begin{tabular}{lcccc}
\toprule
Model & MIST-1.8B (frozen) & MIST-1.8B (unfrozen) & MIST-28M (frozen) & MIST-28M (unfrozen) \\
\midrule
$C_V$                & 0.4297 $\pm$ 0.0164 & 0.1542 $\pm$ 0.0124 & 1.5547 $\pm$ 0.0229 & 0.4023 $\pm$ 0.0063 \\
G298                 & 5.0000 $\pm$ 0.1023 & 0.832  $\pm$ 0.035  & 14.875  $\pm$ 0.1807 & 2.3281 $\pm$ 0.0333 \\
Gap                  & 0.0067 $\pm$ 0.0002 & 0.0029 $\pm$ 0.0001 & 0.0199 $\pm$ 0.0002 & 0.0076 $\pm$ 0.0001 \\
H298                 & 5.4375 $\pm$ 0.1455 & 0.8711 $\pm$ 0.0903 & 14.8125 $\pm$ 0.1846 & 2.1875 $\pm$ 0.0386 \\
HOMO                 & 0.0048 $\pm$ 0.0001 & 0.0019 $\pm$ 0.0001 & 0.0115 $\pm$ 0.0001 & 0.0046 $\pm$ 0.0001 \\
LUMO                 & 0.0058 $\pm$ 0.0001 & 0.0025 $\pm$ 0.0001 & 0.0186 $\pm$ 0.0002 & 0.0063 $\pm$ 0.0001 \\
$\langle R^2 \rangle$& 41.2500 $\pm$ 0.8125 & 16.625 $\pm$ 3.0469 & 122.0   $\pm$ 1.4434 & 34.5   $\pm$ 0.5186 \\
U0                   & 4.9689 $\pm$ 0.1246 & 0.8594 $\pm$ 0.5449 & 14.875  $\pm$ 0.1879 & 2.2344 $\pm$ 0.0363 \\
U298                 & 4.9688 $\pm$ 0.6836 & 1.2031 $\pm$ 0.1332 & 14.8125 $\pm$ 0.1987 & 2.2348 $\pm$ 0.0325 \\
ZPVE                 & 0.0026 $\pm$ 0.0001 & 0.0007 $\pm$ 0.0001 & 0.0103 $\pm$ 0.0001 & 0.0020 $\pm$ 0.0001 \\
$\alpha$             & 3.5625 $\pm$ 0.0146 & 0.3744 $\pm$ 0.0182 & 3.4688 $\pm$ 0.0447 & 0.9578 $\pm$ 0.0151 \\
$\mu$                & 0.3535 $\pm$ 0.0095 & 0.2637 $\pm$ 0.0095 & 0.8594 $\pm$ 0.0110 & 0.5156 $\pm$ 0.0064 \\
\hline
Avg. MAE             & 5.4992 $\pm$ 0.1985 & 1.7659 $\pm$ 0.3242 & 15.6099 $\pm$ 0.1895 & 3.7818 $\pm$ 0.0573 \\
\bottomrule
\end{tabular}
}
\caption{
    \label{tab:qm9_benchmark} Comparing performance of MIST variants on the QM9 dataset.
    The per target \ac{MAE} is reported for models trained on individual targets as well as the \ac{MAE} averaged across single target models.
}
\end{table}

\begin{table}[ht!]
\resizebox{\textwidth}{!}{
\begin{tabular}{lccccc}
\toprule
{Property}        & {GNN} & {GNN-3D} & {DiffMix} & {DiffMix-3D} & {MIST} \\
\midrule
{Ionic Conductivity ($ln(mS/cm)$)} & 0.045        & 0.043           & 0.044            & 0.043               &  \textbf{$0.031 \pm 0.02$}         \\
{Excess Molar Volume ($cm^3/mol$)} &  0.039       &     0.035      &       0.033    &    0.031           &  \textbf{$0.011 \pm 0.002$}         \\
{Excess Molar Enthalpy ($J/mol$)} &     9.88   &   9.68      &    5.10      &     5.52       &  \textbf{$4.24 \pm 0.2$}    \\
\bottomrule
\end{tabular}
}
\caption{
    \label{tab:ionic_cond} MIST outperforms (lower \ac{MAE}) existing state-of-the-art mixture property prediction models on tasks relevant to electrolyte design, including those trained specifically for mixture property prediction (DiffMix\cite{ZRA+DifferentiableModelingOptimization2024}) and those which have full 3D molecular structural information (DiffMix-3D\cite{ZRA+DifferentiableModelingOptimization2024}).
    \ac{MAE} Errors for other models are as reported by Zhu et al.\cite{ZRA+DifferentiableModelingOptimization2024}.
}
\end{table}

\begin{table}[ht!]
\centering
\resizebox{\linewidth}{!}{
\begin{tabular}{l|ccccc}
\toprule
Model & \(\alpha\) & \(\beta\) & \begin{tabular}[c]{@{}c@{}}Coeff. a\\ \(N_{\mathrm{opt}} \propto C^{a}\)\end{tabular} & \begin{tabular}[c]{@{}c@{}}Coeff. b\\ \(D_{\mathrm{opt}} \propto C^{b}\)\end{tabular} & \begin{tabular}[c]{@{}c@{}}Scaling Balance\\ \(\alpha/\beta\)\end{tabular} \\
\midrule
Kaplan et al.\cite{KMH+ScalingLawsNeural2020} & 0.076 & 0.103 & 0.58* & 0.42* & 0.74 \\
Hoffmann et al.\cite{HBM+TrainingComputeOptimalLarge2022}, ``Approach 1'' & -- & -- & \ci{0.50}{0.488}{0.502} & \ci{0.50}{0.501}{0.512} & 1* \\ 
Hoffmann et al.\cite{HBM+TrainingComputeOptimalLarge2022}, ``Approach 3'' & 0.34 & 0.28 & \ci{0.46}{0.454}{0.455} & \ci{0.54}{0.542}{0.543} & 1.2* \\ 
DeepSeek LLM\cite{BCC+DeepSeekLLMScaling2024}, ``Early Data'' & -- & -- & 0.450 & 0.550 & 1.22* \\
DeepSeek LLM\cite{BCC+DeepSeekLLMScaling2024}, ``Current Data'' & -- & -- & 0.524 & 0.478 & 0.908* \\
\midrule
MIST, Penalized Scaling \cref{eq:penalized_neural_scaling} & \ci{0.932}{0.774}{1.09} & \ci{0.537}{0.464}{0.631} & \ci{0.366}{0.312}{0.428} & \ci{0.634}{0.572}{0.688} & \ci{1.75}{1.34}{2.2} \\ 
MIST, No Penalty Terms & \ci{0.841}{0.622}{1.06} & \ci{0.751}{0.536}{0.955} & \ci{0.471}{0.38}{0.556} & \ci{0.529}{0.444}{0.62} & \ci{1.14}{0.798}{1.63} \\ 
\bottomrule
\end{tabular}
}
\caption{
    \label{tab:scaling_coeffs}
    Fitted scaling exponents and compute-optimal coefficients for MIST compared with published values from prior work in \ac{NLP}.
    \(\alpha\) and \(\beta\) control how loss scales with model size and dataset size, respectively.
    The coefficients \(a = \beta / (\alpha + \beta)\) and \(b = \alpha/(\alpha + \beta)\) describe the compute-optimal scaling of model \(N_{\mathrm{opt}}\) and data \(D_{\mathrm{opt}}\) with compute \(C\).
    The ratio \(\alpha/\beta\) quantifies the trade-off between parameter and data scaling for compute efficiency with \(D_{\mathrm{opt}} \propto N_{\mathrm{opt}}^{\alpha/\beta}\).
    Confidence intervals (95\%) are shown for Bayesian estimates.
    Values marked with an asterisk (\(*\)) were derived from values reported in the original publications.
}
\end{table}

\begin{table}[ht!]
\resizebox{\linewidth}{!}{%
\begin{tabular}{lcccccccccc}
  \toprule
  Model & \acs{WAIC} & \acs{MAPE} & \makecell{Pearson's\\\(r\)} & \makecell{Spearman's\\\(\rho\)} & $A$ & $B$ & $\alpha$ & $\beta$ & $E \cdot 1000$ & $\frac{\alpha}{\beta}$ \\
  \midrule
  Penalized, \(\eta_{\star} = f(d_{\text{model}})\) & -580 & 20.0\% & 0.928 & 0.837 & \cistacked{\sn{1.53}{4}}{\sn{5.18}{2}}{\sn{8.18}{4}} & \cistacked{\sn{6.88}{2}}{\sn{1.81}{2}}{\sn{1.98}{3}} & \cistacked{0.849}{0.673}{1.02} & \cistacked{0.516}{0.455}{0.588} & \cistacked{1.57}{0.0673}{5.8} & \cistacked{1.65}{1.23}{2.11} \\
  Penalized, Baseline & -517 & 29.1\% & 0.877 & 0.812 & \cistacked{\sn{8.1}{4}}{\sn{4.58}{3}}{\sn{3.89}{5}} & \cistacked{\sn{1.14}{3}}{\sn{2.14}{2}}{\sn{4.2}{3}} & \cistacked{0.932}{0.774}{1.09} & \cistacked{0.537}{0.464}{0.631} & \cistacked{1.99}{0.0755}{7.73} & \cistacked{1.75}{1.34}{2.2} \\
  \makecell{Penalized, Harmonic \\ Shape Penalty} & -517 & 28.2\% & 0.874 & 0.814 & \cistacked{\sn{1.05}{5}}{\sn{5.62}{3}}{\sn{5.42}{5}} & \cistacked{\sn{1.33}{3}}{\sn{2.66}{2}}{\sn{4.85}{3}} & \cistacked{0.933}{0.773}{1.1} & \cistacked{0.539}{0.464}{0.628} & \cistacked{2.19}{0.0807}{8.42} & \cistacked{1.74}{1.33}{2.19} \\
  Penalized, Additive & -464 & 38.1\% & 0.787 & 0.785 & \cistacked{\sn{1.14}{5}}{\sn{2.91}{3}}{\sn{6.74}{5}} & \cistacked{\sn{3.39}{4}}{\sn{1.56}{3}}{\sn{2.13}{5}} & \cistacked{0.905}{0.714}{1.11} & \cistacked{0.679}{0.551}{0.834} & \cistacked{4.4}{0.111}{17.1} & \cistacked{1.35}{0.951}{1.82} \\
  No Penalties & -453 & 38.5\% & 0.772 & 0.776 & \cistacked{\sn{5.69}{4}}{\sn{8.57}{2}}{\sn{3.69}{5}} & \cistacked{\sn{1.87}{5}}{\sn{1.29}{3}}{\sn{1.32}{6}} & \cistacked{0.841}{0.622}{1.06} & \cistacked{0.751}{0.536}{0.955} & \cistacked{29.9}{9.7}{44.3} & \cistacked{1.14}{0.798}{1.63} \\
  \bottomrule
\end{tabular}

}
\caption{
    \label{tab:scaling_law_parameters}
    Performance metrics and 95\% credible intervals for the neural scaling law parameters from \cref{eq:penalized_neural_scaling} are shown for the baseline model (\cref{fig:compute_optimal_frontier}), Chinchilla-style scaling\cite{HBM+TrainingComputeOptimalLarge2022}, and two additional variations described in \cref{sec:si:bayesian_neural_scaling}.
    All performance metrics are defined in \cref{sec:si:metrics}.
    Overall, all fits support our assessment in \cref{sec:scaling} that the ratio \(\alpha/\beta\) deviates markedly from values reported for current \ac{NLP} models (\cref{tab:scaling_coeffs}).
    \ac{MCMC} chains for all models are provided openly in our data release, and include posterior estimates for all model parameters.
    The source code to replicate this analysis is also available, as detailed in \cref{sec:code_availability}.
}
\end{table}

\begin{table}[ht]
 
  \centering

  \begin{subtable}{\textwidth}
    \centering
  \resizebox{\textwidth}{!}{
  \begin{tabular}{cccccccc}
    \toprule
    \multirow{2}{*}{\makecell{Molecular\\rep.}} & \multirow{2}{*}{\makecell{Mixture\\rep.}} & \multicolumn{3}{c}{Miscible Solvents} & NIST-full \\
    \cmidrule(r){3-5}
    & &  $\rho$ & $\Delta H_{mix}$ & $\Delta H_{vap}$ &  $\ln(\eta)$ \\
    \midrule

     \multirow{2}{*}{GNN} & Attention & 0.018 ± 0.020 &  {0.158 ± 0.002} &  0.087 ± 0.006  & 0.136 ± 0.010\\
     & Deepsets &  {0.003 ± 0.000} &  {0.159 ± 0.002} & 0.406 ± 0.668 &  0.131 ± 0.010 \\

     \midrule
     \multirow{3}{*}{MolT5}& XGB & 0.009 ± 0.000 & 0.269 ± 0.004 & 0.306 ± 0.003 &   0.148 ± 0.001 \\
     & Attention & 0.005 ± 0.001 &  {0.157 ± 0.002} & 0.125 ± 0.077 &  0.076 ± 0.004 \\
     & Deepsets &  {0.008 ± 0.005} &  {0.157 ± 0.003} &  {0.071 ± 0.002} & 0.162 ± 0.009  \\

     \midrule
     \multirow{3}{*}{RDKit}& XGB & 0.009 ± 0.000  & 0.225 ± 0.005 & 0.295 ± 0.002 &   {0.055 ± 0.000} \\
     & Attention & 0.006 ± 0.001 & 0.167 ± 0.002 & 0.199 ± 0.030 &  0.069 ± 0.006 \\
     & Deepsets & 0.005 ± 0.000 & 0.207 ± 0.008 & 0.079 ± 0.005 &  0.137 ± 0.005 \\
     \midrule
     MIST & Linear Combination & \textbf{0.003 ± 0.000} & 0.163 ± 0.002 & \textbf{0.067 ±  0.001}  & 0.309 ± 0.002 \\

  \end{tabular}
}
\end{subtable}

  \begin{subtable}{\textwidth}
    \centering
  \resizebox{\textwidth}{!}{
  \begin{tabular}{ccccccc}
    \midrule
    \multirow{2}{*}{\makecell{Molecular\\rep.}} & \multirow{2}{*}{\makecell{Mixture\\rep.}} & \multicolumn{2}{c}{IlThermo} & MON & NIST & Olfaction \\
    \cmidrule(r){3-4}
    & & $\ln(\kappa)$ & $\ln(\eta)$ & MON & $\ln(\eta)$ & Mixture Similarity$^{*}$\\
    \midrule
     \multirow{2}{*}{GNN}& Attention & 0.276 ± 0.044 & 0.154 ± 0.084 & 10.240 ± 1.658 &  {0.035 ± 0.004} & 0.129 ± 0.005 \\
     & Deepsets & 0.226 ± 0.017 & 0.206 ± 0.020 & 5.990 ± 1.382 & 0.091 ± 0.005 & 0.146 ± 0.010 \\
     \midrule
     \multirow{3}{*}{MolT5}& XGB &  {0.071 ± 0.001} &  {0.078 ± 0.002} & 5.002 ± 0.538 & 0.059 ± 0.001 & 0.128 ±  0.006\\
     & Attention & 0.244 ± 0.011 &  {0.083 ± 0.035} &  {4.660 ± 0.603} &  {0.030 ± 0.001} & 0.123 ± 0.005 \\
     & Deepsets & 0.196 ± 0.013 & 0.132 ± 0.003 & 5.296 ± 0.585 & 0.056 ± 0.004 & 0.121 ± 0.006\\
     \midrule
     \multirow{3}{*}{RDKit}& XGB &  {0.073 ± 0.002} &  {0.076 ± 0.002} &  {4.570 ± 0.348} & 0.048 ± 0.002 & 0.125 ± 0.006 \\
     & Attention & 0.407 ± 0.019 & 0.100 ± 0.003 & 11.297 ± 2.110 & 0.056 ± 0.004 & 0.148 ± 0.010 \\
     & Deepsets & 0.290 ± 0.059 & 0.107 ± 0.007 & 7.625 ± 1.874 & 0.047 ± 0.003 & 0.150 ± 0.008 \\
     \midrule
    MIST & Linear Combination & 0.644 ± 0.014 & 0.433 ± 0.005 &  \textbf{4.028 ± 0.631} & \textbf{0.036 ± 0.001} & 0.112 ± 0.008$^{*}$ \\
    \bottomrule
  \end{tabular}
}
\end{subtable}
  \caption{  Model performances across CheMixHub tasks. All scores except MIST are as reported in \cite{RKM+CheMixHubDatasetsBenchmarks2025}. $^{*}$  Olfaction similarity model uses pooled attention to combine embeddings, not a linear combination. \label{tab:chemixhub}}
\end{table}

\FloatBarrier

\section{End Notes}
\paragraph*{Acknowledgements}
Computational resources for this work were provided by an award for computer time was provided by the U.S. Department of Energy’s (DOE) Innovative and Novel Computational Impact on Theory and Experiment (INCITE) Program. This research used resources from the Argonne Leadership Computing Facility, a U.S. DOE Office of Science user facility at Argonne National Laboratory, which is supported by the Office of Science of the U.S.\ DOE under Contract No.\ DE-AC02-06CH11357.
The authors acknowledge the National Artificial Intelligence Research Resource (NAIRR) Pilot  and  NVIDIA DGX Cloud for contributing to this research result provided under grant No.\ NAIRR240190.
This work was supported by Los Alamos National Laboratory under the grant number AWD026741 at the University of Michigan.
This research was supported in part through computational resources and services provided by Advanced Research Computing at the University of Michigan, Ann Arbor.
This work used the Delta system at National Center for Supercomputing Applications through allocation CTS180061 from the Advanced Cyberinfrastructure Coordination Ecosystem: Services \& Support (ACCESS) program, which is supported by U.S.\ National Science Foundation grants \#2138259, \#2138286, \#2138307, \#2137603, and \#2138296.
Training on the Cerebras Wafer was supported by an allocation on the PSC Neocortex cluster and a Director's Discretionary Grant for access to the Argonne National Lab AI Testbed at the Argonne Leadership Computing Facility, a U.S.\ Department of Energy (DOE) Office of Science user facility at Argonne National Laboratory and is based on research supported by the U.S.\ DOE Office of Science-Advanced Scientific Computing Research Program, under Contract No.\ DE-AC02-06CH11357.
MWM also acknowledges DARPA, NSF, the DOE Competitive Portfolios grant, and the DOE SciGPT grant.
This work was supported in part by the Director, Office of Science, Office of Advanced Scientific Computing Research, of the U.S. Department of Energy under Contract No. DE-AC02-05CH11231.
Additionally, Wadell is grateful for the support of Meta's AR/AV Battery Research Fellowship.
Bhutani is supported by a Catalyst grant from the Michigan Institute for Computational Discovery and Engineering at the University of Michigan.

\paragraph*{Author Contributions}
Wadell, Bhutani and Viswanathan designed research objectives;
Wadell and Bhutani developed training code and recipes with guidance on performance from Brace, Emani, Vishwanath and Ramanathan.
Bhutani and Hegazy diagnosed and corrected model ``spikes'' using random matrix theory with guidance from Mahoney on metric interpretations and Duraisamy on optimization.
Wadell developed Bayesian scaling analysis with input from Bhutani, Ellis-Mohr and Viswanathan on best-practices.
Ellis-Mohr, Nayak and Varshney developed a connection between an information-theoretic framework and the observed super-linear scaling of data with model size.
Bhutani and Wadell identified features in MIST's embedding space with guidance from Viswanathan, Mahoney, Ramsundar and Duraisamy.
Bhutani and Wadell developed mixture property models using datasets curated using software developed by Gering; Kelly, Zhao and Viswanathan provided guidance on physics-aware state-space modeling.
Wadell and Bhutani curated excess mixture property dataset from the literature.
Chen, Lin and Simatos collected experimental excess density data to validate the excess properties model.
High-throughput screening workflow was developed by Wadell and Bhutani using models fine-tuned by Bhutani; target properties and bounds were selected by Azumah and discussed with Kelly, Zhao and Viswanathan;
Wadell and Varshney evaluated creativity metrics for the screening process and implications for molecular discovery.
Wadell and Azumah explored trends in chemical property predictions and MIST's robustness to variations in the molecular encoding.
Bhutani and Azumah explored chemical trends in odour profiles.
Qian, Gerkin, Amorelli and Wiltschko provided experimentally validated labels for discordant triplets.
Bhutani explored discordance in molecular structure and odour perception with guidance from  Qian, Gerkin, Amorelli, Wiltschko and Vishwanath.
Bhutani and Stier validated the hyperbolic geometry of predicted perceptual profiles.
All authors analyzed the data and wrote the paper.
\FloatBarrier

\clearpage
\printbibliography[
    title={References},
    segment=\therefsegment,
    heading=bibintoc,
]
\clearpage
\end{refsegment}

\setcounter{section}{0}
\setcounter{figure}{0}
\setcounter{equation}{0}
\setcounter{page}{1} 
\setcounter{table}{0}

\renewcommand{\tablename}{Table }
\renewcommand{\figurename}{Figure }
\renewcommand{\thefigure}{S\arabic{figure}}
\renewcommand{\theequation}{S\arabic{equation}}
\renewcommand{\thepage}{S\arabic{page}}
\renewcommand{\thetable}{S\arabic{table}}
\appendix
\begin{refsegment}
\title{Supplementary Information for Foundation Models for Discovery and Exploration in Chemical Space}
\maketitle
\clearpage

\renewcommand{\ptctitle}{Supplementary Information Table of Contents}
\part{} 
\parttoc 
\clearpage

\section{Pretraining}
\label{sec:si:pretraining}
In this section,
we discuss the dataset used for pretraining;
we discuss the objective and optimizer used to pretrain \ac{MIST} models;
we discuss training instabilities encountered when scaling \ac{MIST} models to larger datasets and how these were diagnosed; and
we quantify the advantage of the pretraining process on downstream model performance.

\subsection{Dataset}
\label{sec:si:realspace}
We use the Enamine REAL Space dataset to pretrain \ac{MIST} models \cite{EnaREALSpace2024}.
At time of writing, Enamine REAL Space is the largest database of commercially available compounds.
The dataset was constructed using ``forward synthetic analysis'': experimentally validated building blocks were converted into synthons annotated with reactivity features.
The synthons were then coupled virtually under well-established reaction templates, enumerating only those products permitted by the synthon annotations.
Reactivity-based exclusion rules were applied to ensure that the resulting library is synthetically accessible.
Enamine REAL Space was selected as the pretraining dataset since it was the largest database of molecular \ac{SMILES} at the time of training, it is easily accessible for academic use and molecules relevant to downstream tasks -- drug candidates, electrolytes, fragrances --- live in synthetically accessible regions of chemical space.

Given the restrictions on the molecules in REAL Space --- generated from a limited set of synthons using a fixed set of reaction pathways (172 ``well-validated parallel synthesis protocols'' and synthons from 181,288 reagents~\cite{EnaREALSpace2024}) --- the diversity of the pretraining dataset was limited.
As discussed in~\cref{sec:scaling,sec:si:data_quality}, we believe this lack of diversity limited the scaling efficiency of the model with respect to dataset size.
However, as we show in~\cref{sec:si:pretraining_adv}, the pretraining stage significantly improved performance on downstream tasks, including with structures very different from those in the pretraining dataset.

\subsection{Pretraining Objective}
\ac{MIST} models are pretrained using a \acf{MLM} pretraining objective which is widely used by encoder-only transformers in \ac{NLP} \cite{DCLTBERTPretrainingDeep2019,LOG+RoBERTaRobustlyOptimized2019} and by prior molecular foundation models \cite{CGRChemBERTaLargeScaleSelfSupervised2020,ASC+ChemBERTa2ChemicalFoundation2022,RBC+LargescaleChemicalLanguage2022}.
\ac{MLM} is a self-supervised learning task.
The training process involves masking, or hiding, a percentage of tokens in a sequence and then training the model to predict the original hidden tokens based on the surrounding context.
The model's objective is to predict the original tokens that were masked, based on the context provided by the non-masked tokens.
\ac{MLM} allows the model to learn from the context of tokens on both the left and right of the masked token.

Given a token sequence $X = x_{1:T}$ over vocabulary $V$, we randomly sample a masked set $S \subset \{1, \dots , T\}$.
For each $t \in S$, the token at the position $t$ is replaced with a masked token \tok{[MASK]}, yielding a corrupted sequence $\tilde{X}$.
The \ac{MIST} encoder predicts logits $\vec{z}_t \in \mathbb{R}^{|V|}$ which are used to calculate probabilities $p_{\theta}(\cdot | \tilde{X}) = \textrm{softmax}(\vec{z}_t)$.
The loss function is the cross-entropy computed only on masked positions:
\[
\mathcal{L}_{\mathrm{MLM}}
= - \sum_{t \in S} \log p_\theta\!\left(x_t \mid \tilde{X}\right) .
\]

\subsection{Large Batch Optimization} \label{sec:lamb_optimizer}

\begin{figure}[ht!]
    \centering
    \includegraphics{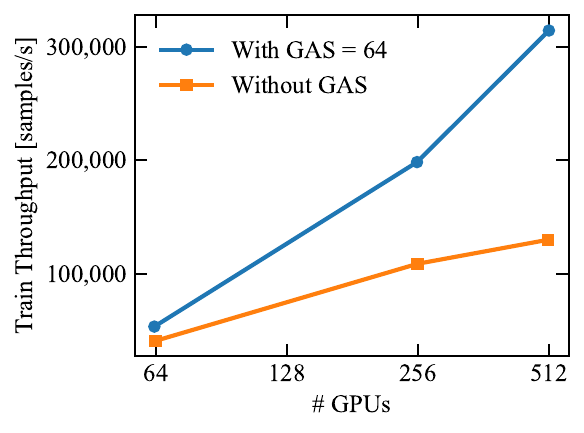}
    \caption{
        \label{fig:gas_scaling}
        Enabling \ac{GAS} was key to efficiently scaling MIST to 128 Polaris nodes or 512 GPUs, by reducing frequency of collective gradient accumulations.
        Without \acs{GAS}, scaling beyond 16 nodes (64 GPUs) incurred a significant performance penalty.
        Our production run used 40 GPUs due to the inefficiency of large batch optimization as discussed below.
    }
\end{figure}

\ac{MIST} models were trained using \acf{DDP}.
Hence, the effective batch size (observations per optimizer step) scales linearly with the number of devices (e.g., GPUs) used for training:
\begin{align*}
\textrm{Effective Batch Size} &= \textrm{Number of Nodes} \\ &\times \textrm{4 GPUs per Node} \\ &\times \textrm{GAS} \times \textrm{512 samples/microbatch} .
\end{align*}
where \acs{GAS} is the number of \acf{GAS}.
The microbatch size was selected to maintain high GPU utilization without triggering out-of-memory errors.

In order to achieve a feasible wall-time, we need to parallelize over a large number of GPUs during pretraining.
We found gradient accumulation to be key to efficiently scaling beyond a limited number of GPUs (\cref{fig:gas_scaling}).
Gradient accumulation involves accumulating gradients over several batches locally on a GPU and only stepping the optimizer after multiple batches.
In a distributed training setup, this helps significantly accelerate training, by reducing the frequency of collective communication operations needed to compute global model gradients.
For our training workflow, we see peak training throughput (i.e., samples per second) when \ac{GAS} is set to eight.
However, increasing \ac{GAS} also increases the effective batch size.

Naively increasing the effective batch size, by increasing the number of GPUs or \ac{GAS}, to accelerate training can result in poor data efficiency and thus negate any gains in computational efficiency \cite{YLR+LargeBatchOptimization2020}.
We observed this when pretraining MIST with the Adam\cite{Adam} and AdamW\cite{LHDecoupledWeightDecay2019} optimizers; the model training loss plateaued for effective batch sizes larger than 16,000 samples.
Hence, we switched to the \ac{LAMB} optimizer\cite{YLR+LargeBatchOptimization2020}.
The \ac{LAMB} optimizer enables better convergence with large batch sizes.
\subsection{Training Instabilities}
\label{sec:si:spikes}
\begin{figure}
    \centering
    \includegraphics[width=0.75\linewidth]{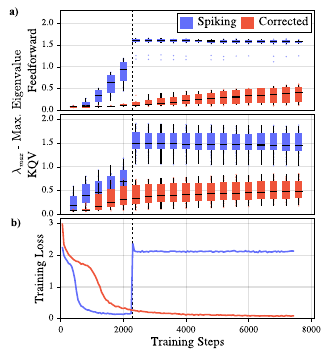}
    \labelphantom{fig:spiking_lambdas}
    \labelphantom{fig:spiking_loss}
    \caption{Instabilities in training loss were diagnosed by analyzing the model's weight matrices.
    (\subref*{fig:spiking_lambdas})~Comparing the change in the distributions of $\lambda_{max}$ (see \cref{sec:si:metrics}) for the key, query, value weights and the feedforward network weights during training.
    The variance in $\lambda_{max}$ for the feedforward layer weights collapses immediately after the spike while the key, query, value $\lambda_{max}$ maintains similar variance to before the spike.
    The magnitude of $\lambda_{max}$ remains similar in both cases.
    (\subref*{fig:spiking_loss})~Comparing loss curves for a run where a spike occurred and a run where the spiking was corrected; spikes were prevented in the latter case by switching to a pre-layernorm architecture and setting the optimizer $\beta_2$ hyperparameter to a low value.
    }
\end{figure}

When training MIST on a large dataset (larger batch sizes for a larger number of training steps), we observed ``spikes'' or ``irreversible steps'' in the training loss as seen in \cref{fig:spiking_loss}.
Similar spiking behaviour has been reported during \ac{LLM} pretraining\cite{takaseSpikeNoMore2024}.
As observed in the ``spiking'' curve (\cref{fig:spiking_loss}), the model's training loss does not recover spontaneously with continued training.
If a spike is not detected and corrected the training run will continue but the model will fail to converge.
Given the large computational cost associated with pretraining, understanding the causes and minimizing the risk of this failure mode is important.

Consistent with prior works\cite{RBC+LargescaleChemicalLanguage2022,takaseSpikeNoMore2024} we found that training could be stabilized by decreasing $\beta_2$, the coefficient used to compute a running average of the square of the gradient in the \ac{LAMB} optimizer\cite{taniguchiADOPTModifiedAdam2024a}.
We hypothesized that the spikes are caused by degradation of the intermediate layer matrices, which is consistent with prior work \cite{takaseSpikeNoMore2024}.
In addition to tuning the $\beta_2$ values, we adopted the \texttt{RoBERTaPreLayerNorm} architecture to help mitigate spikes.
Pre-layer normalization (where the states are normalized before they are passed to the feedforward layers) has been shown to improve training stability\cite{XYH+LayerNormalizationTransformer2020}.
Our ablation studies indicate these two changes were key to stabilizing MIST's pretraining.

Alternative hypothesized causes for the spikes are discussed below and were evaluated by sweeping over model architectures and training hyperparameters:

\begin{itemize}
    \item Precision: We use \texttt{bfloat16} mixed precision for training. Switching to full single precision \texttt{fp32} had no impact on the spikes. 
    \item Hardware: We train across several different supercomputers with different hardware (A100/ V100/ H100 GPUs), and the behaviour was repeatable on all machines.
    \item Learning Rate: Spikes can be prevented by significantly reducing the learning rate, but this slows convergence and leads to a higher final loss.
    \item Weight Decay: We use weight decay regularization. Increasing the weight decay coefficient delays the spikes, but it does not prevent them.
    \item Depth: Spikes did not occur in shallow models (with less than 10 layers), and they appear to be more likely with increasing model depth.
    \item Weight Initialization: Some prior models, e.g., NVIDIA’s Megatron-LM\cite{Megatron-LM}, use a low variance weight initialization to prevent the loss from exploding but this did not help in our case.
    \item Layer Normalization $\epsilon$: This refers to the offset added to the denominator during normalization to prevent numerical instabilities when dividing by a small number. It did not seem to impact spikes.
    \item Tokenizers: Spiking behaviour is reproducible across different tokenizers, and hence it is not an artifact of the data tokenization process.
    \item Batch size: Spikes occurred earlier for smaller batches.
    \item Bad Data: We could not trace spiking behaviour to particular training datum. Spikes occurred for multiple seeds for dataset shuffling.
\end{itemize}

\subsection{Advantage of Pretraining}
\label{sec:si:pretraining_adv}
\begin{figure}[ht!]
    \centering
    \labelphantom{fig:transfer_unfrozen}
    \labelphantom{fig:transfer_frozen}
    \includegraphics{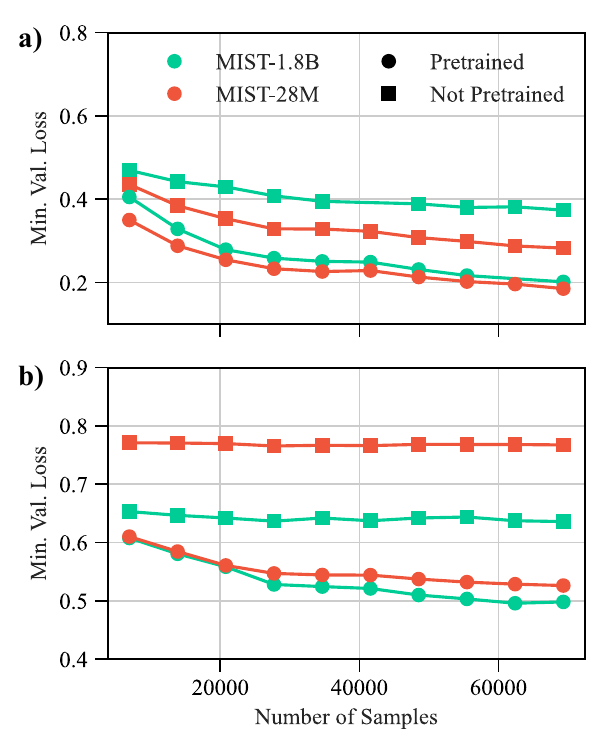}
    \caption{
        \label{fig:pretraining_adv}
        \textbf{Quantifying the Advantage of Pretraining.}
        (\subref*{fig:transfer_unfrozen}) shows traces for models where both the encoder and task network weights were updated during fine-tuning; (\subref*{fig:transfer_frozen}) shows traces for models where only the task network weights were updated during fine-tuning. In both cases, we observe that pretraining significantly improves model prediction accuracy for a complex downstream regression task.
        When only the task network weights are updated (\subref*{fig:transfer_frozen}) we see that performance of models which were not pretrained does not change with increasing dataset size; this is not unexpected since they are severely under-parameterized.
    }
\end{figure}
In order to quantify the impact of our pretraining process on downstream performance, we fine-tuned both pretrained and randomly initialized models with identical architectures.
The fine-tuning datasets were sampled from the tmQM dataset \cite{BSTmQMDatasetQuantum2020} and ranged in size from $900$ to $70,000$ samples.
We tested the advantage of pretraining in two settings: (\cref{fig:transfer_unfrozen}) both encoder and task network weights are updated during fine-tuning and  (\cref{fig:transfer_frozen}) only the weights of the task network are updated during fine-tuning (i.e the pretrained encoder weights remain frozen).
In both cases, we observe (\cref{fig:pretraining_adv}) that pretraining significantly improves model prediction accuracy for a complex multi-target regression task.
This is notable because the chemical structures in this dataset, organometallic complexes, are significantly different from those in the pretraining dataset and contain many tokens (for example transition metal elements and chirality tags) not seen during pretraining.

\section{Model Design Choices}
\label{sec:si:ablation_studies}

\begin{figure}
    \centering
    \includegraphics{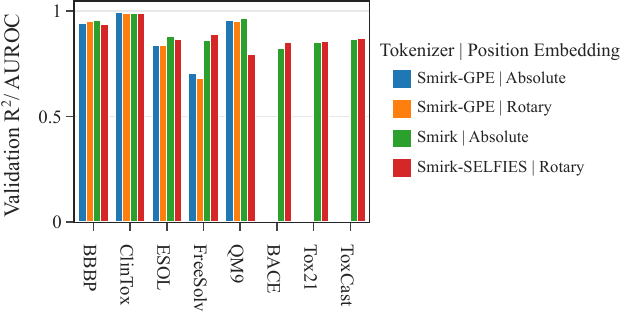}
    \caption{%
    Fine-tuning results on three regression and five classification tasks from MoleculeNet for model trained on different molecular representations (the Smirk-SELFIES tokenizer model uses SELFIES, all others use SMILES), using two different tokenization algorithms (Smirk and Smirk-GPE) and position embeddings (absolute and RoPE \cite{SLP+RoFormerEnhancedTransformer2023}).
    Validation $R^2$ and \ac{AUROC} are reported for the regression and classification tasks, respectively.
    In the case of multitask regression, the average $R^2$ is reported.
    Higher is better, with 1 being a perfect score for both metrics.
    }
\label{fig:ablation_studies}
\end{figure}

Prior to pretraining \ac{MIST} models at scale, extensive ablation studies were conducted to identify the best-performing molecular data representation (\acs{SMILES}\cite{WeiSMILESChemicalLanguage1988} or SELFIES\cite{KHN+SelfreferencingEmbeddedStrings2020}), tokenization strategy (Smirk\cite{WBVTokenizationMolecularFoundation2026} or SmirkGPE\cite{WBVTokenizationMolecularFoundation2026}) and positional encodings (Absolute\cite{VSP+AttentionAllYou2017} or Rotary\cite{SLP+RoFormerEnhancedTransformer2023}) based on MoleculeNet\cite{WRF+MoleculeNetBenchmarkMolecular2018} benchmark task performance.
The baseline architecture used for this study was a 163M parameter RoBERTa\cite{LOG+RoBERTaRobustlyOptimized2019} model with 12 attention heads, 18 hidden layers, a feedforward size of 3,072 and a sequence length of 512.
The rotary position embedding models use RoFormer \cite{SLP+RoFormerEnhancedTransformer2023}, as implemented by HuggingFace's \texttt{transformers} package\cite{HuggingfaceTransformersTransformers2024}.
\Cref{fig:ablation_studies} summarizes results from the full set of ablation studies.
Overall, we do not see a significant sensitivity to these choices across tasks.
All design choices considered are explained below.

The final \ac{MIST} models use the \ac{SMILES} molecular encoding for its broader molecular coverage, absolute position embeddings for implementation simplicity, and the Smirk tokenizer\cite{WBVTokenizationMolecularFoundation2026} for its robust \ac{SMILES} coverage and fine-grained encoding of atom-level information.

\paragraph{Molecular Representation.}
We considered two text based molecular representations -- \ac{SMILES} and SELFIES.
\acf{SMILES} strings are the most widely adopted molecular representation for training transformers on chemical data \cite{CGRChemBERTaLargeScaleSelfSupervised2020}.
Besides being the de-facto standard for data storage in cheminformatics, \ac{SMILES} are a particularly convenient choice for training molecular \acp{FM}, since they require minimal pre-processing and resemble data used in \ac{NLP}.
Hence, existing tokenization algorithms and \ac{LLM} architectures can be employed with little modification.
Some researchers have experimented with using SELFIES (Self-Referencing Embedded Strings) rather than \ac{SMILES} to train \ac{LLM} architectures on molecular data \cite{KAB+SELFIESFutureMolecular2022,ASC+ChemBERTa2ChemicalFoundation2022}.
SELFIES retain the convenience of \ac{SMILES} since they are text-like representations of molecular structure.
However, they can be more robust since physical constraints are encoded in the deriving formal automaton.
Despite the theoretical benefits of using SELFIES rather than \ac{SMILES}, empirical studies to compare their impact on molecular \ac{FM} prediction accuracy remain inconclusive\cite{WBVTokenizationMolecularFoundation2026}.
Our results (\cref{fig:ablation_studies}) agree with this conclusion.

\paragraph{Tokenizer.}
We tested two tokenization strategies described in our prior work \cite{WBVTokenizationMolecularFoundation2026}, which systematically evaluated thirty-four tokenizers for their use in molecular \acp{FM}.
Existing chemistry tokenization algorithms, such as Atom-wise tokenization\cite{SGL+FoundTranslationPredicting2018}, have significant gaps in their coverage of \ac{SMILES}, leading to information loss in the tokenization step\cite{WBVTokenizationMolecularFoundation2026}.
Smirk is a tokenization scheme which fully decomposes bracketed atoms into their constituent parts allowing for lossless tokenization of any OpenSMILES\cite{CraOpenSMILES2016} encoded molecule.
SmirkGPE (Smirk - Glyph Pair Encoding) is a variant of Smirk proposed to lower the fertility (tokens per sequence) of input sequences\cite{WBVTokenizationMolecularFoundation2026}.
While prior work suggests that lower fertility improves model performance \cite{GCE+UnpackingTokenizationEvaluating2024,RPV+HowGoodYour2021,GMFindingOptimalVocabulary2020},
we do not see a significant difference between models trained using Smirk and SmirkGPE (\cref{fig:ablation_studies}).

\paragraph{Positional Encodings.}
The architecture of attention layers is inherently position-invariant, so positional encodings are introduced to provide the model with information about sequence order\cite{VSP+AttentionAllYou2017}.
Many transformer architectures in language and other domains\cite{LOG+RoBERTaRobustlyOptimized2019,DCLTBERTPretrainingDeep2019,CGRChemBERTaLargeScaleSelfSupervised2020} use the absolute sinusoidal embedding proposed by Vaswani et al.\cite{VSP+AttentionAllYou2017}.
Recently, rotary (relative) position encodings have been proposed\cite{RBC+LargescaleChemicalLanguage2022}.
The training stability and convergence time of the MolFormer model improved when using rotary rather than absolute position embeddings \cite{RBC+LargescaleChemicalLanguage2022}.
Despite the theoretical benefits of using rotary attention, we do not observe a significant improvement in downstream performance.

\section{Fine-tuning}
\label{sec:si:fine-tuning}

\subsection{Fine-tuning Datasets and Tasks}
\label{sec:si:datasets}
In this section, we discuss various downstream tasks and datasets on which \ac{MIST} variants were fine-tuned.

\subsubsection{MoleculeNet}
\label{sec:si:molecule_net}
MoleculeNet\cite{WRF+MoleculeNetBenchmarkMolecular2018} is a benchmark suite designed for the systematic evaluation of machine learning models in molecular sciences.
It provides standardized datasets and evaluation metrics across a variety of chemical and biochemical tasks.
Each dataset represents a specific domain, such as quantum mechanics or biophysics, facilitating the development and comparison of cheminformatics models.
We benchmark \ac{MIST} on 11 MoleculeNet datasets, described below using the dataset splits specified in Ref.\cite{WRF+MoleculeNetBenchmarkMolecular2018}:

\begin{description}
  \item[QM8] contains 21,786 small molecules for which time-dependent density functional theory (TDDFT) was used to compute eight excitation energies and four oscillator strengths, resulting in 12 scalar targets per molecule.
  \item[QM9] comprises 133,885 drug-like organic molecules with 13 computed quantum-chemical properties (e.g., dipole moment, polarizability, HOMO/LUMO energies, vibrational frequencies) at the B3LYP/6-31G(2df,p) level.
  \item[ESOL] provides experimental aqueous solubility (log S) measurements for 1,128 small molecules. It is a single-target regression task.
  \item[FreeSolv] contains experimental and calculated hydration free energies for 642 neutral molecules in water. It has one regression target per molecule and benchmarks models’ ability to predict molecule–solvent interaction energies.
  \item[Lipophilicity] provides measured octanol–water distribution coefficients (log D) for 4,200 small molecules.
  \item[BBBP] (Blood–Brain Barrier Penetration) labels compounds as penetrating or non-penetrating (with respect to the blood–brain barrier). It is a single-task binary classification dataset focused on central nervous system delivery, with 2,039 data points.
  \item[Tox21] features 7,831 compounds tested against 12 biological targets related to nuclear receptor signaling and stress response pathways. This is a multi-label binary-classification dataset.
  \item[SIDER] aggregates adverse drug reactions for 1,427 marketed drugs across 27 MedDRA side-effect terms. This is a multi-label binary-classification dataset. 
  \item[ClinTox] provides clinical toxicity data for 1,478 compounds, distinguishing FDA-approved drugs from those withdrawn for toxicity. It comprises two binary-classification tasks (approval status and withdrawal risk).
  \item[HIV] contains 41,127 compounds tested for inhibition of HIV-1 replication in T-cell lines. It is a single-task binary classification. Dataset was split 80/10/10 using a scaffold split~\cite{WRF+MoleculeNetBenchmarkMolecular2018}.
  \item[BACE] includes 1,513 small molecules assayed for inhibition of \(\beta\)-secretase 1 (BACE-1), a target relevant to Alzheimer’s disease. It is a single-task binary classification dataset.
\end{description}

\subsubsection{tmQM}
\label{sec:si:tmQM_dataset}
tmQM\cite{BSTmQMDatasetQuantum2020} is a quantum mechanics dataset of 108k transition-metal organometallic compounds curated from the Cambridge Structural Database\cite{GBLWCambridgeStructuralDatabase2016};
we used the 2024 expanded tmQM release containing 108k complexes; the original publication reported 86k.
tmQM provides quantum mechanical properties (\ac{HOMO}, \ac{LUMO}, \ac{HOMO}/\ac{LUMO} Gap, dipole moment and natural charge) computed using \ac{DFT} at the TPSSh-D3BJ/def2-SVP level using geometries optimized at the GFN2-xTB level\cite{BSTmQMDatasetQuantum2020}.
A complete description of the curation filter used by Balcells and Skjelstad can be found in the dataset's paper, but notably tmQM was restricted to only include structures with a single transition metal atom.
The original dataset provides 3D-coordinates as XYZ files.
However, here we use the SMILES encoding with enhanced stereochemistry, as generated in our prior work using the 2024 release of the tmQM dataset\cite{WBVTokenizationMolecularFoundation2026}.
The dataset is of particular interest due to its expanded elemental and stereochemical diversity relative to other datasets \cite{WBVTokenizationMolecularFoundation2026}.
The dataset with SMILES encodings can be retrieved from \url{https://doi.org/10.5281/zenodo.13761263}, while the original dataset can be found at \url{https://github.com/uiocompcat/tmQM}.

\subsubsection{Kamlet-Taft Solvatochromic Parameters}
\label{sec:si:kt}
\acf{KT} parameters are used to characterize solvent properties, particularly their ability to participate in hydrogen bonding and their polarizability.
These parameters are used in chemistry, pharmacology, and materials science to help to understand and predict solvent effects in processes such as chemical reactions and solution equilibria.
For example, Krishnamurthy et al.\cite{Krishnamurthy2021} used the \ac{KT} parameters as descriptors for predicting nitrogen reduction activity in lithium mediated ammonia synthesis.
The three \ac{KT} parameters are $\alpha$ (hydrogen bond acidity, the solvent's ability to donate a hydrogen bond), $\beta$ (hydrogen bond basicity, the solvent's ability to accept a hydrogen bond) and $\pi^{*}$ (polarizability, the solvent's overall polarity).
The fine-tuning dataset consisted of 182 \ac{KT} values measured using \ac{NMR} from Ref.~\cite{Krishnamurthy2021}.

\subsubsection{BF$_3$ Affinity}
\label{sec:si:donor_number}
The Gutmann \acf{DN} was first proposed as a measure of a molecule's Lewis basicity; and it was later applied to electrolyte design for analyzing lithium-ion salt solubility, and to solvation environment to guide solvent choice\cite{LGLewisBasicityAffinity2009}.
BF$_3$ affinity is often used as a donor-strength proxy for the Gutmann \ac{DN}.
Through choosing solvents and salt anions with desired BF$_3$ affinity, past researchers have successfully designed Li-metal compatible electrolytes with anion-rich \ac{SEI} compositions\cite{ZXLUnderstandingApplyingDonor2024}.
BF$_3$ affinity has also been proven to be a rational design metric for \ac{LHCE}, where solvents ideally possess high BF$_3$ affinity to ensure salt-dissociation, and diluents should have low BF$_3$ affinity to not coordinate with lithium-ions in the solvation shell.
Chen et al.\cite{CZF+DesignLocalizedHighConcentration2023} used BF$_3$ affinity to successfully identify a novel diluent choice that reduces viscosity and lowers the cost.
Xu et al.\cite{XZP+ElectrolyteDesignLiion2023} used BF$_3$ affinity to help guide solvent choice which balances weak lithium-ion/solvent interactions with sufficient lithium salt dissociation.
The \ac{MIST} variant was fine-tuned on a dataset of 344 $BF_3$ affinity measurements of Lewis bases in dichloromethane at 298 K and 1 atm\cite{LGLewisBasicityAffinity2009}.

\subsubsection{Characteristic Temperatures}
\label{sec:si:datasets_ct}
The boiling point, melting point and flash point of a substance determine its thermal operating window and flammability; and the accurate prediction of these properties is desired across a very broad range of applications. 
The \ac{MIST} variant was fine-tuned on a dataset of 3,969 boiling-point, 5,734 melting-point and 10,090 flash-point data entries curated from the literature by Shang Zhu et al. for an in-preparation manuscript.
Melting points and boiling points were curated from the PubChem database\cite{KCC+PubChem2025Update2025}, while flash points were obtained from the literature\cite{gaudinMixtureDescriptorsDevelopment2015}.
The dataset was filtered to chemicals with fewer than 20 heavy atoms (not including hydrogen) and zero radical electrons.

\subsubsection{Dimroth--Reichardt Solvent Polarity Parameters}
\label{sec:si:dimroth_eth}
The Dimroth--Reichardt solvent polarity parameters are used to characterize solvent ionizing power and empirical polarity using solvatochromic measurements of Reichardt's betaine dye.
Specifically, the wavelength of the dye's absorption maximum is converted into the empirical polarity parameter $E_T(30)$, and a normalized form, $E_T^{N}$, is obtained by scaling the values relative to tetramethylsilane and water.
This scale is used widely to compare solvents across a broad polarity range and is particularly sensitive to hydrogen-bond donation, making it complementary to other solvent descriptor sets such as the Kamlet--Taft parameters.
The dataset used here was retrieved from \url{https://www.stenutz.eu/chem/solv20.php?sort=3}~\cite{DimrothReichardt} and consists of 379 listed solvents and solvent-like media with tabulated $E_T^{N}$ values compiled in the Stenutz Dimroth--Reichardt solvent table.

\subsubsection{Dissociation Constants (pKa)}
\label{sec:si:pKa_dmso}
The $pK_a$ is a measure of the acidity of a molecule and is defined as the negative logarithm of the acid dissociation constant ($K_a$). 
It is an important property for understanding chemical reactivity, solubility, and biological activity.
A dataset of $pK_a$ values in dipolar non-hydrogen-bond-donor solvents curated by~Leito et al.\cite{LeitoKaljurandPiirsaluTshepelevitshZhengRos} was used to fine-tune \ac{MIST} on $pK_a$ in \ac{DMSO}. 
The dataset was filtered to include only measurements which included a value for the measurement in \ac{DMSO}.
Duplicates were averaged over based on InChI keys and two measurements for mixtures were excluded, resulting in a final training dataset of 3254 pKa values for small molecules. 

\subsubsection{PubChemQC}
\label{sec:si:pubchem_qc}
PubChemQC\cite{NMPubChemQCB3LYP631G2023,NSPubChemQCProjectLargeScale2017} (\href{https://nakatamaho.riken.jp/pubchemqc.riken.jp/b3lyp_2017.html}{JCIM2017 release}) is a dataset of electronic properties for 3 million molecules extracted from PubChem\cite{KCC+PubChem2025Update2025}.
The dataset provides orbital energies (\ac{HOMO}, \ac{LUMO} and \ac{HOMO}/\ac{LUMO} Gap), total energies, dipole moments, atom-level partial charges (Mulliken and L\"owdin) and relaxed coordinates computed using \ac{DFT} at the B3LYP/6-31G* level.
While not used here, time-dependent \ac{DFT} was additionally used to compute 10 low-lying excited states at the B3LYP/6-31+G* level.
We elected to not use later releases of the dataset due to the lack of a programmatic method to download the raw files directly and our lack of a suitable machine to host a 20TB PostgreSQL database bundled as a Docker image.

\subsubsection{Ionic Conductivity}
\label{sec:si:ic_dataset}
Ionic conductivity is commonly used to characterize electrolytes, as it describes the ability of the electrolyte to transfer ions between the anode and the cathode~\cite{eftekhari2018sodium,XuNonaqueousLiquidElectrolytes2004}.
High ionic conductivity is associated with better battery performance and is often sought for fast-charging electrolytes~\cite{lei2023fast}.
The \ac{MIST} ionic conductivity models were fine-tuned on a dataset of 24,822 mixtures of single-salt ternary-solvent electrolyte solutions generated by Zhu et al.\cite{ZRA+DifferentiableModelingOptimization2024} using the \ac{AEM}~\cite{gering2017prediction}.
The \ac{AEM} generates high-fidelity outputs covering a range of properties such as ion solvation, ion association (ion pairs, triple ions etc.) and viscosity; it enables the generation of large synthetic datasets which capture a wide range of physics relevant to electrolyte design.

\subsubsection{Excess Properties}
\label{sec:si:excess_dataset}

\begin{figure}
  \centering
  \labelphantom{fig:mixture_property_distribution}
  \labelphantom{fig:mixture_functional_coverage}
  \labelphantom{fig:mixture_property_coverage}
  \includegraphics[width=\textwidth]{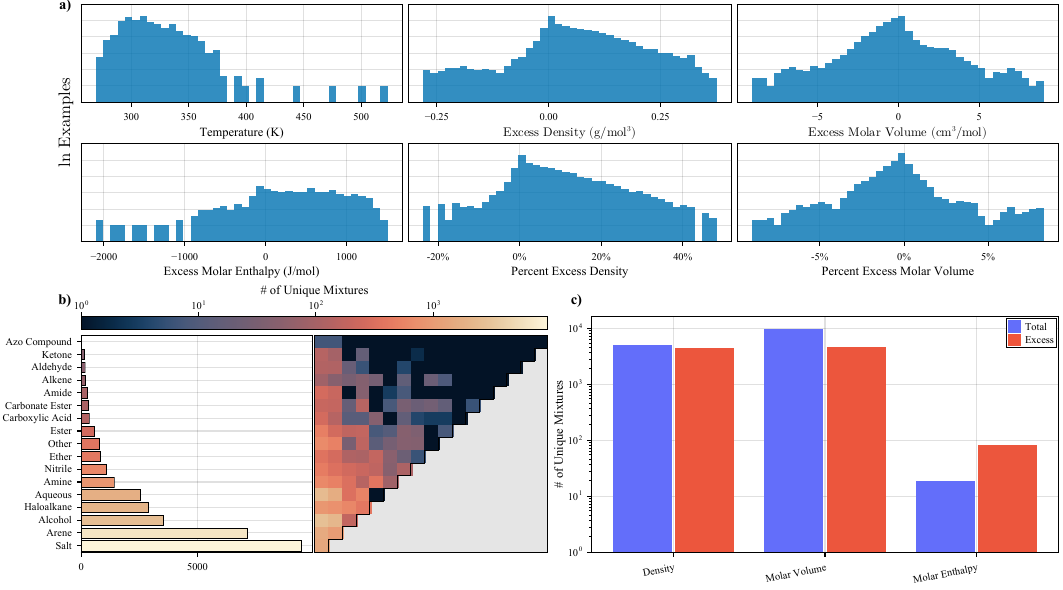}
  \caption{
    \label{fig:mixture_dataset}
    \textbf{We curated a dataset of densities, molar enthalpies, and molar volumes for 1,519 unique binary mixtures.}
    (\subref*{fig:mixture_property_distribution})~Distribution of Mixture Excess Properties (Density, Molar Volume, Molar Enthalpy) and temperatures within our excess properties dataset.
    Due to the limited availability of mixture molar enthalpy data, only the excess molar enthalpy is shown.
    (\subref*{fig:mixture_functional_coverage})~Coverage of functional groups at the component or pair-wise level within the dataset.
    Molecules could be labeled with multiple functional groups; as such, the total example count can exceed the size of the dataset.
    (\subref*{fig:mixture_property_coverage})~Our dataset was collated from the literature and as such contains a high degree of sparsity.
    For example, only a fraction of papers report mixture molar enthalpy, instead opting to report only the excess molar enthalpy.
    When possible, we imputed the missing values from the reported data.
  }
\end{figure}

We curated a dataset of mixture excess properties (density, molar enthalpy, and molar volume) from the literature\cite{DMK+ILThermoFreeAccessWeb2007,KADVExcessDensityDescriptor2025,FCCExcessMolarEnthalpies1999,COFExcessMolarEnthalpies1997,ZWH+DensitySpeedSound2025,LZZ+DensityViscosityExcess2010,FCExcessMolarEnthalpies1995,CFExcessMolarEnthalpies1996,RRK+ExcessEnthalpiesBinary2016,CCDensitiesViscositiesBinary2005,AAMPhysicalPropertiesDensity2008,PSExcessMolarVolumes1998,MMMSDensityExcessProperties2008,Zhang_2018,PSExcessEnthalpyExcess2013,Francesconi_1996,HSExcessVolumesExcess1996,Ottani_2000,SLGSDensityViscosityExcess2019,VMCDensityViscosityExcess2023,RZZ+DensityExcessMolar2011,ZCTExcessPropertiesTetrabutylammonium2024,KTDensityViscosityBinary1998,OGP+ExcessMolarVolumes2004,VPC+ExcessEnthalpyDensity2004,JSDensityViscosityExcess2004,CYC+DensityViscositySpeed2015,TMDensityRefractiveIndex2012,MGP+ExcessMolarEnthalpies2011,Lugo_2002,MFA+DensityViscositiesExcess2014,DKDensitySpeedSound2016,VPM+ExcessEnthalpyDensity2006,Letcher_1999,Zhao_2000,PKExcessMolarVolumes1999,CGMSDensityViscosityVaporLiquid2010,RKLExcessPropertiesBinary1999,CIPropertiesPure1Butyl23dimethylimidazolium2012,Fan_2009,YXMExcessMolarVolumes2004,Comelli_1998,Comelli_2006,Muhuri_1996,TMSDensitySurfaceTension2006,Roy_2006,Francesconi_1994,YSB+DensityExcessMolar1997,ACM+DensityViscosityExcess2006,WWC+ExcessPropertiesIntermolecular2024,ARExcessMolarVolumes2018,LWZ+ExcessMolarVolumes2001,PKExcessMolarVolumes1999,Zaitseva_2016,Shafaati_2017}.
As not all data sources reported all measurements of interest, our data set is sparse (\cref{fig:mixture_property_coverage}).
Where possible, we calculated the relevant properties from the reported values, e.g., computing excess molar volume from the provided density measurements, or vice versa.
The resulting dataset is a superset of the dataset developed previously by~Zhu et al.\cite{ZRA+DifferentiableModelingOptimization2024} for excess property prediction.
The expanded dataset contains 888,045 sparse observations spanning 715 molecules and 1,519 unique binary mixtures.
We have included the dataset within our data release.

\paragraph{Excess Value Imputation.}
When possible, we computed the excess mixture properties \(P_E\) directly from the reported mixture \(P_{mix}\) and pure compound properties \(P_i\) using 
\begin{equation}
\label{eq:si:excess_property}
  P_{E} = P_{mix} - \sum_i x_i P_i .
\end{equation}
If the excess property was provided, but not the mixture property, we would calculate \(P_{mix}\) from the reported pure compound properties \(P_i\).
We only used pure compound properties as reported within the paper; we did not impute pure compound properties from elsewhere in the literature.

To reduce our data entry burden, we did not record the reported molar volume when mixture densities were reported.
Instead, molar volumes \(V_m = M / \rho\) were computed from the reported densities \(\rho\) with molecular weights \(M\) computed using RDKit\cite{GreRDKitOpensourceCheminformatics2024}.

\subsubsection{Odor Profiles}
\label{sec:si:olf_dataset}
The fine-tuning dataset consisted of 4,983 molecules annotated with 135 odour descriptors, curated by combining the GoodScents and Leffingwell PMP 2001 odour databases as distributed in the OpenPOM \texttt{curated\_GS\_LF\_merged\_4983} dataset~\cite{OpenPOM}.
The OpenPOM is a replication of the Principal Odor Map by~Lee et al.\cite{leePrincipalOdorMap2023}.

\subsubsection{Mixture Odor Profiles}
\label{sec:si:olf_mix_dataset}
The fine-tuning dataset was a curated version of the dataset originally released by the \ac{DREAM} Olfactory Mixtures Prediction Challenge which consisted of four datasets collected from three separate psychophysical studies.
The dataset consists of pairs of mixtures with varying number of components with experimentally measured distances ranging from 0 (indistinguishable) to 1 (maximally distinct).
The curated version was compiled for the OlfBoost~\cite{CWYKOlfboostDataProcessed} submission to the second \ac{DREAM} challenge and retrieved from \url{https://github.com/Satarifard/CWYK-Olfboost.git}.
During model development, the OlfBoost team identified parsing errors in the Snitz~1 and Snitz~2 datasets, as well as mislabeling issues and missing compound identifiers in the Bushdid dataset, motivating a re-curation of the benchmark from the original publications.
They further augmented the training resource by incorporating three additional experiments which were not included in the initial challenge release.
The resulting curated OlfBoost dataset comprised 850 unique mixtures and 780 experimentally measured similarity annotations.

\subsubsection{Isotopes}
We fine-tuned \ac{MIST} on the Radioactive Isotope Half Lives dataset from the Wolfram Data Repository\cite{marshallRadioactiveIsotope}.
The data set contains the half lives for 1,175 radioactive isotopes in seconds.
Metastable isotopes were removed from the dataset, as they cannot be represented in \ac{SMILES}, resulting in 925 samples.

\subsection{Data Augmentation}
\label{sec:si:data_augmentation}
\begin{figure}
    \centering
    \labelphantom{fig:alkene_baseline}
    \labelphantom{fig:alkene_augmented}
    \labelphantom{fig:smi_perm_cv}
    \labelphantom{fig:smi_perm_rel}
    \includegraphics[width=\textwidth]{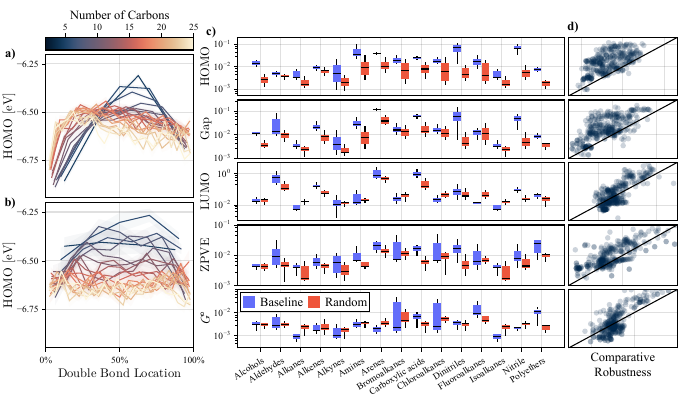}
    \caption{
        \label{fig:smi_permutation}
        \ac{SMILES} encodings are syntactically diverse, lacking a well-defined encoding for a molecule.
        Pragmatically, equivalent encodings should yield identical predictions --- an invariance not explicitly enforced by the \ac{MIST} architecture or training.
        (\subref*{fig:alkene_baseline})~Fine-tuning on an unaugmented dataset does not recover this invariance, as evidenced by the asymmetry in the predicted \ac{HOMO} energies, when the terminal double bond in the \ac{SMILES} string for an alkene moves is written on the left (\smiles{C=CCC}) rather than the right (\smiles{CCC=C}).
        (\subref*{fig:alkene_augmented})~Training on randomized \ac{SMILES} improves symmetry, indicating enhanced robustness.
        (\subref*{fig:smi_perm_cv})~We quantified robustness as the coefficient of variation \(\sigma/\mu\) across permutations.
        (\subref*{fig:smi_perm_rel})~Randomized augmentation either improves or has negligible impact on prediction consistency, while restoring symmetry in key cases like alkenes.
    }
\end{figure}
Our architecture and training objective do not explicitly enforce molecular symmetries.
However, we are able to learn these using \ac{SMILES} encoding randomization as a form of data augmentation during fine-tuning.
Notably, the \ac{SMILES} representation for a given molecule is not unique;
despite efforts to formalize \ac{SMILES}, no broad consensus exists\cite{WBVTokenizationMolecularFoundation2026}.
For chemical consistency, it would be desirable for \ac{MIST}’s embeddings to be invariant to these syntactic differences, mapping equivalent encodings to the same vector in the embedding space. 
However, this invariance is neither enforced by \ac{MIST}’s architecture nor explicitly encouraged during pretraining.

A concrete illustration of this phenomenon is provided in \cref{fig:smi_permutation}.
Here, we predicted the \ac{HOMO} energy level of alkenes while shifting the double bond from one end of the chain (\smiles{C=CCC}) to the other (\smiles{CCC=C}).
Ideally, the model should exhibit symmetry in its predictions (i.e., \(f(\smiles{C=CCC}) = f(\smiles{CCC=C})\)) as both encodings are chemically equivalent.
However, without augmentation (Baseline) the model produced significantly asymmetric predictions (\cref{fig:alkene_baseline}).
Augmentation by randomizing the \ac{SMILES} encoding seen by the model during fine-tuning restored much of this symmetry (\cref{fig:alkene_augmented}).
More generally, as shown in \cref{fig:smi_perm_cv}, augmenting with randomized encodings consistently improved robustness and, critically, did not harm performance.
We define robustness as the coefficient of variation \(\sigma/\mu\) of predictions over \ac{SMILES} permutations.
Since augmentation via randomization can be done online and is computationally inexpensive, it presents a low-cost strategy for improving invariance to syntactic variation in \ac{SMILES} encodings.

To assess \ac{MIST}'s robustness to equivalent \ac{SMILES} encodings,
we generated multiple \ac{SMILES} encodings for hydrocarbons of the form \texttt{R-X}, where \texttt{R} is an alkyl group and \texttt{X} is the functional group shown in \cref{fig:smi_perm_cv}.
We then measured the robustness of model predictions as the coefficient of variation -- the ratio of the standard deviation to the mean --- across predictions for different encodings of the same molecule.
As shown in \cref{fig:smi_perm_cv}, after fine-tuning on unaugmented \ac{SMILES} encodings the model was not invariant to equivalent \ac{SMILES} encodings.
However, augmenting the training data with randomly permuted encodings improved encoding robustness (\cref{fig:smi_perm_cv,fig:smi_perm_rel}).
We evaluated robustness to permutation using the coefficient of variation of \ac{MIST}'s prediction when evaluated on multiple \ac{SMILES} encodings for the same molecule.
We evaluated hydrocarbons with up to 25 carbon atoms for each functional group, generating up to 100 \ac{SMILES} encodings for each molecule.
Encodings were generated using RDKit's random \ac{SMILES} encoding feature\cite{LTK+RdkitRdkit2024_09_42024}.
The number of possible encodings varies due to \ac{SMILES} grammar and the stochastic nature of generating random \ac{SMILES}.
For example, alkanes (i.e. \smiles{CCCC}) have only a single unique encoding while alkenes (i.e. \smiles{C=CC}) have many more (e.g. \smiles{CC=C}, \smiles{C(=C)C}, \smiles{C(C)=C}) valid non-canonical encodings.

\begin{table}
\centering
\resizebox{\textwidth}{!}{%
\setlength{\tabcolsep}{4pt}
\renewcommand{\arraystretch}{1.1}
\begin{tabular}{@{}lllll@{}}
\toprule
Model & Figure & Base Encoder & Finetuning Dataset & Targets \\
\midrule
MIST-28M-Olfaction    & \cref{fig:olfaction_panel} & MIST-28M & Olfaction(\cref{sec:si:olf_dataset}) & 135 binary scent descriptors \cite{leePrincipalOdorMap2023} \\
MIST-28M-QM9           & \cref{fig:lio2_screening_pareto,fig:si:screening_creativity} & MIST-28M & QM9 (\cref{sec:si:molecule_net}) & See \cref{tab:qm9_benchmark} \\
MIST-28M-Melting       & \cref{fig:lio2_screening_pareto,fig:hc_trends_size} & MIST-28M & Characteristic Temperatures (\cref{sec:si:datasets_ct}) & Melting point [K] \\
MIST-28M-Boiling       & \cref{fig:lio2_screening_pareto,fig:hc_trends_size} & MIST-28M & Characteristic Temperatures (\cref{sec:si:datasets_ct}) & Boiling point [K] \\
MIST-28M-pKa           & \cref{fig:lio2_screening_pareto} & MIST-28M & $pK_a$ in \ac{DMSO} (\cref{sec:si:pKa_dmso}) & $pK_a$ in \ac{DMSO} \\
MIST-28M-ETN           & \cref{fig:lio2_screening_pareto} & MIST-28M & $E_T^N$ (\cref{sec:si:dimroth_eth}) & Normalized Dimroth–Reichardt solvent polarity\\
MIST-28M-Excess        & \cref{fig:skew_parity,fig:soap_outliers} & MIST-28M & Excess Property Dataset (\cref{sec:si:excess_dataset}) &
\makecell[tl]{Excess molar volume ($V_m^E$)\\ Excess molar enthalpy ($H_m^E$)} \\
MIST-28M-Ionic         & \cref{fig:ionic_prediction} & MIST-28M & Ionic Conductivity & Ionic conductivity [mS/cm] \\
MIST-28M-tmQM          & \cref{fig:tmQM,fig:token_emb_tsne} & MIST-28M & tmQM (\cref{sec:si:tmQM_dataset}) &
\makecell[tl]{Electronic energy ($E_{\mathrm{elec}}$) [Hartree]\\
Dispersion energy ($E_{\mathrm{disp}}$) [Hartree]\\
\acs{HOMO} ($E_{\text{HOMO}}$) [Hartree]\\
\acs{LUMO} ($E_{\text{LUMO}}$) [Hartree]\\
\acs{HOMO-LUMO} gap ($\Delta E_{\text{HL}}$) [Hartree]\\
Dipole moment ($\mu$) [Debye]\\
Metal-center natural charge ($q_{\mathrm{M}}$) [e]\\
Polarizability ($\alpha$) [$a_0^3$]} \\
MIST-28M-Isotopes      & \cref{fig:isotope_clusters} & MIST-28M & Isotopes & Half-life \\
    MIST-28M               & \cref{fig:token_emb_tsne} & N.A. & Pretrained model & N.A. \\
MIST-28M-FreeSolv      & \cref{fig:token_emb_tsne} & MIST-28M & FreeSolv (\cref{sec:si:molecule_net}) & Hydration free energy [kcal/mol] \\
MIST-1.8B              & \cref{fig:interp_emb_pah,fig:interp_emb_rings} & N.A. & Pretrained model & N.A. \\
\bottomrule
\end{tabular}%
}
\caption{\label{tab:model_list} Details of pretrained and fine-tuned \ac{MIST} model variants used to plot figures in main manuscript.}
\end{table}

\section{High-Throughput Screening}
\label{sec:si:screening}

\begin{figure}[ht!]
    \centering
    \includegraphics[width=\linewidth]{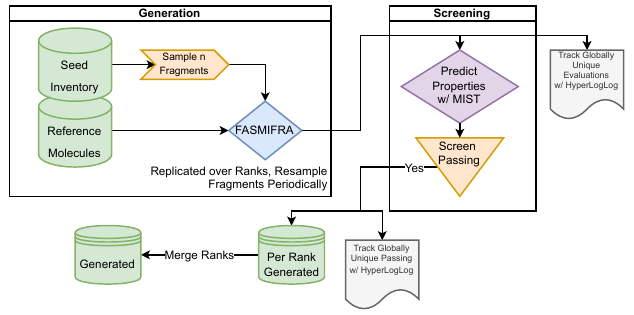}
    \caption{
        \label{fig:screening_workflow}
        High-throughput screening workflow using FASMIFRA\cite{BTMolecularGenerationFast2021} for molecule generation and MIST-28M models for property prediction.
        Each rank generates molecules using both the reference set and \(n\) molecules sampled from the seed inventory (randomly selected molecules from the ChEMBL and ZINC datasets~\cite{ZFH+ChEMBLDatabase20232024,ITY+ZINC20AFreeUltralargeScale2020}).
        The seed inventory provides additional material for generation; without it, we found that uniqueness dropped excessively rapidly (\cref{sec:si:screening_perf_tuning}).
        The generation process is periodically restarted with a new set of seed molecules.
        As each rank operates fully independently, there is a risk of ranks receiving identical feedstock for FASMIFRA; however, this effect is largely mitigated by the size of the seed inventory.
        The stream of generated molecules is then processed by the fine-tuned MIST-28M models to generate property predictions.
        Molecules passing the screening criteria are saved to a per-rank SQLite database along with their predicted properties.
        Per-rank databases are merged as a post-processing step, eliminating the need for expensive synchronization for on-the-fly de-duplication.
        We track global generation and screening statistics on-the-fly using the HyperLogLog~\cite{FFGMHyperLogLogAnalysisNearoptimal2007} algorithm, which estimates the approximate number of unique molecules screened or evaluated,
        thereby avoiding the intractable memory cost and communication costs of an exact cardinality metric.
    }
\end{figure}

In this section, we describe validation and optimization of the high-throughput screening workflow built using fine-tuned \ac{MIST}-28M models (\cref{sec:screening} and Methods~\cref{sec:methods:screen}) on two different multi-objective battery electrolyte solvent screening problems.
First, we target a set of pre-requisite requirements for most lithium batteries --- maximizing the electrochemical and thermal stability of electrolyte solvents. 
Second, we apply the screening workflow to the pertinent and stringent problem of finding electrolyte solvents for lithium-air batteries.
A diagram of the workflow is shown in \cref{fig:screening_workflow}, and the source code is provided in the \texttt{opt/screening} folder in our GitHub repository and code in the \datadrop[data release].

\subsection{Performance Tuning for Screening Workflow}
\label{sec:si:screening_perf_tuning}

\begin{figure}[h!]
    \centering
    \includegraphics{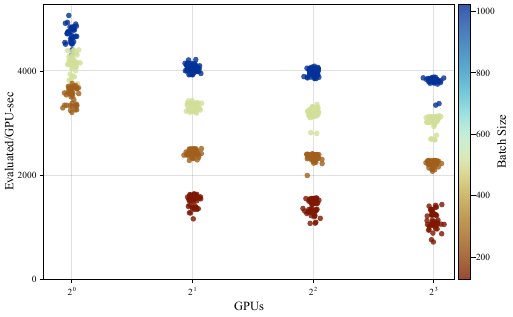}
    \caption{
        \label{fig:screening_eval_scaling}
        Weak scaling of global evaluation throughput (excluding duplicate molecules) as a function of the number of GPUs, based on a limited performance tuning sweep.
        Based on our results (\cref{tab:screening_perf}), a batch size of 822 maximizes global throughput. A batch size of 768 molecules was selected (closest power of 2 to optimum).
        Reasonable scaling efficiency is observed beyond 2 GPUs.
        The drop in throughput when moving from 1 to 2 GPUs is expected, due to the overhead of inter-rank communication.
    }
\end{figure}

We performed a limited sweep over workflow parameters to tune our pipeline to maximize the global (over all ranks) screening throughput (global molecules evaluated per second) and global hit rate (global unique passing molecules per second).
As shown in \cref{fig:screening_eval_scaling},  our evaluation throughput shows good weak scaling with increasing world size (each rank has one GPU for \ac{MIST} model inference).
This is expected as, beyond periodic synchronization to track the global count of unique passing and evaluated molecules, each rank operates independently.
As ranks operate independently and our generation pipeline is inherently stochastic, we do observe a fairly high rate of duplication across ranks, resulting in a median generation efficiency of 54.6\% (the ratio of the globally unique molecules evaluated to the total number of evaluated molecules).

\begin{figure}[h!]
    \centering
    \includegraphics{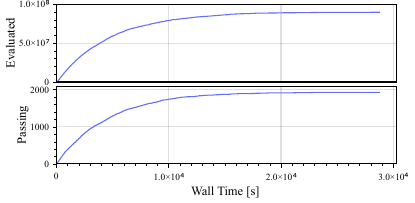}
    \caption{
        \label{fig:screening_gen_trace}
        Trace of the approximate number of globally unique molecules evaluated (top) and accepted (bottom) during a production electrolyte screening run using 8 A100 GPUs.
        At peak, the workflow evaluated 19,687 unique molecules per second, with a hit rate of 0.44 molecules per second.
        We observed an exponential decay in the evaluation rate, asymptotically approaching a limit of approximately 91 million unique molecules --- slightly exceeding our combined fragment inventory of 88 million molecules.
        This highlights the need for scalable, distributed molecular generation methods capable of keeping pace with molecular \acp{FM}.
    }
\end{figure}

Developing scalable, distributed methods for molecular generation that can keep pace with the evaluation throughput of molecular \acp{FM}, such as \ac{MIST}, is key to addressing this limitation\cite{HRPAParallelDistributedThompson2017}.
Ultimately, even with our limited generation efficiency, our pipeline reached a peak screening throughput of 19,687 unique molecules per second with 8 GPUs (8 ranks, 1 GPU per rank).
The generation efficiency decreases rapidly once the number of molecules generated exceeded the number of fragments provided, indicated by the negative coefficient for \texttt{Epoch Size} in \cref{tab:screening_perf}.
To compensate for this we limited FASMIFRA generation to 1.2 times the number of fragments provided.
We also selected a batch size of 768 molecules based on our performance tuning results (\cref{tab:screening_perf}) to maximize the overall throughput of the workflow.

To tune our high-throughput screening pipeline, we fit linear models to the performance sweep results (\cref{tab:screening_perf}).
Results for all 707 runs are provided in our data release.
The following outlines the control parameters evaluated for performance tuning.

\begin{description}
    \item[Batch Size.] The number of molecules evaluated by the MIST models in a single batch per GPU. Each rank has 1 GPU.
    \item[Limit Ref. Fragments.] The fraction of fragments generated from reference electrolytes to use in the generation pipeline.
    \item[Limit Inv. Fragments.] The fraction of fragments generated from the seed inventory to use in the generation pipeline.
    \item[Num. Fragments.] The number of fragments to provide to FASMIFRA\cite{BTMolecularGenerationFast2021} per generation epoch.
    \item[Epoch Size.] The number of molecules, relative to the number of fragments, to generate before reseeding FASMIFRA\cite{BTMolecularGenerationFast2021} with new fragments.
\end{description}

\begin{table}
    \centering
    \begin{tabular}{lrr}
\toprule
                      & \multicolumn{1}{c}{log(global\_throughput)} & \multicolumn{1}{c}{generation\_efficiency} \\
                      &                                         (1) &                                        (2) \\
\midrule
(Intercept)           &                                    7.043*** &                                    1.410** \\
                      &                                     (0.063) &                                    (0.494) \\
\(\log(\mbox{ranks})\) &                                    0.838*** &                                  -0.012*** \\
                      &                                     (0.007) &                                    (0.003) \\
Batch Size            &                                    0.004*** &                                            \\
                      &                                     (0.000) &                                            \\
\((\mbox{Batch Size})^2\)   &                               -0.000*** &                                            \\
                      &                                     (0.000) &                                            \\
Num. Fragments        &                                      -0.000 &                                            \\
                      &                                     (0.000) &                                            \\
Epoch Size            &                                       0.053 &                                     -1.503 \\
                      &                                     (0.060) &                                    (0.993) \\
\((\mbox{Epoch Size})^2\) &                                           &                                      0.644 \\
                      &                                             &                                    (0.496) \\
Limit Ref.\ Fragments &                                             &                                  -0.059*** \\
                      &                                             &                                    (0.007) \\
Limit Inv.\ Fragments &                                             &                                   0.039*** \\
                      &                                             &                                    (0.007) \\
\midrule
$N$                   &                                         707 &                                        707 \\
Degrees of Freedom    &                                         701 &                                        701 \\
$R^2$                 &                                       0.965 &                                      0.206 \\
MAE                   &                                       0.096 &                                      0.043 \\
RMSE                  &                                       0.127 &                                      0.063 \\
\bottomrule
\end{tabular}

    \caption{
        \label{tab:screening_perf}
        Linear models used to guide performance tuning of our high-throughput screening workflow.
        Asterisks denote coefficient significance: * for \(p \leq 0.05\), ** for \(p \leq 0.01\), and *** for \(p \leq 0.001\).
        Descriptions of each parameter are provided in \cref{sec:si:screening_perf_tuning}
    }
\end{table}

\subsection{Maximizing Electrochemical and Thermal Stability Windows}
\begin{figure}
  \centering
  \label{fig:si:screening_electrolytes}
  \includegraphics[width=\textwidth]{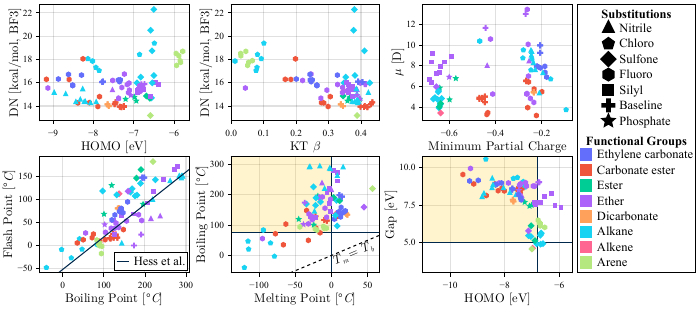}
  \caption{
  \ac{MIST}-28M accurately predicts quantum, chemical, and thermodynamic descriptors for electrolyte design --- including orbital energy levels,  \ac{DN} and \ac{KT} solvatochromic parameters --- when fine-tuned on labelled datasets as small as $\sim$180 samples.
  }
\end{figure}

\begin{figure}
  \centering
  \label{fig:si:screening_pareto}
  \includegraphics[width=\textwidth]{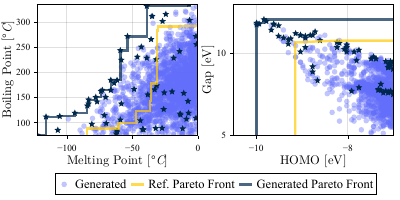}
  \caption{
  Using fine-tuned MIST-28M models, we identified 63 molecules on the Pareto front, improving both electronic and thermal performance relative to our reference set.
  }
\end{figure}

\subsubsection{Pareto Front Molecules}
\label{sec:si:pareto_front_molecules}
\begin{figure}[h!]
    \centering
    \includegraphics[width=\linewidth]{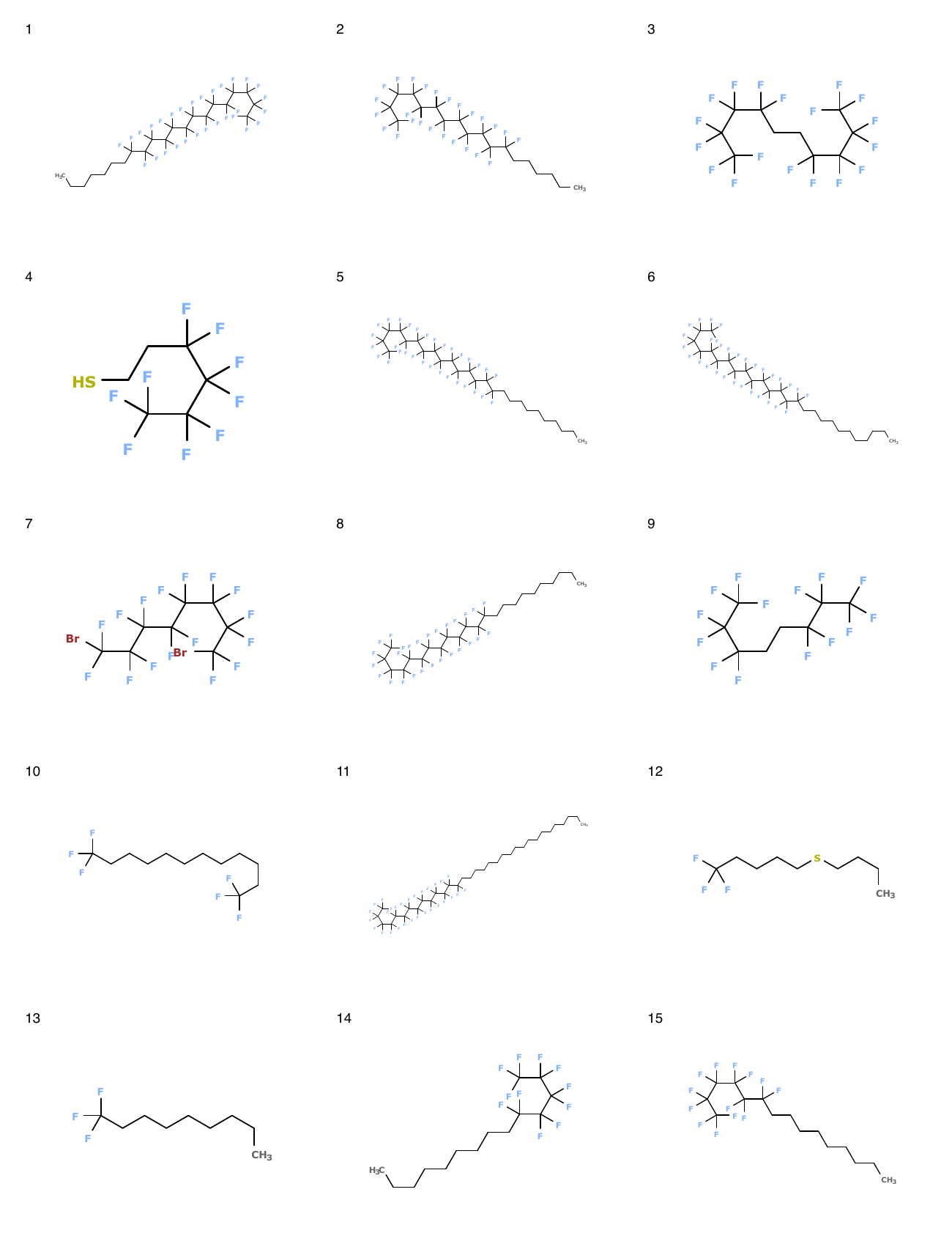}
    \caption{
        \label{fig:pareto_front_p1}
        Screened Pareto-Front molecules identified by the high-throughput screening pipeline presented in the manuscript.
        Figure 1 of 5.
    }
\end{figure}

\begin{figure}[h!]
    \centering
    \includegraphics[width=\linewidth]{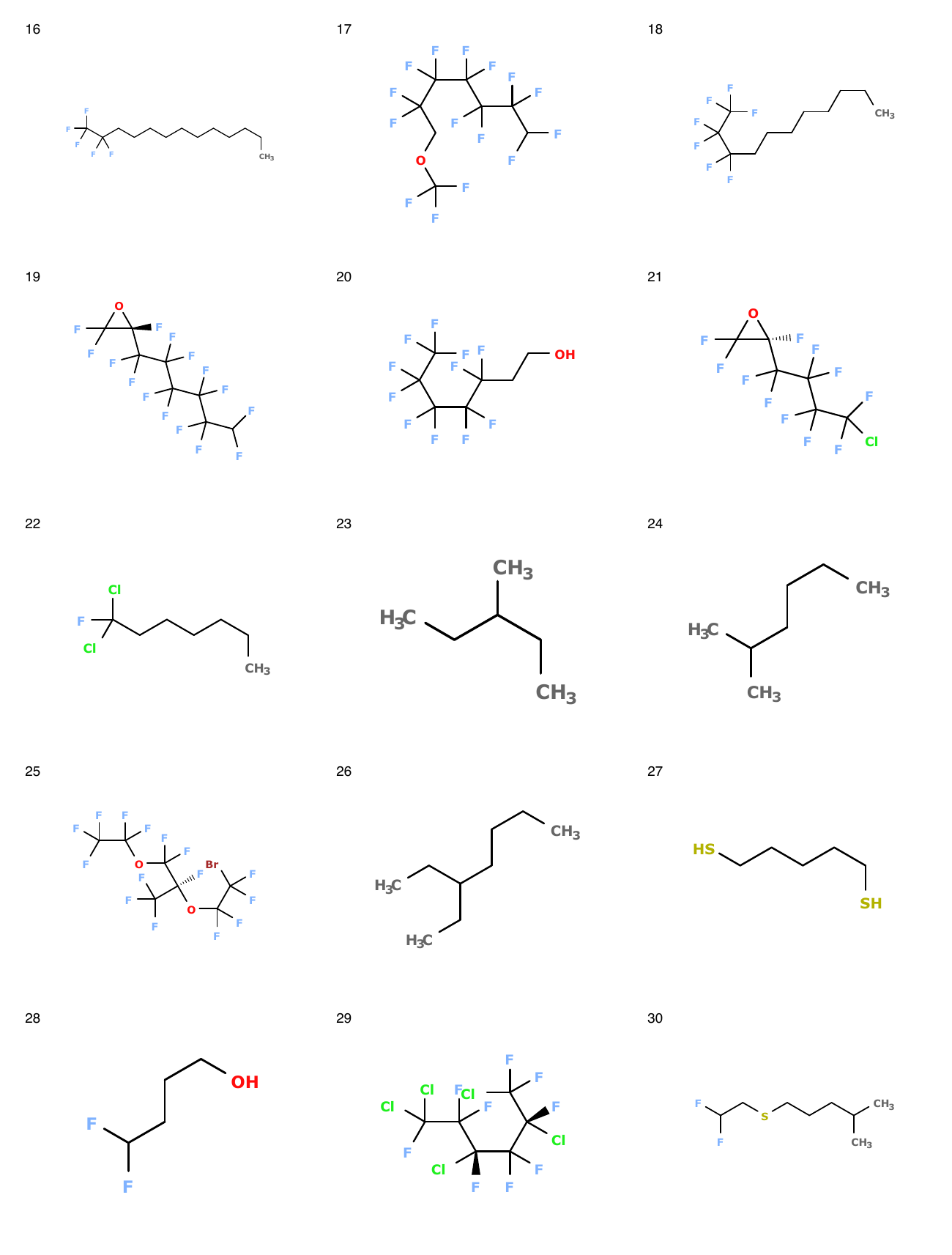}
    \caption{
        \label{fig:pareto_front_p2}
        Screened Pareto-Front molecules identified by the high-throughput screening pipeline presented in the manuscript.
        Estimated properties are shown in \cref{tab:generated_pareto_front}.
        Figure 2 of 5.
    }
\end{figure}

\begin{figure}[h!]
    \centering
    \includegraphics[width=\linewidth]{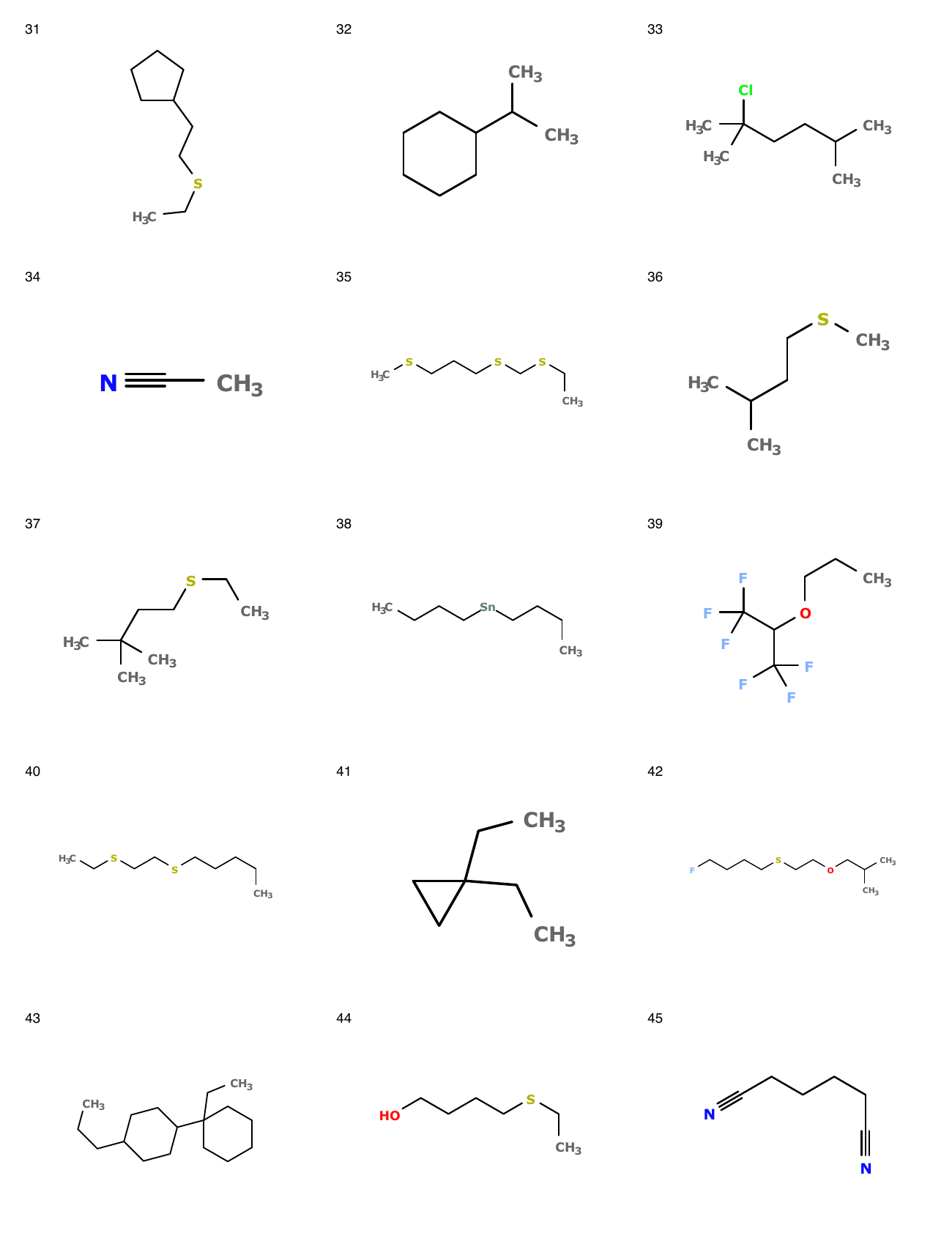}
    \caption{
        \label{fig:pareto_front_p3}
        Screened Pareto-Front molecules identified by the high-throughput screening pipeline presented in the manuscript.
        Estimated properties are shown in \cref{tab:generated_pareto_front}.
        Figure 3 of 5.
    }
\end{figure}

\begin{figure}[h!]
    \centering
    \includegraphics[width=\linewidth]{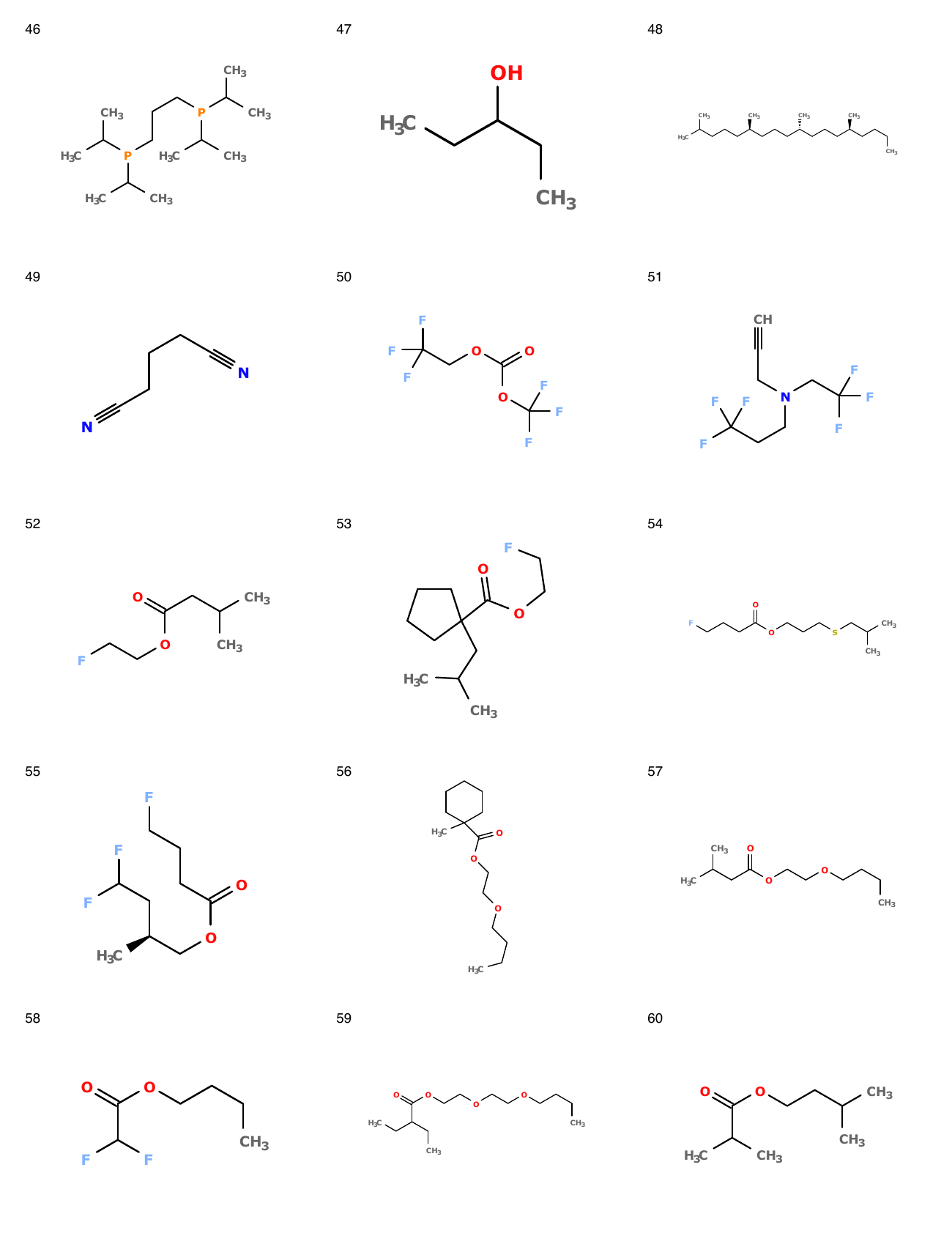}
    \caption{
        \label{fig:pareto_front_p4}
        Screened Pareto-Front molecules identified by the high-throughput screening pipeline presented in the manuscript.
        Estimated properties are shown in \cref{tab:generated_pareto_front}.
        Figure 4 of 5.
    }
\end{figure}

\begin{figure}[h!]
    \centering
    \includegraphics[width=\linewidth]{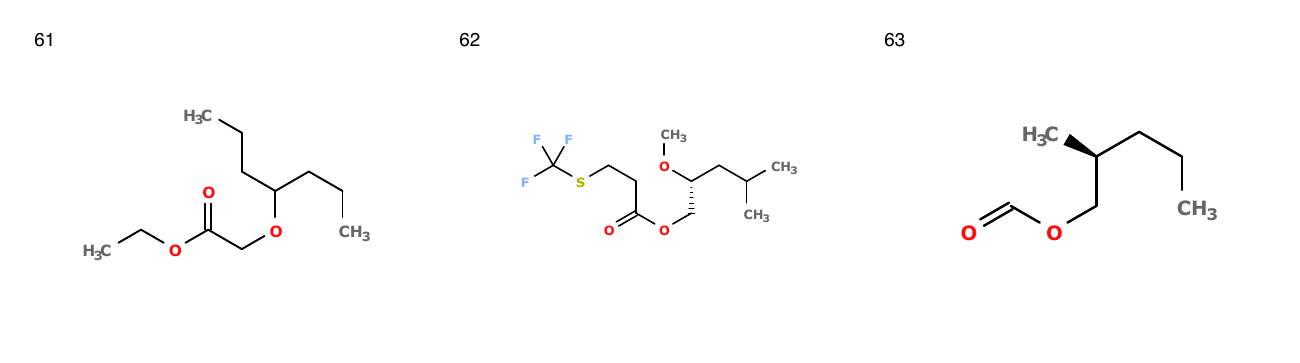}
    \caption{
        \label{fig:pareto_front_p5}
        Screened Pareto-Front molecules identified by the high-throughput screening pipeline presented in the manuscript.
        Estimated properties are shown in \cref{tab:generated_pareto_front}.
        Figure 5 of 5.
    }
\end{figure}

The 63 screened electrolytes on the Pareto front are shown in~\cref{fig:pareto_front_p1,fig:pareto_front_p2,fig:pareto_front_p3,fig:pareto_front_p4,fig:pareto_front_p5}.
Property values computed using fine-tuned \ac{MIST}-28M models for the Pareto front candidates are in \cref{tab:generated_pareto_front}.
\ac{SMILES} encodings for the Pareto front candidates are provided within our data release.

\begin{figure}
    \centering
    \labelphantom{fig:si:scent_pareto_elec}
    \labelphantom{fig:si:scent_pareto_thermal}
    \labelphantom{fig:si:scent_counts}
    \includegraphics[width=4in]{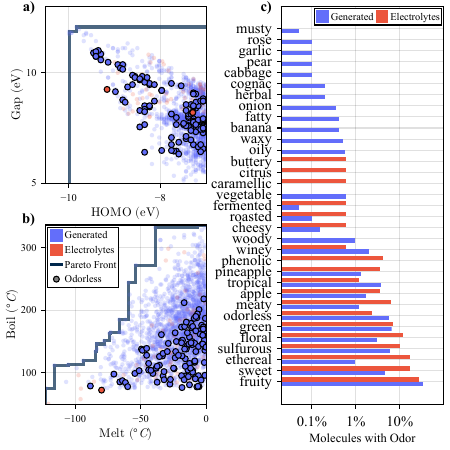}
    \caption{
        \label{fig:screened_scents}
        (\subref*{fig:si:scent_pareto_elec}\&\subref*{fig:si:scent_pareto_thermal})~Using MIST-28M fine-tuned on an olfaction dataset\cite{leePrincipalOdorMap2023}, we further filtered 1,936 electrolyte candidates for odourless compounds.
        (\subref*{fig:si:scent_counts})~Finding electrolytes with a wide range of olfactory profiles,
        demonstrating \ac{MIST}'s ability to further refine candidate molecules from a high-throughput screening workflow.
    }
\end{figure}

\begin{longtable}{cccccc}
  \caption{
    \label{tab:generated_pareto_front}
    Pareto Front Candidate Electrolytes and computed properties using fine-tuned MIST-28M models.
    Molecular structures (Indexed by \#) are shown in \cref{fig:pareto_front_p1,fig:pareto_front_p2,fig:pareto_front_p3,fig:pareto_front_p4,fig:pareto_front_p5}.
  }\\
  \textbf{\#} & \textbf{HOMO} & \textbf{LUMO} & \textbf{Gap} & \textbf{Melt} & \textbf{Boil} \\
              & \textbf{(eV)} & \textbf{(eV)} & \textbf{(eV)} & \(\pmb{(\degree C)}\) & \(\pmb{(\degree C})\) \\\hline
  \endfirsthead
  \multicolumn{6}{c}%
{{\bfseries \tablename\ \thetable{} -- continued from previous page}} \\
\hline
  \textbf{\#} & \textbf{HOMO} & \textbf{LUMO} & \textbf{Gap} & \textbf{Melt} & \textbf{Boil} \\ \hline
  \endhead

  \hline \multicolumn{6}{r}{{Continued on next page}} \\
  \endfoot

  \hline
  \endlastfoot
  1 & -9.8 & 2.2 & 12.1 & -9 & 108 \\
  2 & -9.8 & 2.2 & 12.0 & -2 & 110 \\
  3 & -10.0 & 1.8 & 11.9 & -12 & 88 \\
  4 & -9.9 & 1.9 & 11.8 & -62 & 82 \\
  5 & -9.6 & 2.1 & 11.8 & -13 & 116 \\
  6 & -9.6 & 2.1 & 11.7 & -14 & 118 \\
  7 & -10.0 & 1.8 & 11.7 & -42 & 86 \\
  8 & -9.5 & 2.1 & 11.7 & -7 & 120 \\
  9 & -9.7 & 1.7 & 11.4 & -62 & 78 \\
  10 & -8.7 & 2.5 & 11.3 & -11 & 159 \\
  11 & -9.1 & 2.1 & 11.2 & -8 & 143 \\
  12 & -8.8 & 2.3 & 11.1 & -31 & 174 \\
  13 & -8.8 & 2.3 & 11.1 & -58 & 138 \\
  14 & -9.0 & 2.1 & 11.1 & -38 & 156 \\
  15 & -9.0 & 2.1 & 11.1 & -20 & 162 \\
  16 & -8.9 & 2.1 & 11.1 & -33 & 166 \\
  17 & -9.5 & 1.5 & 11.1 & -42 & 87 \\
  18 & -8.9 & 1.9 & 10.9 & -81 & 98 \\
  19 & -9.4 & 1.5 & 10.9 & -27 & 90 \\
  20 & -9.2 & 1.6 & 10.9 & -26 & 127 \\
  21 & -9.3 & 1.5 & 10.8 & -62 & 84 \\
  22 & -8.6 & 2.2 & 10.8 & -59 & 106 \\
  23 & -8.2 & 2.4 & 10.7 & -122 & 76 \\
  24 & -8.2 & 2.4 & 10.6 & -106 & 113 \\
  25 & -9.3 & 1.2 & 10.6 & -32 & 89 \\
  26 & -8.0 & 2.5 & 10.5 & -66 & 165 \\
  27 & -8.0 & 2.3 & 10.4 & -58 & 225 \\
  28 & -8.2 & 2.1 & 10.4 & -72 & 115 \\
  29 & -8.9 & 1.4 & 10.4 & -44 & 104 \\
  30 & -8.1 & 2.2 & 10.3 & -49 & 167 \\
  31 & -7.8 & 2.4 & 10.2 & -32 & 226 \\
  32 & -7.9 & 2.3 & 10.1 & -74 & 163 \\
  33 & -8.0 & 2.0 & 10.1 & -94 & 120 \\
  34 & -9.0 & 1.0 & 10.1 & -26 & 160 \\
  35 & -7.4 & 2.3 & 9.8 & -1 & 256 \\
  36 & -7.4 & 2.4 & 9.8 & -84 & 134 \\
  37 & -7.5 & 2.1 & 9.7 & -73 & 165 \\
  38 & -7.4 & 2.3 & 9.7 & -77 & 142 \\
  39 & -7.8 & 1.7 & 9.6 & -107 & 86 \\
  40 & -7.1 & 2.5 & 9.5 & -2 & 272 \\
  41 & -7.2 & 2.3 & 9.5 & -116 & 112 \\
  42 & -7.0 & 2.4 & 9.5 & -14 & 242 \\
  43 & -7.1 & 2.2 & 9.4 & -6 & 304 \\
  44 & -7.1 & 2.1 & 9.2 & -18 & 229 \\
  45 & -8.8 & 0.4 & 9.2 & -24 & 316 \\
  46 & -7.1 & 1.9 & 9.1 & -59 & 244 \\
  47 & -7.1 & 2.0 & 9.1 & -84 & 138 \\
  48 & -7.0 & 2.0 & 9.0 & -27 & 302 \\
  49 & -8.8 & 0.1 & 8.9 & -35 & 312 \\
  50 & -8.7 & -0.4 & 8.3 & -70 & 99 \\
  51 & -7.3 & 0.8 & 8.1 & -84 & 146 \\
  52 & -7.8 & 0.2 & 7.9 & -64 & 179 \\
  53 & -7.3 & 0.4 & 7.8 & -36 & 266 \\
  54 & -7.3 & 0.5 & 7.8 & -39 & 229 \\
  55 & -7.3 & 0.5 & 7.8 & -62 & 185 \\
  56 & -7.1 & 0.6 & 7.7 & -7 & 336 \\
  57 & -7.2 & 0.5 & 7.7 & -43 & 251 \\
  58 & -7.9 & -0.2 & 7.7 & -70 & 151 \\
  59 & -7.1 & 0.5 & 7.7 & -39 & 332 \\
  60 & -7.1 & 0.6 & 7.7 & -78 & 151 \\
  61 & -7.1 & 0.5 & 7.6 & -54 & 272 \\
  62 & -7.2 & 0.3 & 7.6 & -46 & 231 \\
  63 & -7.0 & 0.5 & 7.5 & -66 & 185 \\
\end{longtable}

\FloatBarrier

\subsubsection{Replicating QM9 Calculations for Validation}
\label{sec:si:qm9_replication}
To validate the efficacy of \ac{MIST} models for an electrolyte high-throughput screening campaign, we computed \ac{HOMO}, \ac{LUMO} and \ac{HOMO-LUMO} Gap using \ac{DFT} for a randomly selected subset of the passing molecules from our screening run (\cref{fig:si:screening_pareto}).
Our \ac{DFT} calculations followed the procedure used by the QM9\cite{RDRVQuantumChemistryStructures2014} dataset, on which \ac{MIST} models were trained: we performed an initial conformation generation step, followed by an initial relaxation at PM7 semi-empirical level of theory using MOPAC\cite{MSMOPAC2025} with \texttt{precise} convergence thresholds enabled.
PM7 geometries were then further relaxed using Gaussian 09\cite{FTS+Gaussian092009} at the B3LYP/6-31G(2df,p) level of theory.
This procedure deviates from the one used to produce the QM9 dataset\cite{RDRVQuantumChemistryStructures2014} in several ways.
First, we did not invoke their iterative refinement procedure for molecules that failed to converge with the default setting.
Instead, we opted to discard molecules that did not converge.
Second, we used the latest version of MOPAC (v23.1.2) instead of version 13.136L 2012 used by QM9\cite{RDRVQuantumChemistryStructures2014}.
Third, we did not use the proprietary code, Corina, for conformer generation.
Instead, we evaluated the following three different procedures for their ability to replicate the results of the QM9 calculations.

\begin{enumerate}
    \item Generate a single conformer using RDKit's\cite{GreRDKitOpensourceCheminformatics2024} \texttt{ETKDGv3}\cite{WWLRImprovingConformerGeneration2020}.
    \item Embed molecules using OpenBabel\cite{OBJ+OpenBabelOpen2011} and the UFF (Universal force field)\cite{RCC+UFFFullPeriodic1992} to generate a single starting conformer.
    \item Generate 200 conformers using RDKit's\cite{GreRDKitOpensourceCheminformatics2024} \texttt{ETKDGv3}\cite{WWLRImprovingConformerGeneration2020} and select the lowest energy conformer after relaxation with the UFF\cite{RCC+UFFFullPeriodic1992} or MMFF (Merck molecular force field)\cite{TSLBringingMMFFForce2014,HalMerckMolecularForce1996}.
\end{enumerate}

\begin{figure}[h!]
    \centering
    \includegraphics[width=\linewidth]{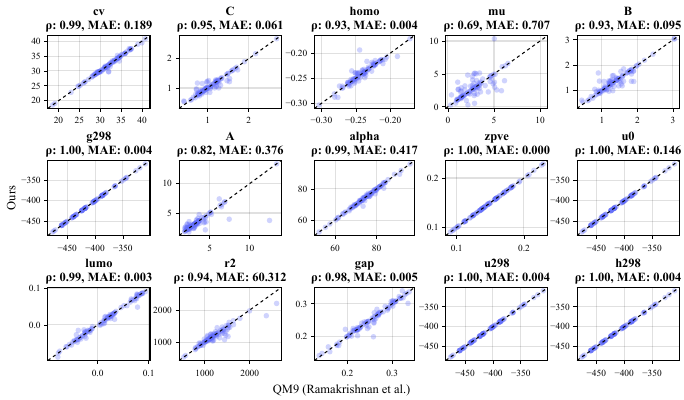}
    \caption{
        \label{fig:qmist_rdkit}
        Parity plots of the QM9\cite{RDRVQuantumChemistryStructures2014} reported property values and our \ac{DFT} calculations using Method \#1 for 95 randomly selected molecules.
        Pearson correlation coefficients (\(\rho\)) and \ac{MAE} values are reported for each property.
    }
\end{figure}

\begin{figure}[h!]
    \centering
    \includegraphics[width=\linewidth]{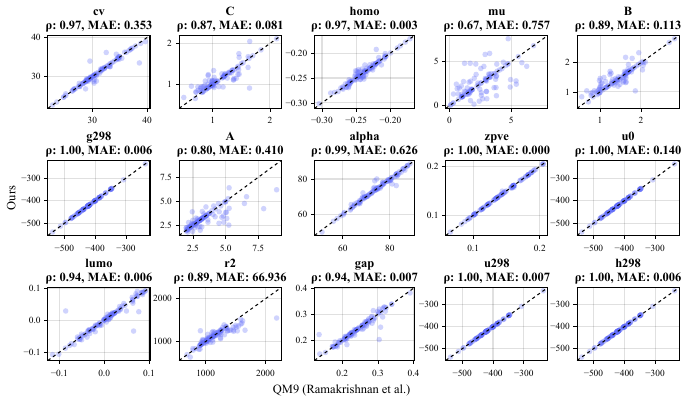}
    \caption{
        \label{fig:qmist_obabel}
        Parity plots of the QM9\cite{RDRVQuantumChemistryStructures2014} reported property values and our \ac{DFT} calculations using Method \#2 for 72 randomly selected molecules.
        Pearson correlation coefficients (\(\rho\)) and \ac{MAE} values are reported for each property.
    }
\end{figure}

\begin{figure}[h!]
    \centering
    \includegraphics[width=\linewidth]{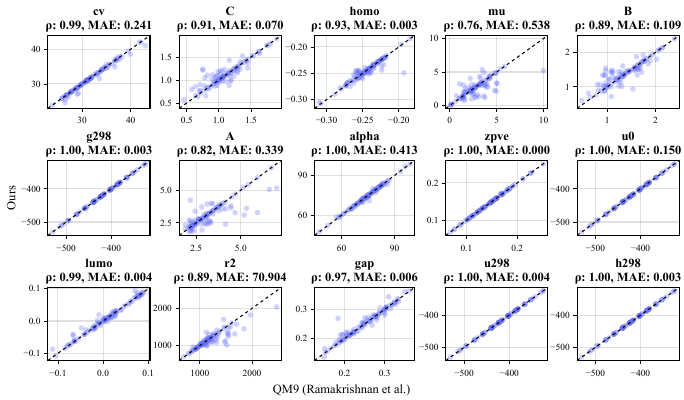}
    \caption{
        \label{fig:qmist_conformer}
        Parity plots of the QM9\cite{RDRVQuantumChemistryStructures2014} reported property values and our \ac{DFT} calculations using Method \#3 for 99 randomly selected molecules.
        Pearson correlation coefficients (\(\rho\)) and \ac{MAE} values are reported for each property.
    }
\end{figure}

\begin{figure}[h!]
    \centering
    \includegraphics[width=\linewidth]{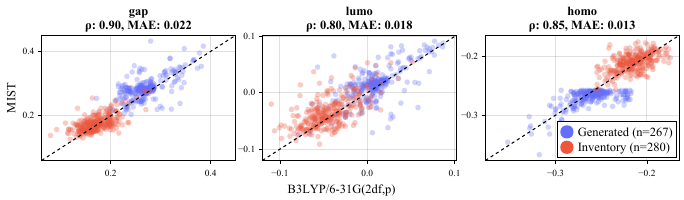}
    \caption{
        \label{fig:screening_parity}
        Parity plots comparing QM9 property values predicted by the fine-tuned \ac{MIST} model with \ac{DFT} calculations at the B3LYP/6-31G(2df,p) level of theory, as detailed in \cref{sec:si:qm9_replication}.
        Results are presented for both generated molecules and inventory molecules.
        \ac{MIST} exhibits consistent error bounds across the full range of values, supporting its utility as a \ac{DFT} surrogate model for early-stage high-throughput screening.
        The observed distributional shift between inventory and generated molecules arises from the screening process, which only retains generated molecules that satisfy predefined criteria.
        Pearson correlation coefficients (\(\rho\)) and \ac{MAE} values are reported for each property. All properties are expressed in units of Hartrees.
    }
\end{figure}

We evaluated each method by computing all QM9 reported properties for up to 100 randomly selected molecules from the QM9 dataset\cite{RDRVQuantumChemistryStructures2014}.
Parity plots of our calculations versus the results published by the QM9 dataset\cite{RDRVQuantumChemistryStructures2014} are shown in \cref{fig:qmist_rdkit,fig:qmist_obabel,fig:qmist_conformer}.
Ultimately, we selected Method \#3 as our protocol, as it provided the lowest \ac{MAE} for the \ac{HOMO}, \ac{LUMO} and Gap properties.

As shown in \cref{fig:screening_parity}, \ac{MIST}-28M fine-tuned on QM9 is predictive over the full range of screened parameters, demonstrating its efficacy for a high-throughput screening application.
We do observe a higher \ac{MAE} than reported in Extended Data~\cref{tab:qm9_benchmark}.
This can partially be explained by differences between our \ac{DFT} calculations and the procedure used by QM9\cite{RDRVQuantumChemistryStructures2014} (\cref{fig:qmist_conformer}).
For example, we were only able to replicate the \ac{HOMO} values to within a \ac{MAE} of 0.003 Hartree of the QM9 published values.
As such, our estimated error for \ac{MIST} (\ac{MAE} 0.013 Hartree) likely partially results from calculation differences between QM9 and our \ac{DFT} calculations.
The \ac{MIST} model's accuracy is sufficient for early-stage high-throughput screening, particularly, when the molecules of interest are far from the decision boundary, as is the case here.
\subsubsection{Mixed Halogenated Electrolytes Discovered From Screening}
\label{sec:si:mixed_halogenation_validation}

In this section, we analyze the Pareto front solvent molecules found by the screening run targeting electrochemically and thermally stable solvent molecules.
Several generated candidates contained mixed halogen substitution patterns, including molecules bearing combinations of fluorine, chlorine, and bromine. 
This observation is noteworthy because the screening was multi-objective: candidate molecules were prioritized to maximize electrochemical stability descriptors, maintain liquidity under room-temperature conditions, and reduce volatility. 
In the screening workflow, these objectives were operationalized by favoring molecules with lower \ac{HOMO} energies, higher \ac{LUMO} energies, wide HOMO--LUMO gaps, boiling points above 75~$^{\circ}$C, and melting points below 25~$^{\circ}$C. 
The appearance of mixed-halogenated molecules on the Pareto front therefore suggested that the model had identified a chemically non-trivial trade-off space not captured by the more familiar fluorination-only design rules~\cite{XuNonaqueousLiquidElectrolytes2004,ZVNotAllFluorination2020}.

Fluorination provides a natural baseline against which to evaluate this hypothesis. 
In lithium-based battery electrolytes, fluorinated solvents and additives have been widely studied because fluorine substitution is often associated with improved electrochemical robustness, altered solvation behaviour, and favorable interphase chemistry~\cite{XuNonaqueousLiquidElectrolytes2004,shkrob2015makes,michan2016fluoroethylene,ZVNotAllFluorination2020,li2019fluorinated,xia2021recent,li2023critical,wang2025fluorine}.
In particular, fluorinated components such as \ac{FEC} are widely used to promote interphases enriched in inorganic fluorides, especially LiF, which are often associated with improved cycling stability and suppressed parasitic decomposition, although the precise mechanistic role of LiF can depend strongly on its origin, morphology, and coupling to the rest of the interphase~\cite{shkrob2015makes,michan2016fluoroethylene,ZVNotAllFluorination2020,li2019fluorinated,he2020intrinsic,li2023critical}.
Since fluorination is already an established and heavily studied design axis in electrolyte science, we used fully fluorinated analogues as the reference state for evaluating whether partial replacement of fluorine with heavier halogens produces a meaningful and chemically interpretable trade-off.

To keep the validation set computationally tractable while preserving structural diversity, we selected all generated mixed-halogenated molecules identified in the screening campaign and supplemented them with a matched comparison set of non-mixed halogenated analogues. 
The comparison set was restricted to molecules containing between two and five total halogen substituents and no more than 20 atoms in total. 
These filters were chosen to limit the cost of quantum-chemical follow-up calculations while still spanning a sufficiently broad range of substitution motifs. 
This allowed us to test whether the mixed-halogenation signal reflected a robust trend. 
We additionally included the fluorinated cyclic carbonate (tetrafluoroethylene carbonate --TFEC) as an anchor molecule because Zhang and Viswanathan\cite{ZVNotAllFluorination2020} used fluorinated ethylene-carbonate derivatives to demonstrate how fluorination pattern can strongly influence electrolyte reduction chemistry and interphase formation on lithium metal~\cite{ZVNotAllFluorination2020}.

For each parent scaffold, we generated two families of comparison molecules. 
In the first family, all halogen sites were assigned to a single halogen type, yielding fully fluorinated, fully chlorinated, fully brominated, and fully iodinated analogues where chemically feasible. 
In the second family, we generated all pairwise mixed-halogen variants, \emph{i.e.}, F/Cl, F/Br, F/I, Cl/Br, Cl/I, and Br/I substitutions, again subject to chemical validity. 
This construction allowed us to distinguish between two conceptually different questions: whether mixed halogenation behaves as a qualitatively distinct design strategy relative to single-halogen substitution, and whether trends across the halogen series follow an interpretable ordering as fluorine is progressively replaced by heavier halogens.

The Pareto front molecules (\cref{sec:si:pareto_front_molecules}) contained fluorinated molecules and esters, consistent with established electrolyte design heuristics, but it also contained multiple examples of mixed halogenation, including F/Br and F/Cl combinations. 
Our goal here is to test whether the emergence of these molecules on the Pareto front encoded a chemically meaningful hypothesis. 
Specifically, we asked whether mixed halogenation systematically perturbs frontier orbital energies (\ac{HOMO} and \ac{LUMO}), gap, charge distribution, and relevant thermal descriptors in a way that is both statistically significant and interpretable relative to the fully fluorinated baseline.

To validate the effect of mixed-halogen substitution on molecular electronic structure, all molecules in the mixed-halogenation set were evaluated in the gas phase with \textsc{Gaussian 16} using a common three-step workflow\cite{frisch2016gaussian}. First, each structure was geometry-optimized at the B3LYP/def2-TZVPP level of theory\cite{becke1993density,lee1988development,stephens1994ab,weigend2005balanced} with Grimme's D3 dispersion correction\cite{grimme2010consistent}, using tight self-consistent-field convergence and the XQC procedure to improve SCF robustness. Harmonic frequency calculations were performed at the same level on the optimized geometries to characterize the resulting stationary points. Second, a natural population analysis (NPA) was performed as a single-point calculation on the converged optimized geometry read from the checkpoint file\cite{reed1985natural,glendening2013nbo}, from which the halogen-site charges were extracted. For each molecule, both the minimum halogen NPA charge and the mean halogen NPA charge across all halogen atoms were computed. An additional single-point calculation with full population analysis was performed on the same checkpoint geometry to obtain molecular orbital coefficients for frontier-orbital analysis. HOMO and LUMO energies were taken directly from this wavefunction, and the HOMO--LUMO gap was defined as the energy difference between the two frontier orbitals. To quantify halogen participation in the frontier states, atom-resolved halogen fractions of the HOMO and LUMO were obtained by projecting the corresponding molecular orbital coefficients onto basis functions centered on halogen atoms and summing the resulting contributions. This workflow provided a consistent electronic-structure description across the full mixed-halogenation validation set and enabled direct comparison between halogen-dependent charge metrics, frontier-orbital properties, and experimentally relevant observables.

The resulting analysis, therefore, serves as a validation of the \ac{MIST} models in a broader sense than simple property prediction. 
Rather than only reproducing known fluorination heuristics, the Pareto optimization found a neighboring design space---mixed halogenation---that is uncommon in presently deployed battery electrolytes, but chemically plausible.
By constructing matched families of fully halogenated and mixed-halogenated analogues and evaluating them with quantum-chemical and statistical analyses, we test whether the screening workflow can do more than recover established design rules; namely, whether it can generate experimentally and mechanistically actionable hypotheses for further electrolyte discovery.

To assess whether the observed trends in electronic, charge, and physical descriptors were systematically associated with halogen substitution pattern rather than parent-scaffold identity alone, we fit linear models of the form
\begin{equation}
y = \beta_0 + \beta_1 \times \mathrm{HalogenType} + \beta_2 \times \mathrm{Parent} + \varepsilon ,
\end{equation}
where \texttt{HalogenType} was treated as a categorical predictor and \texttt{Parent} was included as a fixed effect to account for parent-specific baseline differences. 
For each response variable, we report the coefficient of determination ($R^2$), adjusted coefficient of determination (adjusted $R^2$), and the joint significance of the halogen-type terms after controlling for parent effects.

Across all eleven response variables, the joint effect of halogen type remained statistically significant after accounting for parent identity (\cref{tab:si:mixed_halogenation_model_summary}).
Model fit was strongest for the halogen charge descriptors, particularly the minimum and mean halogen \ac{NPA} charges, for which the combined halogen-type plus parent model explained 98.85\% and 96.91\% of the observed variance, respectively.
Strong explanatory power was also observed for the frontier orbital energies and gap, with $R^2$ values of 0.8408 for \ac{HOMO} energy, 0.8834 for \ac{LUMO} energy, and 0.8715 for the HOMO--LUMO gap. Among the physical-property descriptors, boiling point, flash point, and melting point were all significantly described by the same low-dimensional model, with $R^2$ values between 0.8271 and 0.9186. 
By contrast, the halogen-localization fractions of the frontier orbitals showed comparatively lower, though still substantial, explanatory power ($R^2 = 0.7146$ for \ac{HOMO} halogen fraction and $R^2 = 0.7288$ for \ac{LUMO} halogen fraction), consistent with greater scaffold-specific variation in orbital localization than in the corresponding orbital energies themselves.

Importantly, the reported $R^2$ values quantify variance explained by the full model, \emph{i.e.}, by halogen type together with parent fixed effects.
The statistical significance of halogen substitution pattern specifically is therefore best captured by the mixed-halogen type test reported in~\cref{tab:si:mixed_halogenation_model_summary}.
These results indicate that halogen identity and mixing pattern remain significant sources of variation even after controlling for parent-scaffold effects.

\begin{table}[htbp]
\centering
\caption{Summary of linear fixed-effect models including halogen type and parent fixed effects. The reported $p$ values correspond to the joint significance test for the halogen-type terms after accounting for parent effects.}
\label{tab:si:mixed_halogenation_model_summary}
\begin{tabular}{lcccccc}
\hline
Property & $n$ & $R^2$ & Adjusted $R^2$ & $F_{\mathrm{halo}}$ & $p_{\mathrm{halo}}$ \\
\hline
Halogen \ac{NPA} minimum charge    & 60 & 0.9885 & 0.9821 & 2409.88 & $3.37\times10^{-19}$ \\
Halogen \ac{NPA} mean charge       & 60 & 0.9691 & 0.9519 & 1042.91 & $7.73\times10^{-17}$ \\
Flash point                   & 60 & 0.9186 & 0.8736 & 9.07    & $3.90\times10^{-4}$ \\
Heat Capacity (Cv)         & 60 & 0.8947 & 0.8274 & 53.83  & $1.96\times10^{-8}$ \\
\ac{LUMO} energy                   & 60 & 0.8834 & 0.8189 & 51.28   & $1.84\times10^{-8}$ \\
HOMO--LUMO gap                & 60 & 0.8715 & 0.8004 & 63.33   & $4.95\times10^{-9}$ \\
Boiling point                 & 60 & 0.8585 & 0.7803 & 673.74  & $1.31\times10^{-15}$ \\
HOMO energy                   & 60 & 0.8408 & 0.7528 & 93.93   & $4.14\times10^{-10}$ \\
Melting point                 & 60 & 0.8271 & 0.7315 & 5.35    & $4.62\times10^{-3}$ \\
LUMO halogen fraction         & 60 & 0.7288 & 0.5789 & 2327.30 & $4.23\times10^{-19}$ \\
HOMO halogen fraction         & 60 & 0.7146 & 0.5569 & 221.03  & $1.75\times10^{-12}$ \\
\hline
\end{tabular}
\end{table}

\begin{figure}[h!]
    \centering
    \includegraphics[width=.9\textwidth]{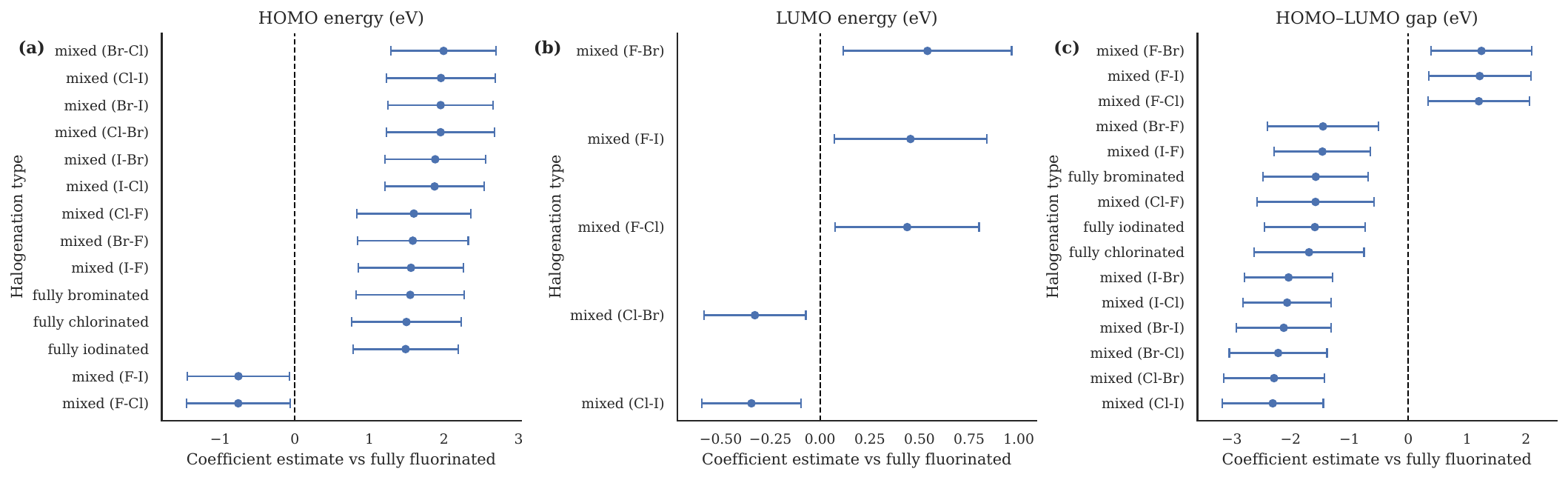}
    \caption{
        Coefficient estimates from parent-fixed-effects linear models for electrochemical descriptors of the mixed-halogenation validation set, shown relative to the fully fluorinated reference.
        a) \ac{HOMO} energy b) \ac{LUMO} energy and c) HOMO--LUMO gap.
        Points denote coefficient estimates for halogenation types whose effects were statistically significant in the fitted models, and horizontal bars indicate 95\% confidence intervals.
        Positive and negative coefficients indicate increases and decreases, respectively, relative to the fully fluorinated analogue of the same parent scaffold.
        For the mixed-halogenation cases, the halogen named first is the dominant substituent.
        For example, Br-Cl has more Bromine than Chlorine while Cl-Br is the inverse.
        \label{fig:si:mixed_halogenation_electrochemical}
    }
\end{figure}

We next examined how halogenation pattern influences electronic stability descriptors using the parent-fixed-effects linear models summarized in~\cref{fig:si:mixed_halogenation_electrochemical}. In each panel, coefficients are reported relative to the fully fluorinated analogue, so that the plotted values isolate the directional effect of halogenation type after accounting for scaffold-specific baseline differences. We display only those halogenation types for which the fitted coefficient was statistically significant.

For \ac{HOMO} energy (\cref{fig:si:mixed_halogenation_electrochemical}a), the clearest stabilizing effects are observed for mixed halogenation patterns in which fluorine remains the majority halogen, particularly mixed (F--Cl) and mixed (F--I), both of which shift the \ac{HOMO} downward relative to the fully fluorinated reference. 
In contrast, halogenation patterns lacking a fluorine-majority character generally shift the \ac{HOMO} upward, often substantially, indicating reduced resistance to electrooxidation. 
~\Cref{fig:si:mixed_halogenation_electrochemical} therefore suggests that partial substitution of fluorine can, in selected cases, preserve or even improve oxidative robustness without changing the underlying molecular scaffold, but that this effect is not general to all mixed-halogenation patterns. 
Once fluorine ceases to dominate the halogen environment, the trend reverses and oxidative stability deteriorates relative to the fully fluorinated baseline. This provides an important nuance: mixed halogenation is not a universal route to lowering the HOMO, but rather a conditional strategy whose success depends strongly on the retained fluorine content.

For \ac{LUMO} energy (\cref{fig:si:mixed_halogenation_electrochemical}b), fewer halogenation types remain significant after parent correction, but the same broad bifurcation persists. Mixed halogenation patterns with fluorine as the dominant halogen, namely mixed (F--Cl), mixed (F--Br), and mixed (F--I), exhibit positive coefficients relative to the fully fluorinated reference, whereas mixed systems without fluorine, such as mixed (Cl--Br) and mixed (Cl--I), exhibit negative coefficients. Under the conventional interpretation of a higher \ac{LUMO} as greater resistance to reduction, these results indicate that partially retaining fluorine while introducing a second halogen can, for some scaffolds, further increase reductive robustness beyond the fully fluorinated case, whereas complete loss of fluorine drives the opposite behaviour. 
This distinction is especially relevant for battery chemistries operating near highly reducing potentials, where even modest upward shifts in the \ac{LUMO} may expand the accessible electrolyte design space.

The HOMO--LUMO gap (\cref{fig:si:mixed_halogenation_electrochemical}c) integrates these directional trends into a single measure of the electronic stability window. Here again a clear separation emerges. 
Mixed halogenation patterns in which fluorine remains dominant produce positive coefficients, corresponding to wider gaps than the fully fluorinated analogue, whereas fully chlorinated, fully brominated, fully iodinated, and mixed systems lacking fluorine all decrease the gap. 
The gap analysis therefore reinforces the picture obtained from the \ac{HOMO} and \ac{LUMO} models separately: modest or asymmetric replacement of fluorine can in some cases improve the electronic stability window, but progressive or complete defluorination systematically narrows it.

Taken together, the three panels show that the performance penalty associated with departing from complete fluorination is highly structured rather than monotonic. Among the halogenation types that underperform the fully fluorinated analogue, those that still retain fluorine generally remain closer to the fluorinated baseline than those in which fluorine is absent altogether. 
Conversely, the only statistically supported improvements over the fully fluorinated reference arise in mixed systems where fluorine remains the majority halogen. 
In this sense, the models identify mixed halogenation not as a blanket replacement for fluorination, but as a chemically constrained tuning strategy that can preserve or enhance electrochemical robustness when applied within a fluorine-dominant regime.

\begin{figure}[h!]
    \centering
    \includegraphics[width=.9\textwidth]{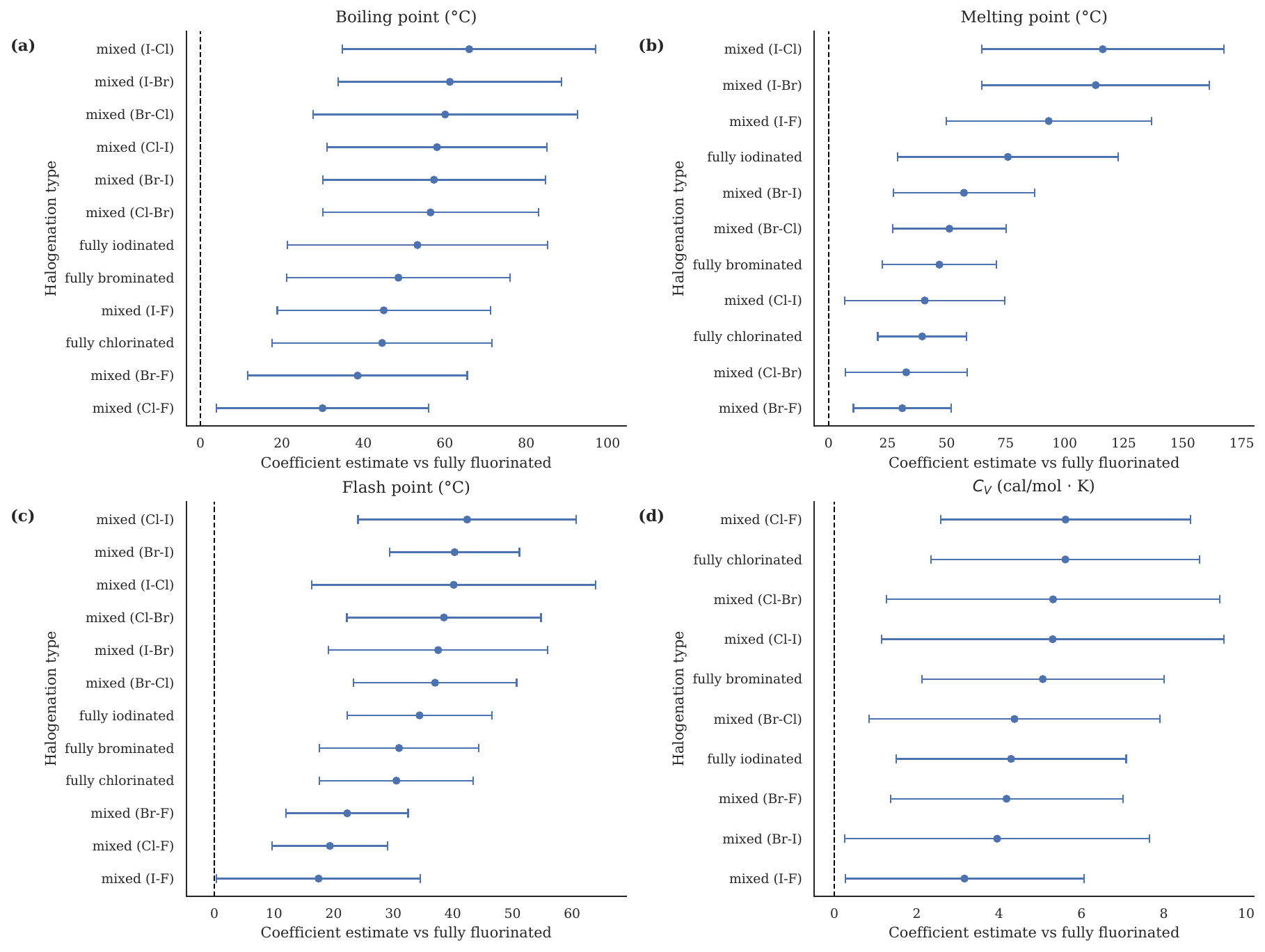}
    \caption{
        Coefficient estimates from parent-fixed-effects linear models for thermophysical descriptors of the mixed-halogenation validation set, shown relative to the fully fluorinated reference. a) Boiling point b) Melting point c) Flash point and d) Constant-volume heat capacity, $C_V$. Points denote coefficient estimates for halogenation types whose effects were statistically significant in the fitted models, and horizontal bars indicate 95\% confidence intervals. Positive and negative coefficients indicate increases and decreases, respectively, relative to the fully fluorinated analogue of the same parent scaffold.
        \label{fig:si:mixed_halogenation_thermochemical}
}
\end{figure}

We next examined how halogenation pattern affects thermophysical descriptors using the parent-fixed-effects linear models shown in \cref{fig:si:mixed_halogenation_thermochemical}. 
As in the electrochemical analysis, coefficients are reported relative to the fully fluorinated analogue, so that the plotted values isolate the directional effect of halogenation type after accounting for scaffold-specific baseline differences. 
Only statistically significant halogenation coefficients are shown.

For boiling point (\cref{fig:si:mixed_halogenation_thermochemical}a), all statistically significant coefficients are positive, indicating that departure from complete fluorination is consistently associated with higher boiling points in the analyzed set. 
Notably, the largest boiling-point increases are observed for mixed halogenation patterns composed exclusively of the heavier halogens, particularly Cl-I, I-Br, and Br-Cl containing systems, whereas fluorine-containing mixed systems exhibit more modest increases. 
Fully brominated and fully iodinated analogues also increase boiling point relative to the fully fluorinated baseline, but several mixed heavy-halogen patterns exceed the corresponding fully halogenated cases. 
This suggests that the observed boiling-point elevation is not simply a trivial mass effect, but likely a broader combination of increased molecular polarizability and altered intermolecular interactions. 
From an electrolyte-design perspective, higher boiling points expand the upper-temperature liquid operating window by reducing the tendency toward vaporization.

The melting-point trends in~\cref{fig:si:mixed_halogenation_thermochemical}b show the same sign, but a more structured ordering. 
All significant coefficients are again positive, implying that mixed or heavier halogen substitution generally raises the melting point relative to the fully fluorinated scaffold. 
The largest upward shifts are seen for I-Cl and I-Br containing mixed systems, followed by mixed I-F and fully iodinated analogues. 
In contrast to boiling point, however, mixed halogenation does not universally outperform the corresponding fully halogenated endmember. 
For bromine- and iodine-rich systems, some mixed variants exceed the fully brominated or fully iodinated cases, whereas the chloride-containing series is more heterogeneous. 
This distinction is important for battery applications that rely on liquid electrolytes over broad temperature windows: although the same chemistry that suppresses volatility can also improve high-temperature robustness, increasing the melting point narrows the low-temperature regime over which the electrolyte remains fluid.

Flash point, shown in~\cref{fig:si:mixed_halogenation_thermochemical}c, provides a complementary measure of volatility and ignition resistance. 
Here too, all statistically significant coefficients are positive, indicating that mixed or heavier halogen substitution increases flash point relative to the fully fluorinated baseline. 
The largest increases again occur in the heavier mixed-halogen families, especially Cl-I, Br-I, and I-Cl containing systems, whereas fluorine-containing mixed systems show smaller but still positive shifts. 
Several mixed heavy-halogen patterns also exceed the fully brominated and fully iodinated analogues, consistent with the strong empirical connection between flash point and boiling point in organic liquids~\cite{HWWFlammabilityLiIonBattery2015}. 
In practical terms, these trends suggest that mixed halogenation may offer a route to reduced flammability risk, especially when heavier halogens are introduced without fully abandoning the parent scaffold.

Finally, the constant-volume heat-capacity coefficients in ~\cref{fig:si:mixed_halogenation_thermochemical}d are uniformly positive, indicating that all statistically significant halogenation patterns increase $C_V$ relative to the fully fluorinated reference. 
In contrast to the boiling- and flash-point panels, the largest increases are concentrated in chloride-containing systems, including mixed Cl-F, fully chlorinated, mixed Cl-Br, and mixed Cl-I. 
Brominated and iodinated systems also show positive shifts, but these are generally smaller. 
Higher heat capacity means that a larger thermal input is required to produce a given temperature rise, so these changes are consistent with improved buffering against transient thermal excursions, although the practical benefit in a working battery will depend on the full electrolyte formulation and cell architecture.

Taken together, the thermophysical analysis indicates that mixed halogenation is associated with systematic upward shifts in boiling point, melting point, flash point, and heat capacity relative to fully fluorinated analogues. 
The strongest gains in boiling point and flash point are concentrated in mixed systems built from the heavier halogens, whereas the largest heat-capacity increases are more characteristic of chloride-containing chemistries. 
These results point to a genuine design trade-off: mixed halogenation can improve high-temperature retention and reduce volatility or flammability, but often at the cost of elevated melting point and therefore diminished low-temperature fluidity. 

A holistic interpretation of both the electrochemical and thermophysical results indicates that mixed halogenation is a viable design lever for tuning both the electrochemical and thermophysical properties of electrolyte molecules in ways that may improve battery performance. 

\subsubsection{Creativity Metrics for Generated Molecules}

\begin{figure}[h!]
    \centering
    \includegraphics{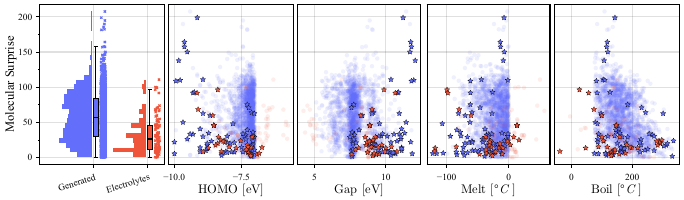}
    \caption{
        \label{fig:si:screening_creativity}
        ``Surprising'' generated molecules yielded improved performance characteristics, for example, a wider electrochemical stability window (\acs{HOMO-LUMO} Gap).
    }
\end{figure}

As shown in \cref{fig:si:screening_creativity}, we found the generated molecules to have a higher average molecular surprise (\cref{eq:molecular_surprise}), or entropy, than our reference set of electrolytes.
This indicates that FASMIFRA\cite{BTMolecularGenerationFast2021} generates molecules that are, on average, more surprising or novel\cite{VarMathematicalLimitTheorems2019} to MIST-28M than those in the reference set.
While a high molecular surprise does not appear to be a prerequisite for advancing the Pareto front, access to surprising molecules does help.
Of the generated molecules, 98.7\% did not occur in our reference set of electrolytes.
This is after de-duplicating generated molecules within and across ranks.

To evaluate the creativity of the generator, we estimated the R\'enyi entropy \(H_{\gamma}\) of the embedding vectors for both generated and reference molecules.
Given \(n\) molecular embedding vectors \(X \in \mathcal{R}^d\) sampled from a probability density \(f\), the estimated R\'enyi entropy is:
\begin{equation}
    H_{\gamma} = \frac{1}{1 - \gamma}\left( \ln L_n  - \gamma \ln n - \ln \beta \right)
\end{equation}
where \(\gamma = (d - 1)/d\), \(\beta\) is a fixed constant, and \(L_n\) is the length of the minimum spanning tree of the embedding vectors\cite{VRSExplainingCreativeArtifacts2020,HBMGApplicationsEntropicSpanning2002}.
As \(\beta\) is defined to be \(L_n / n^{\gamma}\) in the limit of \(n \to \infty\), a numerical estimate of \(H_{\gamma}\) is impractical.
Notably, \(\beta\) is a fixed constant independent of the generating function \(f\)\cite{VRSExplainingCreativeArtifacts2020,HBMGApplicationsEntropicSpanning2002}.
We thus set \(\beta = 1\), dropping it from our estimation of R\'enyi entropy \(\hat{H}_{\gamma}\):
\begin{equation}
    \hat{H}_{\gamma} = \frac{1}{1 - \gamma}\left( \ln L_n  - \gamma \ln n \right)
    \label{eq:renyi_entropy}
\end{equation}
As shown in \cref{tab:creativity_metrics}, the generated molecules have a lower \(\hat{H}_{\gamma}\) than the reference electrolytes.
The ordering of \(H\) and \(\hat{H}_{\gamma}\) reverses because these two metrics measure slightly different phenomena:
\(\hat{H}_{\gamma}\) captures the entropy of the generation process, while \(H\) evaluates the entropy of the generated molecules.
This indicates that although the generated molecules explore regions of embedding space that MIST-28M considers surprising, the generation process itself is low-entropy.
That is, the generated molecules are clumped within embedding space.
This suggests that while fragment-based generators, such as FASMIFRA\cite{BTMolecularGenerationFast2021}, can produce surprising molecules, their generation process has limited exploration of chemical space.
This is consistent with our observation of decaying generation efficiency (\cref{fig:screening_gen_trace}), as the generator rapidly explores and then exhausts a finite region of chemical space.

\begin{table}[ht!]
\centering
\begin{tabular}{@{}lcccc@{}}
\toprule
             & n     & \(L_n\)  & \(\hat{H}_{\gamma}\) & H    \\ \midrule
Generated    & 1,936 & 7,600.45 & 707.8                & 57.8 \\
Electrolytes & 105   & 681.72   & 962.4                & 31.4 \\ \bottomrule
\end{tabular}
\caption{
    \label{tab:creativity_metrics}
    Estimated R\'enyi entropy \(\hat{H}_{\gamma}\) (\cref{eq:renyi_entropy}) and average molecular surprise \(H\), akin to Shannon entropy, for generated molecules and reference electrolytes.
    We observe a reversal in the ordering of \(H\) and \(\hat{H}_{\gamma}\) because the two metrics measure different phenomena:
    \(\hat{H}_{\gamma}\) captures the entropy of the generation process, while \(H\) evaluates the entropy of the generated molecules.
}
\end{table}

\subsection{Screening for Lithium Air Battery Electrolyte Solvents}
\label{sec:si:lithium_air_screening}

Lithium--air (Li--O$_2$) batteries, which generate electricity by coupling lithium oxidation at the anode with oxygen reduction from the air at the cathode, offer exceptionally high theoretical energy density, but impose severe demands on electrolyte performance~\cite{khetan2014identifying}.
These batteries present a stringent problem for electrolyte design as solvent candidates must simultaneously satisfy constraints on phase behaviour, electrochemical stability, and solvation~\cite{khetan2014identifying,khetan2015trade}.
We use lithium--air solvent discovery as a focused test of whether \ac{MIST} can translate literature-derived design rules into a scalable molecular screening workflow. 
In this setting, \ac{MIST}-predicted thermophysical, electronic, and solvation-related descriptors are combined into a small set of mechanistically motivated screening criteria serving as proxies for the various performance demands of the system.

\subsubsection{Bounds}
In this section, we explain the derivation and rational for the bounds used to discover Li--O$_2$ battery electrolyte solvent candidates with \ac{MIST}.

In aprotic Li--O$_2$ batteries, lithium peroxide (\ce{Li2O2}) can form in two ways. 
It can grow directly on the electrode surface, or it can form through a solution-based route where an intermediate species, \ce{LiO2^*}, dissolves into the electrolyte as \ce{Li+} and \ce{O2^-}.
The solution-based pathway is preferred because it leads to lithium peroxide (\ce{Li2O2}) forming as larger particles instead of a thin coating on the electrode. 
This is beneficial because \ce{Li2O2} is electrically insulating, so the thin coating can block the electrode surface and shut down the reaction early. 
By avoiding this surface blockage, the battery can discharge more fully and store more energy~\cite{aetukuri2015solvating,mccloskey2012limitations,luntz2013tunneling}. 
We therefore treat solution-mediated growth as the first screening bound and require the reduced-order free-energy descriptor used in our workflow to satisfy:
\begin{equation}
-6.46~\mathrm{eV} \le \Delta G_{\mathrm{sol}}(\beta,E_T^N) \le -0.35~\mathrm{eV}
\label{eq:li_air_solution_window}
\end{equation}
Where the implemented relation is evaluated from \ac{MIST}-predicted $\beta$ and $E_T^N$. 
This bound retains solvents that lie inside the admissible window for solution-mediated peroxide formation.

The bound (\cref{eq:li_air_solution_window}) is motivated by the solution-mediated discharge mechanism in aprotic Li--O$_2$ batteries, in which the adsorbed superoxide intermediate dissolves into the electrolyte instead of undergoing immediate further reduction on the surface:
\begin{equation}
\ce{LiO2^* -> ^* + Li^+_{(sol)} + O2^-_{(sol)}}.
\end{equation}
The corresponding dissolution free energy is
\begin{equation}
\Delta G_{\mathrm{sol}}
=
G(*) + G(\ce{Li^+_{(sol)}}) + G(\ce{O2^-_{(sol)}}) - G(\ce{LiO2^*}).
\label{eq:li_air_dgsol_def}
\end{equation}
Using the equilibria of the \ce{Li+}||\ce{Li} and \ce{O2}||\ce{O2^-} redox couples, this expression can be rewritten as a difference of solvent-dependent ionic free-energy terms plus a solvent-independent offset associated with the surface superoxide reference state. In reduced-order form:
\begin{equation}
\Delta G_{\mathrm{sol}}
\approx
U_{\ce{Li^+/Li}}^{sol} - U_{\ce{O2/O2^-}}^{sol} + C
\label{eq:li_air_dgsol_reduced}
\end{equation}
where \(C\) collects the non-solvent-dependent contributions from the surface thermodynamics. 
This recovers the same qualitative structure used in prior Li--O$_2$ analyses: solution-mediated growth becomes more favorable when the solvent better stabilizes dissolved \ce{Li+} and \ce{O2^-} relative to the adsorbed \ce{LiO2^*} state.
Aetukuri et al. showed that enhanced stabilization of these dissolved intermediates promotes solution-mediated electrochemistry and toroidal \ce{Li2O2} growth~\cite{aetukuri2015solvating}.
While Khetan et al. used a related framework to analyze the trade-off between solution-driven growth and solvent stability~\cite{khetan2015trade}.

The numerical bounds in~\cref{eq:li_air_solution_window} are chosen to enforce an admissible window for solution-mediated peroxide formation. The upper limit, \(-0.35\) eV, follows the benchmark used by Khetan et al. for onset of favorable solution-mediated growth \cite{khetan2015trade}. The lower limit, \(-6.46\) eV, is taken as the bulk precipitation limit used in our workflow; this prevents the dissolved state from being stabilized so strongly that the formation of the peroxide phase becomes an uphill chemical process, and the solution-mediated mechanism is no longer represented consistently within the screening construction.

\begin{figure}
    \centering
    \includegraphics[width=\linewidth]{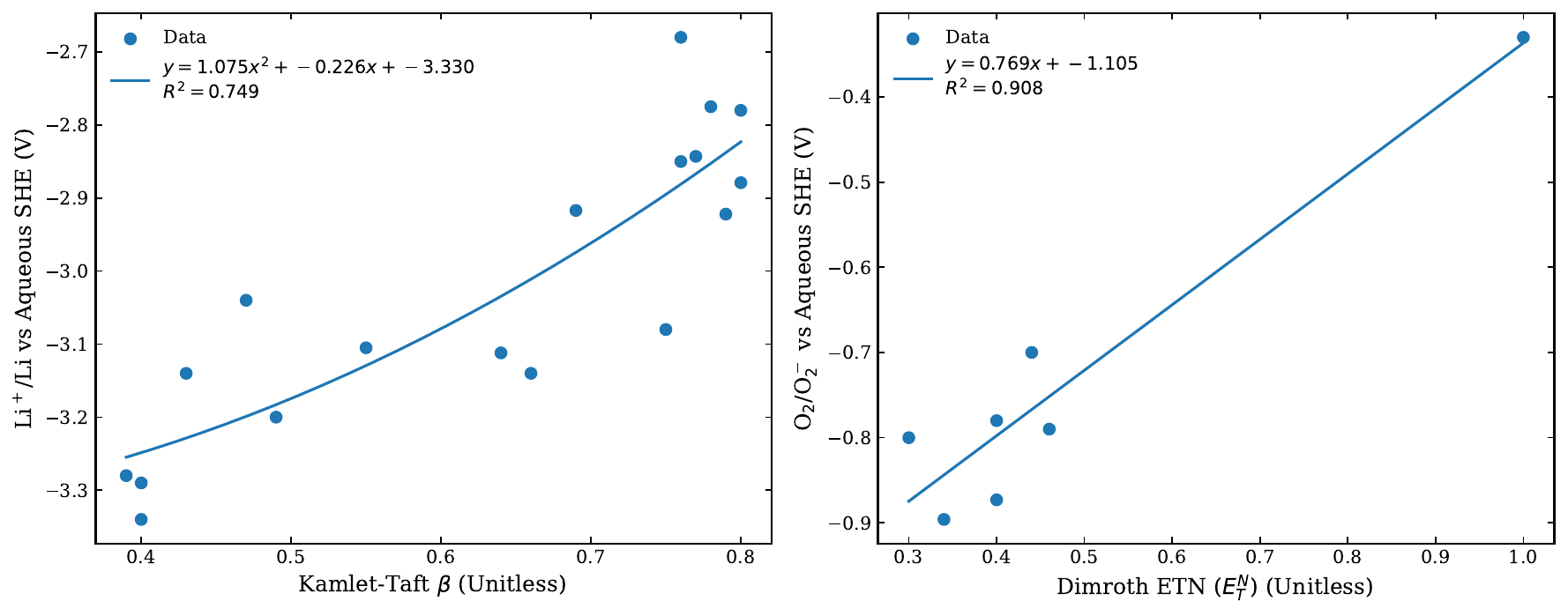}
    \caption{
        (a) $\ce{Li+}||\ce{Li}$ reduction potential fit vs SHE as a function of Kamlet-Taft $\beta$ (b) $\ce{O2}||\ce{O2-}$ reduction potential fit vs SHE as a function of Dimroth-Reichardt $E_T^N$
    }
    \label{fig:Li_air_beta_dimroth_fitting}
    
\end{figure}

To evaluate~\cref{eq:li_air_solution_window} across large molecular libraries, we replace the original donor/acceptor-number picture with MIST-predicted solvent descriptors. 
Khetan et al. showed that the $\ce{Li+}||\ce{Li}$ half-wave potential is strongly correlated with solvent donor/basicity descriptors, whereas the $\ce{O2}||\ce{O2^-}$ couple is strongly correlated with solvent acceptor/acidity descriptors\cite{khetan2015trade}. 
In our workflow, these roles are approximated using Kamlet--Taft \(\beta\) and normalized Dimroth--Reichardt \(E_T^N\), which serve as practical reduced-order surrogates for solvent basicity and polarity/acidity response, respectively. 
We therefore fit the implemented screening relation as:
\begin{equation}
\Delta G_{\mathrm{sol}}(\beta,E_T^N) =  U_{\ce{Li+}||\ce{Li}}^{sol}(\beta) - U_{\ce{O2}\ce{O2^-}}^{sol}(E_T^N)
\label{eq:li_air_dgsol_fit}
\end{equation}
With the respective fitting functions obtained from literature-derived calibration data for the Li--O$_2$ screening model (\cref{fig:Li_air_beta_dimroth_fitting})~\cite{khetan2014solvent,khetan2015trade}. 
We emphasize that \(\beta\) and \(E_T^N\) are not asserted to be uniquely sufficient descriptors of the full electrolyte problem; rather, they are tractable solvent descriptors that capture the dominant reduced-order trends needed to screen for solution-mediated growth at the molecular scale.

The second screening criterion targets solvent stability against hydrogen-abstraction (H-abstraction) driven degradation, a reaction in which a hydrogen atom is pulled off the solvent molecule.
Khetan et al. showed that a solvent's resistance to adsorption-induced H-abstraction can be rationalized using its acidity on a common \ac{DMSO} scale together with frontier-orbital energetics (\ac{HOMO} energies)~\cite{khetan2014solvent,khetan2015trade,khetan2014identifying}. 
Following that framework, we apply a $pK_a$-based stability screen together with a \ac{HOMO} cutoff, requiring the acidity descriptor and orbital-energy criterion used in our implementation to satisfy:
\begin{equation}
pK_a^{\mathrm{DMSO}} \ge 30
\; \textrm{and} \;
E_{\mathrm{HOMO}} \le -7.01 ~\mathrm{eV}
\end{equation}
The \ac{HOMO} bound is selected based on the \ac{HOMO} energy of \ac{DME}, an effective Li--O$_2$ solvent, calculated using \ac{DFT} at the B3LYP/6-31G(2df,p) level of theory (matching the \ac{HOMO} calculation on which the \ac{MIST} model was fine-tuned).
Together, the above two bounds are used to exclude solvents that are electronically or chemically predisposed to H-abstraction during discharge.

Finally, we retain only molecules that satisfy simple ambient liquid-state constraints:
\begin{equation}
T_{\mathrm{m}} \le 25\,^{\circ}\mathrm{C},
\qquad
T_{\mathrm{b}} \ge 75\,^{\circ}\mathrm{C},
\end{equation}
Ensuring that screened candidates remain liquid under near-ambient operating conditions while avoiding excessively volatile solvents.

The resulting lithium--air workflow therefore screens candidates using one bound on solution-mediated peroxide growth, one stability bound derived from the H-abstraction literature, and one liquid-state filter. 
In this form, the lithium--air case study provides a focused demonstration of how fine-tuned \ac{MIST} models can convert literature-grounded molecular design rules into a scalable solvent-screening engine.

\subsubsection{Pareto Front Molecules}
\label{sec:si:lio2_pareto_front_molecules}

\begin{figure}
    \centering
    \includegraphics[width=\linewidth]{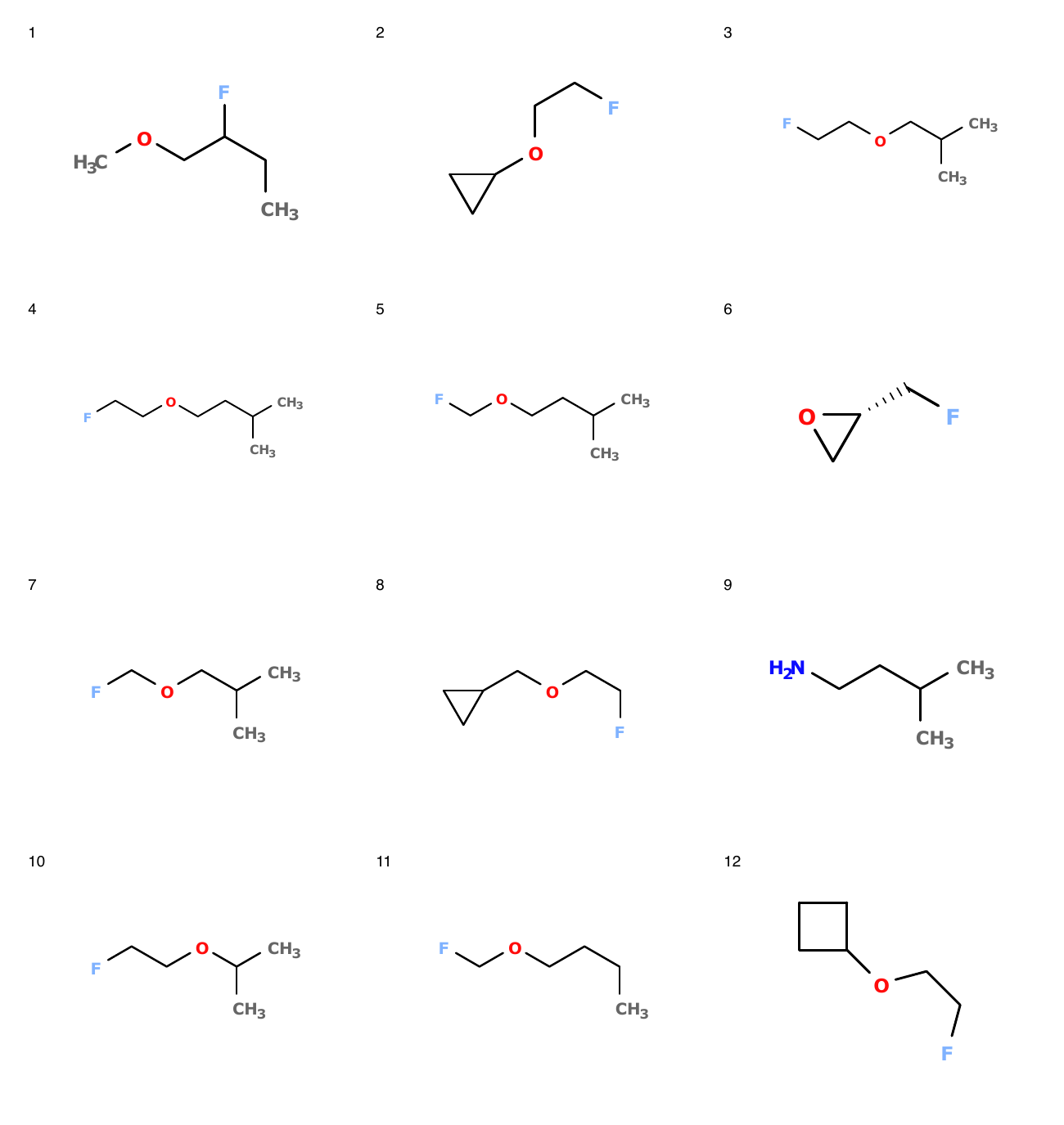}
    \caption{
        \label{fig:lio2_pareto_front_p01}
        Screened Pareto-Front molecules for lithium--air battery solvents identified by the high-throughput screening pipeline presented in the manuscript.
        Figure 1 of 33.
    }
\end{figure}

\begin{figure}
    \centering
    \includegraphics[width=\linewidth]{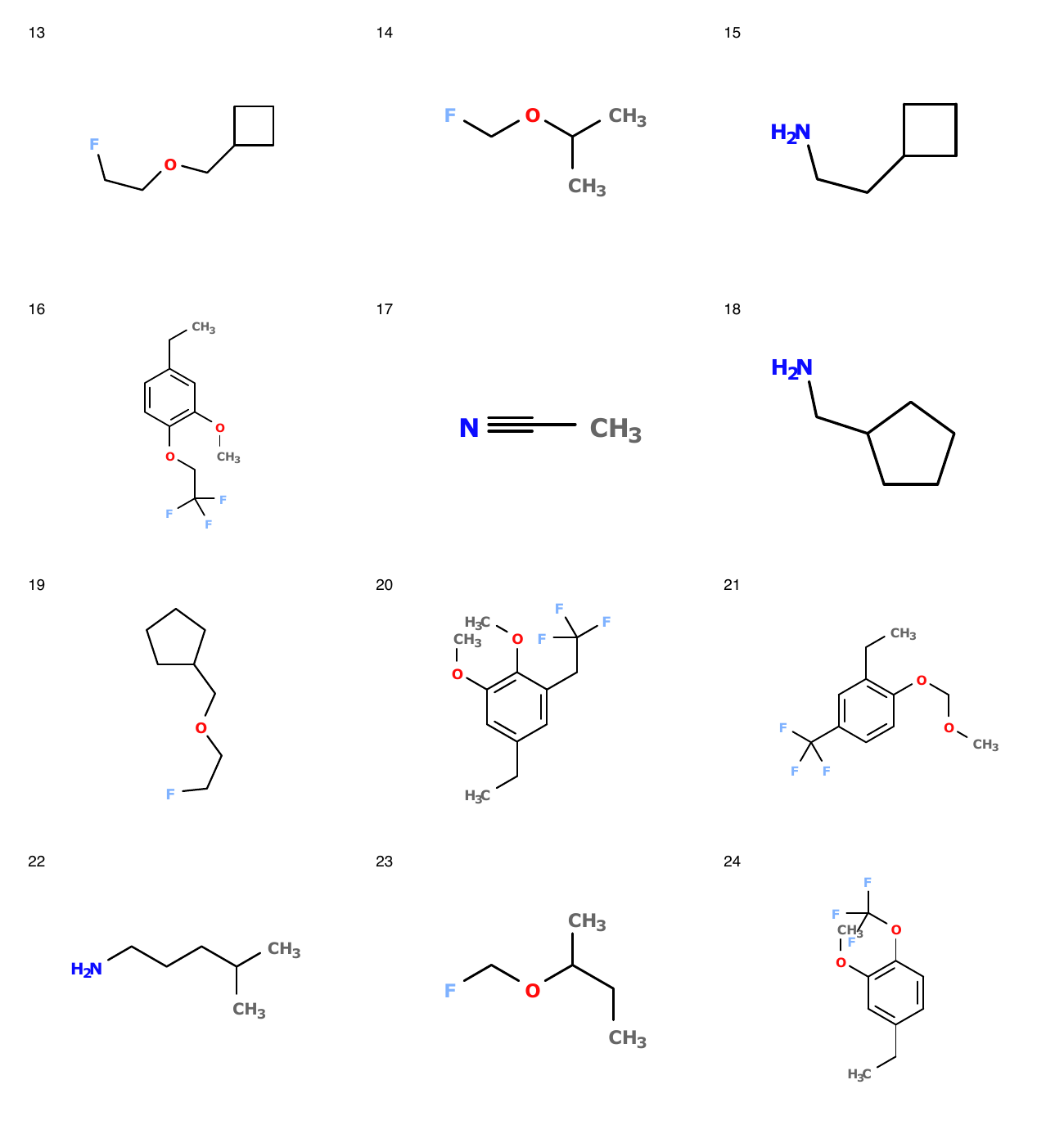}
    \caption{
        \label{fig:lio2_pareto_front_p02}
        Screened Pareto-Front molecules for lithium--air battery solvents identified by the high-throughput screening pipeline presented in the manuscript.
        Figure 2 of 33.
    }
\end{figure}

\begin{figure}
    \centering
    \includegraphics[width=\linewidth]{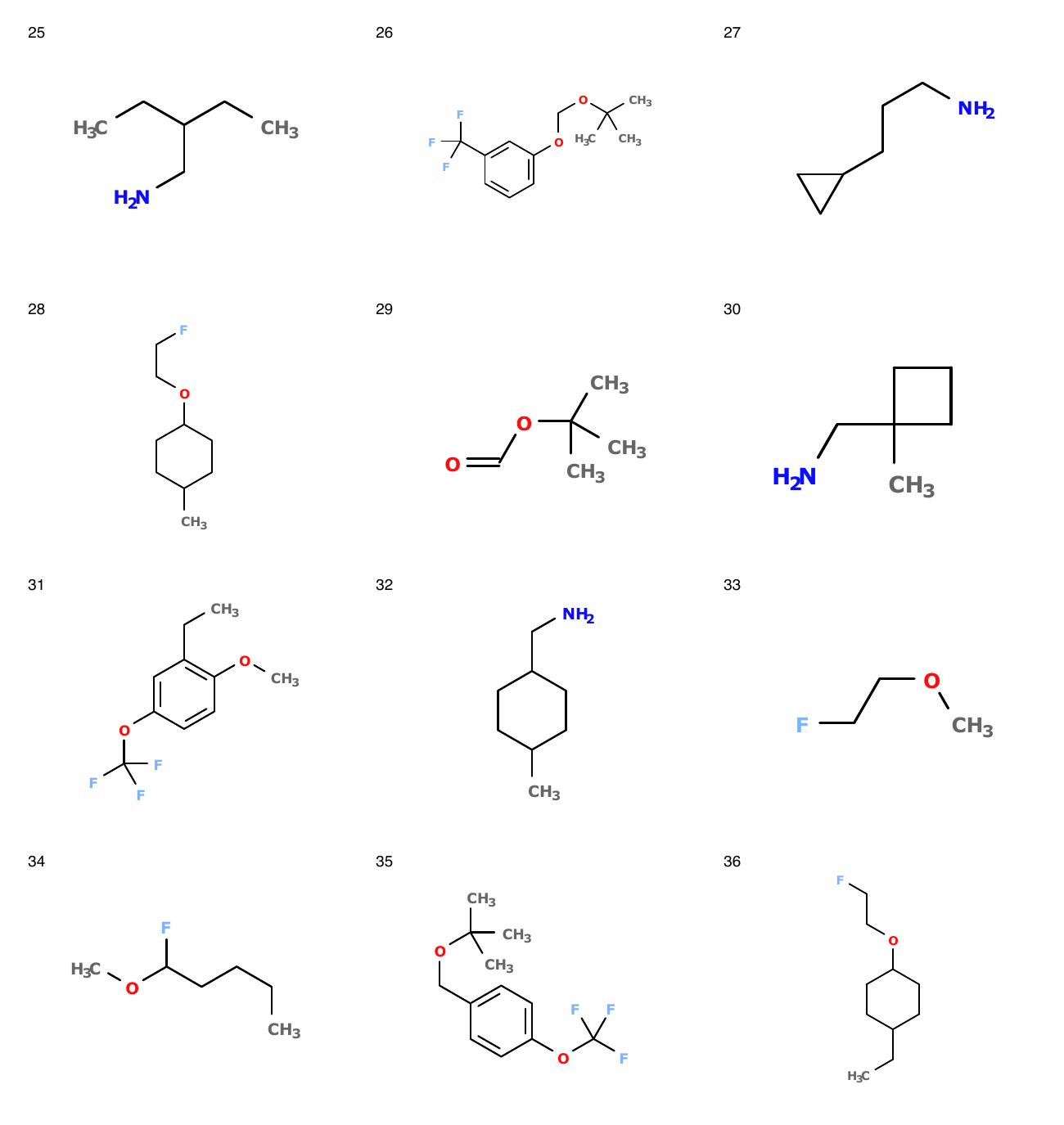}
    \caption{
        \label{fig:lio2_pareto_front_p03}
        Screened Pareto-Front molecules for lithium--air battery solvents identified by the high-throughput screening pipeline presented in the manuscript.
        Figure 3 of 33.
    }
\end{figure}

\begin{figure}
    \centering
    \includegraphics[width=\linewidth]{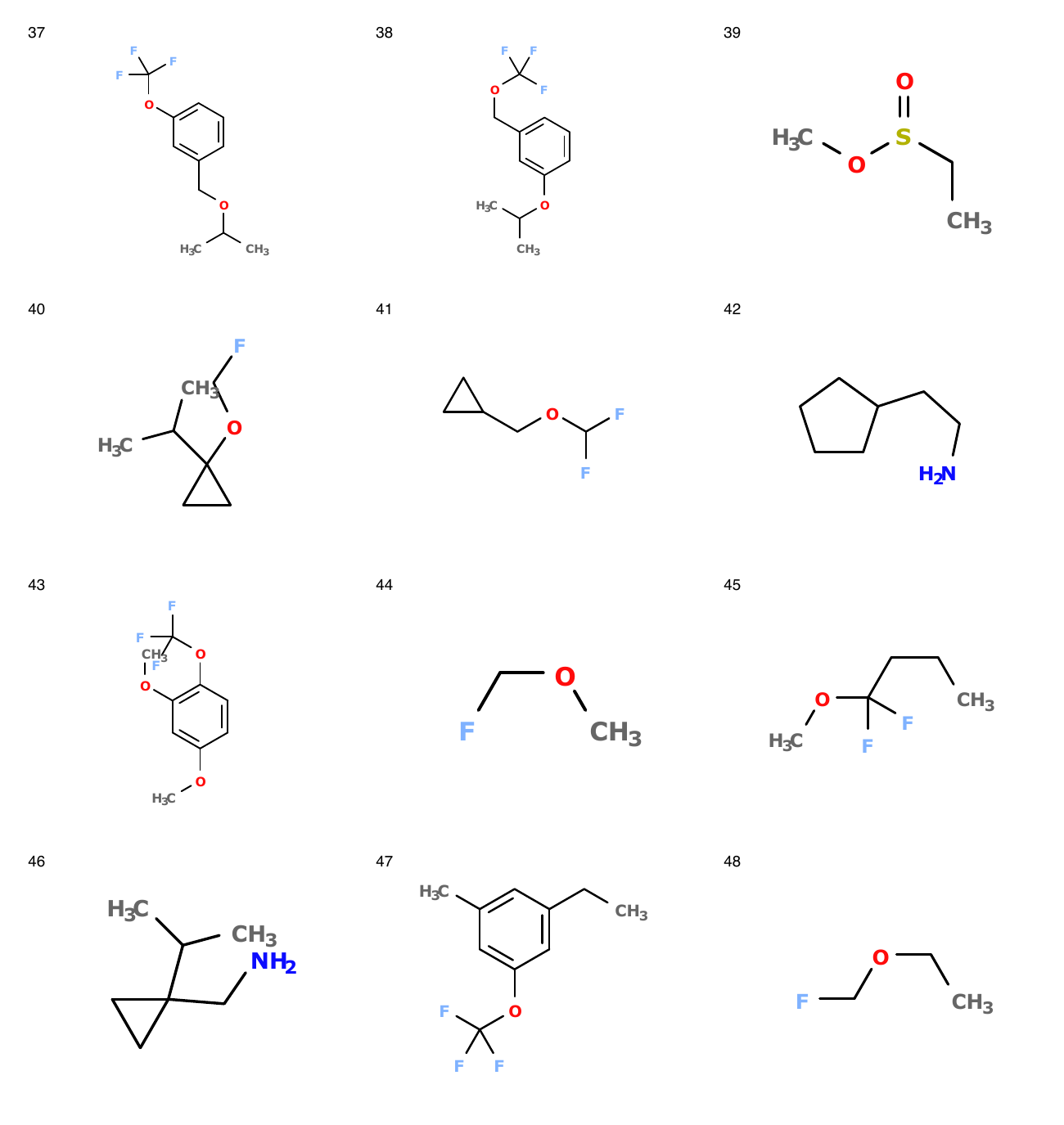}
    \caption{
        \label{fig:lio2_pareto_front_p04}
        Screened Pareto-Front molecules for lithium--air battery solvents identified by the high-throughput screening pipeline presented in the manuscript.
        Figure 4 of 33.
    }
\end{figure}

\begin{figure}
    \centering
    \includegraphics[width=\linewidth]{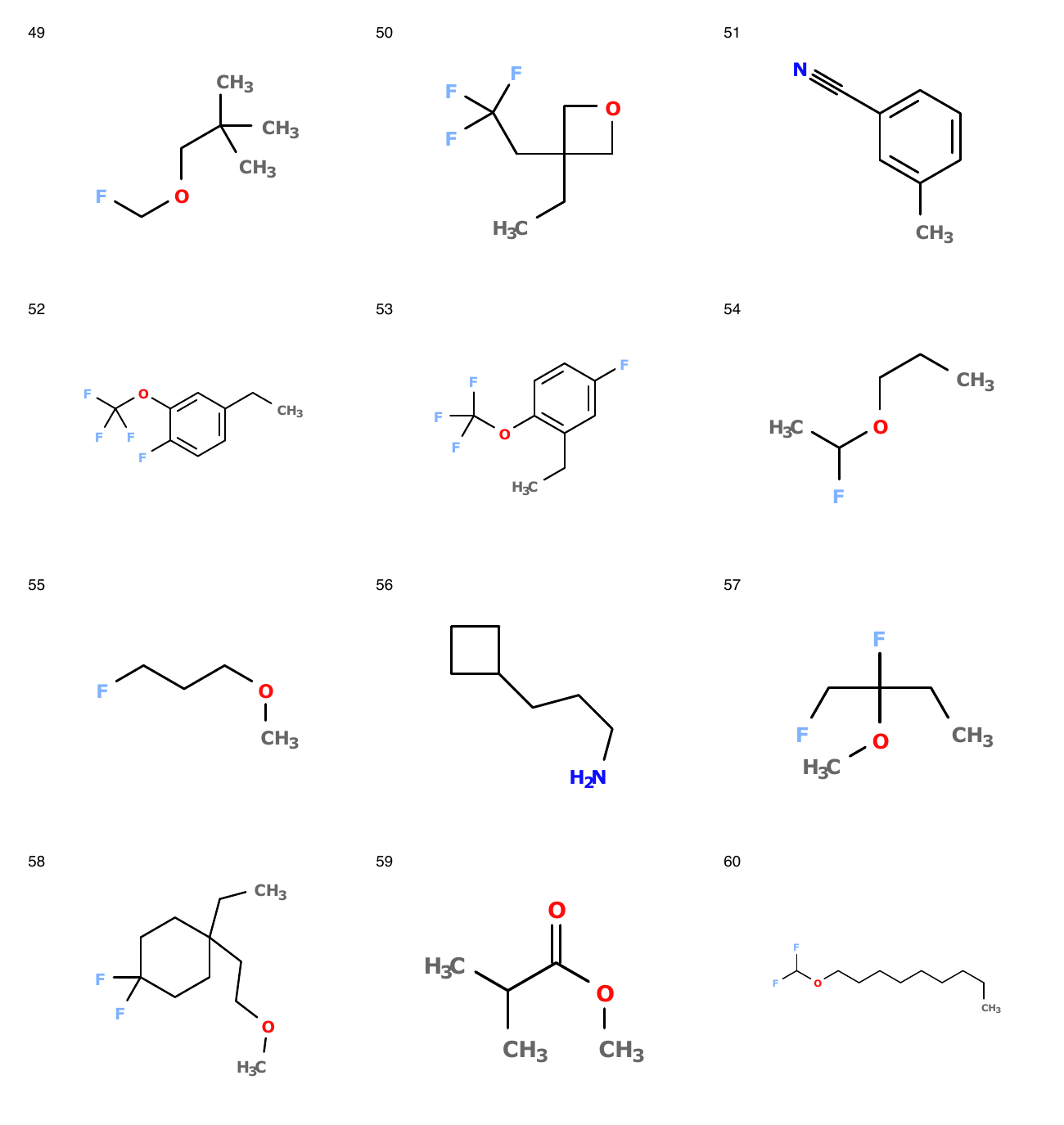}
    \caption{
        \label{fig:lio2_pareto_front_p05}
        Screened Pareto-Front molecules for lithium--air battery solvents identified by the high-throughput screening pipeline presented in the manuscript.
        Figure 5 of 33.
    }
\end{figure}

\begin{figure}
    \centering
    \includegraphics[width=\linewidth]{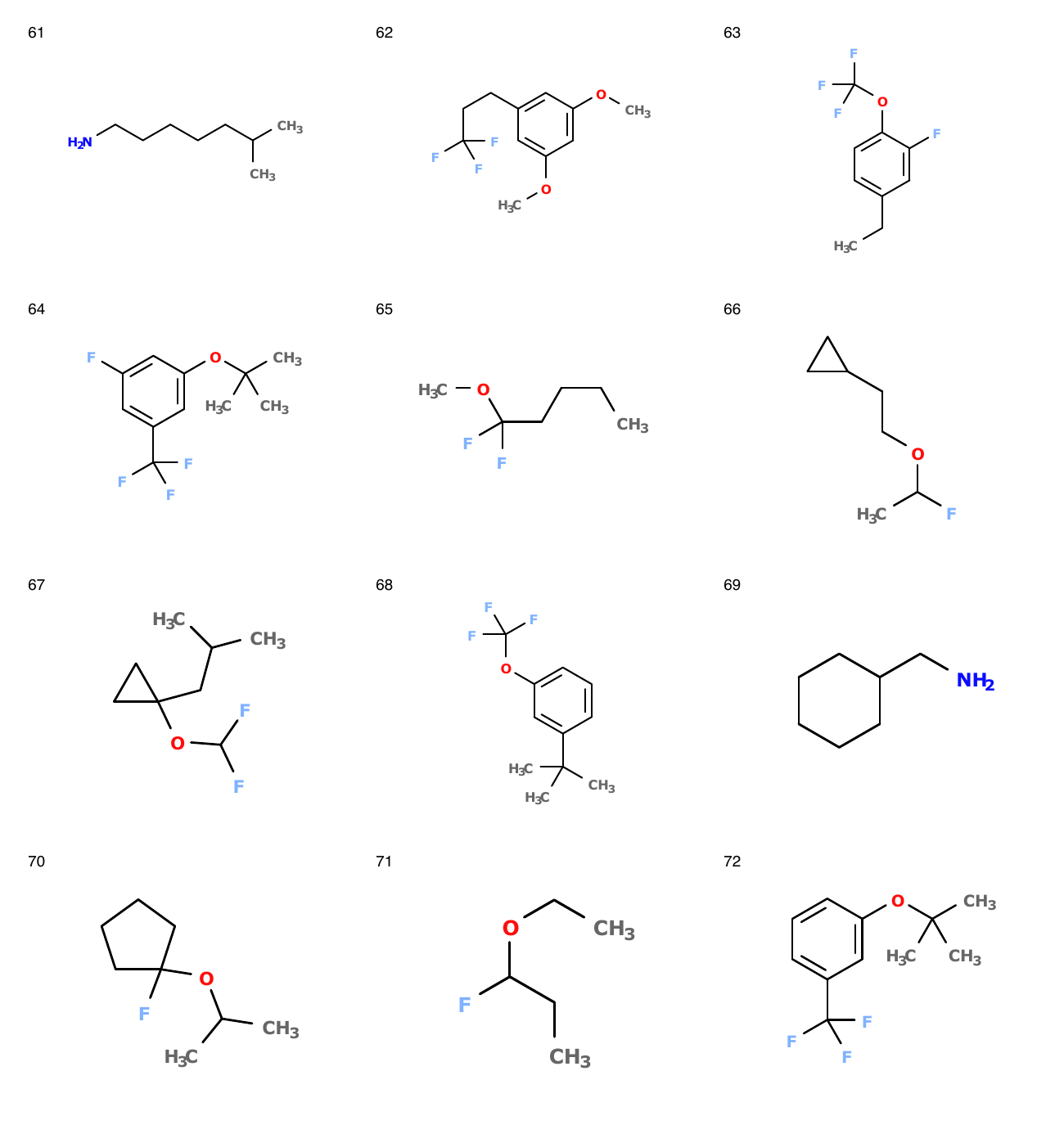}
    \caption{
        \label{fig:lio2_pareto_front_p06}
        Screened Pareto-Front molecules for lithium--air battery solvents identified by the high-throughput screening pipeline presented in the manuscript.
        Figure 6 of 33.
    }
\end{figure}

\begin{figure}
    \centering
    \includegraphics[width=\linewidth]{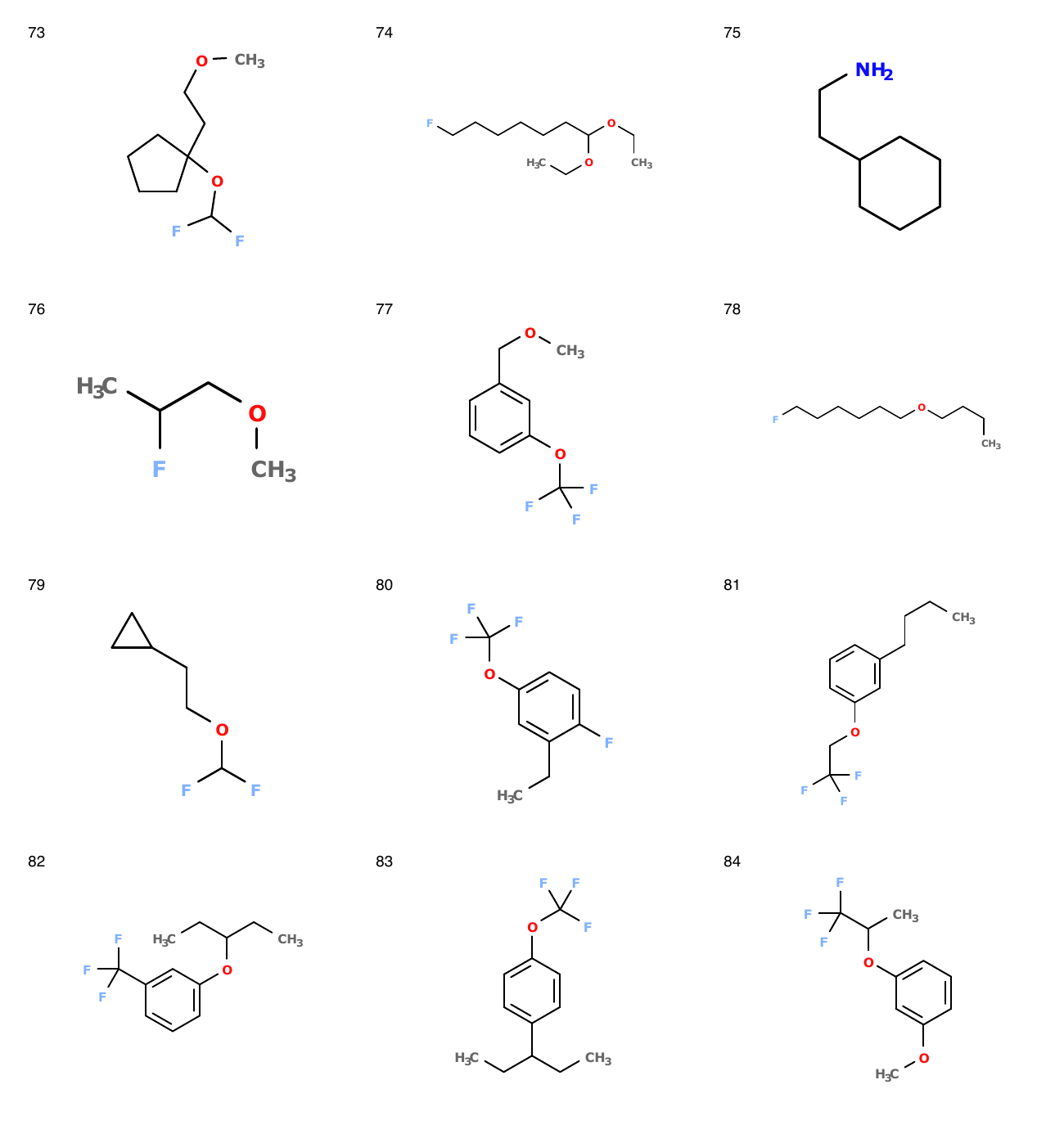}
    \caption{
        \label{fig:lio2_pareto_front_p07}
        Screened Pareto-Front molecules for lithium--air battery solvents identified by the high-throughput screening pipeline presented in the manuscript.
        Figure 7 of 33.
    }
\end{figure}

\begin{figure}
    \centering
    \includegraphics[width=\linewidth]{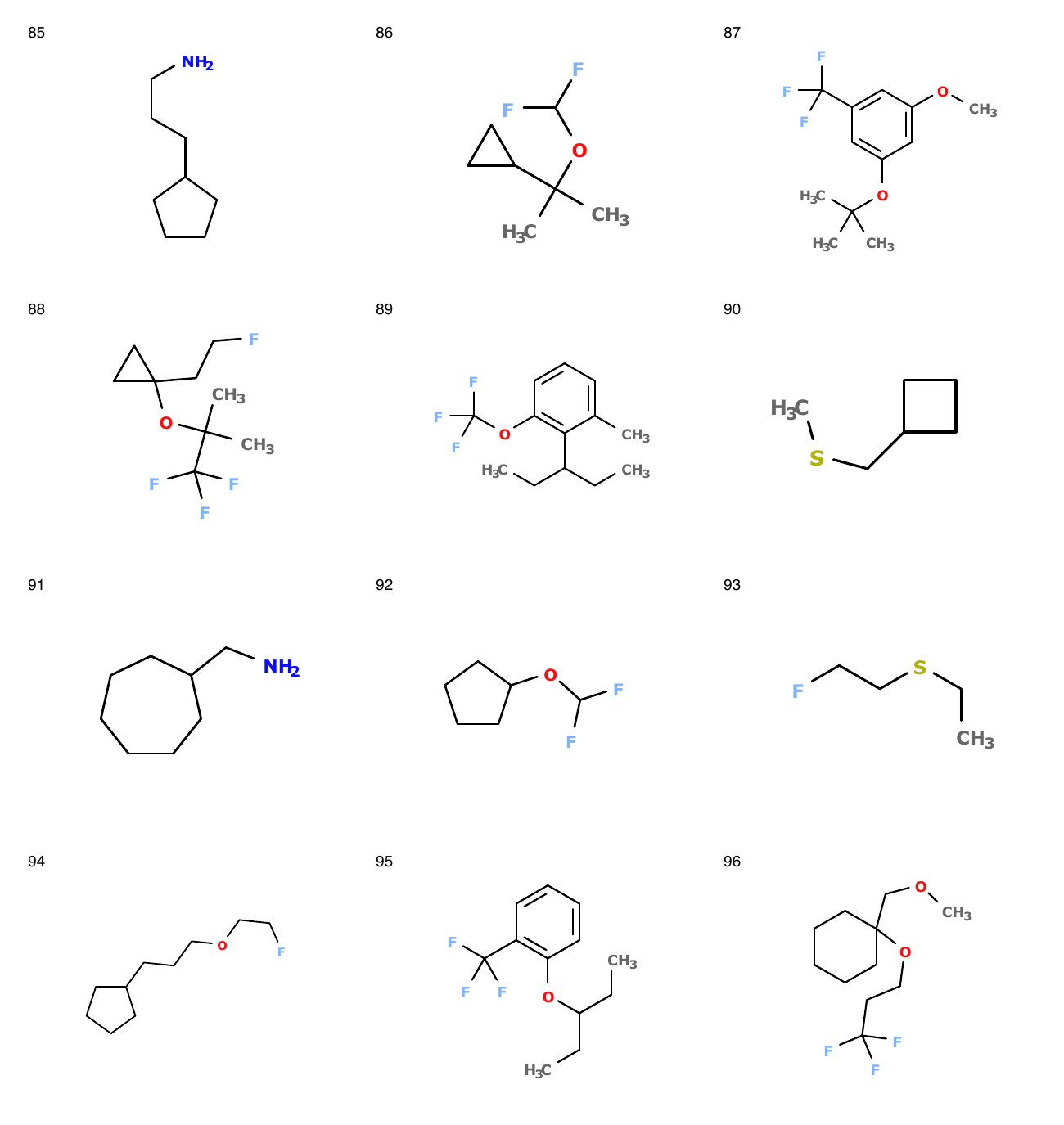}
    \caption{
        \label{fig:lio2_pareto_front_p08}
        Screened Pareto-Front molecules for lithium--air battery solvents identified by the high-throughput screening pipeline presented in the manuscript.
        Figure 8 of 33.
    }
\end{figure}

\begin{figure}
    \centering
    \includegraphics[width=\linewidth]{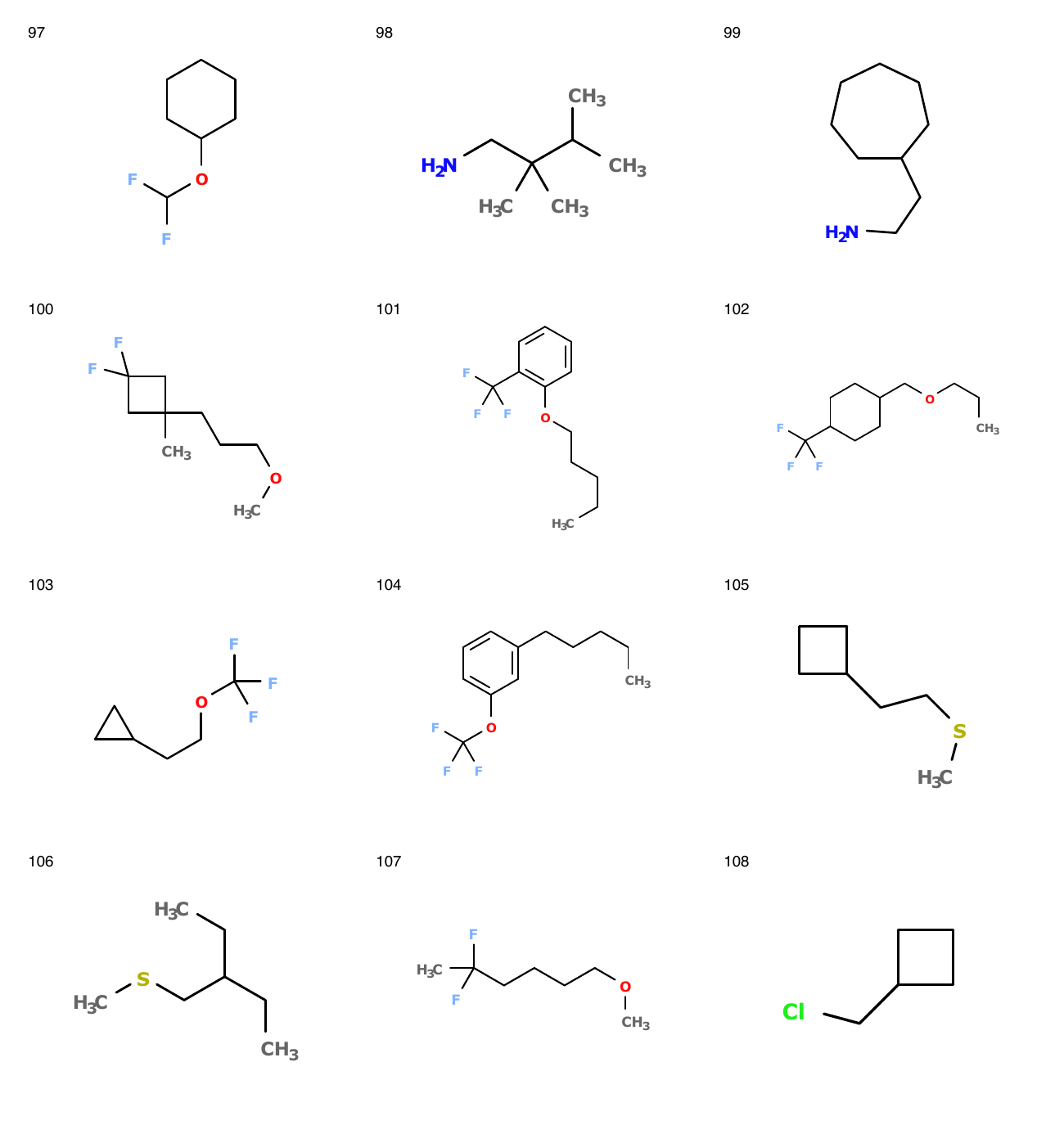}
    \caption{
        \label{fig:lio2_pareto_front_p09}
        Screened Pareto-Front molecules for lithium--air battery solvents identified by the high-throughput screening pipeline presented in the manuscript.
        Figure 9 of 33.
    }
\end{figure}

\begin{figure}
    \centering
    \includegraphics[width=\linewidth]{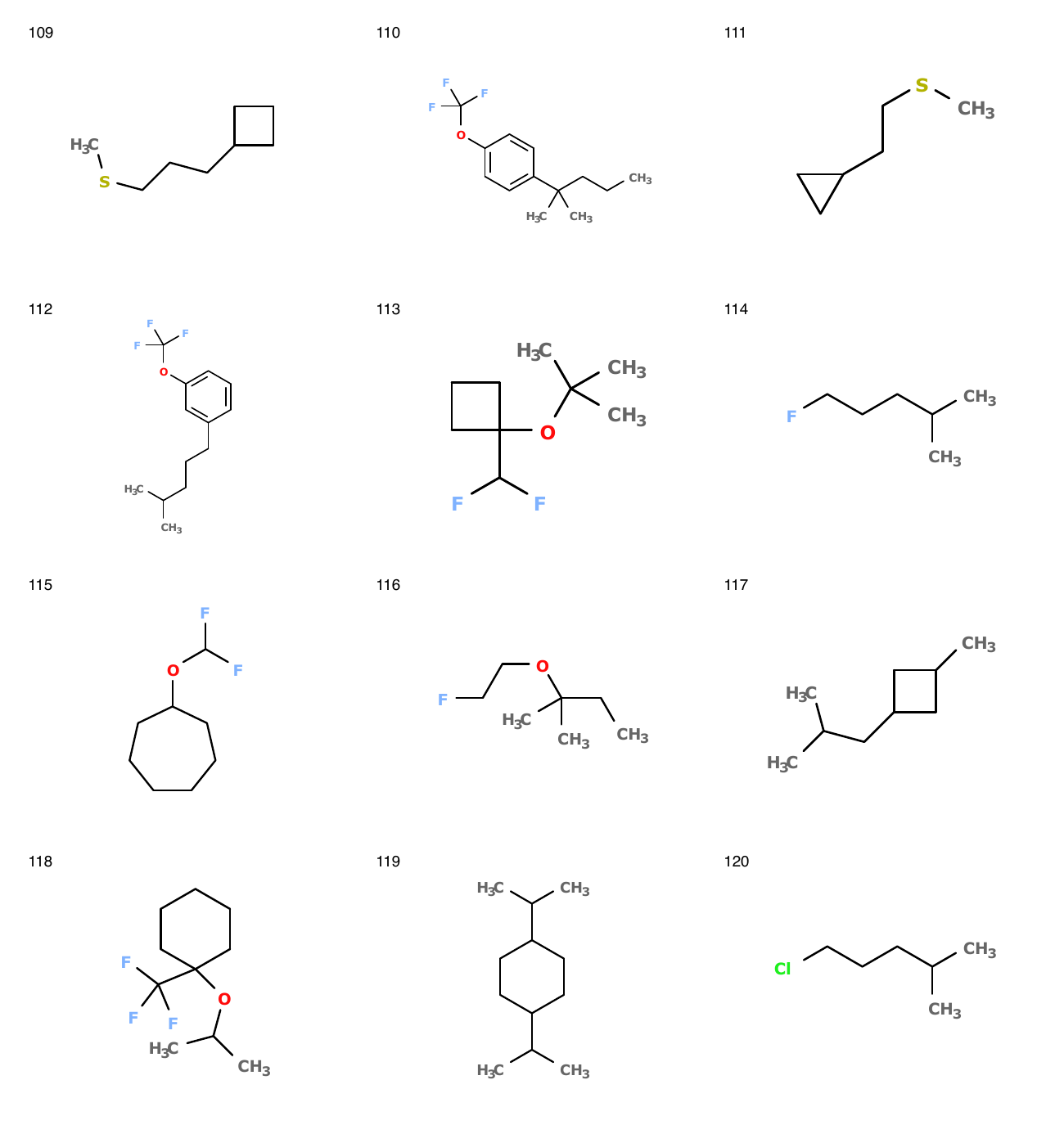}
    \caption{
        \label{fig:lio2_pareto_front_p10}
        Screened Pareto-Front molecules for lithium--air battery solvents identified by the high-throughput screening pipeline presented in the manuscript.
        Figure 10 of 33.
    }
\end{figure}

The 390 screened electrolytes on the Pareto front are shown in~\cref{fig:lio2_pareto_front_p01}--\cref{fig:lio2_pareto_front_p10}.
Property values computed using fine-tuned \ac{MIST}-28M models for the Pareto front candidates are in \cref{tab:lio2_generated_pareto_front}.
\ac{SMILES} encodings for the Pareto front candidates are provided within our data release.

\begin{figure}
    \centering
    \includegraphics[width=\linewidth]{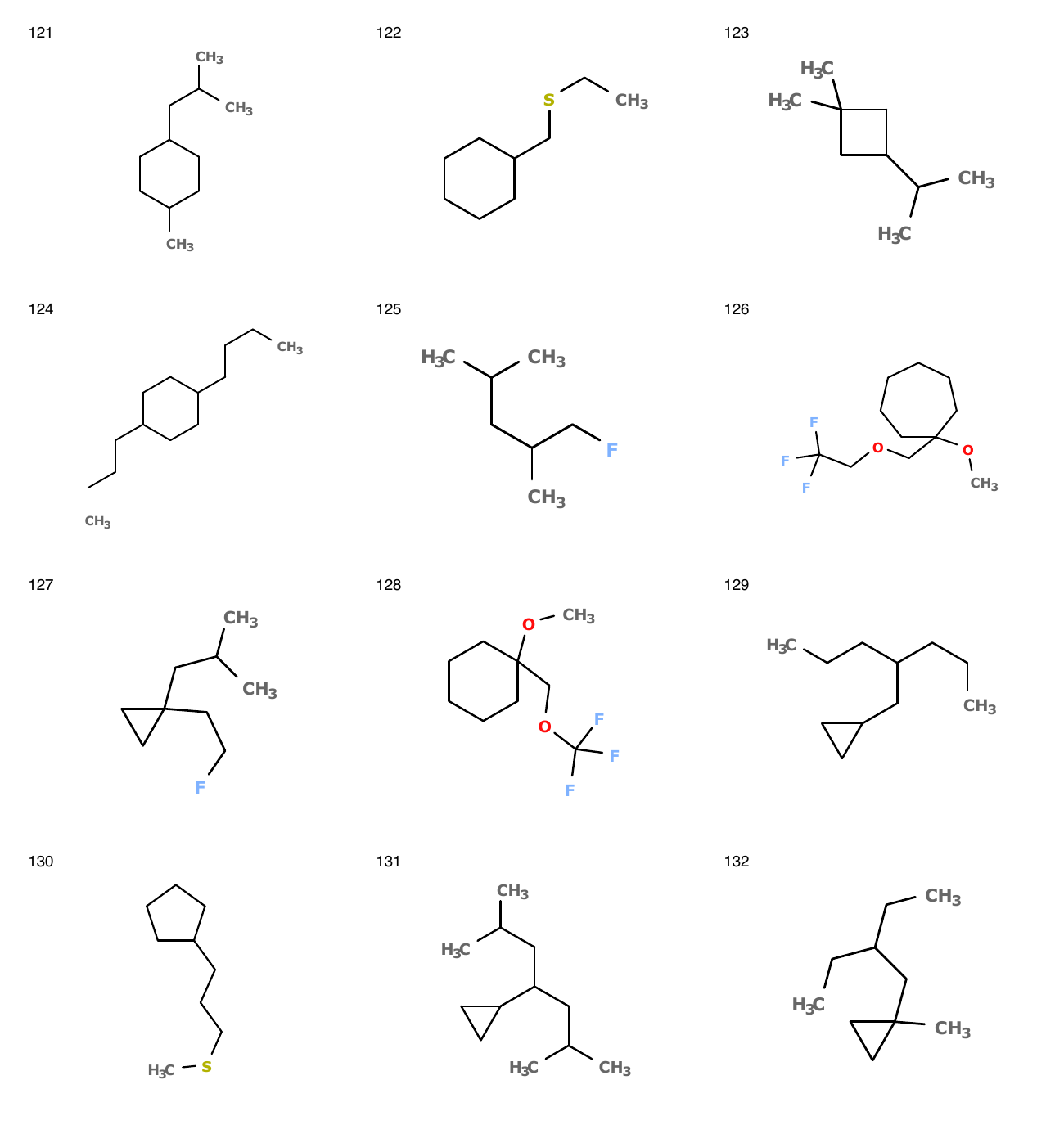}
    \caption{
        \label{fig:lio2_pareto_front_p11}
        Screened Pareto-Front molecules for lithium--air battery solvents identified by the high-throughput screening pipeline presented in the manuscript.
        Figure 11 of 33.
    }
\end{figure}

\begin{figure}
    \centering
    \includegraphics[width=\linewidth]{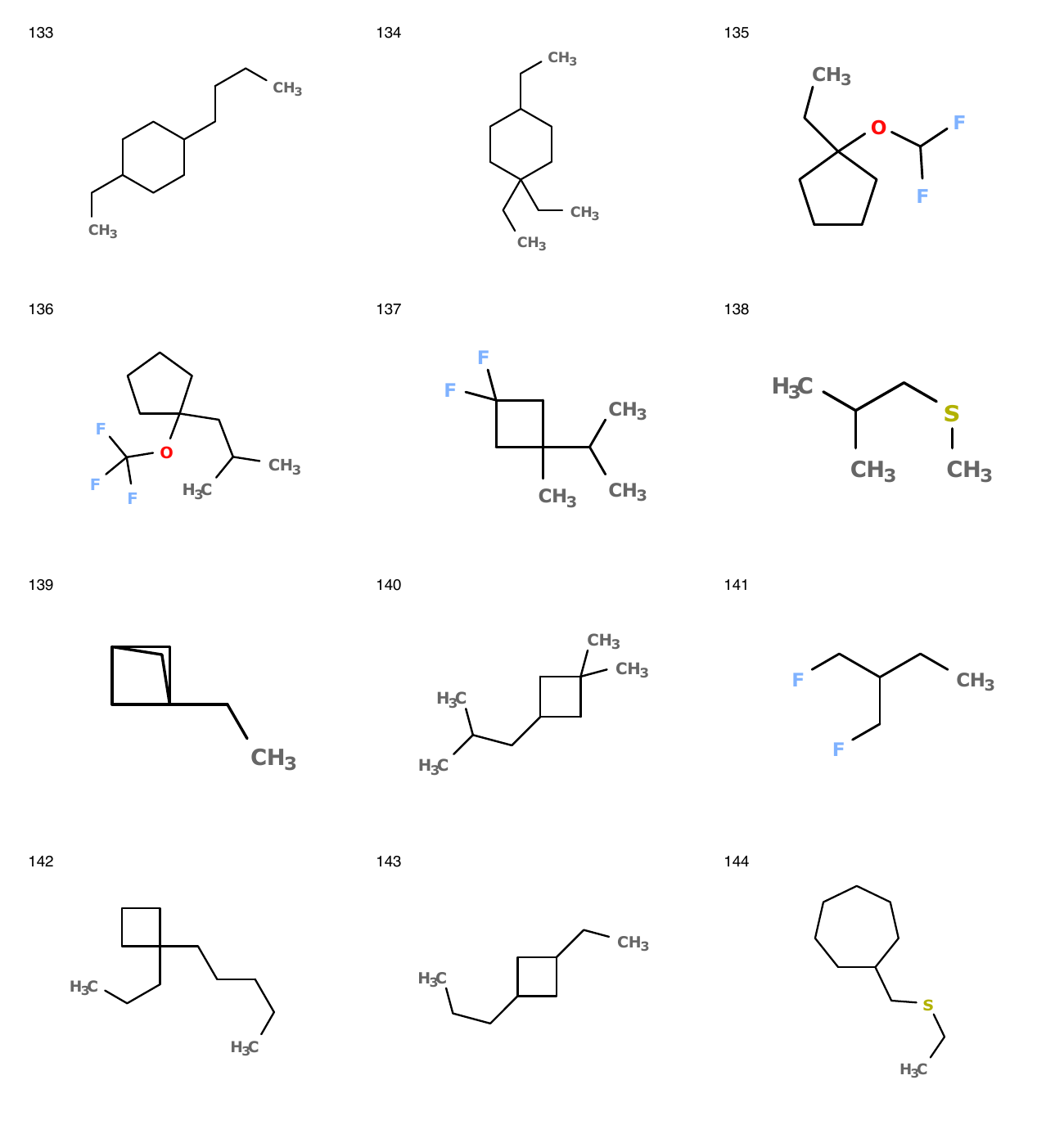}
    \caption{
        \label{fig:lio2_pareto_front_p12}
        Screened Pareto-Front molecules for lithium--air battery solvents identified by the high-throughput screening pipeline presented in the manuscript.
        Figure 12 of 33.
    }
\end{figure}

\begin{figure}
    \centering
    \includegraphics[width=\linewidth]{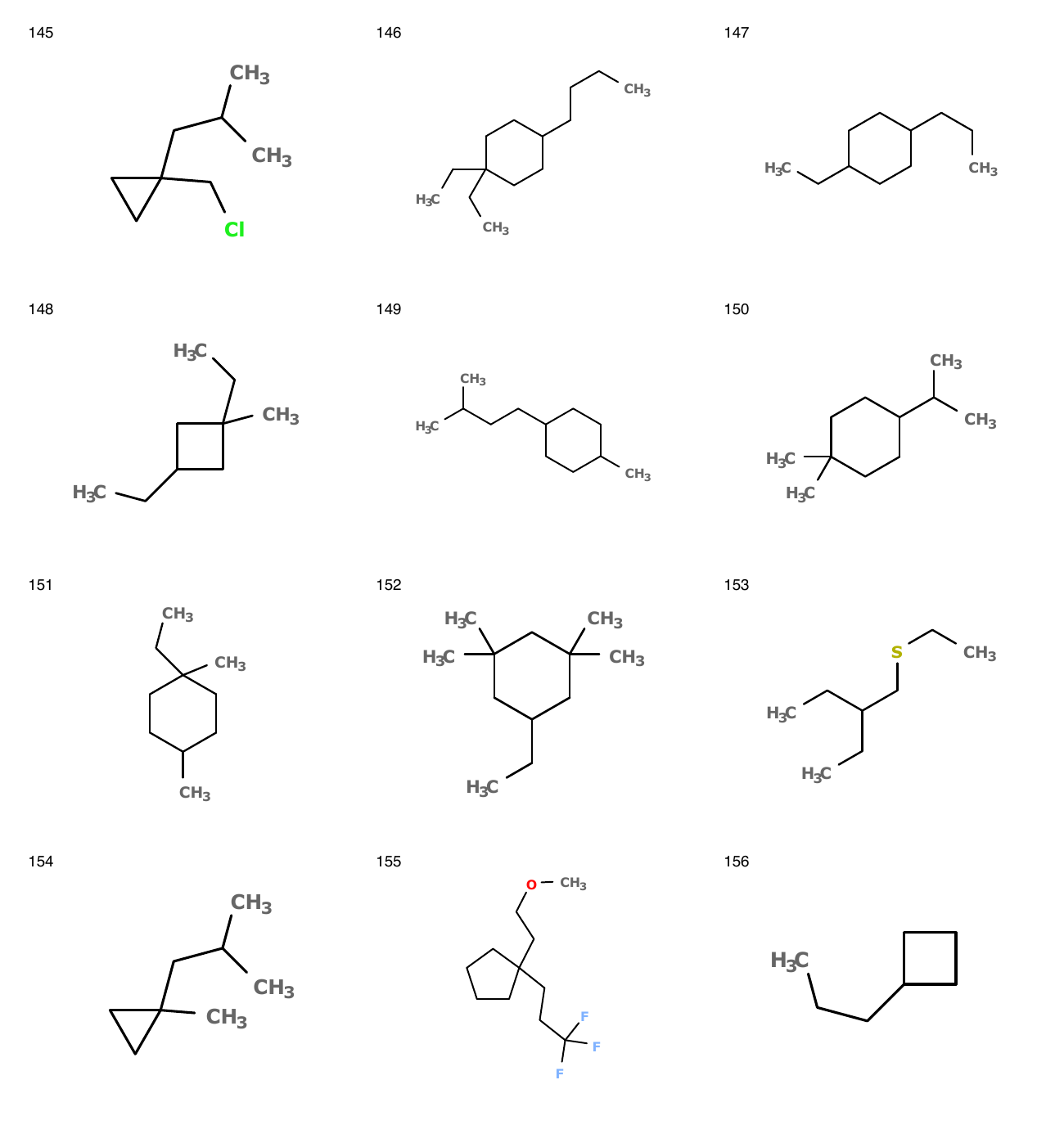}
    \caption{
        \label{fig:lio2_pareto_front_p13}
        Screened Pareto-Front molecules for lithium--air battery solvents identified by the high-throughput screening pipeline presented in the manuscript.
        Figure 13 of 33.
    }
\end{figure}

\begin{figure}
    \centering
    \includegraphics[width=\linewidth]{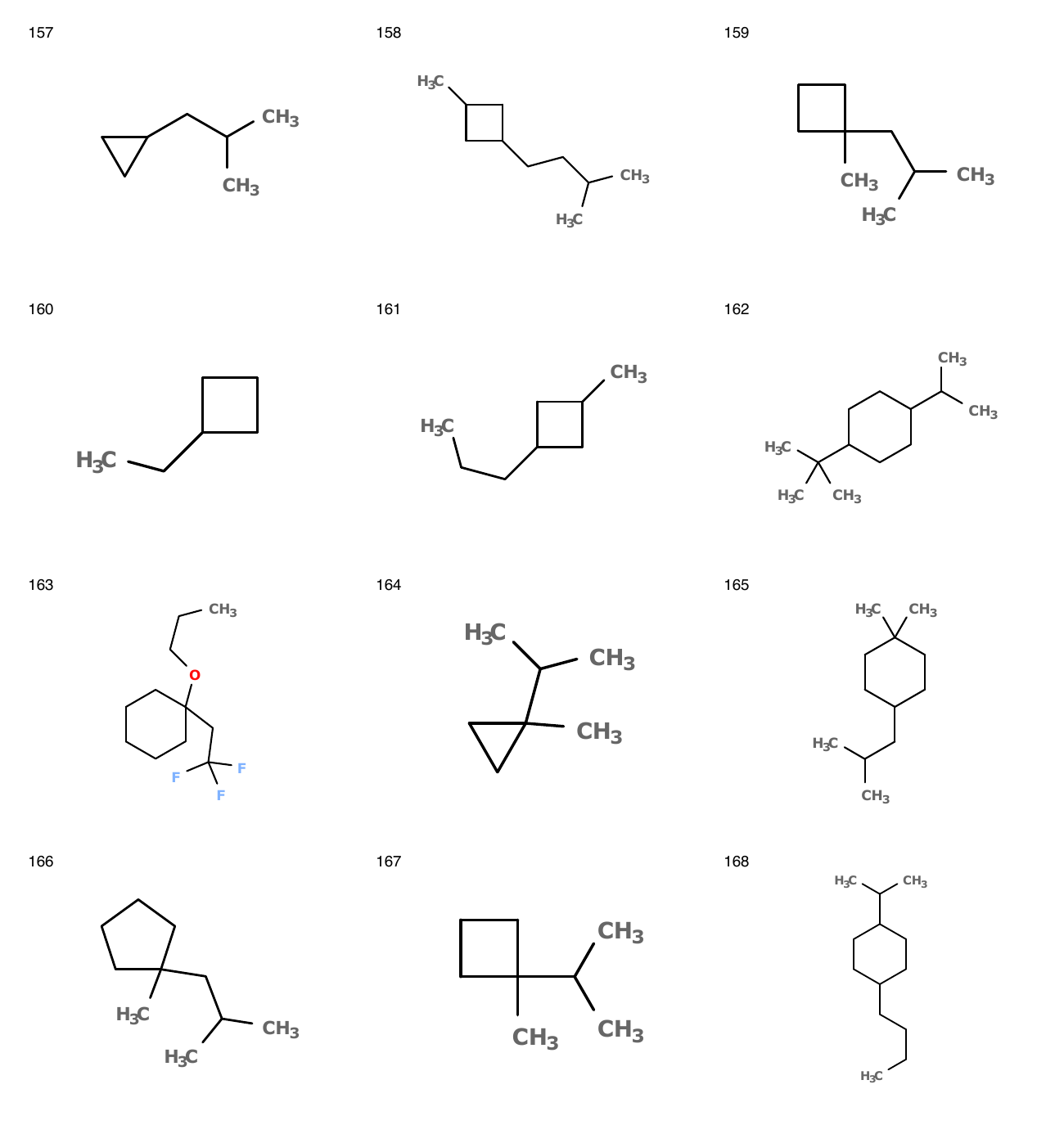}
    \caption{
        \label{fig:lio2_pareto_front_p14}
        Screened Pareto-Front molecules for lithium--air battery solvents identified by the high-throughput screening pipeline presented in the manuscript.
        Figure 14 of 33.
    }
\end{figure}

\begin{figure}
    \centering
    \includegraphics[width=\linewidth]{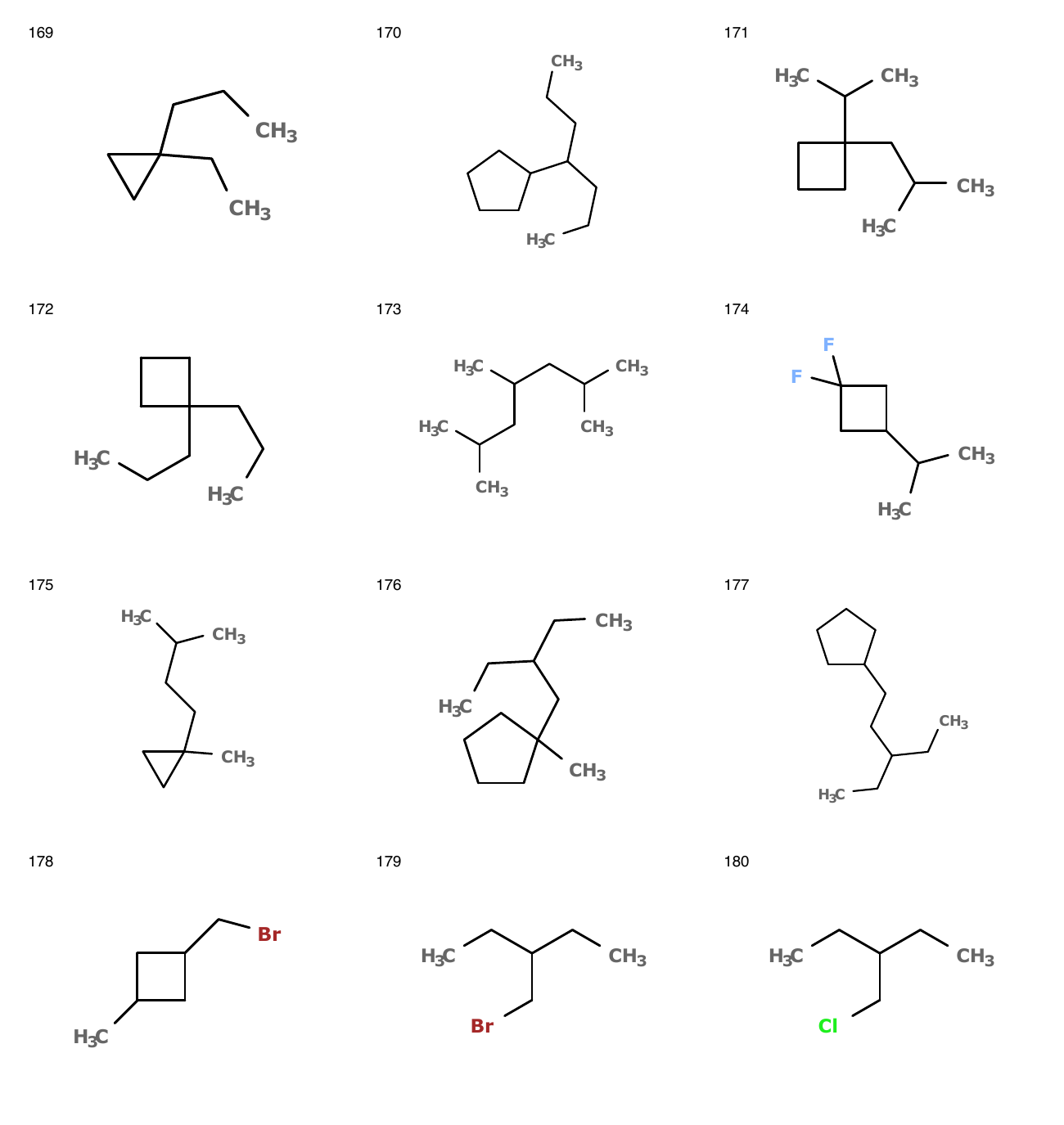}
    \caption{
        \label{fig:lio2_pareto_front_p15}
        Screened Pareto-Front molecules for lithium--air battery solvents identified by the high-throughput screening pipeline presented in the manuscript.
        Figure 15 of 33.
    }
\end{figure}

\begin{figure}
    \centering
    \includegraphics[width=\linewidth]{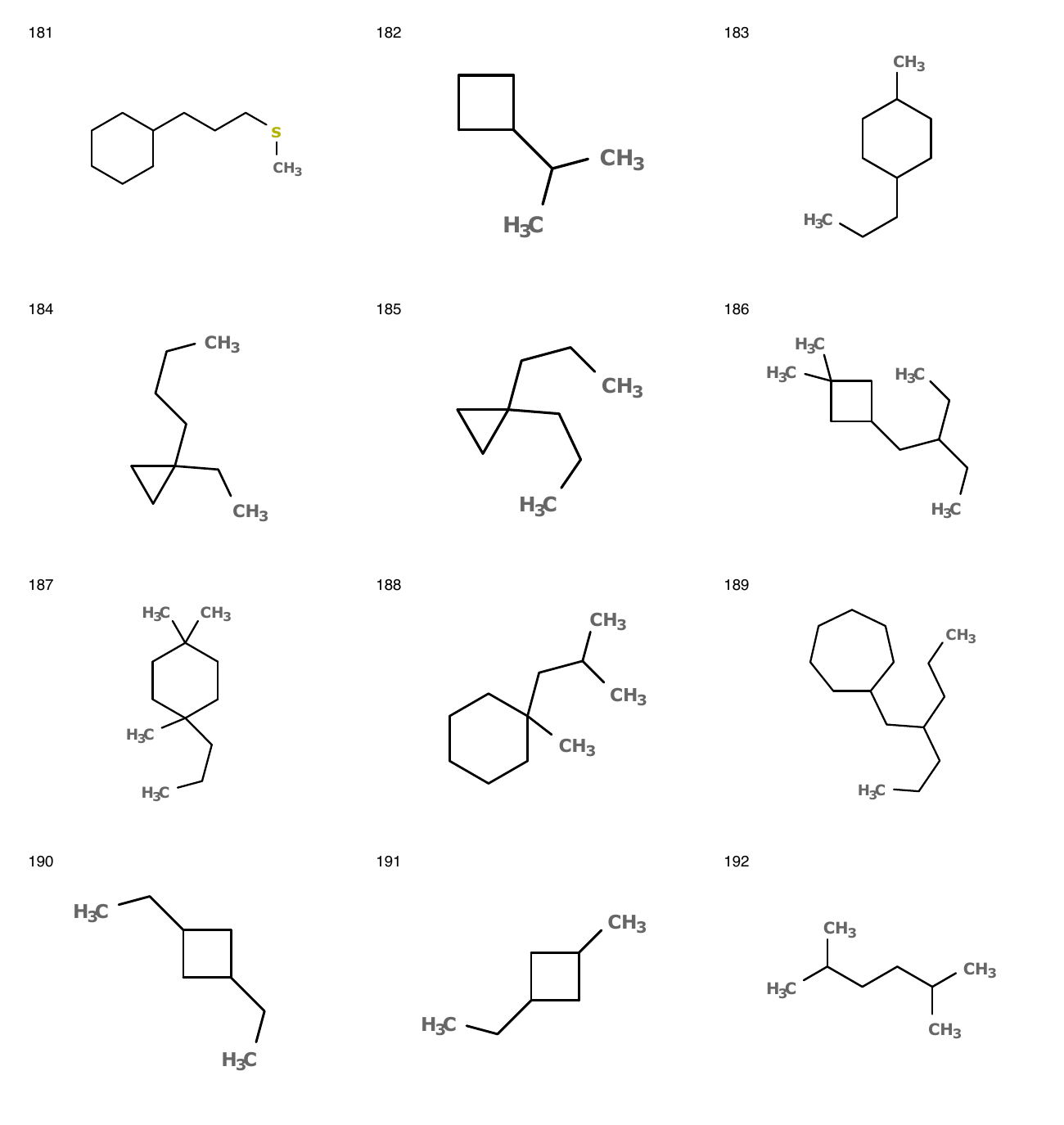}
    \caption{
        \label{fig:lio2_pareto_front_p16}
        Screened Pareto-Front molecules for lithium--air battery solvents identified by the high-throughput screening pipeline presented in the manuscript.
        Figure 16 of 33.
    }
\end{figure}

\begin{figure}
    \centering
    \includegraphics[width=\linewidth]{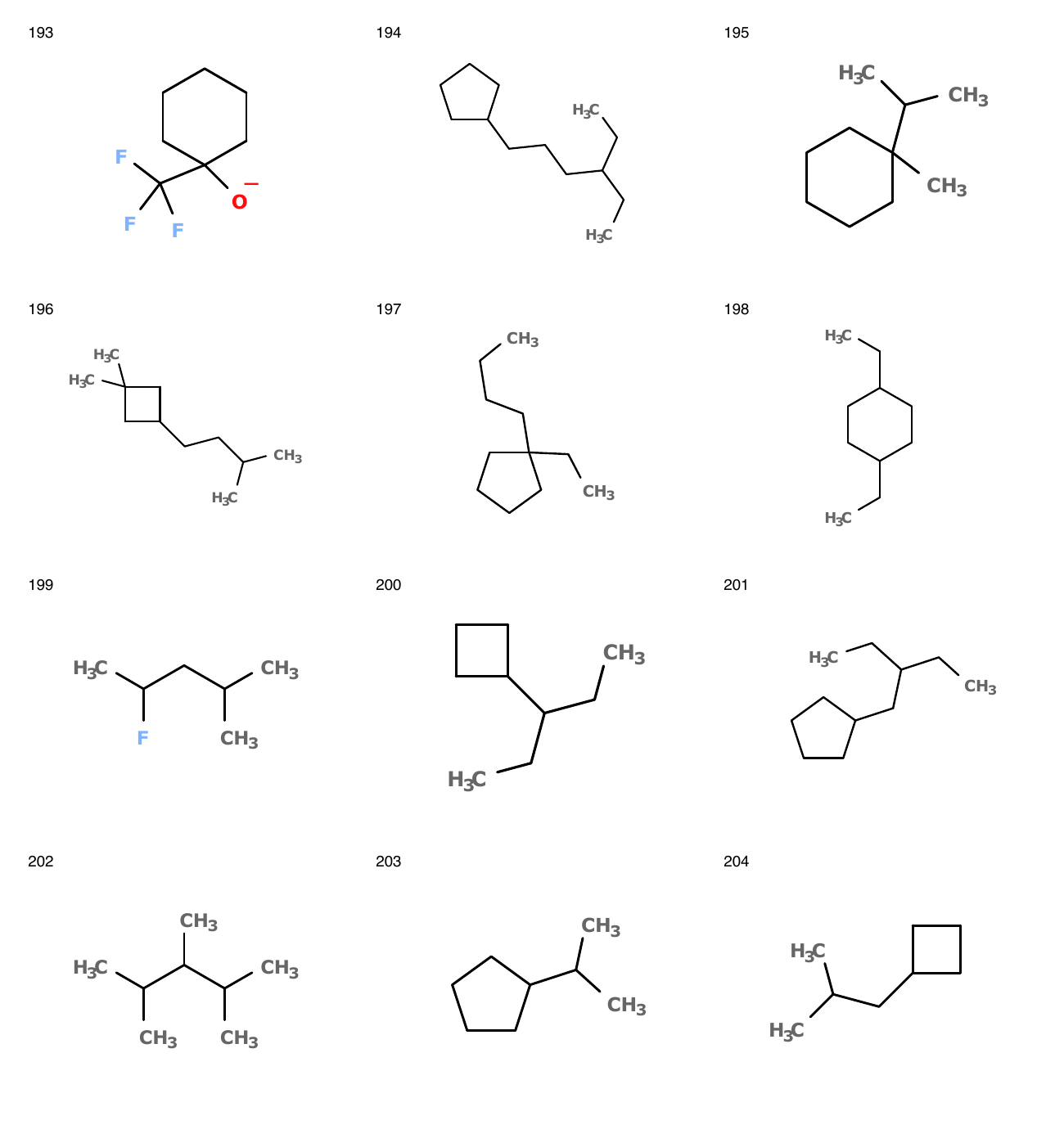}
    \caption{
        \label{fig:lio2_pareto_front_p17}
        Screened Pareto-Front molecules for lithium--air battery solvents identified by the high-throughput screening pipeline presented in the manuscript.
        Figure 17 of 33.
    }
\end{figure}

\begin{figure}
    \centering
    \includegraphics[width=\linewidth]{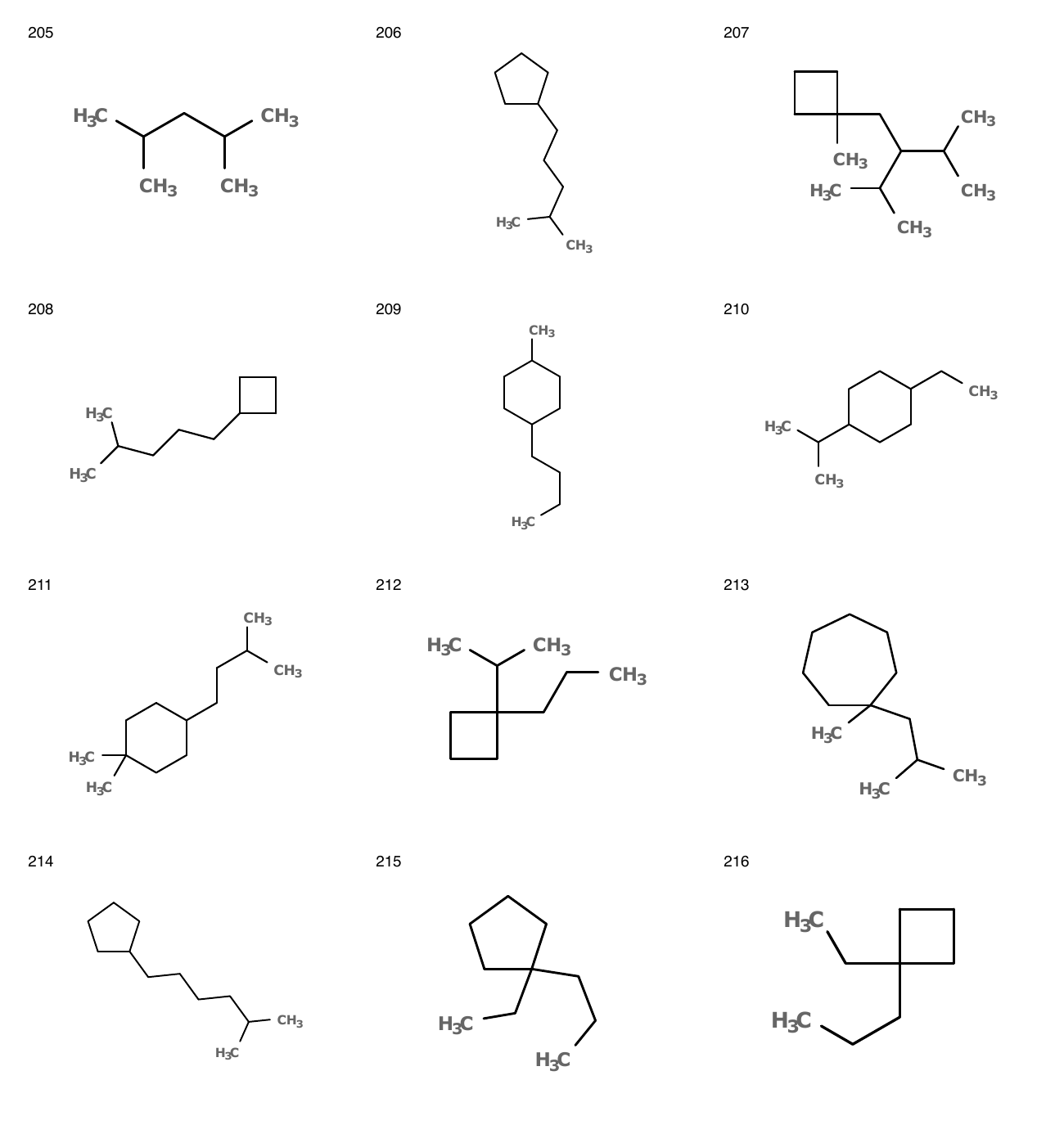}
    \caption{
        \label{fig:lio2_pareto_front_p18}
        Screened Pareto-Front molecules for lithium--air battery solvents identified by the high-throughput screening pipeline presented in the manuscript.
        Figure 18 of 33.
    }
\end{figure}

\begin{figure}
    \centering
    \includegraphics[width=\linewidth]{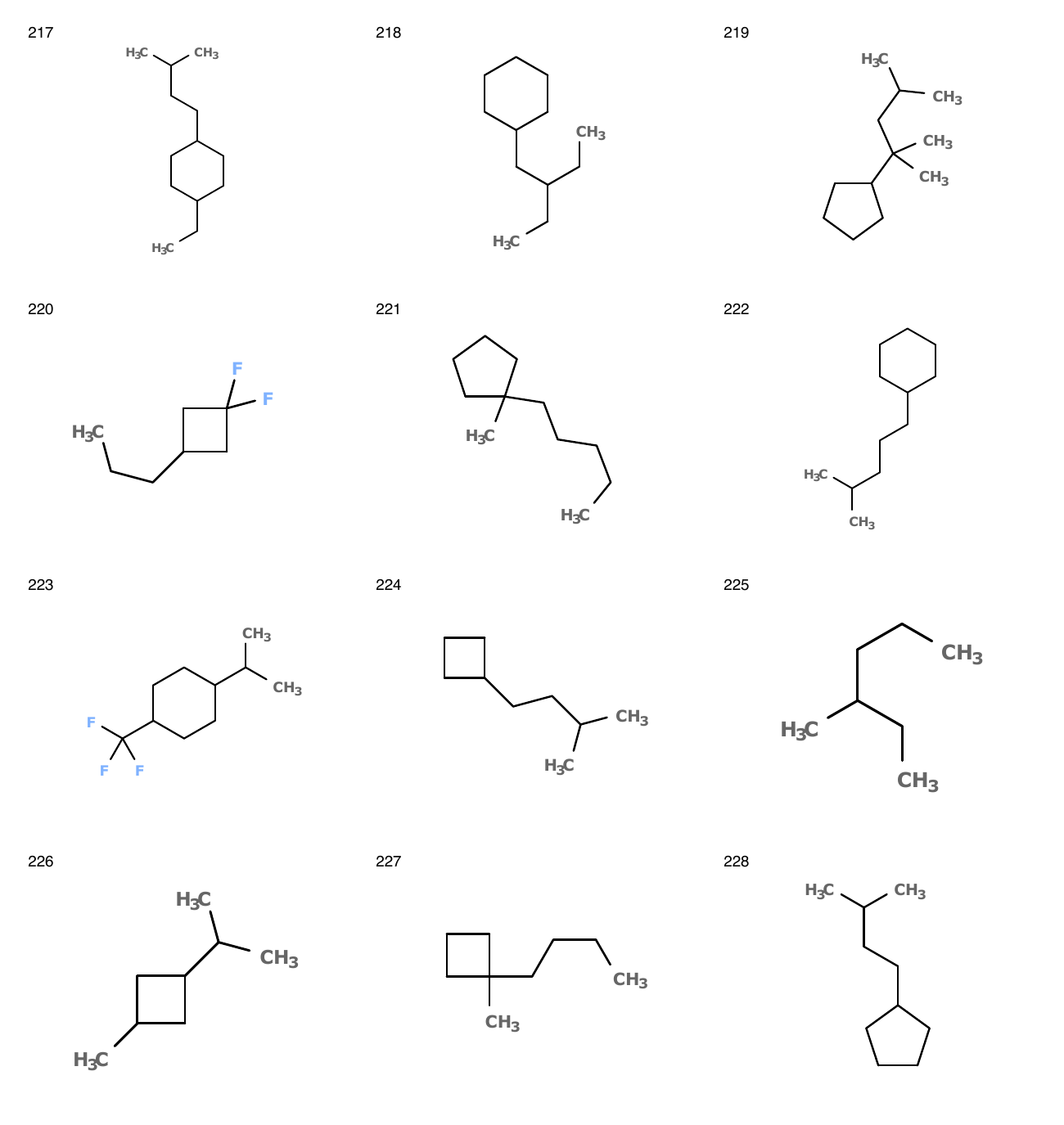}
    \caption{
        \label{fig:lio2_pareto_front_p19}
        Screened Pareto-Front molecules for lithium--air battery solvents identified by the high-throughput screening pipeline presented in the manuscript.
        Figure 19 of 33.
    }
\end{figure}

\begin{figure}
    \centering
    \includegraphics[width=\linewidth]{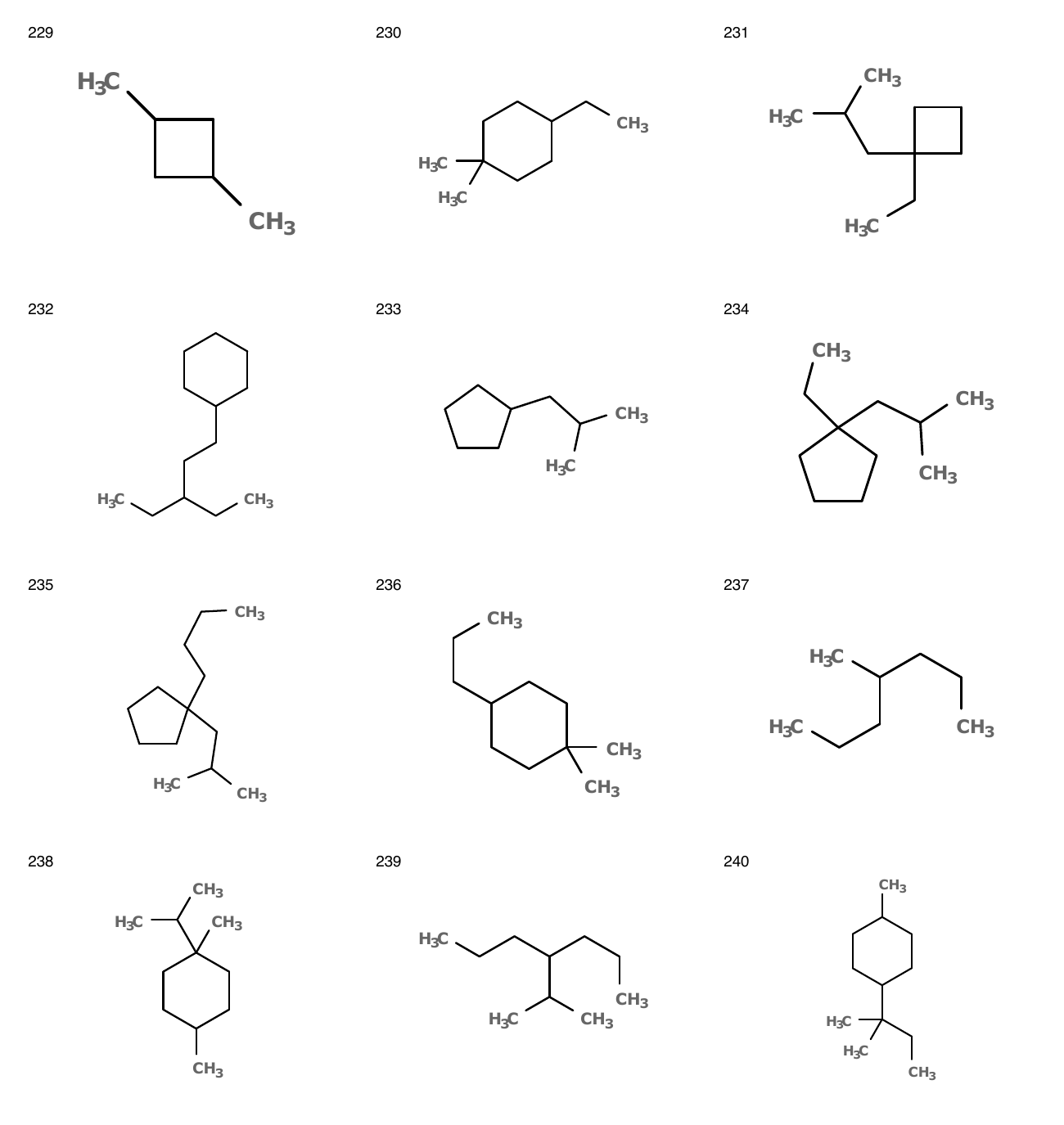}
    \caption{
        \label{fig:lio2_pareto_front_p20}
        Screened Pareto-Front molecules for lithium--air battery solvents identified by the high-throughput screening pipeline presented in the manuscript.
        Figure 20 of 33.
    }
\end{figure}

\begin{figure}
    \centering
    \includegraphics[width=\linewidth]{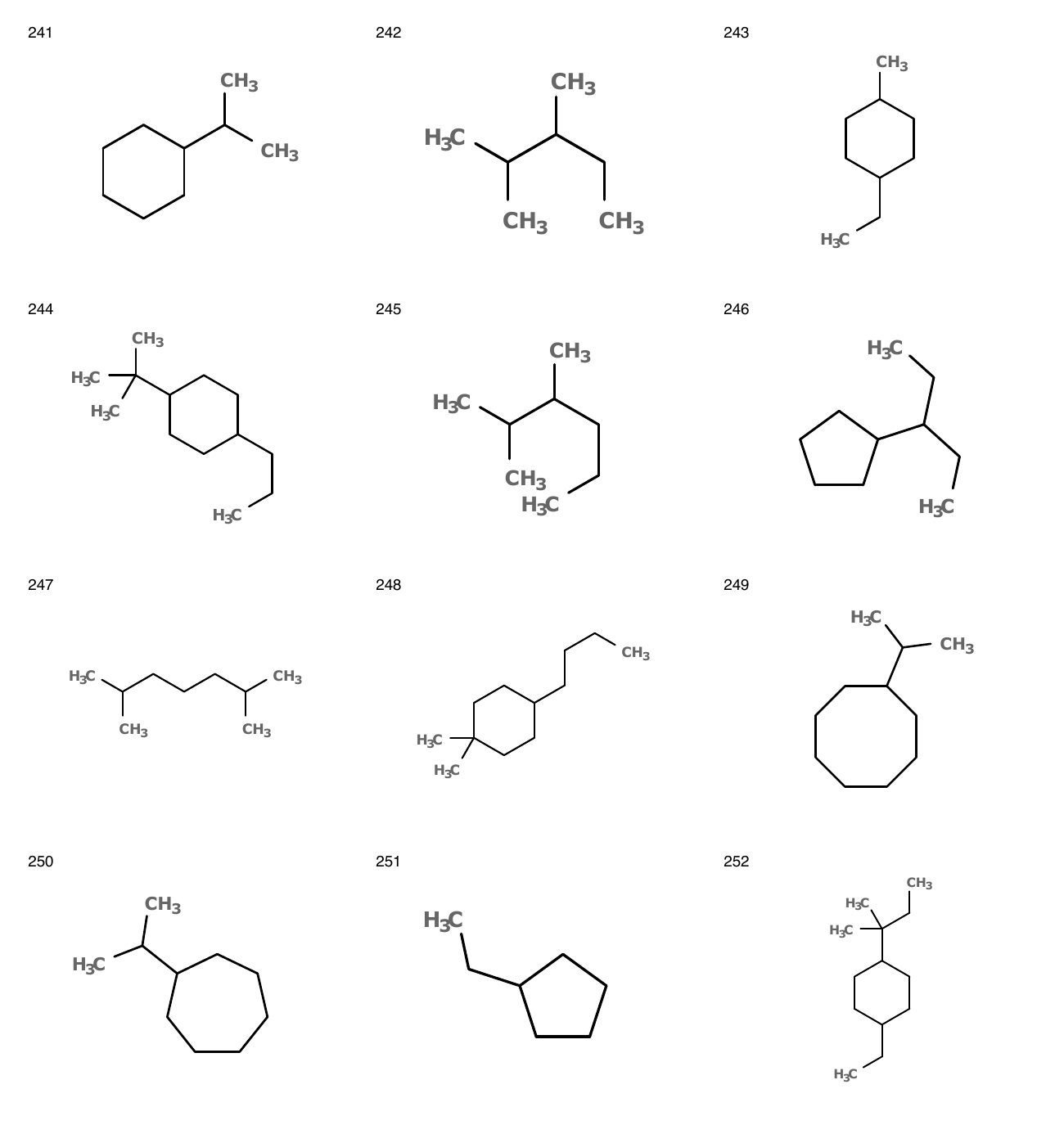}
    \caption{
        \label{fig:lio2_pareto_front_p21}
        Screened Pareto-Front molecules for lithium--air battery solvents identified by the high-throughput screening pipeline presented in the manuscript.
        Figure 21 of 33.
    }
\end{figure}

\begin{figure}
    \centering
    \includegraphics[width=\linewidth]{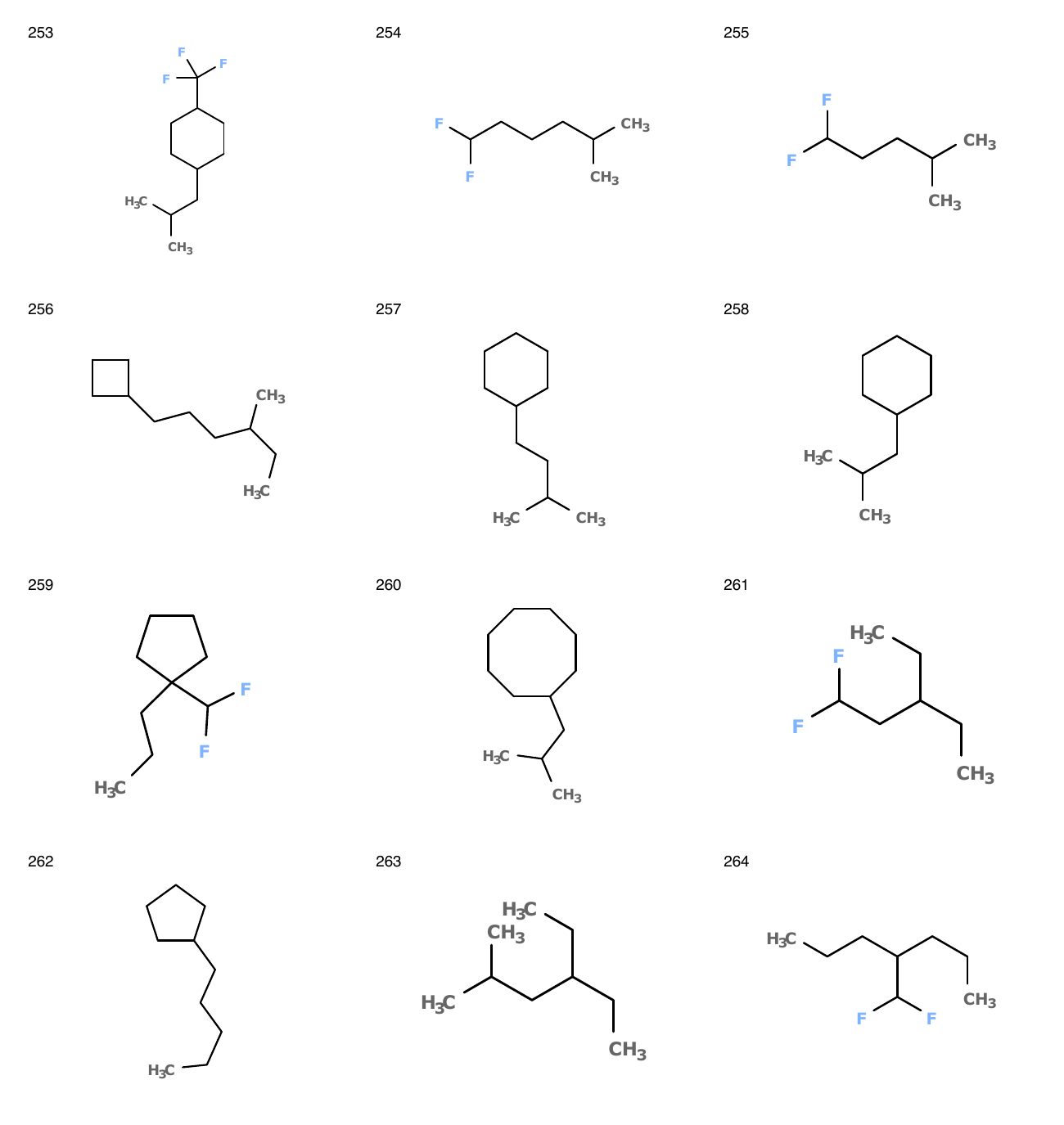}
    \caption{
        \label{fig:lio2_pareto_front_p22}
        Screened Pareto-Front molecules for lithium--air battery solvents identified by the high-throughput screening pipeline presented in the manuscript.
        Figure 22 of 33.
    }
\end{figure}

\begin{figure}
    \centering
    \includegraphics[width=\linewidth]{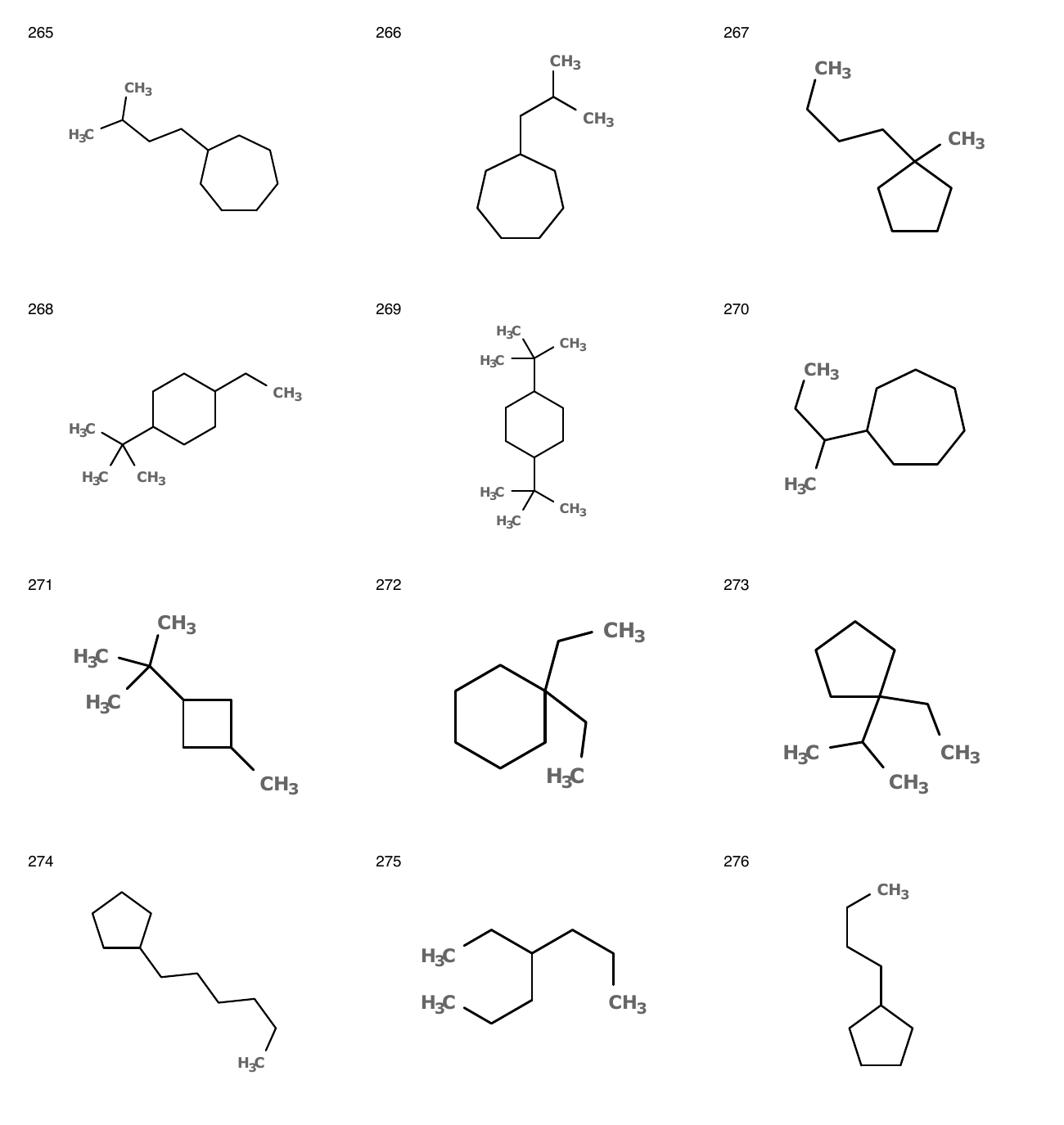}
    \caption{
        \label{fig:lio2_pareto_front_p23}
        Screened Pareto-Front molecules for lithium--air battery solvents identified by the high-throughput screening pipeline presented in the manuscript.
        Figure 23 of 33.
    }
\end{figure}

\begin{figure}
    \centering
    \includegraphics[width=\linewidth]{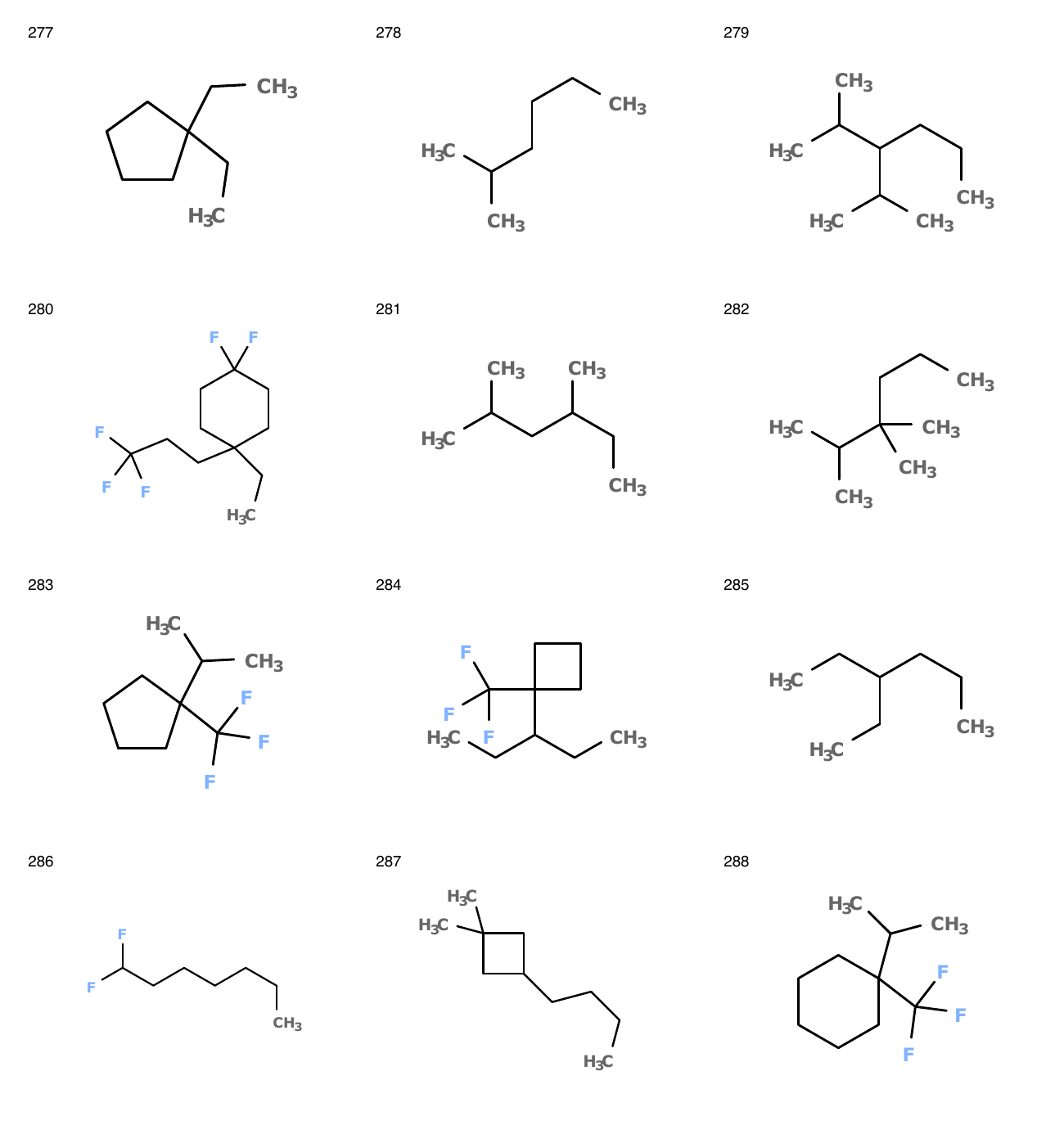}
    \caption{
        \label{fig:lio2_pareto_front_p24}
        Screened Pareto-Front molecules for lithium--air battery solvents identified by the high-throughput screening pipeline presented in the manuscript.
        Figure 24 of 33.
    }
\end{figure}

\begin{figure}
    \centering
    \includegraphics[width=\linewidth]{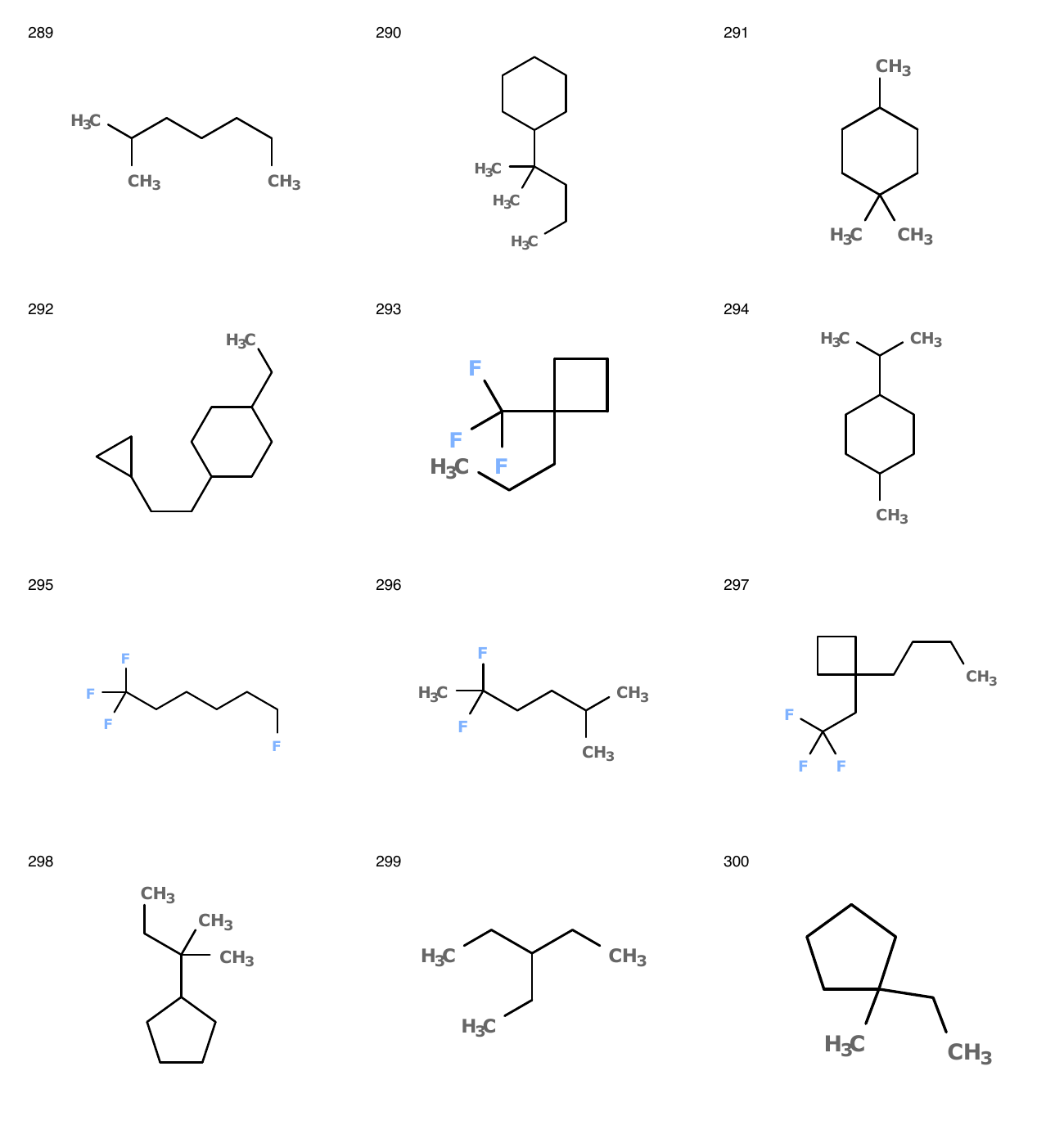}
    \caption{
        \label{fig:lio2_pareto_front_p25}
        Screened Pareto-Front molecules for lithium--air battery solvents identified by the high-throughput screening pipeline presented in the manuscript.
        Figure 25 of 33.
    }
\end{figure}

\begin{figure}
    \centering
    \includegraphics[width=\linewidth]{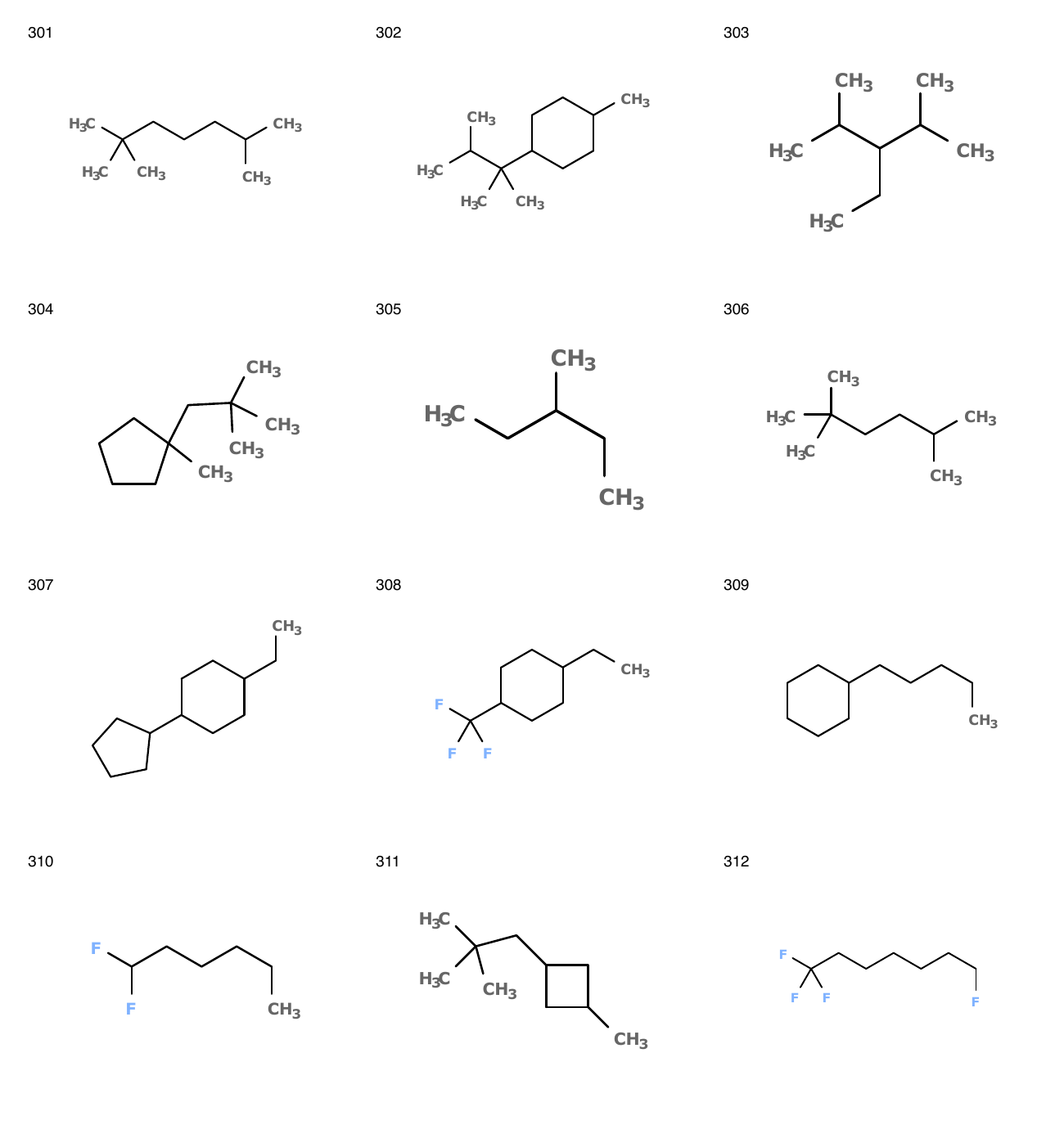}
    \caption{
        \label{fig:lio2_pareto_front_p26}
        Screened Pareto-Front molecules for lithium--air battery solvents identified by the high-throughput screening pipeline presented in the manuscript.
        Figure 26 of 33.
    }
\end{figure}

\begin{figure}
    \centering
    \includegraphics[width=\linewidth]{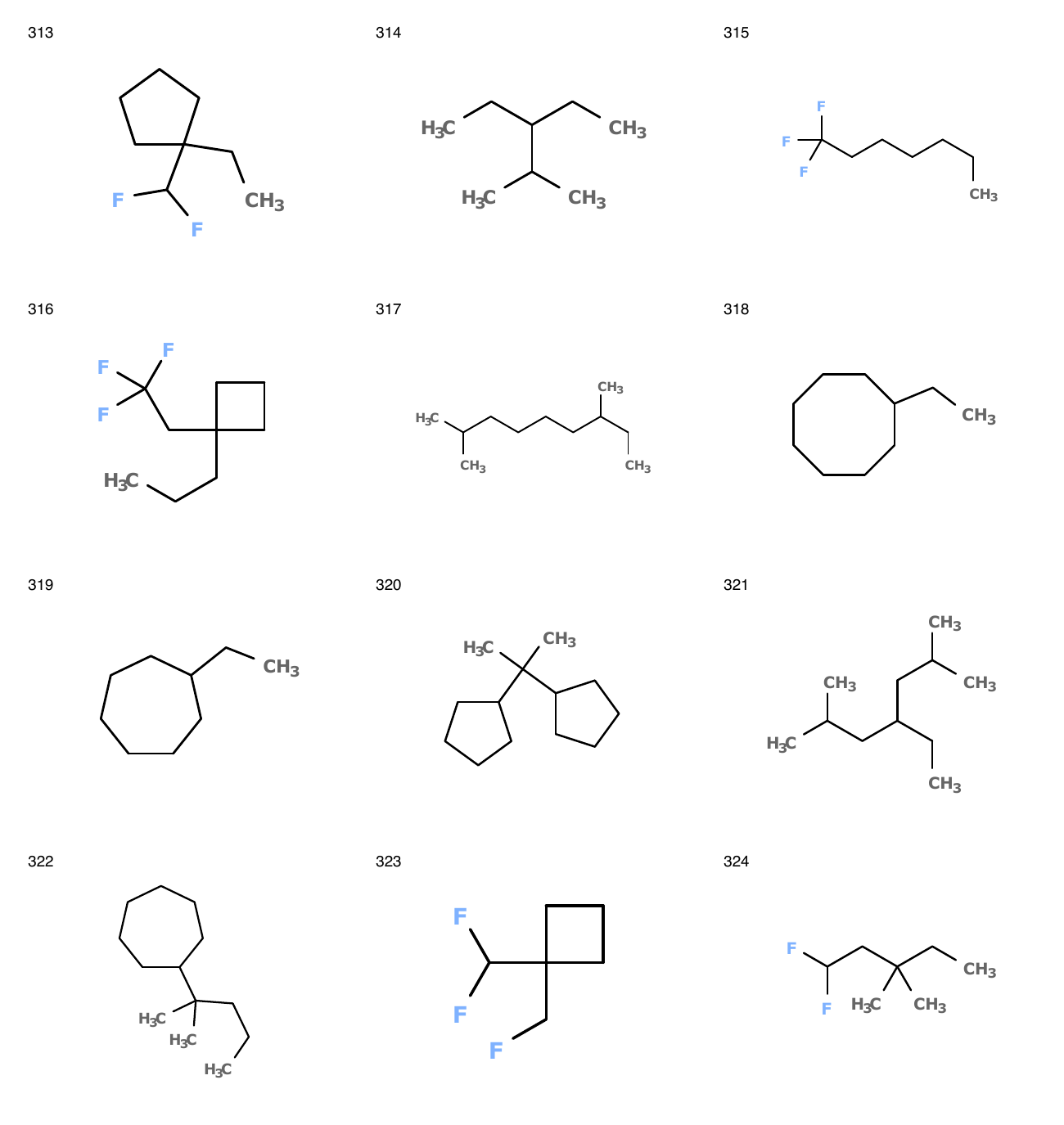}
    \caption{
        \label{fig:lio2_pareto_front_p27}
        Screened Pareto-Front molecules for lithium--air battery solvents identified by the high-throughput screening pipeline presented in the manuscript.
        Figure 27 of 33.
    }
\end{figure}

\begin{figure}
    \centering
    \includegraphics[width=\linewidth]{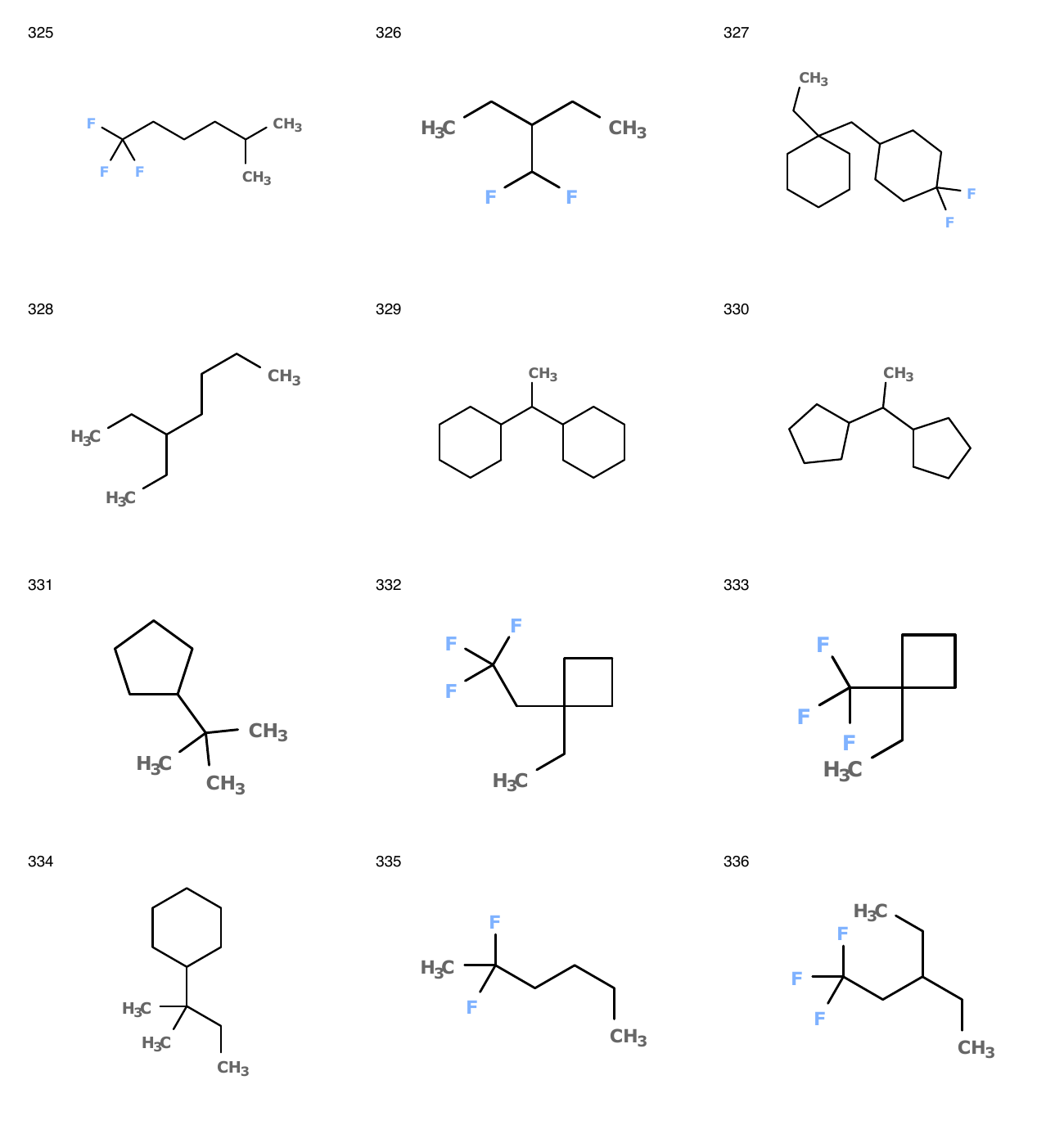}
    \caption{
        \label{fig:lio2_pareto_front_p28}
        Screened Pareto-Front molecules for lithium--air battery solvents identified by the high-throughput screening pipeline presented in the manuscript.
        Figure 28 of 33.
    }
\end{figure}

\begin{figure}
    \centering
    \includegraphics[width=\linewidth]{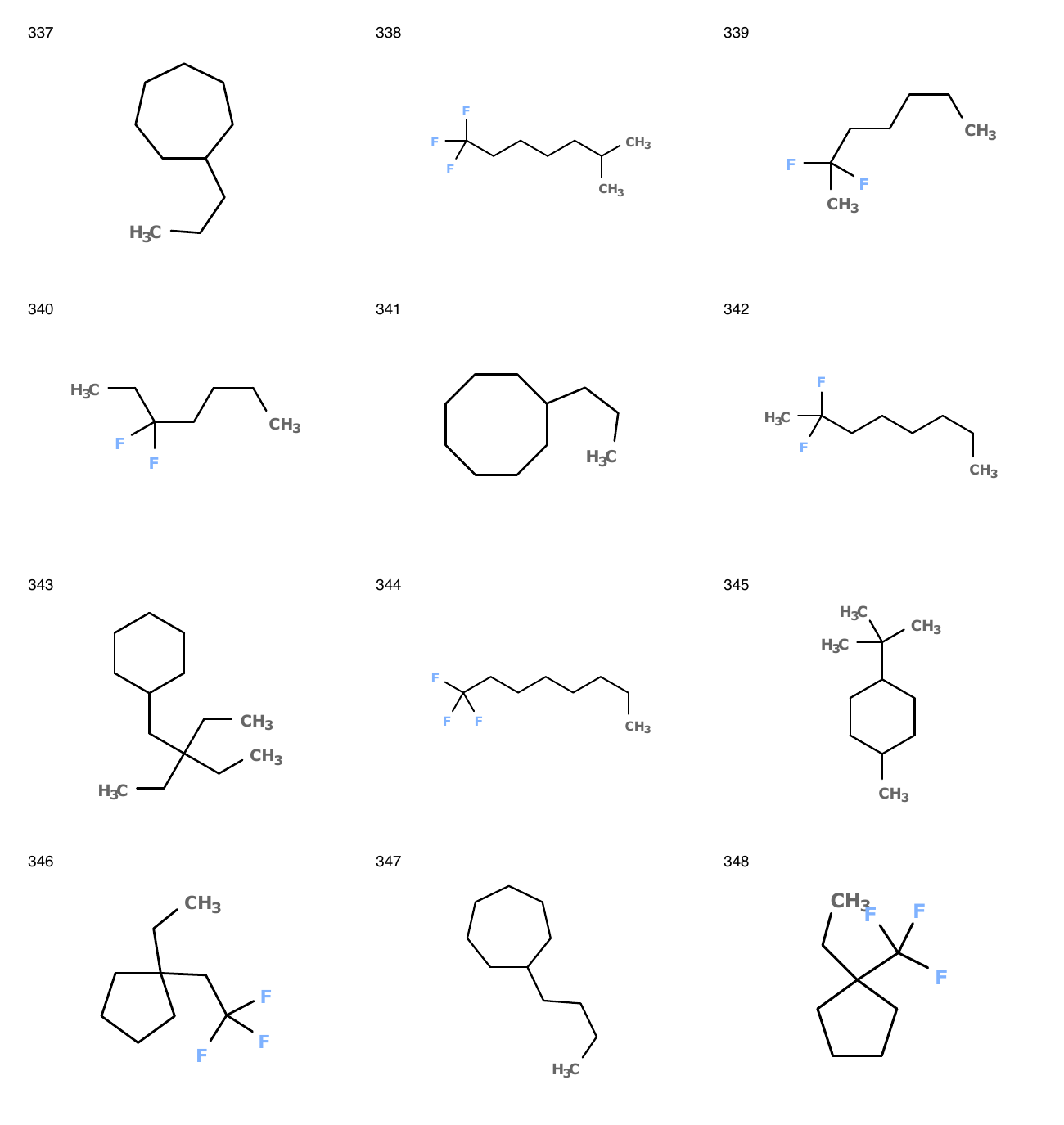}
    \caption{
        \label{fig:lio2_pareto_front_p29}
        Screened Pareto-Front molecules for lithium--air battery solvents identified by the high-throughput screening pipeline presented in the manuscript.
        Figure 29 of 33.
    }
\end{figure}

\begin{figure}
    \centering
    \includegraphics[width=\linewidth]{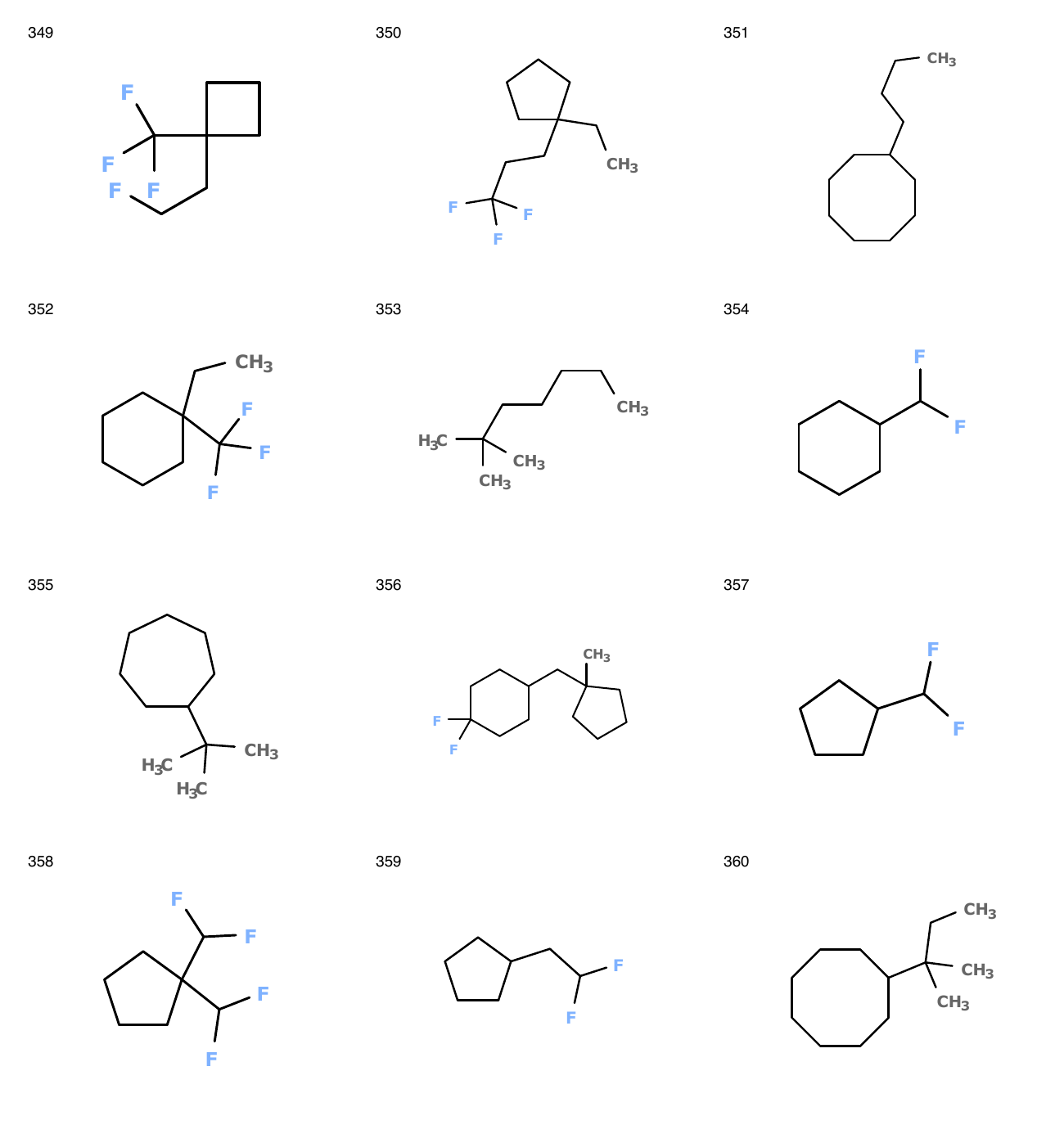}
    \caption{
        \label{fig:lio2_pareto_front_p30}
        Screened Pareto-Front molecules for lithium--air battery solvents identified by the high-throughput screening pipeline presented in the manuscript.
        Figure 30 of 33.
    }
\end{figure}

\begin{figure}
    \centering
    \includegraphics[width=\linewidth]{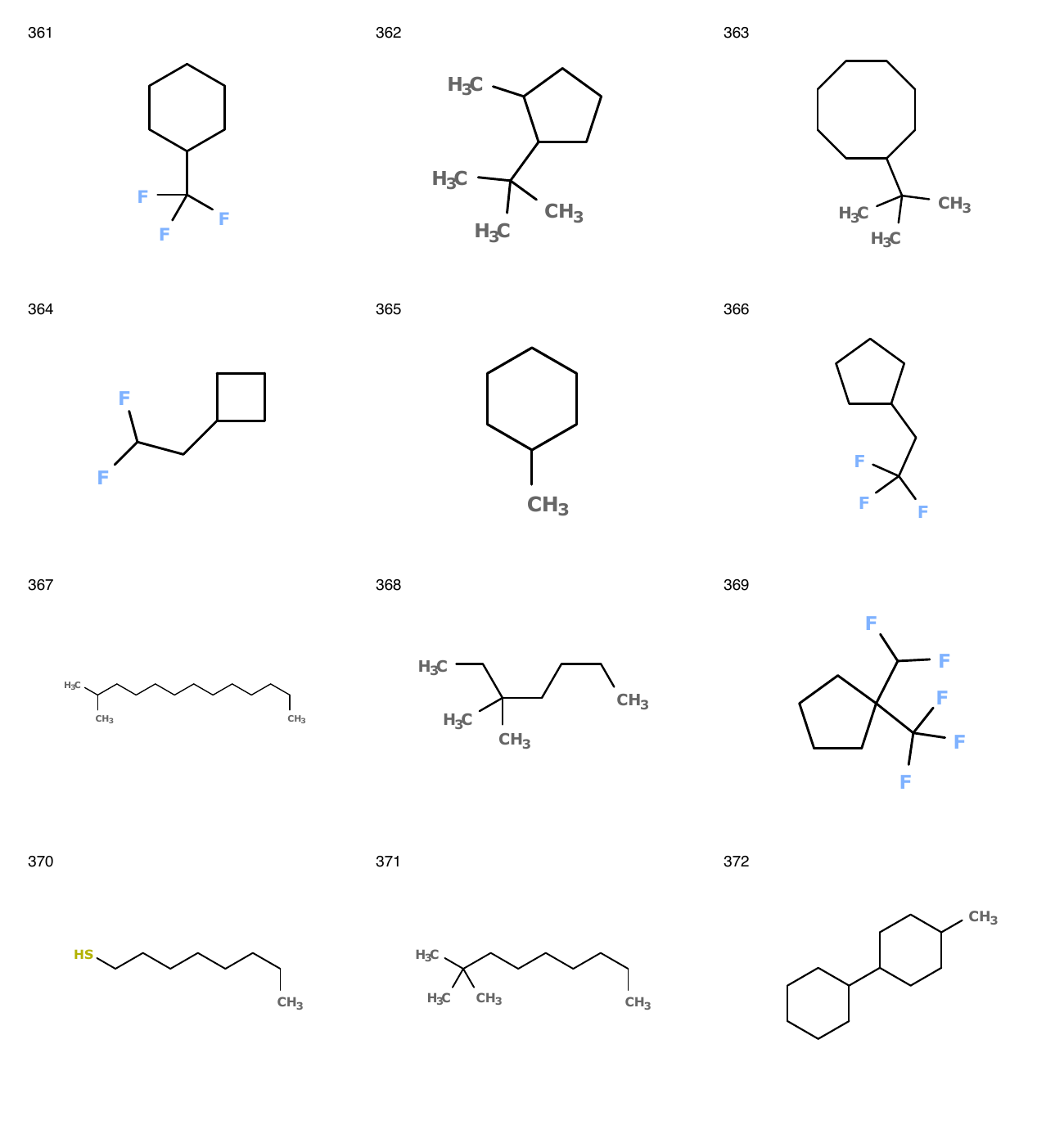}
    \caption{
        \label{fig:lio2_pareto_front_p31}
        Screened Pareto-Front molecules for lithium--air battery solvents identified by the high-throughput screening pipeline presented in the manuscript.
        Figure 31 of 33.
    }
\end{figure}

\begin{figure}
    \centering
    \includegraphics[width=\linewidth]{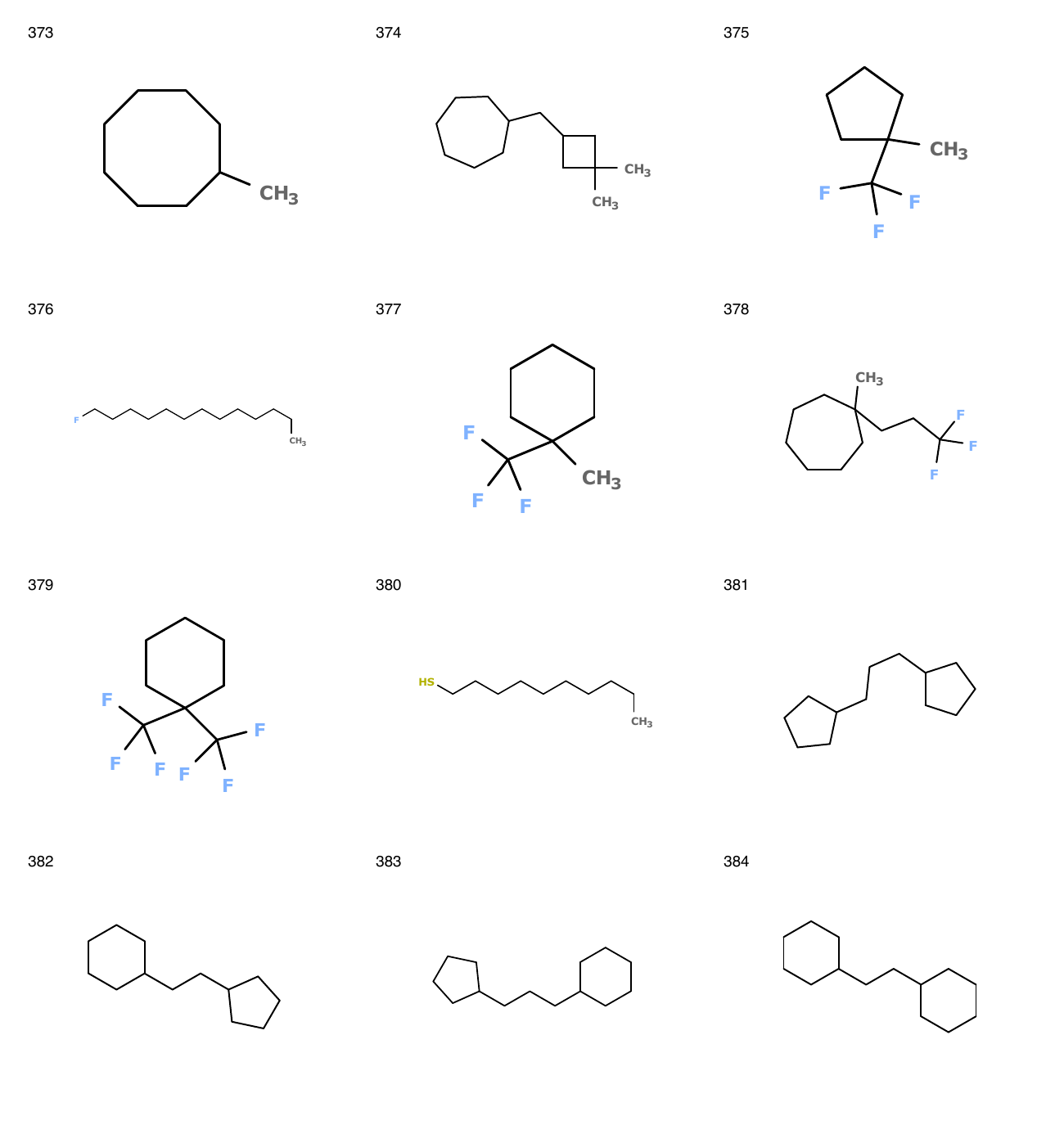}
    \caption{
        \label{fig:lio2_pareto_front_p32}
        Screened Pareto-Front molecules for lithium--air battery solvents identified by the high-throughput screening pipeline presented in the manuscript.
        Figure 32 of 33.
    }
\end{figure}

\begin{figure}
    \centering
    \includegraphics[width=\linewidth]{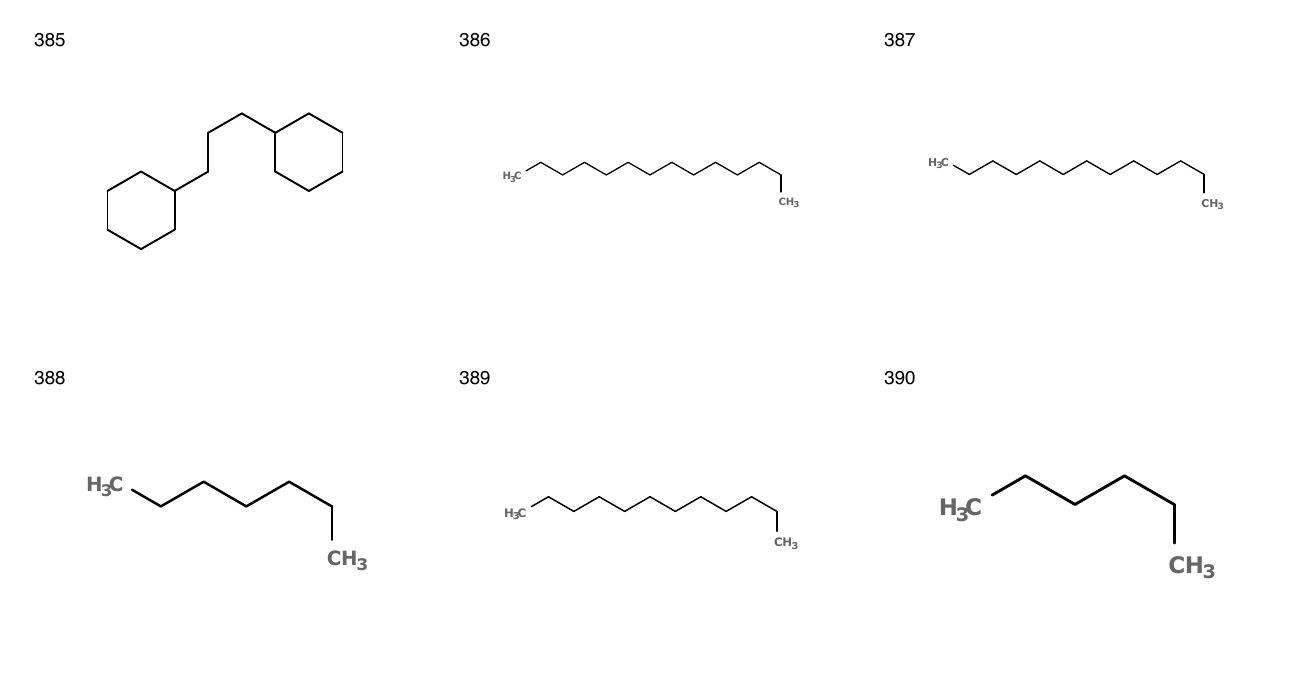}
    \caption{
        \label{fig:lio2_pareto_front_p33}
        Screened Pareto-Front molecules for lithium--air battery solvents identified by the high-throughput screening pipeline presented in the manuscript.
        Figure 33 of 33.
    }
\end{figure}

\begin{longtable}{ccccccc}
  \caption{
    \label{tab:lio2_generated_pareto_front}
    Pareto Front Candidate Electrolytes and computed properties using fine-tuned MIST-28M models.
    Molecular structures (Indexed by \#) are shown in \cref{fig:lio2_pareto_front_p01,fig:lio2_pareto_front_p02,fig:lio2_pareto_front_p03,fig:lio2_pareto_front_p04,fig:lio2_pareto_front_p05,fig:lio2_pareto_front_p06,,fig:lio2_pareto_front_p07,fig:lio2_pareto_front_p08,fig:lio2_pareto_front_p09,fig:lio2_pareto_front_p10,fig:lio2_pareto_front_p11,fig:lio2_pareto_front_p12,fig:lio2_pareto_front_p13,fig:lio2_pareto_front_p14,fig:lio2_pareto_front_p15,fig:lio2_pareto_front_p16,fig:lio2_pareto_front_p17,fig:lio2_pareto_front_p18,fig:lio2_pareto_front_p19,fig:lio2_pareto_front_p20,fig:lio2_pareto_front_p21,fig:lio2_pareto_front_p22,fig:lio2_pareto_front_p23,fig:lio2_pareto_front_p24,fig:lio2_pareto_front_p25,fig:lio2_pareto_front_p26,fig:lio2_pareto_front_p27,fig:lio2_pareto_front_p28,fig:lio2_pareto_front_p29,fig:lio2_pareto_front_p30,fig:lio2_pareto_front_p31,fig:lio2_pareto_front_p32,fig:lio2_pareto_front_p33}.
  }\\
  \textbf{\#} & \textbf{HOMO} & \textbf{pKa (DMSO)} & \textbf{\(\beta\)} & \textbf{\(E_T^N\)} & \textbf{Melt} & \textbf{Boil} \\
              & \textbf{(eV)} &  &  &  & \(\pmb{(\degree C)}\) & \(\pmb{(\degree C)}\) \\\hline
  \endfirsthead
  \multicolumn{7}{c}%
{{\bfseries \tablename\ \thetable{} -- continued from previous page}} \\
\hline
  \textbf{\#} & \textbf{HOMO} & \textbf{pKa (DMSO)} & \textbf{\(\beta\)} & \textbf{\(E_T^N\)} & \textbf{Melt} & \textbf{Boil} \\ \hline
  \endhead

  \hline \multicolumn{7}{r}{{Continued on next page}} \\
  \endfoot

  \hline
  \endlastfoot
  1 & -7.11 & 30.63 & 0.57 & 0.55 & -45 & 113 \\
  2 & -7.02 & 33.62 & 0.58 & 0.45 & -22 & 113 \\
  3 & -7.12 & 31.02 & 0.60 & 0.36 & -45 & 125 \\
  4 & -7.02 & 31.18 & 0.58 & 0.42 & -52 & 159 \\
  5 & -7.11 & 30.81 & 0.57 & 0.39 & -69 & 143 \\
  6 & -7.02 & 31.97 & 0.47 & 0.58 & -81 & 90 \\
  7 & -7.27 & 31.16 & 0.58 & 0.33 & -66 & 102 \\
  8 & -7.16 & 35.84 & 0.51 & 0.48 & -15 & 137 \\
  9 & -7.55 & 33.11 & 0.59 & 0.14 & 2 & 102 \\
  10 & -7.09 & 31.99 & 0.54 & 0.25 & -77 & 101 \\
  11 & -7.07 & 30.50 & 0.44 & 0.50 & -55 & 145 \\
  12 & -7.09 & 33.24 & 0.45 & 0.46 & -20 & 139 \\
  13 & -7.17 & 35.12 & 0.46 & 0.41 & -17 & 160 \\
  14 & -7.26 & 32.73 & 0.50 & 0.29 & -97 & 75 \\
  15 & -7.21 & 45.43 & 0.46 & 0.38 & -47 & 138 \\
  16 & -7.22 & 30.10 & 0.38 & 0.51 & 13 & 238 \\
  17 & -8.97 & 31.77 & 0.41 & 0.45 & -26 & 160 \\
  18 & -7.03 & 43.66 & 0.44 & 0.36 & -39 & 138 \\
  19 & -7.13 & 35.02 & 0.41 & 0.43 & -14 & 185 \\
  20 & -7.15 & 30.39 & 0.38 & 0.48 & 5 & 249 \\
  21 & -7.04 & 30.39 & 0.37 & 0.50 & 8 & 244 \\
  22 & -7.86 & 33.24 & 0.51 & 0.12 & 15 & 139 \\
  23 & -7.19 & 32.85 & 0.49 & 0.21 & -87 & 98 \\
  24 & -7.22 & 31.31 & 0.36 & 0.50 & 5 & 229 \\
  25 & -7.15 & 34.48 & 0.51 & 0.12 & -8 & 179 \\
  26 & -7.16 & 30.36 & 0.33 & 0.53 & 2 & 203 \\
  27 & -7.08 & 46.69 & 0.41 & 0.38 & -46 & 130 \\
  28 & -7.09 & 35.11 & 0.45 & 0.30 & 7 & 203 \\
  29 & -7.46 & 31.55 & 0.37 & 0.46 & -50 & 88 \\
  30 & -7.04 & 42.84 & 0.48 & 0.18 & 12 & 148 \\
  31 & -7.08 & 31.34 & 0.36 & 0.47 & 18 & 237 \\
  32 & -7.73 & 44.83 & 0.47 & 0.18 & 12 & 209 \\
  33 & -7.10 & 30.78 & 0.39 & 0.40 & -81 & 114 \\
  34 & -7.17 & 32.40 & 0.36 & 0.45 & -56 & 130 \\
  35 & -7.28 & 30.31 & 0.30 & 0.54 & 25 & 217 \\
  36 & -7.08 & 35.90 & 0.40 & 0.37 & -5 & 221 \\
  37 & -7.29 & 30.17 & 0.34 & 0.48 & 17 & 224 \\
  38 & -7.20 & 30.05 & 0.33 & 0.48 & 8 & 236 \\
  39 & -7.35 & 30.40 & 0.43 & 0.28 & -26 & 113 \\
  40 & -7.03 & 38.28 & 0.46 & 0.20 & -61 & 139 \\
  41 & -7.59 & 31.06 & 0.19 & 0.64 & -71 & 83 \\
  42 & -7.26 & 47.14 & 0.39 & 0.36 & -19 & 165 \\
  43 & -7.15 & 30.28 & 0.33 & 0.45 & 25 & 238 \\
  44 & -7.19 & 30.45 & 0.42 & 0.26 & -92 & 94 \\
  45 & -7.80 & 30.70 & 0.22 & 0.58 & -75 & 86 \\
  46 & -7.06 & 40.73 & 0.45 & 0.13 & -2 & 168 \\
  47 & -7.08 & 32.36 & 0.32 & 0.45 & 0 & 231 \\
  48 & -7.25 & 30.26 & 0.42 & 0.21 & -90 & 83 \\
  49 & -7.24 & 30.81 & 0.41 & 0.24 & -67 & 106 \\
  50 & -7.21 & 30.98 & 0.27 & 0.50 & -30 & 128 \\
  51 & -7.06 & 32.22 & 0.31 & 0.44 & 18 & 245 \\
  52 & -7.17 & 30.51 & 0.29 & 0.46 & -4 & 190 \\
  53 & -7.22 & 30.86 & 0.27 & 0.50 & -10 & 187 \\
  54 & -7.22 & 30.41 & 0.37 & 0.31 & -70 & 119 \\
  55 & -7.05 & 31.57 & 0.41 & 0.22 & -40 & 138 \\
  56 & -7.28 & 48.05 & 0.35 & 0.34 & -7 & 164 \\
  57 & -7.27 & 30.77 & 0.24 & 0.51 & -77 & 95 \\
  58 & -7.12 & 32.07 & 0.37 & 0.30 & -21 & 208 \\
  59 & -7.27 & 30.53 & 0.40 & 0.22 & -77 & 121 \\
  60 & -7.38 & 30.04 & 0.27 & 0.48 & -17 & 196 \\
  61 & -7.94 & 38.24 & 0.43 & 0.10 & 23 & 176 \\
  62 & -7.12 & 30.39 & 0.34 & 0.36 & 7 & 249 \\
  63 & -7.31 & 30.57 & 0.28 & 0.46 & 9 & 170 \\
  64 & -7.32 & 30.36 & 0.26 & 0.49 & -36 & 181 \\
  65 & -7.50 & 30.89 & 0.20 & 0.55 & -60 & 107 \\
  66 & -7.06 & 34.47 & 0.25 & 0.49 & -68 & 126 \\
  67 & -7.51 & 31.02 & 0.33 & 0.37 & -43 & 160 \\
  68 & -7.25 & 30.13 & 0.27 & 0.47 & -29 & 204 \\
  69 & -7.02 & 45.49 & 0.35 & 0.33 & -11 & 166 \\
  70 & -7.04 & 33.43 & 0.39 & 0.22 & -39 & 144 \\
  71 & -7.26 & 31.13 & 0.36 & 0.29 & -30 & 107 \\
  72 & -7.16 & 30.25 & 0.26 & 0.47 & -39 & 193 \\
  73 & -7.29 & 30.12 & 0.35 & 0.31 & -5 & 203 \\
  74 & -7.03 & 30.05 & 0.39 & 0.20 & -28 & 225 \\
  75 & -7.38 & 47.92 & 0.34 & 0.31 & -41 & 188 \\
  76 & -7.20 & 31.29 & 0.38 & 0.23 & -101 & 78 \\
  77 & -7.05 & 30.71 & 0.30 & 0.39 & 17 & 233 \\
  78 & -7.03 & 32.18 & 0.27 & 0.43 & 6 & 241 \\
  79 & -7.46 & 31.41 & 0.13 & 0.57 & -55 & 107 \\
  80 & -7.25 & 30.77 & 0.23 & 0.47 & -14 & 201 \\
  81 & -7.04 & 30.74 & 0.26 & 0.43 & 12 & 275 \\
  82 & -7.10 & 30.37 & 0.29 & 0.38 & -6 & 213 \\
  83 & -7.07 & 31.39 & 0.28 & 0.39 & 20 & 239 \\
  84 & -7.19 & 30.37 & 0.30 & 0.35 & 23 & 243 \\
  85 & -7.52 & 48.66 & 0.33 & 0.30 & -35 & 188 \\
  86 & -7.46 & 31.33 & 0.28 & 0.37 & -46 & 121 \\
  87 & -7.16 & 31.81 & 0.30 & 0.34 & 20 & 234 \\
  88 & -7.55 & 30.08 & 0.27 & 0.39 & -75 & 128 \\
  89 & -7.08 & 30.34 & 0.27 & 0.38 & 0 & 253 \\
  90 & -7.10 & 54.26 & 0.39 & 0.10 & -16 & 160 \\
  91 & -7.08 & 46.49 & 0.31 & 0.30 & -35 & 194 \\
  92 & -7.60 & 31.58 & 0.11 & 0.55 & -66 & 106 \\
  93 & -7.07 & 33.10 & 0.37 & 0.16 & -86 & 110 \\
  94 & -7.02 & 37.52 & 0.30 & 0.31 & 1 & 228 \\
  95 & -7.09 & 30.02 & 0.27 & 0.36 & -2 & 226 \\
  96 & -7.27 & 31.30 & 0.29 & 0.31 & 18 & 262 \\
  97 & -7.58 & 31.44 & 0.11 & 0.53 & -42 & 127 \\
  98 & -7.66 & 31.15 & 0.38 & 0.07 & -28 & 120 \\
  99 & -7.22 & 48.85 & 0.29 & 0.31 & -27 & 217 \\
  100 & -7.06 & 31.62 & 0.33 & 0.22 & -43 & 163 \\
  101 & -7.06 & 30.09 & 0.24 & 0.38 & 15 & 255 \\
  102 & -7.18 & 32.37 & 0.24 & 0.39 & 8 & 234 \\
  103 & -7.61 & 30.80 & -0.01 & 0.58 & -59 & 88 \\
  104 & -7.11 & 31.34 & 0.23 & 0.40 & -3 & 266 \\
  105 & -7.17 & 56.40 & 0.37 & 0.07 & 1 & 179 \\
  106 & -7.07 & 42.81 & 0.37 & 0.09 & -53 & 175 \\
  107 & -7.11 & 31.05 & 0.29 & 0.30 & -71 & 136 \\
  108 & -7.52 & 56.18 & 0.33 & 0.21 & -95 & 88 \\
  109 & -7.13 & 57.65 & 0.37 & 0.08 & -2 & 200 \\
  110 & -7.16 & 31.34 & 0.26 & 0.34 & 2 & 236 \\
  111 & -7.10 & 54.61 & 0.36 & 0.08 & -10 & 156 \\
  112 & -7.20 & 30.28 & 0.26 & 0.33 & -15 & 246 \\
  113 & -7.22 & 30.13 & 0.27 & 0.30 & -51 & 146 \\
  114 & -7.38 & 49.54 & 0.35 & 0.09 & -110 & 94 \\
  115 & -7.54 & 31.49 & 0.08 & 0.50 & -37 & 155 \\
  116 & -7.08 & 31.82 & 0.34 & 0.11 & -80 & 134 \\
  117 & -7.49 & 63.99 & 0.36 & 0.04 & -105 & 160 \\
  118 & -7.52 & 32.41 & 0.24 & 0.33 & -21 & 206 \\
  119 & -7.54 & 65.23 & 0.36 & 0.05 & -36 & 212 \\
  120 & -8.04 & 51.86 & 0.34 & 0.08 & -119 & 89 \\
  121 & -7.60 & 64.41 & 0.36 & 0.03 & -72 & 200 \\
  122 & -7.03 & 52.80 & 0.34 & 0.09 & -25 & 221 \\
  123 & -7.56 & 64.49 & 0.36 & 0.02 & -87 & 140 \\
  124 & -7.12 & 63.89 & 0.33 & 0.12 & -1 & 276 \\
  125 & -7.36 & 51.18 & 0.33 & 0.11 & -116 & 111 \\
  126 & -7.28 & 30.36 & 0.25 & 0.29 & 11 & 253 \\
  127 & -7.24 & 60.82 & 0.33 & 0.12 & -90 & 167 \\
  128 & -7.50 & 30.90 & 0.22 & 0.35 & 14 & 214 \\
  129 & -7.14 & 63.41 & 0.34 & 0.05 & -38 & 218 \\
  130 & -7.17 & 57.68 & 0.34 & 0.07 & -1 & 224 \\
  131 & -7.10 & 65.06 & 0.35 & 0.03 & -42 & 208 \\
  132 & -7.13 & 63.92 & 0.35 & 0.03 & -85 & 171 \\
  133 & -7.46 & 63.68 & 0.33 & 0.09 & -10 & 241 \\
  134 & -7.41 & 64.51 & 0.34 & 0.03 & -40 & 232 \\
  135 & -7.63 & 33.31 & 0.21 & 0.34 & -27 & 144 \\
  136 & -7.68 & 32.82 & 0.22 & 0.33 & -49 & 199 \\
  137 & -7.63 & 37.58 & 0.31 & 0.14 & -79 & 102 \\
  138 & -7.45 & 46.89 & 0.33 & 0.08 & -107 & 107 \\
  139 & -7.74 & 46.13 & 0.33 & 0.08 & -58 & 131 \\
  140 & -7.30 & 64.69 & 0.34 & 0.01 & -93 & 168 \\
  141 & -7.34 & 39.50 & 0.28 & 0.20 & -121 & 82 \\
  142 & -7.17 & 63.51 & 0.32 & 0.11 & -34 & 240 \\
  143 & -7.58 & 63.18 & 0.33 & 0.07 & -81 & 181 \\
  144 & -7.02 & 54.56 & 0.32 & 0.09 & -9 & 244 \\
  145 & -7.04 & 61.37 & 0.32 & 0.11 & -112 & 140 \\
  146 & -7.14 & 64.46 & 0.32 & 0.10 & -11 & 258 \\
  147 & -7.63 & 63.51 & 0.32 & 0.08 & -24 & 224 \\
  148 & -7.56 & 63.95 & 0.34 & 0.02 & -76 & 167 \\
  149 & -7.44 & 64.63 & 0.33 & 0.03 & -55 & 223 \\
  150 & -7.62 & 64.86 & 0.34 & 0.01 & -68 & 185 \\
  151 & -7.74 & 64.11 & 0.34 & 0.02 & -61 & 177 \\
  152 & -7.23 & 65.07 & 0.33 & 0.02 & -59 & 198 \\
  153 & -7.02 & 44.99 & 0.32 & 0.06 & -75 & 188 \\
  154 & -7.11 & 63.30 & 0.33 & 0.04 & -123 & 113 \\
  155 & -7.23 & 33.59 & 0.25 & 0.25 & -9 & 245 \\
  156 & -7.79 & 62.01 & 0.33 & 0.03 & -47 & 129 \\
  157 & -7.40 & 62.29 & 0.32 & 0.06 & -136 & 101 \\
  158 & -7.32 & 64.34 & 0.33 & 0.04 & -95 & 174 \\
  159 & -7.51 & 63.78 & 0.32 & 0.04 & -115 & 145 \\
  160 & -7.82 & 61.74 & 0.33 & 0.02 & -70 & 96 \\
  161 & -7.62 & 63.08 & 0.31 & 0.10 & -100 & 147 \\
  162 & -7.23 & 65.45 & 0.32 & 0.04 & -11 & 224 \\
  163 & -7.26 & 32.52 & 0.18 & 0.34 & -12 & 238 \\
  164 & -7.15 & 62.62 & 0.32 & 0.04 & -126 & 83 \\
  165 & -7.45 & 64.95 & 0.32 & 0.02 & -65 & 207 \\
  166 & -7.63 & 64.04 & 0.32 & 0.05 & -94 & 169 \\
  167 & -7.52 & 63.38 & 0.32 & 0.03 & -109 & 117 \\
  168 & -7.31 & 64.71 & 0.30 & 0.11 & -21 & 240 \\
  169 & -7.14 & 62.78 & 0.31 & 0.06 & -116 & 141 \\
  170 & -7.40 & 63.59 & 0.31 & 0.06 & -27 & 234 \\
  171 & -7.18 & 65.13 & 0.31 & 0.06 & -83 & 189 \\
  172 & -7.43 & 63.22 & 0.31 & 0.07 & -76 & 198 \\
  173 & -7.84 & 61.77 & 0.32 & 0.02 & -98 & 166 \\
  174 & -7.92 & 42.05 & 0.27 & 0.17 & -72 & 92 \\
  175 & -7.12 & 63.90 & 0.31 & 0.05 & -111 & 136 \\
  176 & -7.29 & 64.48 & 0.31 & 0.04 & -71 & 222 \\
  177 & -7.42 & 63.63 & 0.31 & 0.03 & -26 & 239 \\
  178 & -7.65 & 58.24 & 0.29 & 0.12 & -108 & 104 \\
  179 & -7.88 & 53.35 & 0.30 & 0.08 & -122 & 101 \\
  180 & -7.86 & 53.60 & 0.30 & 0.07 & -120 & 104 \\
  181 & -7.16 & 58.80 & 0.30 & 0.06 & 16 & 244 \\
  182 & -7.79 & 62.35 & 0.30 & 0.06 & -121 & 103 \\
  183 & -7.84 & 63.36 & 0.30 & 0.07 & -31 & 201 \\
  184 & -7.10 & 62.92 & 0.30 & 0.07 & -101 & 168 \\
  185 & -7.09 & 62.93 & 0.30 & 0.06 & -99 & 173 \\
  186 & -7.11 & 65.12 & 0.31 & 0.03 & -72 & 208 \\
  187 & -7.37 & 65.00 & 0.31 & 0.03 & -50 & 200 \\
  188 & -7.57 & 64.29 & 0.30 & 0.05 & -73 & 190 \\
  189 & -7.04 & 63.79 & 0.30 & 0.04 & -6 & 294 \\
  190 & -7.63 & 63.00 & 0.30 & 0.06 & -111 & 159 \\
  191 & -7.75 & 62.71 & 0.29 & 0.07 & -111 & 124 \\
  192 & -8.05 & 60.99 & 0.31 & 0.01 & -114 & 121 \\
  193 & -7.90 & 30.55 & 0.09 & 0.40 & 3 & 171 \\
  194 & -7.26 & 63.80 & 0.30 & 0.03 & -19 & 259 \\
  195 & -7.72 & 63.95 & 0.29 & 0.04 & -71 & 169 \\
  196 & -7.24 & 65.01 & 0.30 & 0.02 & -84 & 178 \\
  197 & -7.51 & 63.31 & 0.28 & 0.08 & -68 & 220 \\
  198 & -7.85 & 63.43 & 0.30 & 0.02 & -51 & 205 \\
  199 & -7.47 & 44.82 & 0.26 & 0.13 & -127 & 77 \\
  200 & -7.63 & 63.00 & 0.29 & 0.04 & -69 & 171 \\
  201 & -7.59 & 63.43 & 0.30 & 0.02 & -50 & 216 \\
  202 & -8.05 & 61.49 & 0.30 & 0.02 & -80 & 119 \\
  203 & -7.94 & 62.81 & 0.28 & 0.07 & -105 & 136 \\
  204 & -7.67 & 62.88 & 0.29 & 0.05 & -117 & 135 \\
  205 & -8.11 & 60.27 & 0.30 & 0.02 & -133 & 97 \\
  206 & -7.60 & 63.45 & 0.29 & 0.03 & -75 & 212 \\
  207 & -7.08 & 65.51 & 0.29 & 0.04 & -56 & 202 \\
  208 & -7.54 & 63.32 & 0.29 & 0.02 & -81 & 187 \\
  209 & -7.69 & 63.56 & 0.28 & 0.07 & -28 & 221 \\
  210 & -7.68 & 64.60 & 0.28 & 0.05 & -44 & 206 \\
  211 & -7.21 & 65.10 & 0.29 & 0.03 & -64 & 227 \\
  212 & -7.41 & 64.05 & 0.27 & 0.09 & -78 & 179 \\
  213 & -7.44 & 64.42 & 0.28 & 0.05 & -51 & 216 \\
  214 & -7.47 & 63.61 & 0.29 & 0.02 & -42 & 231 \\
  215 & -7.72 & 63.21 & 0.28 & 0.06 & -79 & 193 \\
  216 & -7.56 & 62.98 & 0.28 & 0.05 & -100 & 169 \\
  217 & -7.42 & 64.87 & 0.27 & 0.07 & -24 & 243 \\
  218 & -7.45 & 63.61 & 0.29 & 0.02 & -20 & 236 \\
  219 & -7.32 & 64.78 & 0.28 & 0.03 & -68 & 206 \\
  220 & -8.17 & 47.52 & 0.21 & 0.22 & -61 & 108 \\
  221 & -7.51 & 63.34 & 0.26 & 0.09 & -56 & 216 \\
  222 & -7.48 & 63.62 & 0.28 & 0.03 & -43 & 232 \\
  223 & -7.94 & 31.97 & 0.23 & 0.18 & -33 & 179 \\
  224 & -7.64 & 63.09 & 0.28 & 0.02 & -88 & 161 \\
  225 & -8.19 & 59.49 & 0.29 & -0.04 & -83 & 108 \\
  226 & -7.65 & 63.32 & 0.28 & 0.01 & -90 & 141 \\
  227 & -7.55 & 63.07 & 0.26 & 0.08 & -95 & 162 \\
  228 & -7.77 & 63.34 & 0.28 & 0.03 & -68 & 189 \\
  229 & -7.81 & 61.73 & 0.27 & 0.05 & -112 & 89 \\
  230 & -7.83 & 64.12 & 0.28 & 0.02 & -54 & 177 \\
  231 & -7.42 & 64.23 & 0.27 & 0.04 & -78 & 178 \\
  232 & -7.31 & 63.81 & 0.28 & 0.03 & -12 & 254 \\
  233 & -7.95 & 63.06 & 0.27 & 0.04 & -96 & 166 \\
  234 & -7.51 & 64.49 & 0.27 & 0.05 & -58 & 198 \\
  235 & -7.17 & 64.69 & 0.25 & 0.12 & -44 & 236 \\
  236 & -7.61 & 64.29 & 0.27 & 0.05 & -49 & 197 \\
  237 & -8.12 & 59.98 & 0.29 & -0.04 & -57 & 137 \\
  238 & -7.64 & 64.88 & 0.28 & -0.01 & -22 & 190 \\
  239 & -7.87 & 61.48 & 0.27 & 0.03 & -57 & 178 \\
  240 & -7.46 & 64.95 & 0.27 & 0.03 & -52 & 202 \\
  241 & -8.00 & 62.93 & 0.27 & 0.05 & -74 & 163 \\
  242 & -8.17 & 60.55 & 0.27 & 0.04 & -116 & 102 \\
  243 & -7.96 & 63.21 & 0.26 & 0.04 & -69 & 173 \\
  244 & -7.28 & 65.21 & 0.25 & 0.08 & -16 & 230 \\
  245 & -8.13 & 61.01 & 0.26 & 0.05 & -94 & 124 \\
  246 & -7.79 & 63.32 & 0.26 & 0.04 & -51 & 195 \\
  247 & -7.98 & 61.11 & 0.27 & 0.00 & -106 & 141 \\
  248 & -7.39 & 64.46 & 0.26 & 0.07 & -41 & 226 \\
  249 & -7.82 & 63.31 & 0.26 & 0.04 & -38 & 213 \\
  250 & -7.92 & 63.19 & 0.26 & 0.04 & -52 & 189 \\
  251 & -8.03 & 62.11 & 0.27 & 0.02 & -69 & 131 \\
  252 & -7.34 & 65.21 & 0.26 & 0.04 & -10 & 233 \\
  253 & -7.77 & 34.54 & 0.21 & 0.17 & -61 & 202 \\
  254 & -8.32 & 34.78 & 0.22 & 0.16 & -105 & 105 \\
  255 & -8.56 & 33.41 & 0.20 & 0.19 & -106 & 91 \\
  256 & -7.48 & 63.54 & 0.26 & 0.05 & -48 & 215 \\
  257 & -7.72 & 63.45 & 0.26 & 0.03 & -41 & 212 \\
  258 & -7.88 & 63.32 & 0.26 & 0.03 & -64 & 190 \\
  259 & -8.07 & 46.67 & 0.17 & 0.23 & -51 & 156 \\
  260 & -7.60 & 63.60 & 0.26 & 0.02 & -25 & 235 \\
  261 & -8.29 & 36.13 & 0.22 & 0.14 & -103 & 112 \\
  262 & -7.76 & 62.33 & 0.26 & 0.02 & -14 & 215 \\
  263 & -8.00 & 61.46 & 0.27 & -0.01 & -95 & 151 \\
  264 & -8.32 & 39.02 & 0.20 & 0.18 & -79 & 138 \\
  265 & -7.54 & 63.61 & 0.26 & 0.03 & -23 & 233 \\
  266 & -7.78 & 63.43 & 0.26 & 0.03 & -43 & 215 \\
  267 & -7.75 & 63.28 & 0.24 & 0.08 & -79 & 189 \\
  268 & -7.53 & 65.19 & 0.25 & 0.04 & -5 & 215 \\
  269 & -7.06 & 65.54 & 0.25 & 0.03 & 3 & 210 \\
  270 & -7.81 & 63.56 & 0.24 & 0.06 & -41 & 219 \\
  271 & -7.57 & 64.18 & 0.26 & 0.00 & -75 & 141 \\
  272 & -7.80 & 63.26 & 0.25 & 0.03 & -61 & 191 \\
  273 & -7.63 & 64.05 & 0.24 & 0.05 & -66 & 178 \\
  274 & -7.56 & 62.50 & 0.25 & 0.02 & 16 & 239 \\
  275 & -7.97 & 60.19 & 0.26 & -0.05 & -50 & 165 \\
  276 & -7.94 & 62.31 & 0.25 & 0.02 & -16 & 188 \\
  277 & -7.80 & 63.03 & 0.24 & 0.04 & -91 & 169 \\
  278 & -8.14 & 58.80 & 0.26 & -0.04 & -105 & 113 \\
  279 & -7.83 & 62.42 & 0.23 & 0.06 & -71 & 168 \\
  280 & -7.96 & 34.95 & 0.16 & 0.20 & -49 & 188 \\
  281 & -8.14 & 60.89 & 0.24 & 0.02 & -104 & 127 \\
  282 & -7.98 & 61.71 & 0.24 & 0.03 & -83 & 138 \\
  283 & -8.17 & 35.32 & 0.17 & 0.19 & -51 & 152 \\
  284 & -7.89 & 36.97 & 0.18 & 0.17 & -56 & 185 \\
  285 & -8.08 & 60.06 & 0.26 & -0.05 & -57 & 143 \\
  286 & -8.46 & 38.25 & 0.17 & 0.18 & -89 & 108 \\
  287 & -7.34 & 64.25 & 0.22 & 0.06 & -76 & 176 \\
  288 & -8.05 & 36.44 & 0.16 & 0.19 & -33 & 180 \\
  289 & -8.08 & 58.97 & 0.25 & -0.03 & -61 & 136 \\
  290 & -7.47 & 64.32 & 0.21 & 0.09 & -33 & 217 \\
  291 & -7.90 & 63.63 & 0.24 & -0.01 & -61 & 154 \\
  292 & -7.06 & 62.02 & 0.21 & 0.08 & -5 & 260 \\
  293 & -8.24 & 40.27 & 0.12 & 0.24 & -61 & 131 \\
  294 & -7.81 & 63.98 & 0.23 & 0.01 & -54 & 187 \\
  295 & -8.81 & 30.24 & 0.01 & 0.33 & -81 & 77 \\
  296 & -8.49 & 31.77 & 0.18 & 0.15 & -106 & 86 \\
  297 & -7.95 & 39.35 & 0.13 & 0.22 & -52 & 186 \\
  298 & -7.75 & 64.05 & 0.21 & 0.06 & -64 & 173 \\
  299 & -8.15 & 60.04 & 0.24 & -0.06 & -102 & 114 \\
  300 & -7.83 & 62.93 & 0.21 & 0.07 & -102 & 134 \\
  301 & -7.85 & 60.71 & 0.23 & 0.01 & -97 & 152 \\
  302 & -7.31 & 65.35 & 0.23 & -0.01 & 8 & 227 \\
  303 & -7.95 & 62.70 & 0.22 & 0.03 & -98 & 147 \\
  304 & -7.33 & 64.58 & 0.23 & 0.01 & -69 & 166 \\
  305 & -8.31 & 59.39 & 0.24 & -0.04 & -122 & 76 \\
  306 & -8.00 & 61.04 & 0.23 & 0.00 & -103 & 129 \\
  307 & -7.31 & 62.41 & 0.20 & 0.08 & -13 & 252 \\
  308 & -8.19 & 38.57 & 0.13 & 0.21 & -10 & 169 \\
  309 & -7.62 & 62.40 & 0.23 & 0.01 & 8 & 236 \\
  310 & -8.75 & 35.85 & 0.13 & 0.22 & -92 & 95 \\
  311 & -7.41 & 64.61 & 0.23 & 0.00 & -70 & 163 \\
  312 & -8.63 & 30.52 & 0.00 & 0.32 & -62 & 97 \\
  313 & -8.16 & 47.01 & 0.13 & 0.21 & -53 & 136 \\
  314 & -8.08 & 61.33 & 0.22 & 0.01 & -97 & 131 \\
  315 & -8.75 & 32.86 & 0.09 & 0.25 & -77 & 99 \\
  316 & -8.20 & 40.29 & 0.12 & 0.22 & -44 & 163 \\
  317 & -7.73 & 61.77 & 0.22 & 0.00 & -73 & 193 \\
  318 & -7.94 & 62.49 & 0.21 & 0.03 & -16 & 215 \\
  319 & -8.04 & 62.34 & 0.22 & 0.02 & -30 & 190 \\
  320 & -7.15 & 63.89 & 0.19 & 0.09 & -37 & 240 \\
  321 & -7.69 & 62.63 & 0.22 & 0.02 & -66 & 193 \\
  322 & -7.32 & 64.32 & 0.19 & 0.09 & -26 & 235 \\
  323 & -8.40 & 38.60 & 0.09 & 0.24 & -71 & 88 \\
  324 & -8.34 & 30.82 & 0.17 & 0.13 & -109 & 91 \\
  325 & -8.66 & 30.60 & 0.13 & 0.20 & -95 & 81 \\
  326 & -8.50 & 35.31 & 0.12 & 0.20 & -108 & 91 \\
  327 & -7.15 & 45.48 & 0.16 & 0.13 & 8 & 281 \\
  328 & -7.99 & 59.96 & 0.23 & -0.05 & -67 & 166 \\
  329 & -7.06 & 63.57 & 0.16 & 0.13 & -8 & 280 \\
  330 & -7.27 & 62.98 & 0.17 & 0.12 & -32 & 237 \\
  331 & -7.86 & 63.67 & 0.19 & 0.06 & -64 & 140 \\
  332 & -8.24 & 39.11 & 0.11 & 0.21 & -56 & 131 \\
  333 & -8.35 & 38.24 & 0.10 & 0.22 & -63 & 109 \\
  334 & -7.69 & 64.29 & 0.19 & 0.06 & -42 & 190 \\
  335 & -8.74 & 35.43 & 0.09 & 0.23 & -99 & 76 \\
  336 & -8.69 & 30.87 & 0.12 & 0.18 & -93 & 90 \\
  337 & -7.91 & 62.38 & 0.20 & 0.02 & -18 & 214 \\
  338 & -8.56 & 31.09 & 0.13 & 0.18 & -69 & 120 \\
  339 & -8.53 & 35.85 & 0.12 & 0.18 & -92 & 93 \\
  340 & -8.38 & 39.49 & 0.09 & 0.22 & -62 & 110 \\
  341 & -7.76 & 62.42 & 0.20 & 0.02 & 1 & 234 \\
  342 & -8.30 & 37.74 & 0.11 & 0.19 & -82 & 121 \\
  343 & -7.08 & 64.34 & 0.19 & 0.05 & -12 & 261 \\
  344 & -8.53 & 32.98 & 0.07 & 0.23 & -60 & 121 \\
  345 & -7.68 & 64.84 & 0.20 & 0.00 & -10 & 189 \\
  346 & -8.25 & 39.72 & 0.09 & 0.21 & -47 & 163 \\
  347 & -7.71 & 62.50 & 0.20 & 0.02 & 4 & 236 \\
  348 & -8.52 & 38.56 & 0.07 & 0.22 & -44 & 133 \\
  349 & -8.48 & 35.33 & 0.05 & 0.24 & -61 & 106 \\
  350 & -8.16 & 40.28 & 0.10 & 0.20 & -17 & 190 \\
  351 & -7.56 & 62.69 & 0.18 & 0.02 & 25 & 258 \\
  352 & -8.43 & 39.17 & 0.06 & 0.22 & -31 & 159 \\
  353 & -8.01 & 59.74 & 0.18 & 0.02 & -81 & 143 \\
  354 & -8.17 & 48.31 & -0.02 & 0.26 & -33 & 112 \\
  355 & -7.70 & 64.15 & 0.16 & 0.06 & -31 & 183 \\
  356 & -7.26 & 44.04 & 0.14 & 0.09 & 16 & 254 \\
  357 & -8.27 & 47.00 & -0.01 & 0.25 & -60 & 89 \\
  358 & -8.59 & 31.51 & 0.00 & 0.24 & -61 & 108 \\
  359 & -8.24 & 47.73 & -0.01 & 0.24 & -53 & 116 \\
  360 & -7.42 & 64.38 & 0.14 & 0.07 & -20 & 231 \\
  361 & -8.57 & 39.58 & -0.13 & 0.26 & -37 & 91 \\
  362 & -7.72 & 64.29 & 0.16 & 0.03 & -74 & 181 \\
  363 & -7.56 & 64.24 & 0.14 & 0.06 & -20 & 207 \\
  364 & -8.23 & 45.96 & -0.01 & 0.23 & -70 & 91 \\
  365 & -8.04 & 61.81 & 0.17 & -0.02 & -24 & 125 \\
  366 & -8.52 & 38.15 & -0.13 & 0.25 & -48 & 97 \\
  367 & -7.30 & 59.35 & 0.15 & 0.02 & 17 & 264 \\
  368 & -8.02 & 60.25 & 0.15 & 0.01 & -69 & 148 \\
  369 & -8.61 & 31.56 & -0.05 & 0.23 & -53 & 102 \\
  370 & -7.67 & 36.71 & 0.12 & 0.08 & -3 & 240 \\
  371 & -7.70 & 59.83 & 0.14 & 0.02 & -66 & 194 \\
  372 & -7.38 & 62.01 & 0.13 & 0.04 & 13 & 252 \\
  373 & -7.97 & 62.15 & 0.15 & -0.03 & -31 & 187 \\
  374 & -7.08 & 61.90 & 0.12 & 0.04 & -10 & 260 \\
  375 & -8.61 & 38.11 & 0.00 & 0.18 & -33 & 98 \\
  376 & -7.24 & 52.03 & 0.01 & 0.17 & 16 & 271 \\
  377 & -8.53 & 38.59 & -0.02 & 0.18 & -19 & 122 \\
  378 & -8.04 & 38.49 & 0.03 & 0.15 & -9 & 199 \\
  379 & -8.43 & 31.19 & -0.11 & 0.19 & -66 & 123 \\
  380 & -7.40 & 43.11 & 0.05 & 0.09 & 1 & 266 \\
  381 & -7.35 & 60.22 & 0.04 & 0.01 & -17 & 246 \\
  382 & -7.35 & 60.65 & 0.04 & 0.00 & 11 & 248 \\
  383 & -7.18 & 61.18 & 0.03 & -0.01 & 4 & 263 \\
  384 & -7.21 & 61.16 & 0.02 & 0.00 & 11 & 264 \\
  385 & -7.05 & 61.35 & 0.01 & -0.01 & 13 & 282 \\
  386 & -7.24 & 57.24 & -0.02 & 0.01 & 20 & 276 \\
  387 & -7.34 & 57.21 & -0.02 & 0.00 & 15 & 259 \\
  388 & -8.16 & 57.06 & 0.02 & -0.06 & -51 & 112 \\
  389 & -7.46 & 56.96 & -0.02 & -0.02 & 13 & 240 \\
  390 & -8.24 & 57.50 & 0.01 & -0.06 & -64 & 78 \\
\end{longtable}
\FloatBarrier

\subsubsection{Model Based Solvent Design Rules}
\label{sec:si:lio2_design_rules}

\begin{figure}[h!]
    \centering
    \includegraphics[width=.9\textwidth]{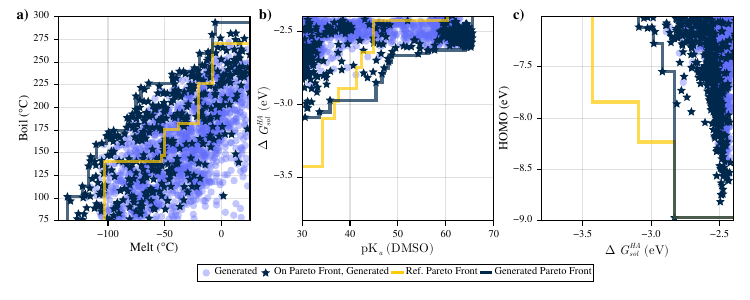}
    \labelphantom{fig:si:lio2_boil_melt}
    \labelphantom{fig:si:lio2_g_pKa}
    \labelphantom{fig:si:lio2_homo_g}
    \caption{
         \label{fig:si:lithium_air_pareto}
        \textbf{Screening outcomes for the lithium--air electrolyte design workflow.}
        Reference molecules taken from prior lithium--air electrolyte studies are shown as yellow crosses, generated molecules as faded blue circles, and Pareto-optimal candidates as blue stars. The orange line traces the non-dominated molecules across all objectives. 
        (\subref*{fig:si:lio2_boil_melt})~Melting point versus boiling point, showing advancement toward broader liquid windows characterized by lower melting points and higher boiling points. 
        (\subref*{fig:si:lio2_g_pKa})~$pK_a$ (DMSO) versus $\Delta G_{\mathrm{sol}}^{HA}$, highlighting the trade-off between resistance to H-abstraction and retention of an admissible solution-mediated growth window. 
        (\subref*{fig:si:lio2_homo_g})~HOMO versus $\Delta G_{\mathrm{sol}}^{HA}$, showing simultaneous improvement in oxidative stability while preserving favorable solution-mediated thermodynamics.
    }
\end{figure}

In~\cref{fig:si:lio2_boil_melt}, the generated set spans a wider thermal window than most of the reference molecules, and the Pareto frontier advances toward the low-melting, high-boiling region. 
This is precisely the direction favored by the liquid-state screen: solvents that remain fluid near ambient conditions while also avoiding excessive volatility. 
Rather than improving a single thermophysical property in isolation, the frontier shows that \ac{MIST} can identify molecules that jointly widen the accessible liquid range.

\Cref{fig:si:lio2_g_pKa} shows the trade-off between $pK_a$ in \ac{DMSO} and $\Delta G_{\mathrm{sol}}^{HA}$. 
In our workflow, $pK_a$ serves as the stability descriptor against H-abstraction, whereas $\Delta G_{\mathrm{sol}}^{HA}$ measures whether the solvent remains inside the admissible solution-mediated window. 
The generated frontier shifts toward substantially larger $pK_a$ values while keeping $\Delta G_{\mathrm{sol}}^{HA}$ in the desired range, indicating that the workflow can identify solvents that are less prone to abstraction-driven degradation without sacrificing the thermodynamic conditions needed for solution-mediated peroxide growth. 
Since solution-mediated growth has been linked to particulate or toroidal \ce{Li2O2} formation and higher discharge capacity, this panel captures a chemically meaningful balance between stability and capacity-enabling discharge chemistry.

\Cref{fig:si:lio2_homo_g} shows an analogous trade-off between \ac{HOMO} level and $\Delta G_{\mathrm{sol}}^{HA}$. 
Here, the Pareto candidates reach deeper \ac{HOMO} energies than most of the reference set, while maintaining favorable values of $\Delta G_{\mathrm{sol}}^{HA}$. 
In the context of the present search, this means that oxidative stability can be improved without moving the solvent outside the solution-mediated regime. 
Taken together,~\cref{fig:si:lio2_g_pKa} and~\cref{fig:si:lio2_homo_g} the workflow populates the narrow region where the solvent simultaneously supports solution-mediated growth, resists H-abstraction, and becomes harder to oxidize.

\begin{figure}[h!]
    \centering
    \includegraphics[width=.7\textwidth]{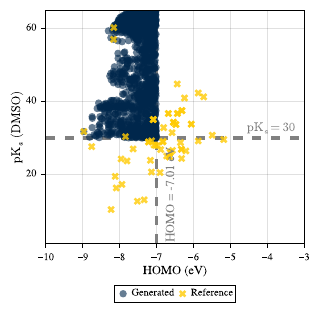}
    \caption{
        Joint solvent-stability screen for lithium--air electrolyte candidates. Generated molecules are shown as blue circles and literature reference molecules as yellow crosses. The dashed lines mark the applied stability thresholds, HOMO $=-7.01$ eV and $pK_a(\mathrm{DMSO})=30$. Candidates in the upper-left region satisfy both criteria simultaneously, combining deeper HOMO levels with larger $pK_a$ values.
        \label{fig:si:lithium_air_pka_sol}
    }
\end{figure}

We summarize the stability screen in~\cref{fig:si:lithium_air_pka_sol}. 
The generated molecules cluster strongly in the upper-left quadrant defined by deeper HOMO levels and larger $pK_a$ values, whereas the reference molecules are more broadly dispersed and often violate one or both thresholds. 

Combined with the liquid-state filter and the enforced solution-mediated bound, these results show that the workflow can identify solvents that remain liquid under practical conditions, resist the two principal degradation modes targeted here, and preserve the thermodynamic conditions associated with solution-mediated \ce{Li2O2} growth. 
More broadly, this case study shows how \ac{MIST} models can translate literature-derived electrolyte design rules into a multi-parameter molecular screening engine, enabling the efficient discovery of candidate solvents within the narrow design windows characteristic of battery electrolyte selection.

\begin{figure}[h!]
    \centering
    \labelphantom{fig:si:lio2_func_groups}
    \labelphantom{fig:si:lio2_flourination}
    \includegraphics[width=.9\textwidth]{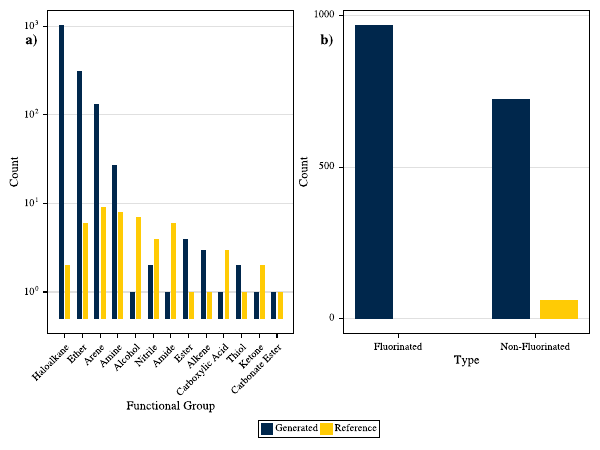}
    \caption{
        Structural motifs represented among the lithium--air screening candidates. 
        (\subref*{fig:si:lio2_func_groups})~Functional-group occurrence in the generated candidate set (blue) and the literature reference set (yellow), shown on a logarithmic count scale. 
        (\subref*{fig:si:lio2_flourination})~Distribution of fluorinated and non-fluorinated molecules in the generated and reference sets. Together, these panels show that the screened candidates are enriched in haloalkanes, ethers, arenes, and amines, with strong representation of fluorinated scaffolds.
        \label{fig:si:lithium_air_functional_group_distribution}
    }
\end{figure}
Considering the functional-group distribution of the screened molecules in~\cref{fig:si:lithium_air_functional_group_distribution}, we find that the generated set is dominated by haloalkanes, ethers, arenes, and amines, with haloalkanes appearing most frequently. 
This enrichment is chemically sensible in the context of aprotic Li--O$_2$ electrolyte design. 
Ethers, in particular, have been widely used in Li--O$_2$ batteries because they are aprotic, comparatively resistant to hydrogen abstraction, and lack carbonyl-type motifs that can promote parasitic pathways leading to passivating \ce{Li2CO3} formation during operation \cite{zhang2011partially,wijaya2015gamma,burke2015enhancing,zhao2018binary,schwenke2013stability,kwak2020optimized,luntz2013tunneling}. 
Arenes \cite{zaidi2024fluorobenzene} and amines \cite{jiang2022quenching,petit2019dabconium} have also been explored in Li--O$_2$ systems, although more commonly as diluents or additives in localized high-concentration electrolytes rather than as stand-alone base solvents.

A second clear feature of the screened set is the prominence of fluorinated motifs, shown in~\cref{fig:si:lio2_flourination}. 
Haloalkanes are themselves abundant, and fluorinated derivatives of other functional classes also appear frequently. 
This trend is again consistent with prior lithium--air and broader battery-electrolyte literature, where fluorinated solvents and co-solvents are often used either as base electrolyte components \cite{chen2022building} or as additives in \ac{LHCE} formulations \cite{wen2025boosting,wijaya2015gamma,rigoni2024use,zhang2011partially}. 
Fluorination has been associated with improved stabilization of reduced oxygen species in solution~\cite{zhang2011partially} and with the formation of LiF-rich interphases at the lithium-metal electrode, which can suppress further electrolyte decomposition and stabilize the anode~\cite{chen2022building,nasybulin2013effects}. 
Nasybulin et al.\cite{nasybulin2013effects}, for example, showed that fluorine-containing solvents can support \ce{LiO2} and \ce{Li2O2} discharge chemistry while simultaneously generating protective LiF, with failure emerging primarily when other electrolyte components are not sufficiently oxidation-resistant.

The prevalence of haloalkanes and other fluorinated scaffolds suggests that MIST is identifying structural motifs that align with established Li--O$_2$ electrolyte design principles while remaining inside the narrow thermodynamic and stability windows enforced by the screening protocol. 
More broadly,~\cref{fig:si:lithium_air_functional_group_distribution} shows that the model is not converging on a single narrow chemical family, but instead populating a chemically interpretable region of solvent space enriched in motifs already associated with favorable Li--O$_2$ behavior.

\section{Molecular Surprise}
\label{sec:si:synth_access}

We investigated whether molecular surprise (\cref{eq:molecular_surprise}) correlates with existing metrics for molecular complexity, specifically synthetic accessibility\cite{ESEstimationSyntheticAccessibility2009,CRGJSCScoreSyntheticComplexity2018,SZS+ModelingCrowdsourcedDefinition2014} and the molecular assembly index\cite{MMC+IdentifyingMoleculesBiosignatures2021}.
Synthetic accessibility aims to estimate the difficulty of synthesizing a molecule, but it lacks a universally agreed-upon definition or ground truth\cite{SZS+ModelingCrowdsourcedDefinition2014}.
Heuristic estimators are typically rule-based\cite{ESEstimationSyntheticAccessibility2009} or derived from reaction pathways\cite{CRGJSCScoreSyntheticComplexity2018}, and expert assessments vary widely\cite{SZS+ModelingCrowdsourcedDefinition2014}.
Unexpectedly, we found the Smirk tokenizer sequence length to be highly predictive of chemist-annotated synthetic accessibility, outperforming existing cheminformatic heuristics (\cref{fig:sa_score_grid_crowd}).

\subsection{Summary of Evaluated Heuristics for Molecular Complexity}

Here, we detail the molecular complexity metrics evaluated, including their retrieval information.

\paragraph{SAScore.}
SAScore\cite{ESEstimationSyntheticAccessibility2009} uses hand-selected and tuned molecular descriptors (e.g., the number of rings) to estimate synthetic accessibility on a scale from 1 (easy) to 10 (hard).
It is integrated into RDKit\cite{GreRDKitOpensourceCheminformatics2024}, making it a widely adopted synthetic accessibility metric in cheminformatics.

\paragraph{BR-SAScore.}
BR-SAScore\cite{CJEstimatingSyntheticAccessibility2024} extends SAScore\cite{ESEstimationSyntheticAccessibility2009} by incorporating additional factors reflecting the complexity of molecular building blocks and reaction pathways.

\paragraph{SyBA.}
SyBA\cite{VKCSSYBABayesianEstimation2020} is a fragment-based method that uses a na\"ive Bayesian classifier to assess molecular complexity.
It was trained on molecules from ZINC15\cite{SIZINC15Ligand2015} (easy class) and molecules generated using Nonpher\cite{VSNonpherComputationalMethod2017} (hard class).
Model parameters were obtained from \url{https://anaconda.org/LICH/syba/1.0.1/download/noarch/syba-1.0.1-py_0.tar.bz2}, as referenced in the project's \href{https://github.com/lich-uct/syba/issues/6}{GitHub issues}.

\paragraph{SCScore.}
SCScore\cite{CRGJSCScoreSyntheticComplexity2018} uses a neural network trained on reaction pathways to compute synthetic accessibility from Morgan fingerprints.
It is trained with a hinge loss that assigns higher scores (less accessible) to products than to reactants.
We retrieved SCScore's source code and model weights from the GitHub at \ghcommit{connorcoley/scscore}{37090a6aa8220408f572de20224c33c221312d16}.
Minor edits were made for compatibility with our Python 3 environment.
The modified code is available at \texttt{opt/synth\_access/vendor/scscore}.

\paragraph{Molecular Assembly Index.}
The Molecular Assembly Index, introduced by Marshall et al.\cite{MMC+IdentifyingMoleculesBiosignatures2021}, was computed using the implementation by Seet et al.\cite{SPS+RapidComputationAssembly2024}, available at \ghcommit{DaymudeLab/assembly-theory}{4116c7bd0a2d590fa4843aa5448d3485a268dfba}.
As exact computation of molecular assembly is NP-complete\cite{SPS+RapidComputationAssembly2024}, we imposed a 10-second time limit per molecule.
This allowed us to compute scores for 55.7\% of the Sheridan et al.\cite{SZS+ModelingCrowdsourcedDefinition2014} dataset and 81.6\% of the Chen and Jung\cite{CJEstimatingSyntheticAccessibility2024} dataset.
Increasing the timeout to 30 seconds resulted in over 2,304 CPU-hours of computation before timing out.

\paragraph{Molecular Weight.}
As a baseline metric, molecular weights were computed using RDKit\cite{GreRDKitOpensourceCheminformatics2024}.

\paragraph{Smirk.}
Using the Smirk tokenizer\cite{WBVTokenizationMolecularFoundation2026}, we computed tokenized sequence lengths for molecules in original, canonical, and Kekul\'e \ac{SMILES} formats.
Canonical and Kekul\'e forms were generated from the original \ac{SMILES} using RDKit\cite{GreRDKitOpensourceCheminformatics2024}.
Smirk fully decomposes \ac{SMILES} strings into constituent glyphs, as defined by the OpenSMILES\cite{CraOpenSMILES2016} specification.
For example, \href{https://pubchem.ncbi.nlm.nih.gov/compound/3672}{Ibuprofen} is tokenized as \tok{C,C,(,C,),C,c,1,c,c,c,(,c,c,1,),[,C,@@,H,],(,C,),C,(,=,O,),O}, yielding a score of 30.

\paragraph{MoLFormer.}
The 44M parameter molecular foundation model developed by IBM\cite{RBC+LargescaleChemicalLanguage2022} was retrieved from IBM's HuggingFace repository (\hfrepo{ibm-research/MoLFormer-XL-both-10pct}).
MoLFormer was trained using \acf{MLM} on 1.1 billion \ac{SMILES} encoded molecules from the PubChem and ZINC dataset\cite{RBC+LargescaleChemicalLanguage2022}.
MoLFormer was not explicitly trained to predicted synthetic accessibility.

\paragraph{ChemBERTa.}
We retrieved the pretrained ChemBERTa model\cite{CGRChemBERTaLargeScaleSelfSupervised2020} from the HuggingFace repository at \hfrepo{seyonec/ChemBERTa-zinc-base-v1}, as indicated in the original paper.
ChemBERTa uses the RoBERTa architecture\cite{LOG+RoBERTaRobustlyOptimized2019} and was trained using \ac{MLM} on molecules from the PubChem dataset\cite{CGRChemBERTaLargeScaleSelfSupervised2020}.
ChemBERTa was not explicitly trained to predict synthetic accessibility.

\subsection{Summary of Molecular \acp{FM} Evaluated for Molecular Complexity}

We computed molecular surprise using both publicly available models (MoLFormer\cite{RBC+LargescaleChemicalLanguage2022} and ChemBERTa\cite{CGRChemBERTaLargeScaleSelfSupervised2020}) and \ac{MIST}.
All \acp{FM} were evaluated without explicit fine-tuning for synthetic accessibility.
Each model was tested in 12 configurations, varying across the following.

\begin{description}
    \item[Encoding.] Original, canonical, and Kekul\'e \ac{SMILES}.
    \item[Trained/Untrained.] Pretrained weights vs.\ randomly initialized weights.
    \item[Normalization.] Scores computed per molecule vs.\ normalized per token.
\end{description}

\begin{table}
    \centering
    \begin{tabular}{lrr}
\toprule
                                            & \multicolumn{2}{c}{AUROC} \\ 
\cmidrule(lr){2-3} 
                                            &         (1) &         (2) \\ 
\midrule
(Intercept)                                 &    0.582*** &    0.539*** \\ 
                                            &     (0.014) &     (0.009) \\ 
untrained: true                             &    0.143*** &     0.026** \\ 
                                            &     (0.014) &     (0.009) \\ 
per\_token: true                            &   -0.130*** &   -0.039*** \\ 
                                            &     (0.014) &     (0.009) \\ 
encoding: canonical                         &      -0.018 &      -0.000 \\ 
                                            &     (0.019) &     (0.013) \\ 
encoding: kekule                            &       0.031 &    0.056*** \\ 
                                            &     (0.019) &     (0.013) \\ 
model: ibm-research/MoLFormer-XL-both-10pct &       0.060 &       0.044 \\ 
                                            &     (0.033) &     (0.023) \\ 
model: models/mist-1.8B-dh61satt            &      -0.050 &     -0.054* \\ 
                                            &     (0.033) &     (0.023) \\ 
model: models/mist-28znv46w                 &      -0.013 &      -0.024 \\ 
                                            &     (0.033) &     (0.023) \\ 
model: models/mist-4yzwys2z                 &      -0.021 &      -0.017 \\ 
                                            &     (0.033) &     (0.023) \\ 
model: models/mist-n2dkcidc                 &       0.007 &       0.013 \\ 
                                            &     (0.033) &     (0.023) \\ 
model: seyonec/ChemBERTa-zinc-base-v1       &     0.097** &    0.085*** \\ 
                                            &     (0.033) &     (0.023) \\ 
\midrule
$N$                                         &          84 &          84 \\ 
Degrees of Freedom                          &          73 &          73 \\ 
$R^2$                                       &       0.751 &       0.510 \\ 
MAE                                         &       0.114 &       0.087 \\ 
RMSE                                        &       0.117 &       0.079 \\ 
\bottomrule
\end{tabular}

    \caption{
        \label{tab:molecular_surprise_fe}
        Linear Fixed-Effects Model for molecular foundation model \ac{AUROC} scores when predicting Chemist annotated (1\cite{SZS+ModelingCrowdsourcedDefinition2014}) or Easy/Hard (2\cite{CJEstimatingSyntheticAccessibility2024}) synthetic accessibility.
        Overall, molecular \acp{FM} scored 58.2\% and 53.4\% on the datasets respectively, indicating limited discriminative capacity.
        Notably, normalizing molecular surprise by the length of the tokenized sequence results in a significant reduction in predictive power.
    }
\end{table}

\begin{figure}[h!]
    \centering
    \labelphantom{fig:model_score_grid_crowd_heuristic}
    \labelphantom{fig:model_score_grid_crowd_mfm}
    \includegraphics[width=\linewidth]{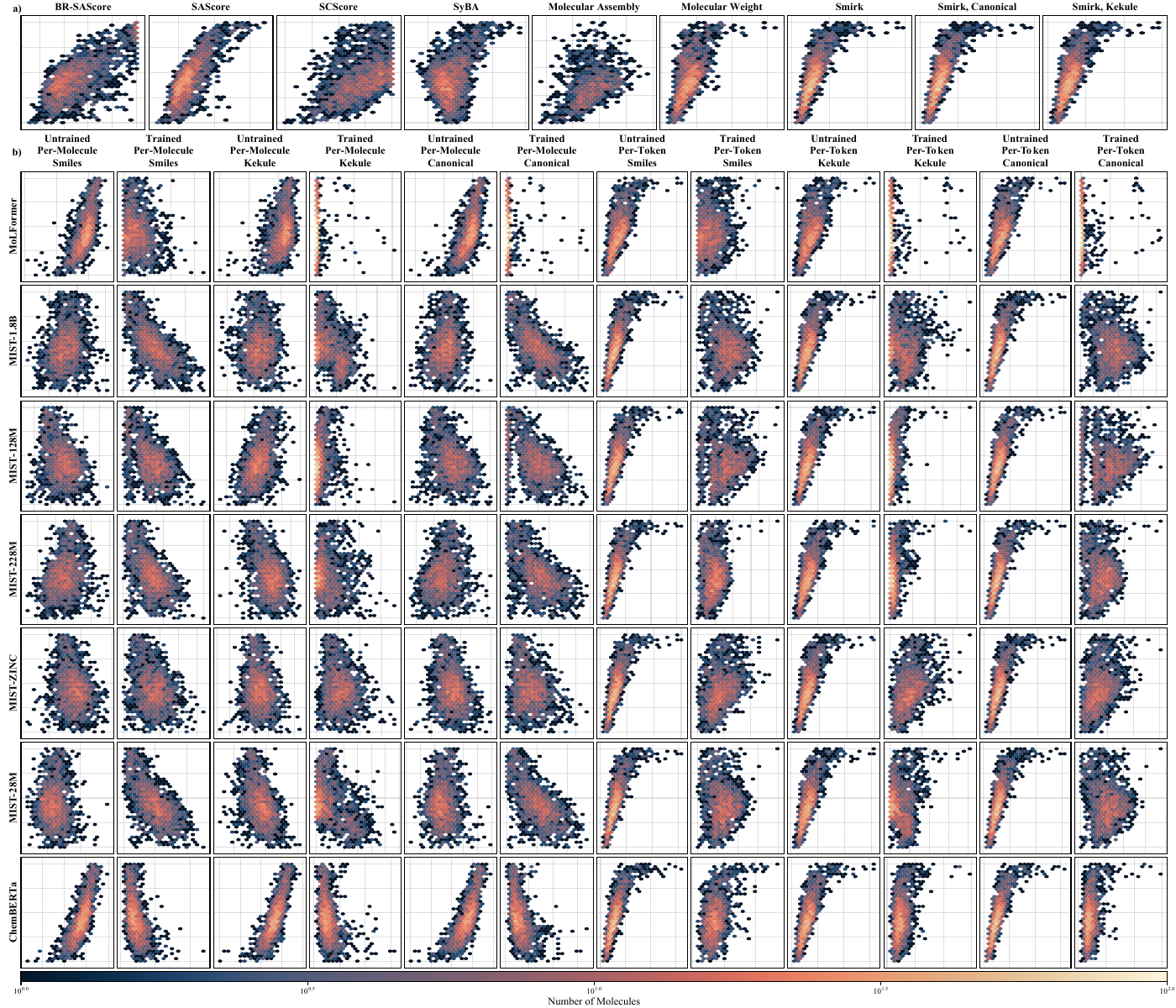}
    \caption{
        \label{fig:sa_score_grid_crowd}
        Comparison of chemist-annotated molecular complexity (y-axis) with various computational metrics (x-axis).
        \subref*{fig:model_score_grid_crowd_heuristic}: heuristic metrics.
        \subref*{fig:model_score_grid_crowd_mfm}: molecular \acp{FM}, grouped by architecture (rows) and treatment (columns).
        Molecular surprise from untrained models shows a weak positive correlation with chemist annotations.
        However, this correlation diminishes—or even becomes weakly negative—after training (e.g., MIST-28M, Trained, Per-Molecule, \ac{SMILES}).
        Notably, both Molecular Weight and Smirk sequence length exhibit a consistent positive correlation with chemist annotations.
    }
\end{figure}

\begin{figure}[h!]
    \centering
    \includegraphics[width=\linewidth]{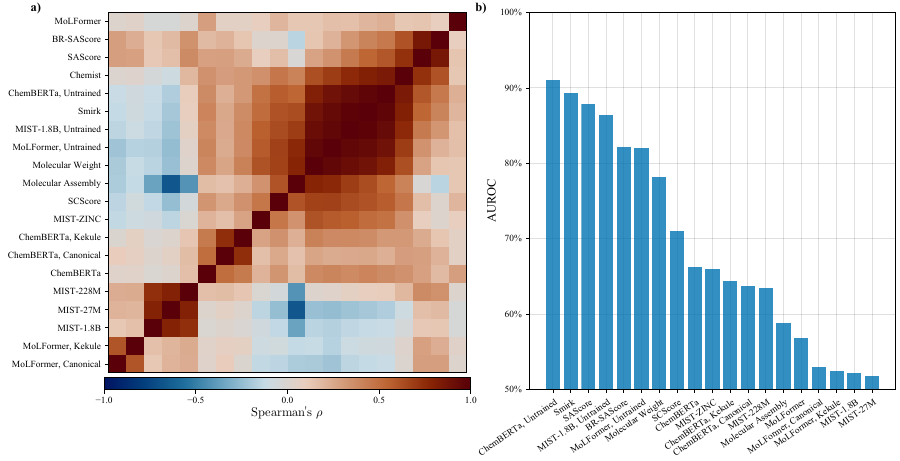}
    \caption{
        \label{fig:molecular_surprise_crowd_all}
        Synthetic accessibility scores across all evaluated models for the dataset from Sheridan et al.\cite{SZS+ModelingCrowdsourcedDefinition2014}.
    }
\end{figure}

\begin{figure}[h!]
    \centering
    \includegraphics[width=\linewidth]{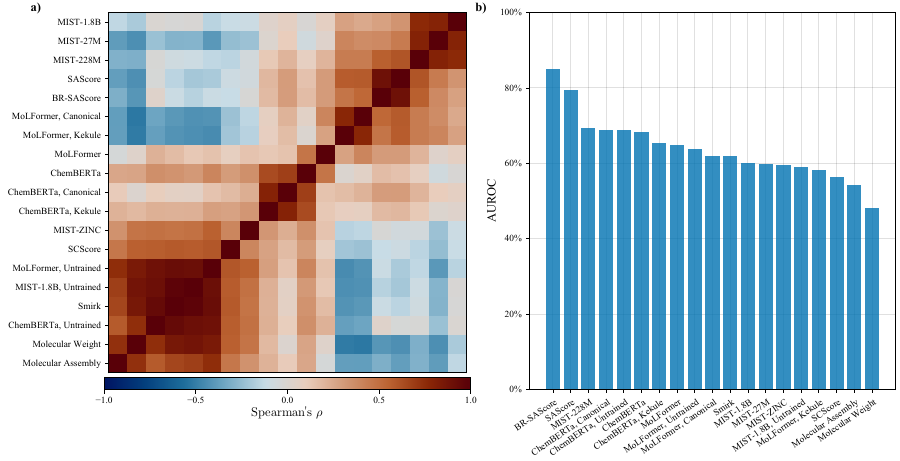}
    \caption{
        \label{fig:molecular_surprise_ba_all}
        Synthetic accessibility scores across all evaluated models for the dataset used by BR-SAScore\cite{CJEstimatingSyntheticAccessibility2024}.
    }
\end{figure}

We fit linear fixed-effects models (\cref{tab:molecular_surprise_fe}) to estimate treatment effects.
Untrained models produced significantly higher molecular surprise scores than trained models, suggesting that pretraining does not align with chemist-assessed synthetic accessibility.
The encoding choice had negligible impact, while per-token normalization reduced model performance (Lower \ac{AUROC}).
Interestingly, Smirk sequence length (\cref{fig:molecular_surprise_crowd_all,fig:molecular_surprise_ba_all}) was among the strongest predictors, suggesting untrained models may serve as noisy proxies for sequence length.
Hexbin plots comparing chemist annotations from Sheridan et al.\cite{SZS+ModelingCrowdsourcedDefinition2014} to evaluated metrics are shown in \cref{fig:sa_score_grid_crowd}.
Pairwise correlations and \ac{AUROC} scores for both the Sheridan et al.\cite{SZS+ModelingCrowdsourcedDefinition2014} and Chen and Jung\cite{CJEstimatingSyntheticAccessibility2024} datasets are shown in \cref{fig:molecular_surprise_crowd_all,fig:molecular_surprise_ba_all}, respectively.

\section{Olfaction}
\label{sec:si:olfaction}

\begin{figure}[ht!]
    \centering
    \labelphantom{fig:num_samples_auroc}
    \labelphantom{fig:functional_group_auroc}
    \labelphantom{fig:num_molecules_auroc}
    \labelphantom{fig:discontinuous_odour}
    \labelphantom{fig:logit_correlation}
    
    \includegraphics[width=\linewidth]{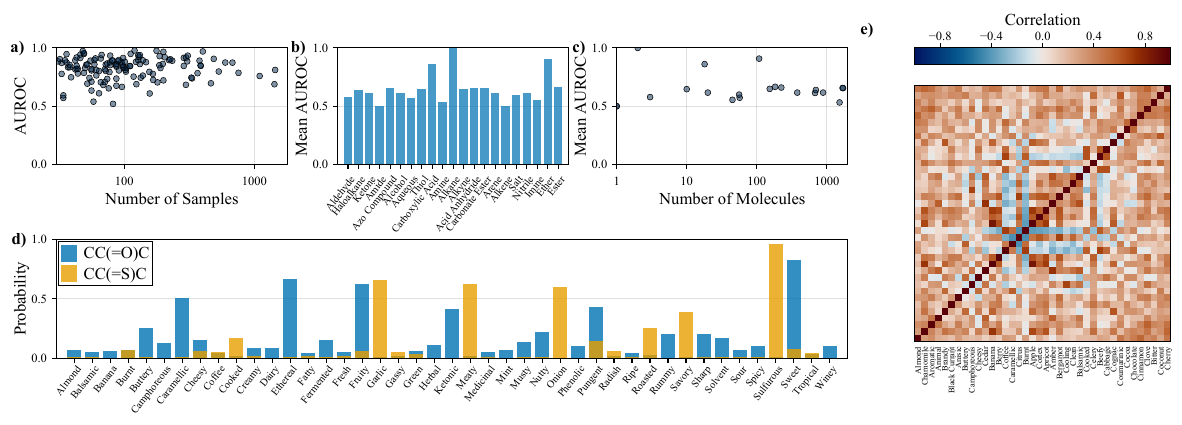}
    \caption{\textbf{A closer look at the MIST-28M Olfaction model.}
        \label{fig:si:olfactory}
        (\subref*{fig:num_samples_auroc})~Despite the sparsity of positive labels for individual scents in the dataset, we find that the model's ability to classify the presence of an odour is uncorrelated with the number of positive examples.
        (\subref*{fig:functional_group_auroc})~Model performance varies by functional group. (\subref*{fig:num_molecules_auroc}) The difference in performance is not explained by the number of examples of the functional group in the training data.
        (\subref*{fig:discontinuous_odour}) The model is able to successfully predict drastic changes in olfactory profile caused by single atom changes in molecular structure, hence capturing the discontinuous nature of olfactory space. For example, it accurately labels acetone \smiles{CC(=O)C} as ``ethereal'' while predicting thioacetone \smiles{CC(=S)C} (single-atom substitution) to be ``sulfurous'' and ``meaty''.
        (\subref*{fig:logit_correlation}) Analyzing correlations (using Pearson's correlation coefficient) between the model's predicted logit distribution across all molecules in the dataset, we find that the model learns correlations consistent with human perception. For example, ``chocolate'' is most correlated with ``cocoa,'' and fruit scents (``apple'', ``apricot'', ``cherry'') are all highly correlated.
    }
\end{figure}

In this section, we discuss the performance of the \ac{MIST}-28M model fine-tuned on a multi-label binary classification olfactory perception dataset of 4,983 non-stereochemical \ac{SMILES} with 135 scent labels~\cite{OpenPOM}.
This OpenPOM dataset was curated to replicate the performance of the \ac{POM} model by~Lee et al.\cite{leePrincipalOdorMap2023}.
We analysed the model's performance across different scents and functional groups, as well as the predicted logits, to better understand the model's capabilities.
Contrary to expectation, we found the model to be robust to the class imbalance (\cref{fig:num_samples_auroc}) and sparsity of the dataset (\cref{fig:num_molecules_auroc}).
We observed that the model successfully learns the discontinuous nature of the molecular structure to olfactory perception mapping (\cref{fig:discontinuous_odour}) and human interpretable odour relationships (\cref{fig:logit_correlation}).

\subsection{Discordant Triads}
\label{sec:si:discordance}
\begin{figure}[ht!]
    \centering
    \labelphantom{fig:si:discordance_structural}
    \labelphantom{fig:si:perceptual_discord}
    \labelphantom{fig:si:gnn_discord}
    \labelphantom{fig:si:mist_cosine_discord}
    \labelphantom{fig:si:mist_euclidean_discord}
    \label{fig:si:discordance}
    \includegraphics[width=\linewidth]{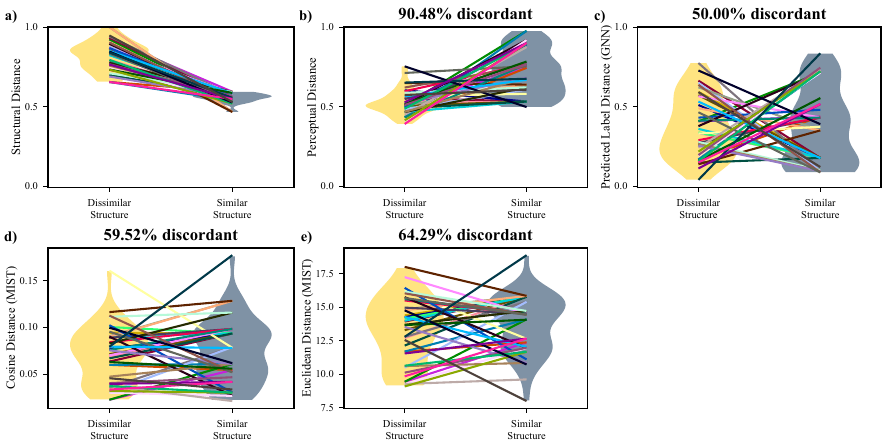}
    \caption{\textbf{Discovering discordant triads.}
        Prior work identified 41 ``discordant'' molecular triplets, in which the structurally closest pair is not the perceptually closest pair. 
        Violin plots show the distributions of structural and perceptual distances for the structurally similar and structurally dissimilar pairs within each triplet; coloured lines connect pairs sharing the same anchor molecule.
        (\subref*{fig:si:discordance_structural})~Tanimoto similarities for the pairs labelled structurally similar and structurally dissimilar.
        (\subref*{fig:si:perceptual_discord})~Experimentally measured explicit perceptual distance ratings in the same triplets show high discordance with structural distance, that is, the molecule that is more structurally similar to the anchor is usually (90\% of the time) less perceptually similar.
        (\subref*{fig:si:gnn_discord})~The \ac{GNN} developed by Lee et al.\cite{leePrincipalOdorMap2023} predicts the empirical discordance only 50\% of the time.
        (\subref*{fig:si:mist_cosine_discord})~Using the cosine distance between \ac{MIST} predicted odour profiles of molecules we correctly predict 59.52\% of the discordance. 
        (\subref*{fig:si:mist_euclidean_discord})~Using the Euclidean distance between \ac{MIST} predicted odour profiles of molecules we correctly predict 64.3\% of the discordance. 
    }
\end{figure}

In this section, we discuss the details of the discordance triad identification task described in~\cref{sec:olfaction}.
This is a prospective validation task which experimentally validates the \ac{MIST} olfaction model's predictions and demonstrates its ability to generalize;
the triads used were not publicly available at the time of training, fine-tuning or benchmarking \ac{MIST}.
Lee et al.\cite{leePrincipalOdorMap2023} curated a set of 41 discordant triplets.
Each triplet contains an ``anchor'' (a reference molecule), a ``structural cliff'' (structurally different from the anchor but perceptually similar to it) and a ``label cliff'' (structurally similar to the anchor but perceptually different from it).
Hence, these triads test whether a model can distinguish perceptual similarity from structural similarity.

For each triad and each distance metric, we compute the perceptual distance between anchor and structural cliff, as well as, the perceptual distance between anchor and label cliff.
The five distance metrics (one structural, four perceptual) shown in~\cref{fig:si:discordance} are:
\begin{description}
    \item[Structural Distance] Tanimoto distance computed from molecular fingerprints, quantifying structural dissimilarity between molecules based on their chemical features.
    \item[Perceptual Distance] Empirical perceptual distance as measured by~Lee et al.\cite{leePrincipalOdorMap2023}, derived from human ratings of odour similarity across multiple perceptual dimensions.
    \item[Predicted Label Distance (\ac{GNN})] Cosine distance between predicted olfactory label profiles, where each molecule is represented by a 54-dimensional binary vector indicating the presence or absence of specific odour descriptors (e.g., floral, fruity, woody) as determined by the \ac{GNN} in Ref~\cite{leePrincipalOdorMap2023}.
    The 54 labels are the overlap between the \ac{POM}~\cite{leePrincipalOdorMap2023} and OpenPOM (\cref{sec:si:olf_dataset}) dataset to allow a comparison between \ac{MIST} and the \ac{GNN}.
    \item[Cosine Distance (MIST)] Cosine distance between \ac{MIST}-predicted olfactory profiles, represented by the model's 54-dimensional logit outputs containing predicted unnormalized probabilities for each odour descriptor.
    \item[Euclidean Distance (MIST)] Euclidean ($L_2$) distance between MIST-predicted logit vectors, providing an alternative distance metric in the same 54-dimensional odour descriptor space.
\end{description}

Following~Lee et al.\cite{leePrincipalOdorMap2023}, we compare the ability of the different distance metrics to identify discordant triplets using the discordance rate:
\begin{align*}
    \textrm{Discordance Rate}= \frac{\textrm{Discordant Count}}{\textrm{Total Discordant Count}} \times 100\%
\end{align*}
where the discordant count is the number of triads identified as discordant by the perceptual distance metrics and the total discordant count is 41 (the total number of discordant triads assembled by Lee et al.\cite{leePrincipalOdorMap2023} for this task).
This percentage represents how often the model correctly predicts that the structurally similar but perceptually different pair is perceptually farther from the anchor than the structurally different but perceptually similar pair. 
Higher percentages indicate better model performance in capturing the discordance between structural and perceptual similarity.

The experimental labels and structures for the molecules in the discordant triplets used here are available on Osmo's GitHub (\url{https://github.com/osmoai/publications}).

\subsection{Chain-length dependent organization of olfactory profiles across homologous series}
\label{sec:si:homologous_series_odour_trends}

\begin{figure}[ht!]
    \centering
    \includegraphics[width=\linewidth]{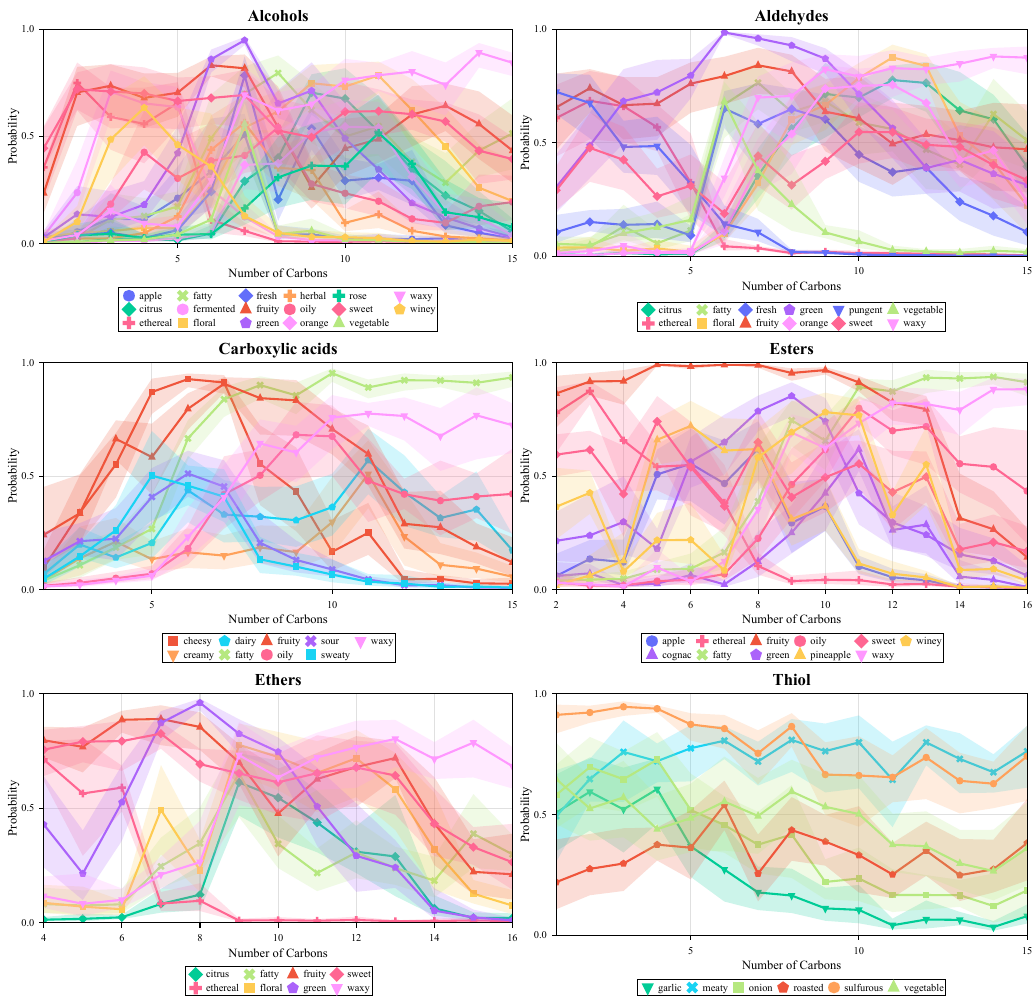}
    \caption{Changes in scent profile with carbon chain length across functional group.
    \label{fig:scent_trends}
    }
\end{figure}

To examine how molecular size modulates perceptual character within shared chemical scaffolds, we compared odour profiles predicted by \ac{MIST} across homologous series of alcohols, aldehydes, carboxylic acids, esters, ethers, and thiols. 
Across all six series, the dominant trend is that odour probability varies non-monotonically with carbon number, with many descriptors exhibiting maxima over relatively narrow chain-length windows rather than monotonic scaling. 
Prior work shows that molecular structure systematically organizes perceived odour quality, and recent large-scale models further show that perceptual hierarchies can be inferred from molecular structure alone~\cite{kellerPredictingHumanOlfactory2017,leePrincipalOdorMap2023}. 
This interpretation is also mechanistically consistent with evidence that the concentration achieved at the receptor binding site depends on transport-related molecular properties, including volatility and hydrophobicity, as well as with structural and functional studies showing that odourant chain-length selectivity is determined by the composition of the olfactory receptor binding pocket~\cite{choiUnderstandingMolecularMechanisms2023}.

Across the oxygen-containing homologous series (all show in Supplementary~\cref{fig:scent_trends} except thiols), increasing carbon number generally shifts odour profiles away from sharp, fresh notes and toward waxy, oily, and fatty character.
Alcohols show a broad progression from fresh, citrus, and ethereal character at lower carbon number toward sweeter, greener, and eventually waxier profiles at larger size, with floral and rose-like notes peaking in an intermediate regime. 
This pattern is consistent with prior homologous-series studies showing that increasing chain length in unsaturated alcohols shifts odour quality from green/grassy character toward more fatty, citrus-like, and soapy notes, as well as with broader reviews emphasizing that alcohol odour depends strongly on carbon number within a fixed functional class~\cite{zarzoEffectFunctionalGroup2012a}.
The fact that floral-type descriptors emerge most strongly at intermediate chain lengths in our profiles further supports the view that pleasant top-note character is often optimized within a restricted size window rather than increasing monotonically.

Aldehydes display one of the clearest size-dependent perceptual trajectories. 
Lower-carbon aldehydes are predicted to smell fresh, green, citrus, orange-like, and pungent, whereas longer-chain aldehydes increasingly acquire waxy and fatty character. 
This agrees with long-standing empirical rules assigning green, leaf-like odours to aldehydes as a class, while also matching homologous-series studies showing that increasing chain length shifts unsaturated aldehydes from grassy/green character toward more fatty, citrus-like, and soapy impressions~\cite{genvaItPossiblePredict2019}. 
The model's predictions for aldehyde therefore support a model in which functional group preserves a recognizable perceptual identity, while chain length continuously reweights the balance between fresh-green and heavier waxy-fatty submodes.

Carboxylic acids exhibit the strongest segregation between short-chain and long-chain perceptual regimes. 
Short-chain acids are dominated by sour, sweaty, cheesy, creamy, and dairy-like notes, whereas longer-chain acids shift toward fatty, oily, and waxy descriptors. 
This trend is strongly supported by prior experimental work: volatile fatty acids are classically associated with sour-to-rancid odours, and homologous unsaturated acids have been reported to transition successively from cheesy and sweaty to plastic-like and finally waxy with increasing chain length~\cite{genvaItPossiblePredict2019}. 
\ac{MIST}'s predicted profiles confirm this trend, indicating that carbon number acts as a strong determinant of where a carboxylic acid lies along the sour/cheesy to fatty/waxy perceptual axis.

Esters remain concentrated in the fruity-sweet odour space, but still show substantial internal structure. 
Apple, pineapple, sweet, winey, and cognac-like descriptors are strongest in the short-to-mid chain regime, while waxy, oily, and fatty notes become more competitive as carbon number increases. 
This is consistent with the classical assignment of esters to fruity and floral odour classes~\cite{genvaItPossiblePredict2019}, while also agreeing with flavour literature showing that esters are major contributors to apple- and pineapple-like aroma and that chain composition influences whether the resulting impression remains bright and fruity or becomes heavier and more fatty. 
Our results therefore suggest that the canonical ``fruity ester'' rule is best interpreted as an intermediate-window effect rather than a uniform property of the entire homologous series.

Ethers show a similar but somewhat smoother progression, transitioning from lighter ethereal, fruity, and fresh descriptors at low carbon number toward sweeter, waxier, and more oily profiles at larger size. 
Although structure--odour rules for ethers are generally less rigid than for aldehydes, esters, or acids, the observed behaviour is still consistent with the broader homologous-series principle that increasing carbon number shifts odour quality from more volatile, top-note character toward heavier and less diffusive notes within a fixed functional-group scaffold \cite{zarzoEffectFunctionalGroup2012a}.

Thiols are the principal exception to the oxygenated trend. 
Across the full chain-length range, sulfur-associated descriptors---including garlic, onion, sulfurous, meaty, roasted, and vegetable---remain dominant, indicating that functional-group identity overwhelmingly constrains the perceptual space of this class.
This strong class specificity is well supported by prior literature: thiols are widely recognized as having alliaceous or rotten/sulfurous character, unusually low odour thresholds, and receptor interactions that can involve metal-assisted recognition, including for onion-related thiol odourants \cite{genvaItPossiblePredict2019}. 
The relative stability of the thiol perceptual manifold in \ac{MIST}'s predictions therefore agrees with the established view that organosulfur odours are more strongly determined by functional group than by gradual variation in carbon number.

Taken together, these results support a hierarchical picture of structure--odour organization. 
Functional groups establish the dominant odour family---for example, fruity for esters, green for aldehydes, sour/rancid for acids, and sulfurous/alliaceous for thiols---while carbon chain length tunes the specific descriptor mixture within that family.

\subsection{Topological analysis of olfactory geometry}
\label{sec:si:tda_olfaction}

\begin{figure}[ht!]
    \centering
    \label{fig:si:euclidean_betti}
    \includegraphics[width=\linewidth]{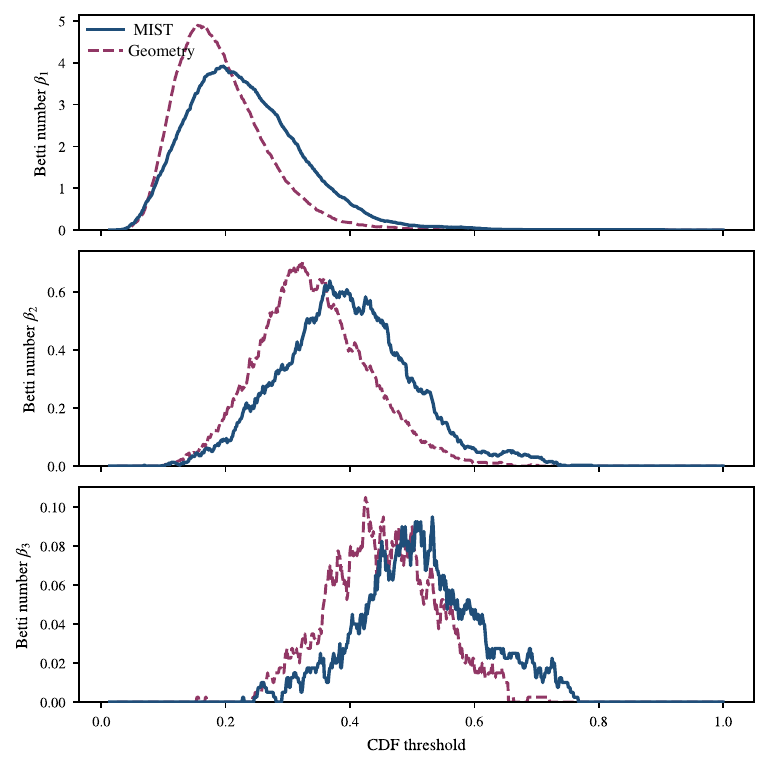}
    \caption{
        Comparison of empirical and best-fit hyperbolic Betti curves for the MIST olfactory logit similarity structure. 
        The hyperbolic model closely reproduces the empirical \(\beta_1\), \(\beta_2\), and \(\beta_3\) curves, with small relative differences in integrated Betti values, consistent with a hyperbolic organization of the learned olfactory representation.
    }
\end{figure}

\begin{figure}[ht!]
    \centering
    \label{fig:si:hyperbolic_betti}
    \includegraphics[width=\linewidth]{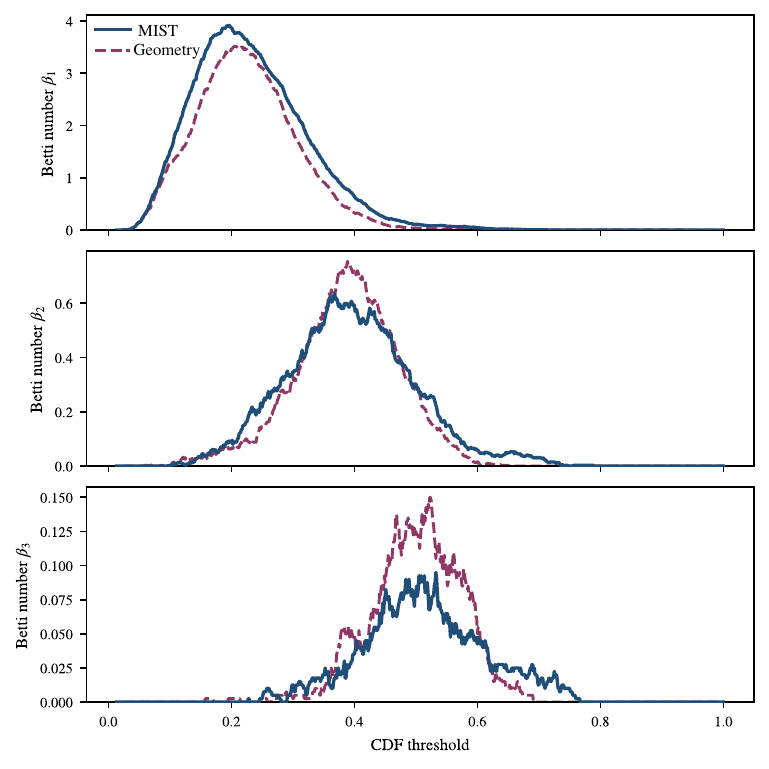}
    \caption{
    Comparison of empirical and best-fit Euclidean Betti curves for the MIST olfactory logit similarity structure. 
    Although the 3D Euclidean model partially captures the \(\beta\) curves, it has a higher  \(\chi^2\) value than the hyperbolic model (5.05 and 0.00 respectively).
    }
\end{figure}

\ac{TDA} provides a multiscale description of structure in pairwise similarity or dissimilarity data without assuming that the underlying space is flat. 
In the context of finding structure in the perceptual profiles predicted by \ac{MIST}, the input is a weighted matrix of pairwise dissimilarities between molecular odour profiles. 
Rather than choosing a single threshold at which to define a graph, \ac{TDA} considers all thresholds simultaneously by constructing a filtered clique complex from the weighted network. 
This approach is well suited to olfaction, where prior work has argued that both natural odour co-occurrence statistics and human perceptual similarity are more consistent with a low-dimensional hyperbolic geometry than with a Euclidean one, and where structured cortical representations of odour space have also been observed~\cite{zhouHyperbolicGeometryOlfactory2018,sagarHighprecisionMappingReveals2023}.

Persistent homology summarizes the topology of the filtered complex as the threshold is varied. 
At any given threshold, the Betti numbers count topological features of different dimensions, including connected components (\(\beta_0\)), loops (\(\beta_1\)) and higher-order cavities (\(\beta_2\) and above). 
Evaluating these quantities across thresholds yields Betti curves, \(\beta_k(t)\), whose shapes summarize how topological features appear and disappear over scale. 
In olfactory data, these curves provide a compact description of whether odour relationships are organized as a flat cloud, a curved manifold or a continuous hierarchy. 
Following previous work on olfaction, however, Betti curves alone are not sufficient to establish an underlying geometry; rather, they must be evaluated against explicit geometric model spaces and interpreted together with biological motivation and explanatory utility~\cite{zhouHyperbolicGeometryOlfactory2018,zhangHippocampalSpatialRepresentations2023}.

To make this comparison computationally tractable for large matrices, we used ALBATROSS~\cite{albatross,stierALBATROSSCheapFiltration2025}. 
ALBATROSS estimates empirical Betti curves by repeatedly drawing random \(n\times n\) submatrices from the full dissimilarity matrix, computing the first three Betti curves on each subsample, and averaging across iterations.
Candidate Euclidean and hyperbolic geometries are then generated by sampling the same number of points from model spaces, computing their pairwise distances, and subjecting the resulting model matrices to the same topological pipeline. 
In the hyperbolic case, points are sampled from a shell with radii bounded between \(r_{\min}\) and \(r_{\max}\), with radial density proportional to \(\sinh^{d-1}(r)\), reflecting the exponential expansion of negatively curved space. 
Best-fit model parameters are identified by minimizing a combined objective based on both the integrated Betti values and the \(L_1\) distances between empirical and model-derived curves. 
Statistical consistency is then assessed using these summary statistics.
We ran ALBATROSS on 40 data samples with 400 iterations and 400 \ac{LHS} samples.

For visualization of the inferred geometry (\cref{fig:olfaction_hyperbolic}), we used the shell-constrained HYDRA embedding implemented in ALBATROSS~\cite{kellerressel2019hydramethodstrainminimizinghyperbolic,albatross}. 

HYDRA performs hyperbolic multidimensional scaling in the Poincar'e ball. The ALBATROSS implementation modifies the original HYDRA procedure in two ways. First, the empirical dissimilarities are linearly transformed onto the geodesic scale of the inferred model,
\[
\tilde{A}_{ij} = 2r_{\max}|1 + A_{ij}|,
\]
where ($r_{\max}$) is the best-fit shell radius from the Betti-curve analysis and ($A_{ij}$) are the empirical distances. This map preserves the rank ordering of pairwise relationships while expressing them on the scale of the inferred model from the Betti-curve analysis. Second, for datasets with large ($r_{\max}$), the curvature parameter is adaptively reduced ($\kappa = (r_{\mathrm{target}} / r_{\max})^2$) to keep the effective embedding radius in the numerically tractable range, avoiding imprecision of Poincar'e ball coordinates near the boundary. 
In olfaction, such embeddings are particularly useful because they yield an interpretable map in which local neighborhoods reflect perceptual similarity, whereas radial position captures hierarchical depth in the learned odour space~\cite{zhouHyperbolicGeometryOlfactory2018,kellerressel2019hydramethodstrainminimizinghyperbolic,albatross}.
The HYDRA embeddings for \ac{MIST} preserve correlation between the embedding and input distance (Spearman's correlation = 0.907), indicating good embedding quality.
\section{Isotope Stability Heuristics}\label{sec:si:isotope_stability}

In this section, we discuss the labels used to classify the stability of each isotope in \cref{fig:isotope_clusters}.
The primary decay channel for each isotope was assigned based on their atomic ($Z$) and mass ($A$) numbers using the following heuristics \cite{wileyDecayModes}:

\paragraph{Heavy-nucleus rules.}
Spontaneous fission and alpha decay are quantum tunneling processes that compete in very heavy nuclei\cite{wileyDecayModes}.
Alpha decay is a common decay mode for very heavy nuclei with $Z > 83$ (heavier than bismuth).
Spontaneous fission is a rarer decay mode that becomes a more significant competitor in extremely heavy nuclei. 
Nuclei with $Z > 91$ and a high mass number $A > 239$ are considered to undergo spontaneous fission.

\paragraph{Beta-stability curve from \ac{SEMF}.}
The \ac{SEMF} is used to identify stable nuclei and those that undergo $\beta^\pm$ decay \cite{weizsaeckerZurTheorieKernmassen1935}.
Stable nuclei lie on or close to the beta-stability curve. 
The curve is defined by the expression:
$$
Z_{\beta}(A) = \frac{A}{2 + kA^{2/3}} ,
$$
where $A$ is the mass number, $Z_{\beta}(A)$ is the proton number which maximizes the binding energy for a given mass number, and $k \approx 0.015$ is a dimensionless constant derived from \ac{SEMF} coefficients.
Nuclei with $Z < Z_{\beta}$ (neutron rich) are labeled $\beta^-$.
Those with $Z > Z_{\beta}$ (proton rich) are labeled $\beta^+$.
Those within a Z-dependent tolerance ($tol(Z) = c + mZ$, where $c$ is set to 0.8 and $m$ to 0.2) are considered ``stable''.
This Z-dependent widening reflects the empirically observed ``valley of stability''\cite{wileyDecayModes}.
\section{Excess Property Models Validation}
\label{sec:si:mix_validation}
\begin{figure}[ht!]
    \centering
    \includegraphics[width=0.7\linewidth]{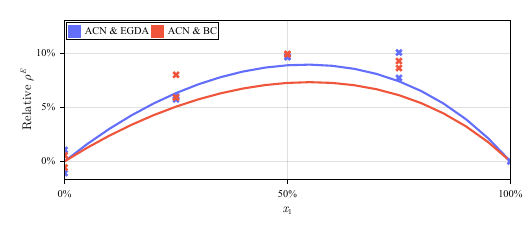}
    \caption{\textbf{Experimentally validating mixture property prediction.}
        \label{fig:mix_validation}
        We ran experiments to validate the prediction accuracy of fine-tuned \ac{MIST} models for binary excess property prediction on two compounds with high relative excess density (Relative $\rho^E$).
        Lines are values predicted by fine-tuned \ac{MIST}-28M excess models.
        Markers are experimental data points (two repeats for every composition).
    }
\end{figure}
The fine-tuned \ac{MIST} models for binary excess property prediction were validated against experimental data for two mixtures with high excess density --- \ac{ACN} and \ac{EGDA},  \ac{ACN} and \ac{BC}.
We additionally validated the model on mixtures whose properties are not well-modeled by existing similarity-based descriptors for excess properties.
In both cases, the model was able to correctly predict the excess properties of high-excess mixtures, which are of interest for electrolyte design\cite{SVArtificialIntelligenceElectrolyte2025}.

\paragraph{Experimental Details.}

All chemicals used in this study were commercially available and further purified prior to use.
\ac{ACN} (anhydrous, $\geq$99.8\%) and \ac{BC} (anhydrous, $\geq$99.9\%) were purchased from Sigma-Aldrich, and \ac{EGDA} ($\geq$98\%) was obtained from BLD Pharm.
The solvents were dried on activated 3\AA{} molecular sieves for at least 24 hours prior to use in the preparation of the mixture.

Density measurements were performed using a flow-through robotic setup adapted from our previous work~\cite{DMB+AutonomousOptimizationNonaqueous2022}.
The mixture components were first combined and homogenized by mechanical stirring for 1~minute in the mixing vessel.
A fixed volume of 0.5~mL of the prepared solution was then delivered via a precision peristaltic pump (Longer, L100-1S-2) directly into the vial on a high precision analytical balance (Radwag, AS~220.R2~PLUS, 220~g~$\times$~0.1~mg).
The delivered volume was controlled through precalibrated pump timing and tubing geometry to ensure volumetric consistency.
The density was calculated as the ratio of the measured mass to the known volume of 0.5~mL, and each measurement was repeated twice to confirm the reproducibility.
All steps were carried out in an enclosed environment to minimize evaporation and environmental interference.

\subsection{Testing Generalization Performance}
\label{sec:si:soap_excess}
\begin{figure}[ht!]
    \centering
    \labelphantom{fig:soap_corr}
    \labelphantom{fig:si:soap_predictions}
    \includegraphics[width=\linewidth]{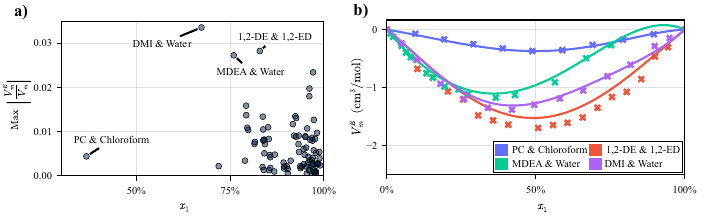}
    \caption{
        \label{fig:soap_validation}
        \textbf{Challenging excess property prediction.}
        (\subref*{fig:soap_corr})~\acs{SOAP} fingerprint similarity has been observed to be negatively correlated with the magnitude of excess quantities.
        We identify mixtures which disagree with this heuristic --- have low similarity and low excess molar volume or vice versa: PC (Propylene carbonate) and Chloroform,  MDEA (N-Methyldiethanolamine) and Water, DMI (1,3-Dimethyl-2-imidazolidinone) and Water, and 1,2-DE (1,2-Diaminoethane) and 1,2-ED (1,2-Ethanediol).
        (\subref*{fig:si:soap_predictions})~The \ac{MIST} model is able to accurately predict the excess molar volume of mixtures which disagree with the similarity heuristic (Reprint of~\cref{fig:soap_outliers}).
    }
\end{figure}
In order to test the generalization performance of the excess property prediction models, we identified ``challenging'' binary mixtures.
Mixtures were considered ``challenging'' if the magnitude of their excess density could not be estimated using the design rule formulated by Kelly et al.\cite{KADVExcessDensityDescriptor2025}:
mixtures of two solvents show larger-magnitude excess properties when the molecules are less similar and smaller deviations when they are more similar.
The identified ``challenging'' mixtures and predictions from a fine-tuned \ac{MIST} excess property model are shown in \cref{fig:soap_corr}.
Similarity here is measured using the REMatch\cite{deMappingClassifyingMolecules2017} kernel;
the REMatch kernel computes a global similarity between two molecules by optimally matching their sets of \ac{SOAP} descriptors\cite{Bart_k_2013}.
Following the methodology in Ref.~\cite{KADVExcessDensityDescriptor2025}, the DScribe\cite{himanenDScribeLibraryDescriptors2020} implementations of the \ac{SOAP} fingerprint and REMatch kernel method were used.

\section{Analyzing Trends in Ionic Conductivity}\label{sec:si:conductivity}
\begin{figure}[ht!]
    \centering
    \includegraphics[width=\linewidth]{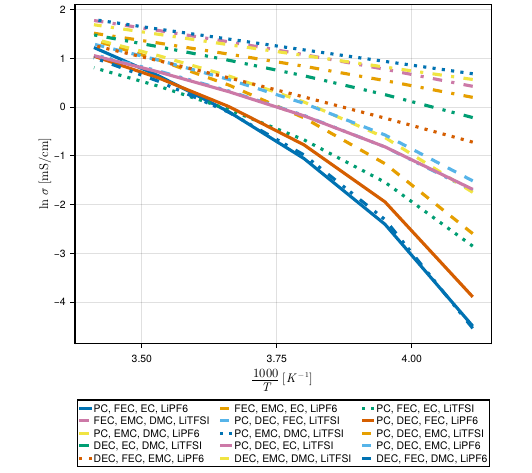}
    \caption{
        \label{fig:non_arr_data}
       Arrhenius plots of conductivity (\(\ln\sigma\) against\ \(1000/T\)) across all compositions are concave-down—demonstrating non-Arrhenius \ac{VFT} behaviour.
    }
\end{figure}
As discussed in~\cref{sec:methods:ionic}, the \ac{MIST} ionic conductivity model uses the \acf{VFT} equation to learn temperature dependence of ionic conductivity.
In this section, we show that a \ac{VFT}, rather than Arrhenius, temperature dependence is appropriate for modelling electrolytes. 
We analyse the \ac{VFT} pseudo-activation energies $E_a$ and Vogel temperatures $T_0$ learnt by the model across a range of compositions.
We examine the differences in $E_a$ between an organic salt, LiTFSI, and an inorganic salt, LiPF$_6$.
We calculate excess $E_a$ and $T_0$ using the \acf{VFT} parameter values learnt by the model, these values can be used to provide insight into the impact of solvent-solvent and solvent-salt interactions on ion mobility.

\subsection{Electrolytes Show \acs{VFT} Temperature Dependence}\label{sec:si:why_vft}
To validate the non-Arrhenius temperature dependence of electrolyte ionic conductivity in our training data, we plot (\cref{fig:non_arr_data}) the inverse temperature $1000/T$ against the log conductivity $\ln \sigma$  for a randomly selected subset of electrolytes in the training set. 
The values plotted here are the labels, as calculated using the \ac{AEM}, rather than model predictions.
Across all compositions and both salts, Arrhenius plots of conductivity (\(\ln \sigma\) versus \ \(1/T\)) are systematically concave-down, with steeper slopes at lower \(T\), indicating a temperature-dependent activation energy and the inadequacy of a single Arrhenius law. 
A \ac{VFT} form naturally produces this curvature and fits the data continuously over the full temperature range.

\subsection{\acs{VFT} Pseudo-Activation Energies for Battery Electrolytes}
\label{sec:si:conductivity_ea}
Here we show the pseudo-activation energies, $E_a$, of electrolytes with a salt (either LiPF$_6$ or LiTFSI) dissolved in an organic carbonates solvent.
The solvents considered are a mixture of \ac{EC}, \ac{PC} and a third solvent --- one of \ac{EMC}, \ac{DMC}, \ac{FEC} or \ac{DEC}.
All predictions are for three salt mole ratios ($x_{Li} \in \{ 0.05, 0.09, 0.16\}$).
We observe physically consistent trends in activation energy ---
activation energies increase with salt concentration,
lowest activation energies are predicted for linear carbonates (for example \ac{DEC} and \ac{DMC}),
and cyclic carbonates increase activation energy.

\begin{figure}[h!]
    \centering
    \includegraphics[width=\linewidth]{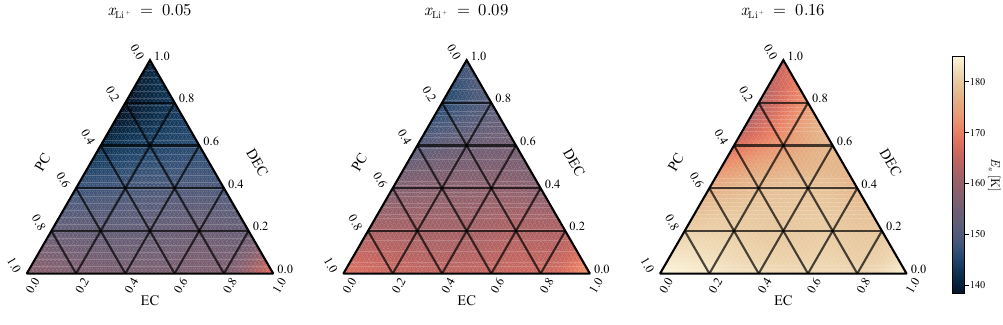}
    \caption{
        \label{fig:ea_dec_pf6}
        Activation energies for LiPF$_6$ in \ac{EC}, \ac{PC} and \ac{DEC}.
    }
\end{figure}

\begin{figure}[h!]
    \centering
    \includegraphics[width=\linewidth]{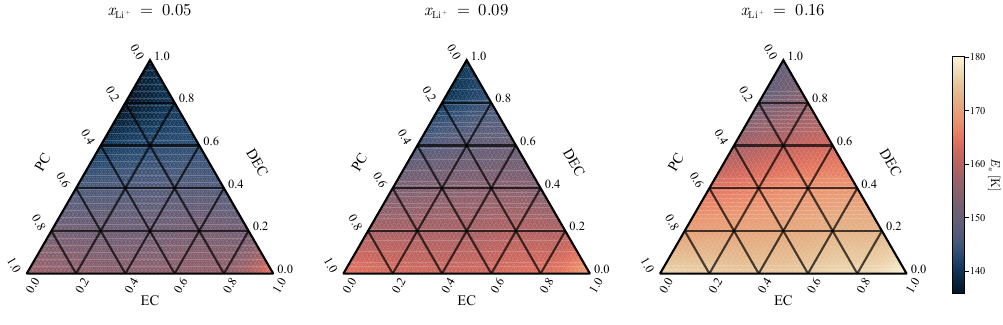}
    \caption{
        \label{fig:ea_dec_tfsi}
        Activation energies for LiTFSI in \ac{EC}, \ac{PC} and \ac{DEC}.
    }
\end{figure}

\begin{figure}[h!]
    \centering
    \includegraphics[width=\linewidth]{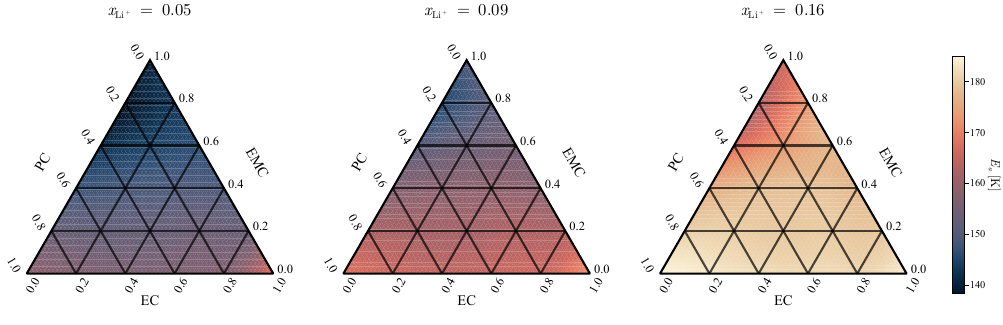}
    \caption{
        \label{fig:ea_emc_pf6}
        Activation energies for LiPF$_6$ in \ac{EC}, \ac{PC} and \ac{EMC}.
    }
\end{figure}

\begin{figure}[h!]
    \centering
    \includegraphics[width=\linewidth]{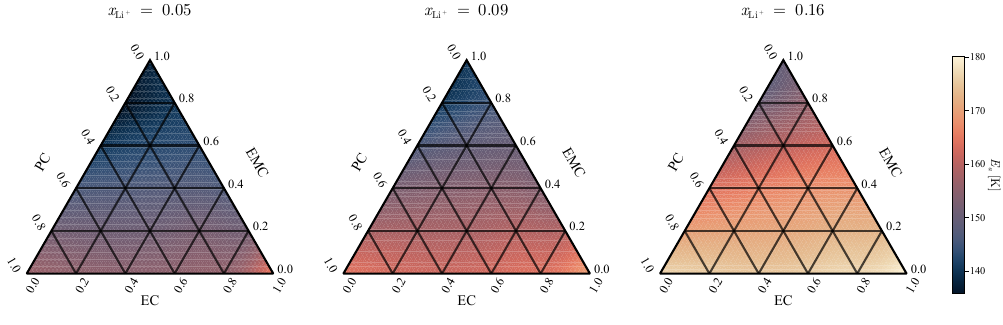}
    \caption{
        \label{fig:ea_emc_tfsi}
        Activation energies for LiTFSI in \ac{EC}, \ac{PC} and \ac{EMC}.
    }
\end{figure}

\begin{figure}[h!]
    \centering
    \includegraphics[width=\linewidth]{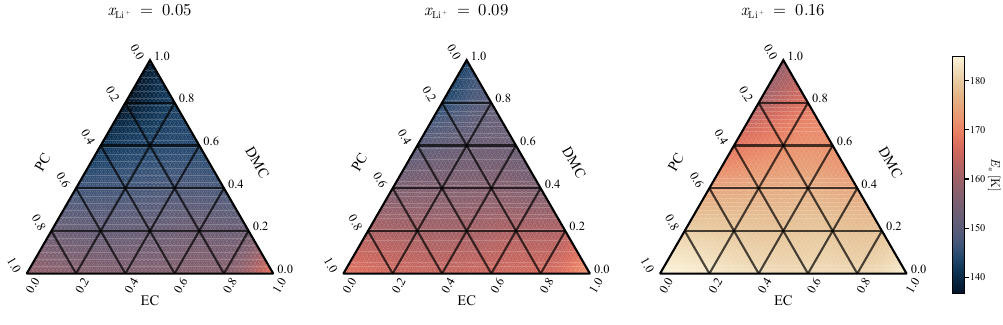}
    \caption{
        \label{fig:ea_dmc_pf6}
        Activation energies for LiPF$_6$ in \ac{EC}, \ac{PC} and \ac{DMC}.
    }
\end{figure}

\begin{figure}[h!]
    \centering
    \includegraphics[width=\linewidth]{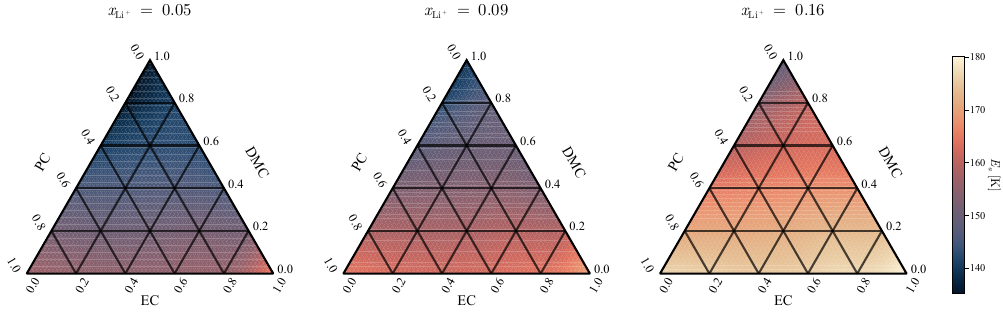}
    \caption{
        \label{fig:ea_dmc_tfsi}
        Activation energies for LiTFSI in \ac{EC}, \ac{PC} and \ac{DMC}.
    }
\end{figure}

\begin{figure}[h!]
    \centering
    \includegraphics[width=\linewidth]{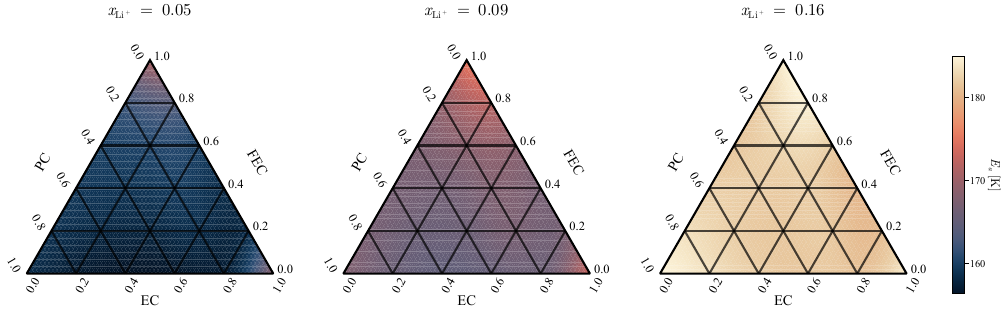}
    \caption{
        \label{fig:ea_fec_pf6}
        Activation energies for LiPF$_6$ in \ac{EC}, \ac{PC} and \ac{FEC}.
    }
\end{figure}

\begin{figure}[h!]
    \centering
    \includegraphics[width=\linewidth]{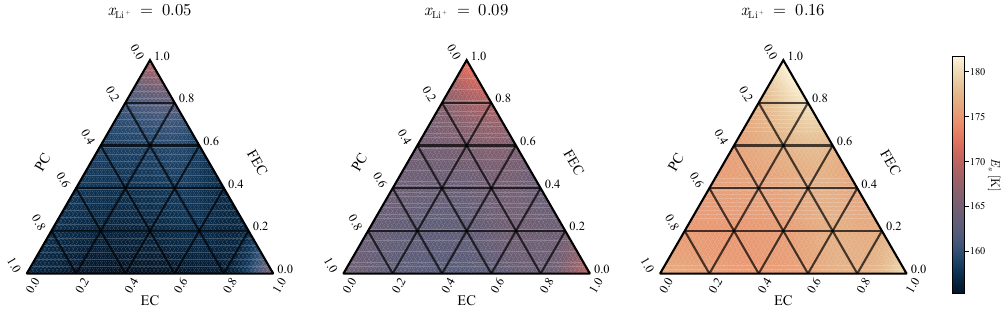}
    \caption{
        \label{fig:ea_fec_tfsi}
        Activation energies for LiTFSI in \ac{EC}, \ac{PC} and \ac{FEC}.
    }
\end{figure}

\clearpage
\subsection{Excess Activation Energy}
\label{sec:si:conductivity_excess_ea}
Here we show the ``excess activation energies'' of the same electrolyte systems discussed above in \cref{sec:si:conductivity_ea}.
The architecture of the \ac{MIST} ionic conductivity model enables excess activation energies to be computed.
These predictions provide rich insight into how salt--solvent and solvent--solvent interactions impact ion mobility in different concentration regimes.
We define the excess activation energy as the difference between the ideal (linear mixing) pseudo-activation energies, $E_a^{ideal}$, and the predicted pseudo-activation energy for the overall mixture, $E_a$:
\begin{align}
    E_a^{excess} &= E_a - E_a^{ideal} \\
    E_a^{ideal} &= \sum_i \frac{x_i}{1 - x_{Li}} E_{a, i} \\
    E_{a, i} &= E_a |_{x_{Li}, x_i = 1 - x_{Li}}
\end{align}
The overall salt to solvent ratio $x_{Li}$ : $ 1 - x_{Li}$ remains fixed for each ternary plot while the ratio of each individual solvent $x_i$ varies.
Hence, the \ac{VFT} pseudo-activation parameter of the electrolyte containing salt and only solvent $i$ is $E_{a, i}$.
We observe that the excess activation energies are lower for LiTFSI as compared to LiPF$_6$.
Additionally,LiTFSI shows largely negative excess while LiPF$_6$ tends to have negative excess $E_a$ in dilute mixtures but positive excess $E_a$ at high salt mole ratios.
These observations are consistent across all solvent systems except those containing \ac{FEC}.
For the \ac{EC}, \ac{PC} and \ac{FEC} solvent system excess $E_a$ is negative for both salts and shows very similar behaviour across both.

\begin{figure}[h!]
    \centering
    \includegraphics[width=\linewidth]{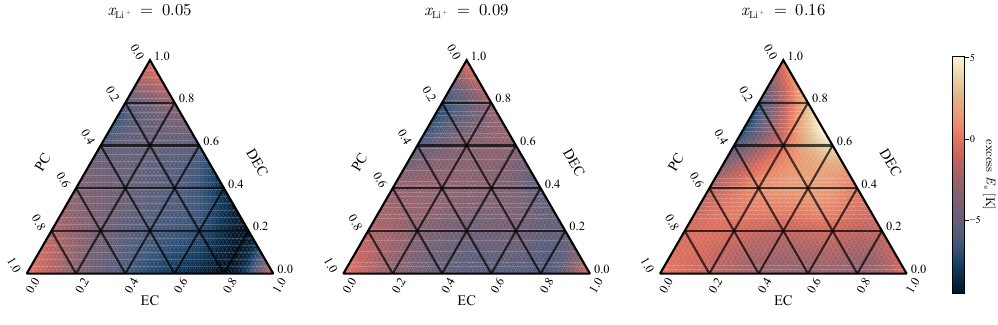}
    \caption{
        \label{fig:excess_dec_pf6}
        Excess activation energies for LiPF$_6$ in \ac{EC}, \ac{PC} and \ac{DEC}.
    }
\end{figure}

\begin{figure}[h!]
    \centering
    \includegraphics[width=\linewidth]{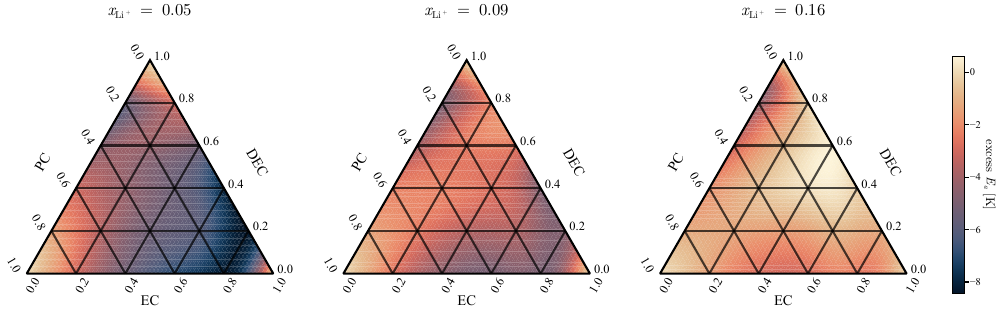}
    \caption{
        \label{fig:excess_dec_tfsi}
        Excess activation energies for LiTFSI in \ac{EC}, \ac{PC} and \ac{DEC}.
    }
\end{figure}

\begin{figure}[h!]
    \centering
    \includegraphics[width=\linewidth]{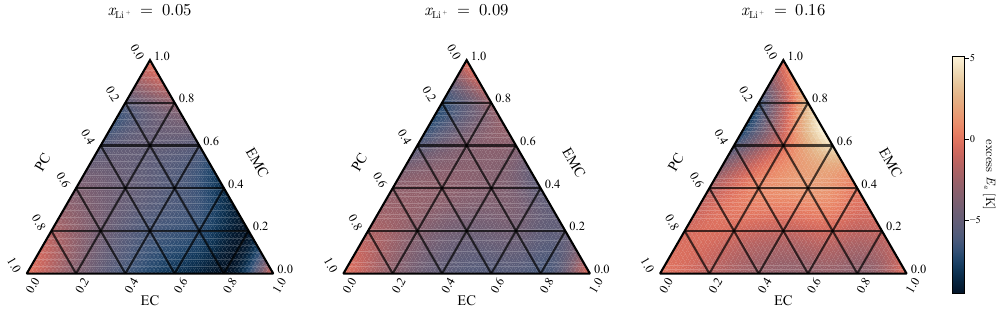}
    \caption{
        \label{fig:excess_emc_pf6}
        Excess activation energies for LiPF$_6$ in \ac{EC}, \ac{PC} and \ac{EMC}.
    }
\end{figure}

\begin{figure}[h!]
    \centering
    \includegraphics[width=\linewidth]{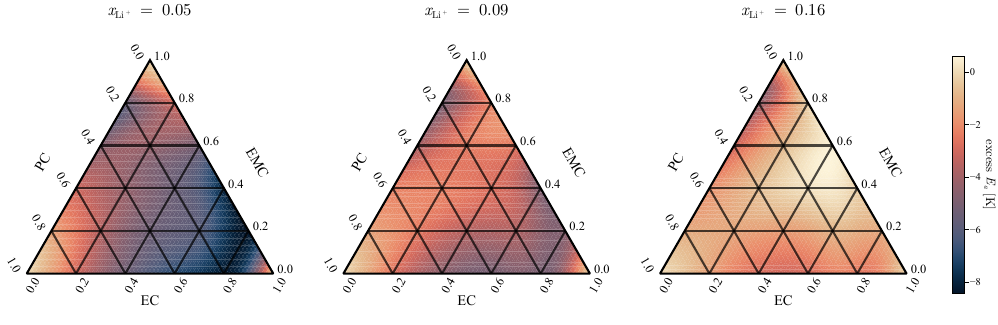}
    \caption{
        \label{fig:excess_emc_tfsi}
        Excess activation energies for LiTFSI in \ac{EC}, \ac{PC} and \ac{EMC}.
    }
\end{figure}

\begin{figure}[h!]
    \centering
    \includegraphics[width=\linewidth]{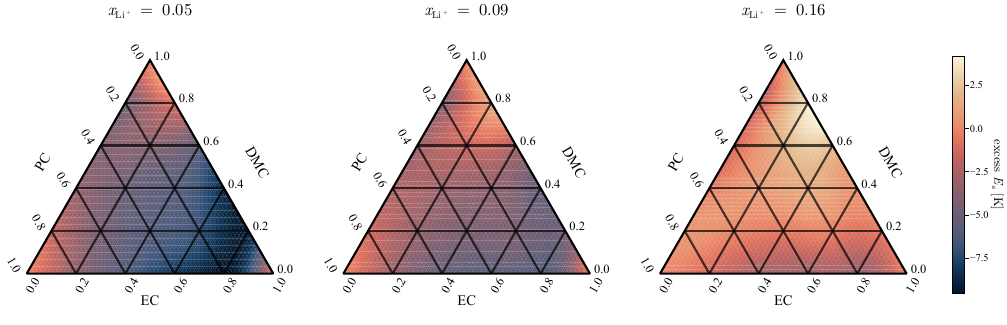}
    \caption{
        \label{fig:excess_dmc_pf6}
        Excess activation energies for LiPF$_6$ in \ac{EC}, \ac{PC} and \ac{DMC}.
    }
\end{figure}

\begin{figure}[h!]
    \centering
    \includegraphics[width=\linewidth]{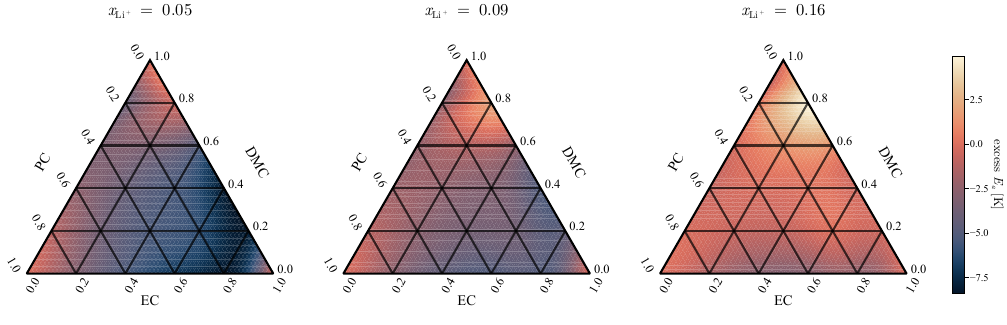}
    \caption{
        \label{fig:excess_dmc_tfsi}
        Excess activation energies for LiTFSI in \ac{EC}, \ac{PC} and \ac{DMC}.
    }
\end{figure}

\begin{figure}[h!]
    \centering
    \includegraphics[width=\linewidth]{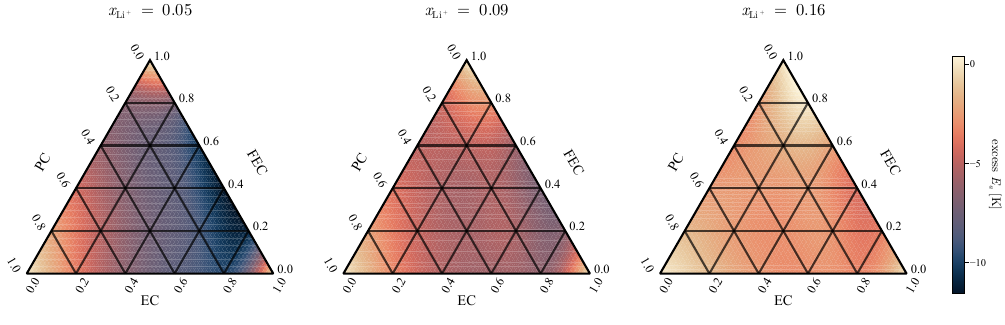}
    \caption{
        \label{fig:excess_fec_pf6}
        Excess activation energies for LiPF$_6$ in \ac{EC}, \ac{PC} and \ac{FEC}.
    }
\end{figure}

\begin{figure}[h!]
    \centering
    \includegraphics[width=\linewidth]{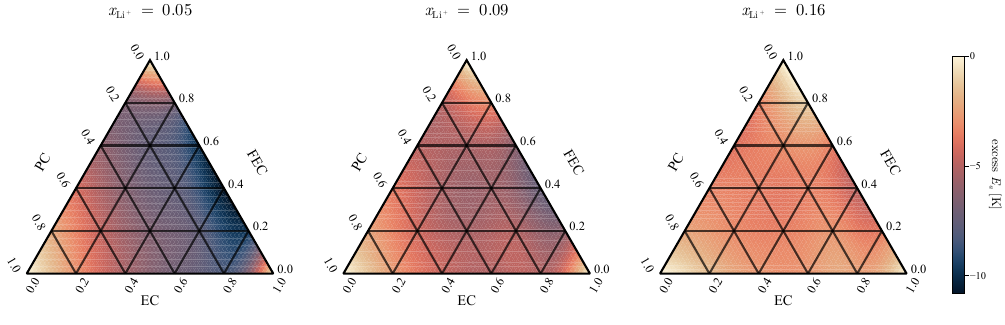}
    \caption{
        \label{fig:excess_fec_tfsi}
        Excess activation energies for LiTFSI in \ac{EC}, \ac{PC} and \ac{FEC}.
    }
\end{figure}
\clearpage

\subsection{Vogel Temperatures and Excess Vogel Temperatures for Battery Electrolytes}
\label{sec:si:conductivity_t0}

The Vogel temperature or ``ideal glass transition temperature'' \(T_{0}\) is a theoretical point where the system's relaxation time becomes infinite and its configurational entropy vanishes.
It is an asymptotic value that is approached as the cooling rate decreases that cannot be measured directly in experiments~\cite{masahiroUnderstandingVogel2013}.
Here we show the Vogel temperatures and ``excess Vogel temperatures'' of the same electrolyte systems discussed above in \cref{sec:si:conductivity_ea}.
Excess Vogel temperatures are computed using the same approach used for excess activation energies in \cref{sec:si:conductivity_ea}.
We observe trends in the excess Vogel temperature match those in the excess pseudo-activation energy.

\begin{figure}[h!]
    \centering
    \includegraphics[width=\linewidth]{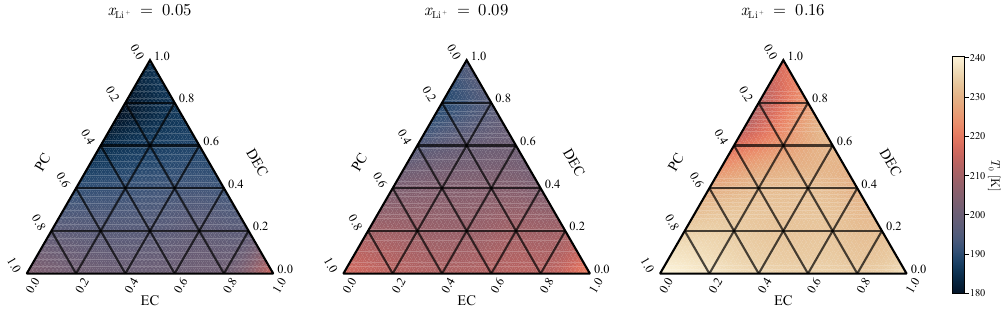}
    \caption{
        \label{fig:t0_dec_pf6}
        Vogel temperatures for LiPF$_6$ in \ac{EC}, \ac{PC} and \ac{DEC}.
    }
\end{figure}

\begin{figure}[h!]
    \centering
    \includegraphics[width=\linewidth]{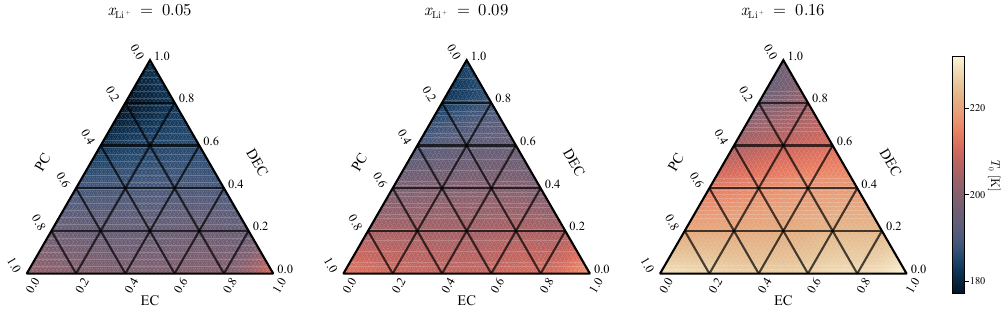}
    \caption{
        \label{fig:t0_dec_tfsi}
        Vogel temperatures for LiTFSI in \ac{EC}, \ac{PC} and \ac{DEC}.
    }
\end{figure}

\begin{figure}[h!]
    \centering
    \includegraphics[width=\linewidth]{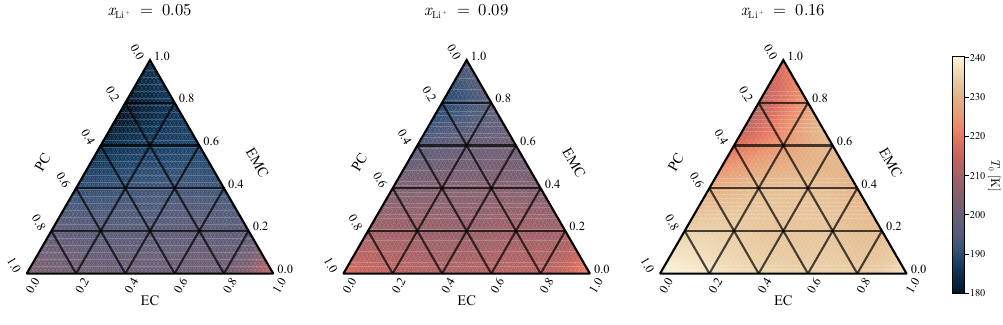}
    \caption{
        \label{fig:t0_emc_pf6}
        Vogel temperatures for LiPF$_6$ in \ac{EC}, \ac{PC} and \ac{EMC}.
    }
\end{figure}

\begin{figure}[h!]
    \centering
    \includegraphics[width=\linewidth]{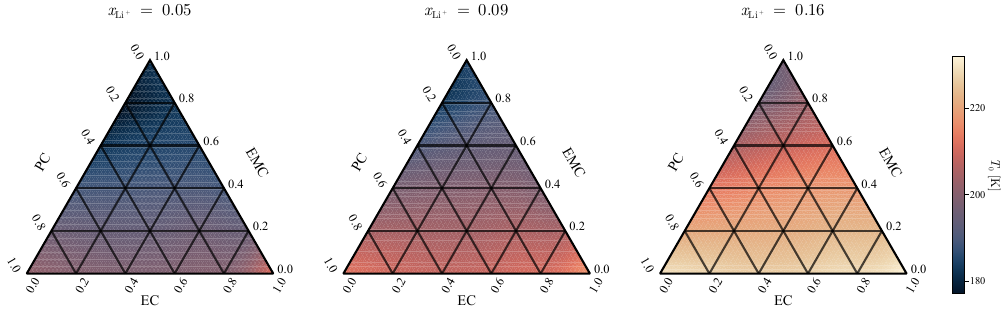}
    \caption{
        \label{fig:t0_emc_tfsi}
        Vogel temperatures for LiTFSI in \ac{EC}, \ac{PC} and \ac{EMC}.
    }
\end{figure}

\begin{figure}[h!]
    \centering
    \includegraphics[width=\linewidth]{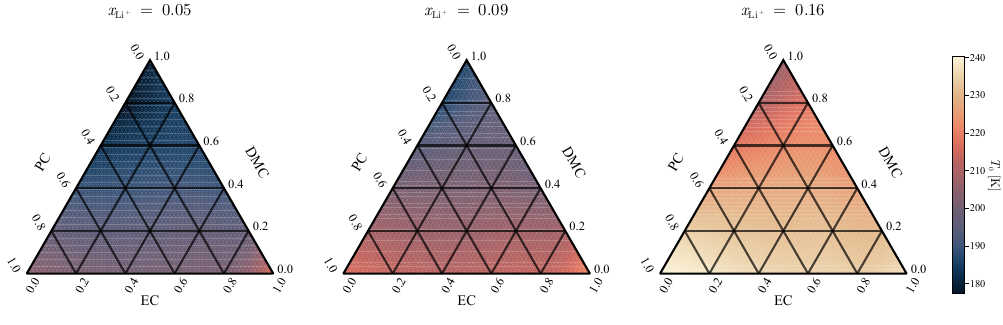}
    \caption{
        \label{fig:t0_dmc_pf6}
        Vogel temperatures for LiPF$_6$ in \ac{EC}, \ac{PC} and \ac{DMC}.
    }
\end{figure}

\begin{figure}[h!]
    \centering
    \includegraphics[width=\linewidth]{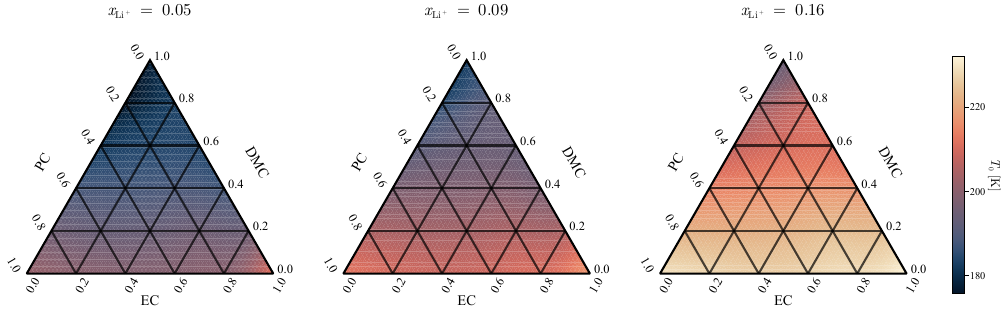}
    \caption{
        \label{fig:t0_dmc_tfsi}
        Vogel temperatures for LiTFSI in \ac{EC}, \ac{PC} and \ac{DMC}.
    }
\end{figure}

\begin{figure}[h!]
    \centering
    \includegraphics[width=\linewidth]{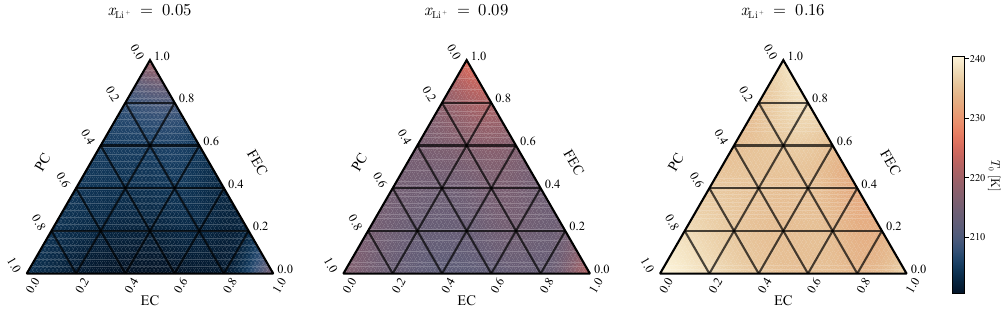}
    \caption{
        \label{fig:t0_fec_pf6}
        Vogel temperatures for LiPF$_6$ in \ac{EC}, \ac{PC} and \ac{FEC}.
    }
\end{figure}

\begin{figure}[h!]
    \centering
    \includegraphics[width=\linewidth]{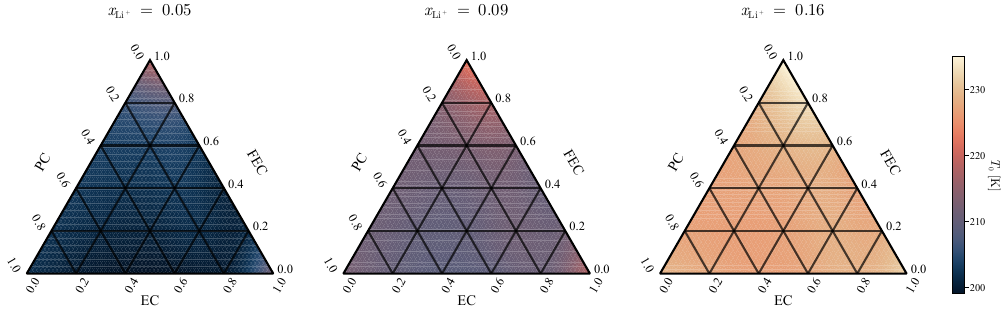}
    \caption{
        \label{fig:t0_fec_tfsi}
        Vogel temperatures for LiTFSI in \ac{EC}, \ac{PC} and \ac{FEC}.
    }
\end{figure}

\begin{figure}[h!]
    \centering
    \includegraphics[width=\linewidth]{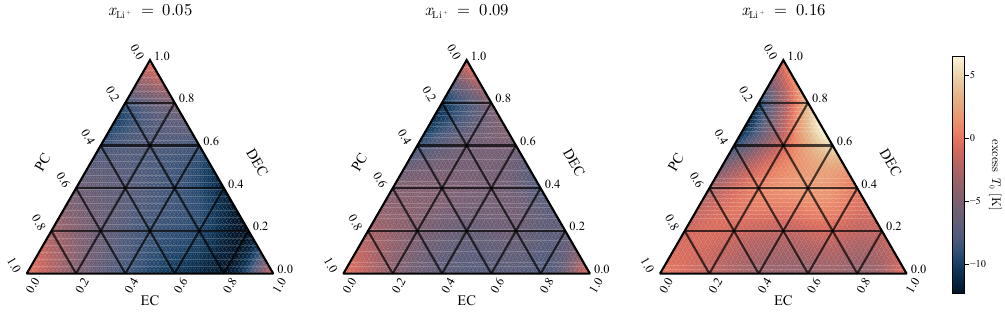}
    \caption{
        \label{fig:excess_t0_dec_pf6}
        Excess Vogel temperatures for LiPF$_6$ in \ac{EC}, \ac{PC} and \ac{DEC}.
    }
\end{figure}

\begin{figure}[h!]
    \centering
    \includegraphics[width=\linewidth]{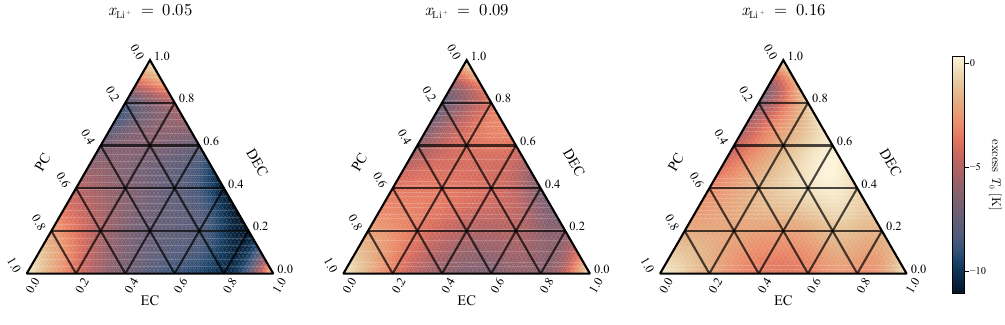}
    \caption{
        \label{fig:excess_t0_dec_tfsi}
        Excess Vogel temperatures for LiTFSI in \ac{EC}, \ac{PC} and \ac{DEC}.
    }
\end{figure}

\begin{figure}[h!]
    \centering
    \includegraphics[width=\linewidth]{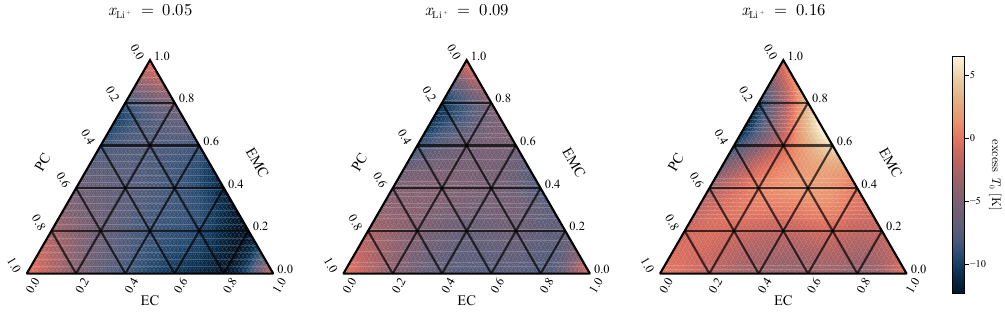}
    \caption{
        \label{fig:excess_t0_emc_pf6}
        Excess Vogel temperatures for LiPF$_6$ in \ac{EC}, \ac{PC} and \ac{EMC}.
    }
\end{figure}

\begin{figure}[h!]
    \centering
    \includegraphics[width=\linewidth]{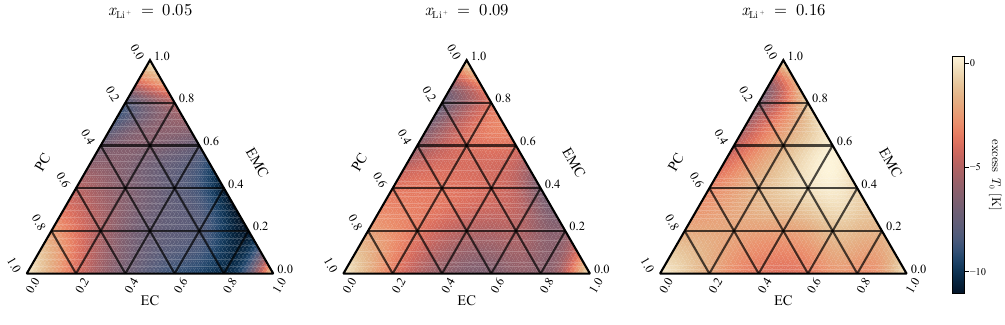}
    \caption{
        \label{fig:excess_t0_emc_tfsi}
        Excess Vogel temperatures for LiTFSI in \ac{EC}, \ac{PC} and \ac{EMC}.
    }
\end{figure}

\begin{figure}[h!]
    \centering
    \includegraphics[width=\linewidth]{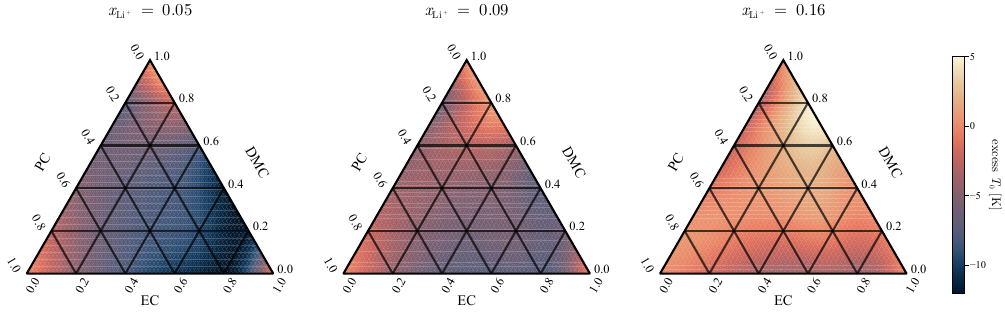}
    \caption{
        \label{fig:excess_t0_dmc_pf6}
        Excess Vogel temperatures for LiPF$_6$ in \ac{EC}, \ac{PC} and \ac{DMC}.
    }
\end{figure}

\begin{figure}[h!]
    \centering
    \includegraphics[width=\linewidth]{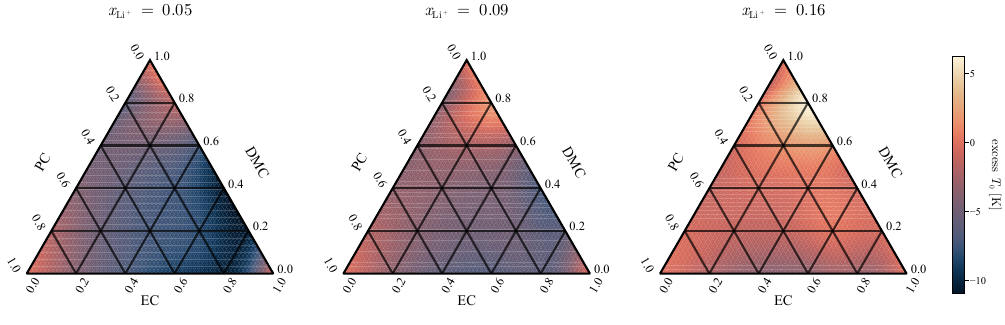}
    \caption{
        \label{fig:excess_t0_dmc_tfsi}
        Excess Vogel temperatures for LiTFSI in \ac{EC}, \ac{PC} and \ac{DMC}.
    }
\end{figure}

\begin{figure}[h!]
    \centering
    \includegraphics[width=\linewidth]{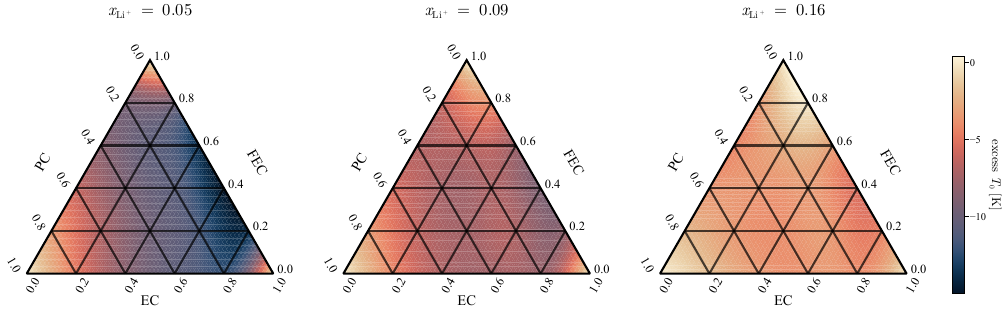}
    \caption{
        \label{fig:excess_t0_fec_pf6}
        Excess Vogel temperatures for LiPF$_6$ in \ac{EC}, \ac{PC} and \ac{FEC}.
    }
\end{figure}

\begin{figure}[h!]
    \centering
    \includegraphics[width=\linewidth]{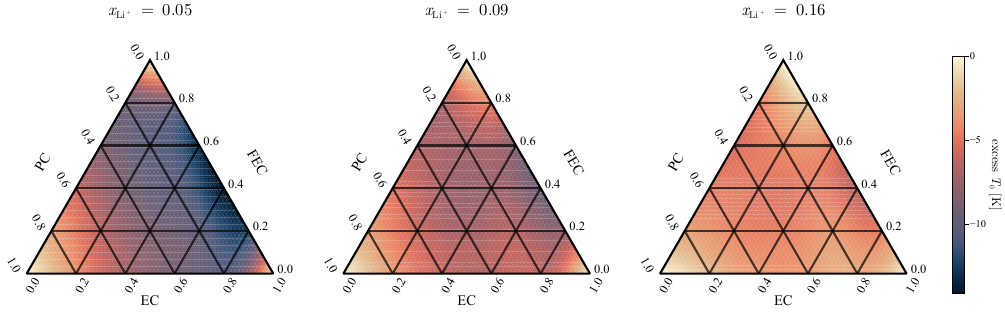}
    \caption{
        \label{fig:excess_t0_fec_tfsi}
        Excess Vogel temperatures for LiTFSI in \ac{EC}, \ac{PC} and \ac{FEC}.
    }
\end{figure}

\clearpage

\subsection{Comparing Temperature Dependence Across Anions}
\label{sec:si:conductivity_comparing}
\begin{figure}[h!]
    \centering
    \includegraphics[width=\linewidth]{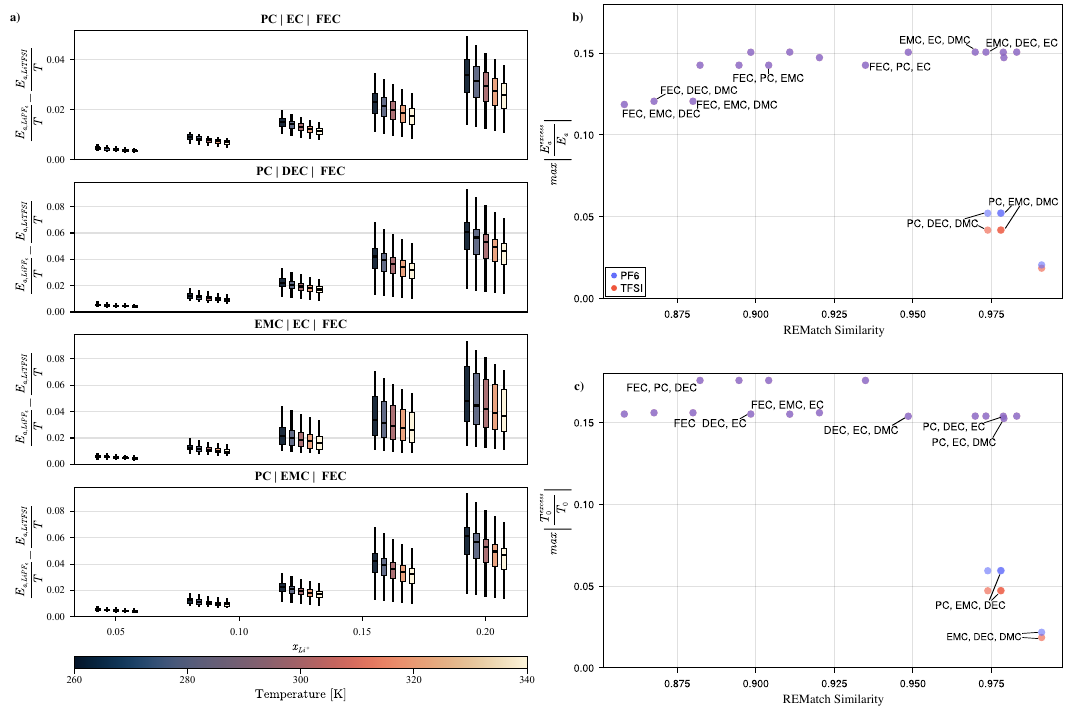}
    \labelphantom{fig:delta_Ea_T}
    \labelphantom{fig:Ea_vs_soap}
    \labelphantom{fig:T0_vs_soap}
    \caption{
        \label{fig:conductivity_comparing}
        \textbf{Comparing \ac{VFT} parameters for LiTFSI and LiPF$_6$.}
        (\subref*{fig:delta_Ea_T})~Differences in the non-dimensionalized pseudo-activation energies for LiTFSI and LiPF$_6$ at various salt mole ratios (0.05 - 0.20) and temperatures (260-340 K).
        (\subref*{fig:Ea_vs_soap})~Maximum absolute relative excess pseudo-activation energy~\cite{ngaiIntroductionProblemsRelaxation2011} versus the structural similarity~\cite{deMappingClassifyingMolecules2017,himanenDScribeLibraryDescriptors2020} for LiTFSI and LiPF$_6$ in various carbonate solvent systems.
        (\subref*{fig:T0_vs_soap})~Maximum absolute relative excess Vogel temperature~\cite{ngaiIntroductionProblemsRelaxation2011} versus the structural similarity~\cite{deMappingClassifyingMolecules2017,himanenDScribeLibraryDescriptors2020} for LiTFSI and LiPF$_6$ in various linear and cyclic carbonate solvent systems.
    }
\end{figure}
In this section, we use the \ac{VFT} parameters learnt by the \ac{MIST} ionic conductivity model, and the associated excess quantities, to compare the behaviour of the LiTFSI and LiPF$_6$ electrolytes at different temperatures.
Insights from this analysis are relevant for designing electrolytes with good performance at extreme operating temperatures.

Overall, we observe that (\cref{fig:delta_Ea_T}) LiPF$_6$ shows a larger pseudo-activation energy than LiTFSI (consistent with ternary plots in~\cref{sec:si:conductivity_ea}).
This is consistent with the fact that PF$_6^-$ often exhibits stronger Li$^+$–anion association (ion pair formation and cluster formation) than highly charge-delocalised TFSI$^-$ in the same carbonate solvents~\cite{dahbiComparativeStudyEC2011}.
A stronger association lowers ionicity and increases the apparent activation barrier for long-range charge transport; LiTFSI remains more dissociated and thus less sensitive to temperature~\cite{rushingTaleNoninteractingAdditive2022}.
The gap between the two salts ($\frac{\Delta E_a}{T}$) widens at low temperatures.
This may be because the ion pairs stabilise as temperature decreases; dissociation increases with increasing temperature.
Since LiPF$_6$ is likely more associated, cooling penalises it disproportionately~\cite{berhautIonicAssociationAnalysis2019}.
In particular, the model captures that solvent choice affects the temperature dependence of ionic conductivity: linear-carbonate-rich blends (\ac{DEC}/\ac{EMC}/\ac{DMC}) have larger gaps compared to solvents with \ac{EC}.
\ac{EC}, a cyclic carbonate, has higher permittivity and strongly binds Li$^+$ ions, favouring solvent-separated structures and higher dissociation.
Linear carbonates, which have lower permittivity, promote contact ion pairing~\cite{gallekankanamgeMolecularStructureChemical2020}.
As with the observations above, assuming that LiPF$_6$ is more prone to association, it would be more affected in mixtures with linear carbonates~\cite{gallekankanamgeMolecularStructureChemical2020}.

We compare the magnitude of the relative excess pseudo-activation energy $\textrm{max} \left|\frac{E^{excess}_a}{E_a} \right|$ and Vogel temperature $\textrm{max} \left|\frac{T^{excess}_0}{T_0} \right|$ with the similarity between the solvent molecules.
As discussed in~\cref{sec:si:soap_excess}, prior work~\cite{KADVExcessDensityDescriptor2025} suggests that mixtures with more dissimilar molecules have larger excess properties.
We compute the relative excess as the ratio between the excess and total parameter values and find where the magnitude of this ratio is maximised across all possible salt and solvent mole ratios; this metric emphasises the deviation from linear mixing.
To quantify similarity, we evaluated the REMatch similarity~\cite{deMappingClassifyingMolecules2017,himanenDScribeLibraryDescriptors2020} between the \ac{SOAP} fingerprints of ternary solvent mixtures as the sum of the pairwise similarities weighted by their normalised mole ratios:
\begin{align}
    \text{sim}_W(\text{mol}_1, \text{mol}_2, \text{mol}_3) = \sum_{1 \leq i \leq j \leq 3} \frac{x_i + x_j}{x_1 + x_2 + x_3} \, \text{sim}_R(\text{mol}_i, \text{mol}_j)
\end{align}
We observe that the absolute relative excess of $E_a$ (\cref{fig:Ea_vs_soap}) and $T_0$ (\cref{fig:T0_vs_soap}) is the same for both salts in mixtures containing \ac{EC} or \ac{FEC};
in mixtures containing only linear carbonates and \ac{PC}, we observe that PF$_6$ has a slightly larger relative excess.
The relative excess is also higher for mixtures containing \ac{FEC} and \ac{EC}, with both high and low similarity.
This could be explained by the high dielectric constants of the two molecules (95.3 for \ac{EC}~\cite{hallDielectricConstantsQuantum2015} and 102 for \ac{FEC}~\cite{leeTuningTwoInterfaces2019}) which leads to stronger salt-solvent and solvent-solvent interactions, which is reflected by the high excess observed.
This effect seems to dominate the impact of structural dissimilarity purported in prior work~\cite{KADVExcessDensityDescriptor2025}.

\subsection{Determinants of Vogel Temperature}
\begin{figure}[h!]
    \centering
    \includegraphics[width=\linewidth]{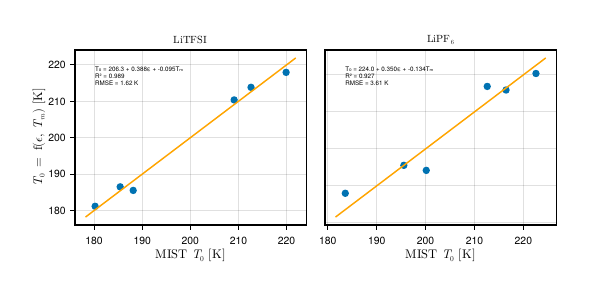}
    \caption{
        \label{fig:gml_fit_1M}
        Parity plots comparing \ac{MIST}-predicted Vogel temperature $T_0$ with linear model predictions based on solvent dielectric constant $\varepsilon$ and melting point $T_m$ for 1M LiTFSI and LiPF$_6$ electrolytes with unary carbonate solvents. Orange line represents perfect agreement between models.
    }
\end{figure}

\begin{figure}[h!]
    \centering
    \includegraphics[width=\linewidth]{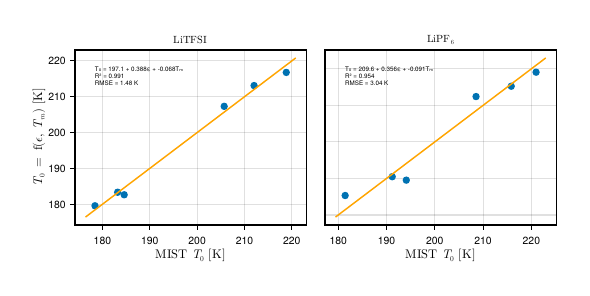}
    \caption{
        \label{fig:gml_fit_0_8M}
        Parity plots comparing \ac{MIST}-predicted Vogel temperature $T_0$ with linear model predictions based on solvent dielectric constant $\varepsilon$ and melting point $T_m$ for 0.8M LiTFSI and LiPF$_6$ electrolytes with unary carbonate solvents. Orange line represents perfect agreement between models.
    }
\end{figure}

\begin{figure}[h!]
    \centering
    \includegraphics[width=\linewidth]{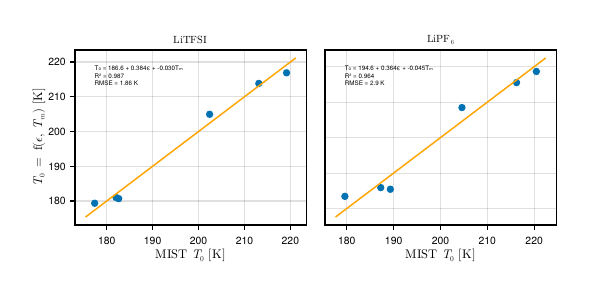}
    \caption{
        \label{fig:gml_fit_0_6M}
        Parity plots comparing \ac{MIST}-predicted Vogel temperature $T_0$ with linear model predictions based on solvent dielectric constant $\varepsilon$ and melting point $T_m$ for 0.6M LiTFSI and LiPF$_6$ electrolytes with unary carbonate solvents. Orange line represents perfect agreement between models.
    }
\end{figure}

\begin{figure}[h!]
    \centering
    \includegraphics[width=\linewidth]{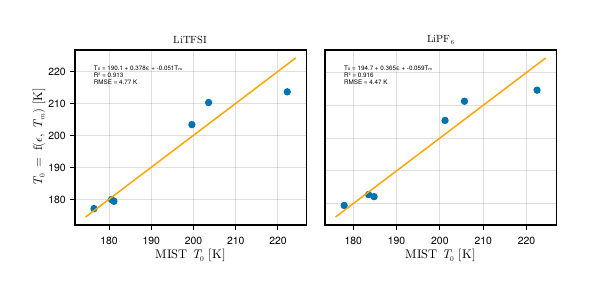}
    \caption{
        \label{fig:gml_fit_0_4M}
        Parity plots comparing \ac{MIST}-predicted Vogel temperatures $T_0$ with linear model predictions based on solvent dielectric constant $\varepsilon$ and melting point $T_m$ for 0.4M LiTFSI and LiPF$_6$ electrolytes with unary carbonate solvents. Orange line represents perfect agreement between models.
    }
\end{figure}

\begin{figure}[h!]
    \centering
    \includegraphics[width=\linewidth]{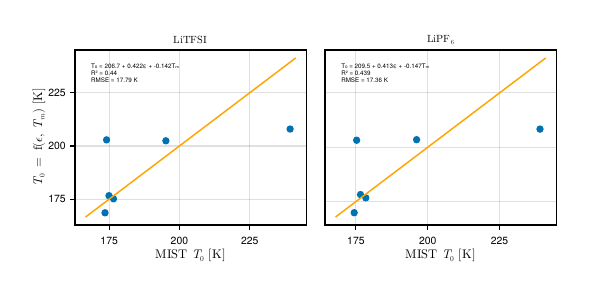}
    \caption{
        \label{fig:gml_fit_0_2M}
        Parity plots comparing \ac{MIST}-predicted Vogel temperature $T_0$ with linear model predictions based on solvent dielectric constant $\varepsilon$ and melting point $T_m$ for 0.2M LiTFSI and LiPF$_6$ electrolytes with unary carbonate solvents. Orange line represents perfect agreement between models.
    }
\end{figure}

To elucidate the physicochemical determinants of Vogel temperature ($T_0$) in the electrolytes studied above, we developed linear models relating $T_0$ to solvent dielectric constant ($\varepsilon$) and melting point ($T_m$) for the six organic carbonate solvents (\ac{EC}, \ac{PC}, \ac{DEC}, \ac{FEC}, \ac{DMC}, and \ac{EMC}) and two salts (LiTFSI and LiPF$_6$) studied above. 
The linear models demonstrate good predictive accuracy at moderate-to-high concentrations. 
For example, for LiTFSI-based electrolytes at 1M molarity, the model $T_0 = 206.3 + 0.388\varepsilon - 0.095T_m$ ($R^2 = 0.989$) demonstrated excellent predictive accuracy, with dielectric constant emerging as the dominant descriptor ($p < 0.001$). 
Each unit increase in dielectric constant corresponds to a 0.39~K increase in $T_0$, consistent with enhanced ion-solvent interactions stabilizing the glassy state at higher temperatures. 
A similar trend was observed for LiPF$_6$ electrolytes ($R^2 = 0.927$), though with slightly reduced accuracy. 
Notably, the melting point contribution exhibited marginal statistical significance ($p = 0.067$ for LiTFSI, $p = 0.174$ for LiPF$_6$), suggesting that solvent polarity, rather than molecular packing efficiency, primarily governs low-temperature transport behaviour~\cite{XuNonaqueousLiquidElectrolytes2004,XuElectrolytesInterphasesLiIon2014}. 
The linear model fits for all systems studied are shown in~\cref{fig:gml_fit_1M,fig:gml_fit_0_8M,fig:gml_fit_0_6M,fig:gml_fit_0_4M,fig:gml_fit_0_2M}.

\begin{figure}[h!]
    \centering
    \includegraphics[width=\linewidth]{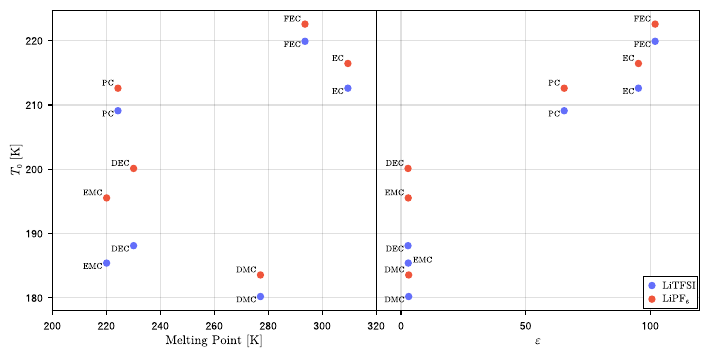}
    \caption{
        \label{fig:descriptors_vs_T0_1M}
       Vogel temperature $T_0$ versus solvent melting point and dielectric constant for 1M LiTFSI and LiPF$_6$ electrolytes. $T_0$ shows strong positive correlation with dielectric constant $\varepsilon$ but weak dependence on melting point $T_m$, confirming the linear model analysis above.
    }
\end{figure}

\begin{figure}[h!]
    \centering
    \includegraphics[width=\linewidth]{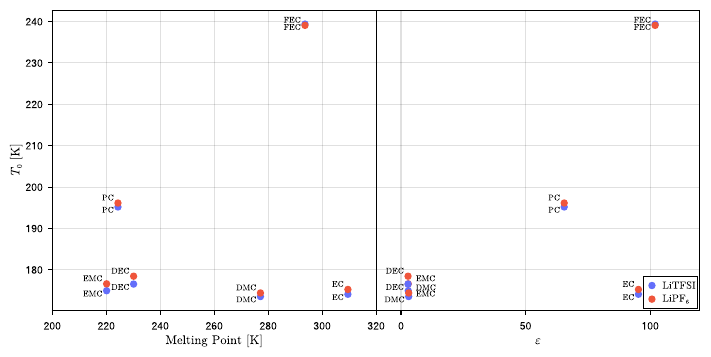}
    \caption{
        \label{fig:descriptors_vs_T0_0.8M}
       At 0.2 M concentration, $T_0$ exhibits increased scatter for both $T_m$ and $\varepsilon$ compared to the more concentrated 1M solution. The two salts show similar but less well-defined trends.
       This could be because dilute electrolytes reflect solvent-dominated behaviour with incomplete solvation shells, reducing the utility of simple molecular descriptors for predicting low-temperature transport properties~\cite{XuElectrolytesInterphasesLiIon2014}.
    }
\end{figure}

\FloatBarrier
\section{Atom-Level Partial Charge Predictions}
\label{sec:si:partial_charge}

The Smirk tokenizer\cite{WBVTokenizationMolecularFoundation2026} provides MIST with tokens that encode atomic-level concepts.
For example, an ammonium ion \smiles{[NH4+]} is tokenized as \tok{[,N,H,4,+,]}, granting \ac{MIST} direct access to the electronic, nuclear, and geometric information contained within bracketed atoms.
As a demonstration of this capability, we fine-tuned a variant of MIST to predict atom-level Mulliken and L\"owdin partial charges.
This model was used to generate the minimum partial charges shown in \cref{fig:si:screening_electrolytes}.

To compute atom-level partial charges, we trained a token-level regression network.
Unlike our other fine-tuned models, which use the final-layer hidden state for the first token (\cref{fig:mist_arch_diagram}), here we feed the hidden state of each token individually through the task network to produce a prediction \(y_i\) for each token \(i\) in the \ac{SMILES}-encoded molecule.
We train the \ac{MIST} models on the PubChemQC dataset (\cref{sec:si:pubchem_qc}) to predict the corresponding partial charge annotation \(y_i\), using the \ac{MSE} loss between predictions and labels to guide optimization.

\begin{figure}[ht!]
    \centering
    \includegraphics[width=0.75\linewidth]{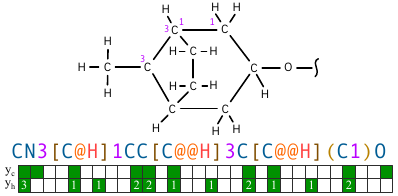}
    \caption{
        \label{fig:partial_charge_encoding}
        Example tokenization and partial charge encoding for a fragment of a larger molecule.
        The structural diagram shows ring closures in purple and all hydrogen atoms explicitly.
        For each token, the model predicts the central atom’s partial charge (\(y_c\)) and the average charge of attached hydrogens (\(y_h\)) for both Mulliken and L\"owdin methods, yielding four values per token (only one charge type is shown here for clarity).
        Green squares indicate tokens with relevant outputs; the rest are left unannotated.
        The numbers in the bottom row indicate the number of hydrogen atoms associated with each \(y_h\) value.
    }
\end{figure}

We use the encoding scheme shown in \cref{fig:partial_charge_encoding} to annotate each token with its associated partial charge.
Specifically, we represent the partial charge as a tuple \(y_i = (y_c, y_h)\), where \(y_c\) denotes the charge on the heavy atom and \(y_h\) is the average charge over its attached hydrogen atoms.
For example, ammonia is encoded in \ac{SMILES} as \smiles{N}, resulting in a single token representing both the nitrogen and its three implicit hydrogens.
When hydrogen atoms are explicitly represented (\cref{fig:partial_charge_encoding}), we annotate the central atom with \((y_c, y_h)\), and the hydrogen tokens with only \(y_h\).
Our regression task then minimizes the masked \ac{MSE} loss between model predictions and reference annotations for both Mulliken and L\"owdin charges.
A similar approach could be used to encode bond-level regression targets or other atom-level information.
Performance metrics for our models are shown in \cref{tab:pubchem_benchmarks}.

\begin{table}[ht!]
\centering
\begin{tabular}{@{}llcc@{}}
\toprule
Model & Dataset & Partial Charge, MAE (e) \\
\midrule
Thaler et al.\cite{TMT+ActiveLearningGraph2024} & CoRE MOF-2019\cite{CHB+AdvancesUpdatesAnalytics2019} & 0.00083  \\
Kancharlapalli et al.\cite{KGHSFastAccurateMachine2021} & CoRE MOF-2019\cite{CHB+AdvancesUpdatesAnalytics2019} & 0.0192 \\
Lehner et al.\cite{LKM+DASHDynamicAttentionBased2023} & ZINC \& ChEMBL & 0.03 (RMSE) \\
\midrule
MIST-28M & PubChemQC & 0.036 \\
MIST-228M & PubChemQC & 0.031 \\
MIST-1.8B & PubChemQC & 0.030 \\
\bottomrule
\end{tabular}%
\caption{
    \label{tab:pubchem_benchmarks}
    Comparison of MIST models trained on PubChemQC with related models.
    To our knowledge, no other models have been benchmarked on the PubChemQC\cite{NMPubChemQCB3LYP631G2023} dataset, limiting direct comparisons.
}
\end{table}

\section{Linear Probes}
\label{sec:si:probes}
In this section, we detail how we trained linear classifier probes to identify \ac{RO5} feature vectors within the internal activations of \ac{MIST} models.

\subsection{Dataset}
Our Lipinski's Rule of Five\cite{LLDFExperimentalComputationalApproaches1997} dataset was curated from MoleculeNet\cite{WRF+MoleculeNetBenchmarkMolecular2018} datasets, specifically: QM8, QM9, HIV, ToxCast, Tox21, ClinTox and BBBP.
The goal of probing a model's activations \(x\) is to identify feature vectors \(f\) that are linearly predictive of the target feature: \(p(\mbox{feat.}|\mbox{mol}) = \sigma(f \cdot x)\);
where \(\sigma\) is the logistic function.
As such, our primary concern when building the probe dataset is to maximize the ``conceptual separation'' between positive and negative classes\cite{GNP+FindingNeuronsHaystack2023}.
We used the following procedure to curate a dataset such that the marginal probability of each of Lipinski's four criteria was as close to 0.5 as possible.

\begin{enumerate}
    \item Remove any molecule that was rejected by \texttt{rdkit}'s \texttt{MolFromSMILES}.
    \item De-duplicate the dataset using \texttt{rdkit}'s computed InChI Key.
    \item Use iterative proportional refitting to randomly sample a balanced dataset.
    \item Use \texttt{scikit-learn}'s \texttt{StratifiedShuffleSplit} to split the dataset into train/validation/test (80/10/10) while preserving the relative frequency of passing molecules.
\end{enumerate}

\begin{table}[ht!]
    \centering
    \begin{tabular}{lcc}
    \toprule
                & Initial & Resampled \\
    \midrule \\
    H-Donor      & 99.2\%   & 84.9\%     \\
    H-Acceptor   & 98.9\%   & 84.2\%     \\
    MWT          & 97.1\%   & 81.6\%     \\
    Log P        & 96.2\%   & 73.6\%     \\ 
    \midrule \\
    Dataset Size & 279,066  & 10,000 \\
    \bottomrule
    \end{tabular}
    \caption{
        \label{tab:lipinski_dataset_si}
        Frequency of molecules passing each of Lipinski's \acs{RO5} criteria, before and after rebalancing.
        Further improving class balance would require reducing the dataset size to the point that linear probes for our largest models (i.e., MIST-1.8B) would become underdetermined (i.e., the hidden size would be greater than the number of samples).
    }
\end{table}

Our iterative proportional refitting procedure consisted of two steps:
(1) assigning sampling weights to each molecule such that the expectation for each criterion was 0.5; and 
(2) randomly sampling \(n\) molecules proportional to these weights to improve the class balance within the dataset.
We sampled 10,000 molecules from the source dataset to balance our need for having a balanced dataset and having sufficient samples to fit linear probes.
Our largest model, MIST-1.8B, has a feed-forward width of 9,128; a smaller, nearly perfectly balanced dataset would be possible, but it would leave the probes underdetermined.

This resampling procedure was necessary for two reasons.
First, the initial dataset was highly unbalanced containing more positive labels.
This imbalance is expected to be detrimental to probe quality by biasing our probes towards predicting positive labels.
Second, the individual features (within the dataset) are highly correlated;
of our initial 279,066 molecules, 262,898 (94.2\%) pass all four of Lipinski's \ac{RO5} criteria, but only 918 (0.3\%) pass only the LogP criteria.
To identify features associated with each criterion separately, as well as an overall \ac{RO5} feature, we needed to re-sample our raw dataset to correct these imbalances.

\subsection{Training Linear Probes}
Encoder activations \(\vec{x}\) were collected during the model's forward pass using PyTorch hooks.
Linear probes were trained using the AdamW\cite{LHDecoupledWeightDecay2019} optimizer to minimize the binary cross entropy loss between target \(y_i\) and predicted \(\hat{y}_i = \sigma(\vec{f}_i \cdot \vec{x} + b_i)\) class labels,
where $i$ indexes over \ac{RO5} criteria and $f_i$ are the feature vectors.
For numerical stability, we used PyTorch's \texttt{binary\_cross\_entropy\_with\_logits} and passed the predicted logits \(\vec{f}_i \cdot \vec{x} + b_i\) directly.
Probes were trained for 10,000 steps with a constant learning rate of \sn{1}{-3}, selecting the probe with the lowest validation loss as the trained probe.
Notably, by construction training linear probes will converge to the global optimum, since a linear classifier trained to minimize the cross-entropy is a convex problem\cite{ABUnderstandingIntermediateLayers2018}.

\subsection{Evaluating Linear Probes}

Once trained, probes were assessed with the following metrics, primarily to study how the identified features mutated across different layers of the encoder.
As with our other classification tasks, our primary metric for classifier efficacy was the \ac{AUROC}.
\paragraph{Additivity of Lipinski's Rule of Five Criteria.}
\begin{figure}
    \centering
    \includegraphics[width=0.7\linewidth]{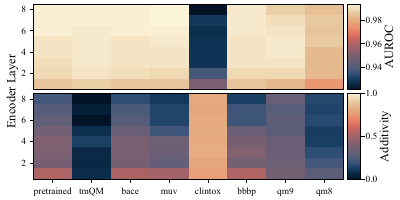}
    \caption{
        \label{fig:lipinski_linear_probes}
        Linear probes recover Lipinski’s \ac{RO5} signals. 
    }
\end{figure}
The \ac{RO5} requires suitable molecules to pass all four criteria.
Here, we explore whether \ac{MIST} models learned an overall feature vector \(f_l\) that is a linear combination of the individual criteria vectors \(f_i\).
This hypothesis would require models to learn orthogonal, or nearly orthogonal\cite{EHO+ToyModelsSuperposition2022}, feature vectors for the individual criteria and to deduce that combining these features linearly is useful.
To assess this, we evaluated the ``Additivity'' of the overall Lipinski feature \(f_{l}\) and the individual criterion features \(f_i\), or the cosine similarity of the overall feature and the sum of the criterion, defined as:
\begin{align*}
    \mbox{Additivity} &= S_C(f_l, \sum_i f_i) \\
    S_C(\vec{a}, \vec{b}) &= \frac{\vec{a} \cdot \vec{b}}{\| \vec{a} \| \| \vec{b}\|} .
\end{align*}

We observe alignment between the aggregate and overall feature directions for some fine-tuned variants (\cref{fig:lipinski_linear_probes}).
The observed alignment between the summed criterion axes and the overall axis is surprising; the overall \ac{RO5} label is a logical conjunction of the individual criteria and need not correspond to a single linear separator.
This behaviour varies by task.
In ClinTox (a clinical toxicity classification dataset), the summed  \ac{RO5} vectors nearly reconstruct the overall probe direction (Additivity near one), suggesting the formation of a fused drug-likeness feature.
In contrast, tmQM (an organometallic quantum mechanical property dataset) retains disentangled features with near-orthogonal orientations (Additivity near zero).

\paragraph{Feature Alignment Across Layers.}
\begin{figure}
    \centering
    \includegraphics[width=\linewidth]{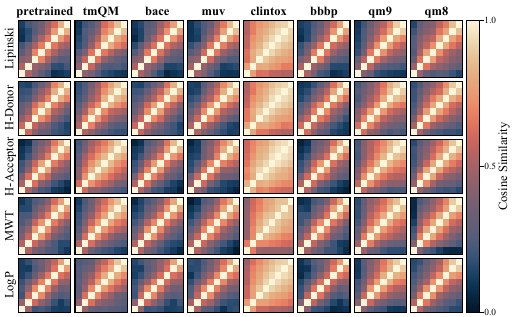}
    \caption{
        \label{fig:lipinski_layer_similarity_output}
        Measured cosine similarity of \ac{RO5} feature vectors identified using linear probes.
    }
\end{figure}
Feature vectors can change across transformer layers as transformer layers rotate, delete or add features to the residual stream\cite{JAJ+SparseCrosscodersCrossLayer2024}.
Our question here was to determine whether the identified features were continuously updated, eventually stabilized or faded out, as they move through the network.
To do this, we computed the cosine similarity \(S_C\) between the identified feature vectors for different layers.
The results for probes at the output of a transformer block are shown in \cref{fig:lipinski_layer_similarity_output}.

\paragraph{Fixed Effect Models for the \ac{AUROC} of Lipinski Feature Vectors.}
\begin{table}[ht!]
\centering
\begin{tabular}{lrrrr}
\toprule
Model                     &        (1) &        (2) &        (3) &        (4) \\\midrule
(Intercept)               &   4.501*** &   3.962*** &   4.333*** &   4.030*** \\
                          &    (0.027) &    (0.137) &    (0.056) &    (0.081) \\
encoder\_dataset: tmQM    &   0.432*** &            &     0.418* &            \\
                          &    (0.071) &            &    (0.177) &            \\
encoder\_dataset: bace    &   0.366*** &            &     0.351* &            \\
                          &    (0.067) &            &    (0.166) &            \\
encoder\_dataset: muv     &   0.411*** &            &     0.397* &            \\
                          &    (0.070) &            &    (0.173) &            \\
encoder\_dataset: clintox &  -1.828*** &            &  -1.750*** &            \\
                          &    (0.024) &            &    (0.060) &            \\
encoder\_dataset: bbbp    &   0.371*** &            &     0.344* &            \\
                          &    (0.068) &            &    (0.165) &            \\
encoder\_dataset: qm9     &    0.167** &            &      0.189 &            \\
                          &    (0.057) &            &    (0.145) &            \\
encoder\_dataset: qm8     &  -0.274*** &            &   -0.289** &            \\
                          &    (0.041) &            &    (0.101) &            \\
location: attention       &   0.261*** &            &            &      0.148 \\
                          &    (0.019) &            &            &    (0.151) \\
location: intermediate    &     -0.015 &            &            &      0.052 \\
                          &    (0.016) &            &            &    (0.141) \\
location: output-dense    &  -0.504*** &            &            &   -0.369** \\
                          &    (0.013) &            &            &    (0.111) \\
layer                     &  -0.016*** &      0.013 &            &            \\
                          &    (0.004) &    (0.034) &            &            \\
\hline
N                         &        256 &        256 &        256 &        256 \\
BIC                       & -2,079.116 & -1,211.592 & -1,623.396 & -1,209.869 \\
Resid. DoF                &        244 &        254 &        248 &        252 \\
Adj. R2                   &      0.971 &     -0.007 &      0.819 &      0.021 \\
RMSD                      &      0.004 &      0.022 &      0.009 &      0.022 \\
\bottomrule
\end{tabular}
\caption{
    \label{tab:lipinski_auroc_regtab}
    Fitted coefficients from generalized linear models predicting the logit of a probe’s \ac{AUROC} score.
    Columns (1)–(4) correspond to different model specifications: (1) includes layer, fine-tuning dataset, and probe location; (2) includes only layer; (3) includes only fine-tuning dataset; and (4) includes only probe location.
    Using all three features (Column 1) provides the highest quality model, as measured by the \(\Delta BIC\) between models, taking \(\Delta BIC > 6\) to be significant\cite{HTFElementsStatisticalLearning2009,leitaoThisScalingNonlinear2016}.
    Fine-tuning effect sizes are positive (higher \ac{AUROC}) for all downstream tasks except ClinTox and QM8.
    Probes at the attention block report significantly higher \ac{AUROC}, while feed-forward probes show negligible or significantly lower \ac{AUROC}, suggesting that \ac{RO5} features may primarily reside within the attention mechanism of \ac{MIST} encoders.
}
\end{table}
We fit fixed effects models to predict the logit of a feature's \ac{AUROC} score.
The pretrained model and output location were used to baseline the effects of fine-tuning and probe location, respectively.
The resulting coefficients are shown in \cref{tab:lipinski_auroc_regtab}, and they can be interpreted as the log-odd-ratio for individual effects relative to the baseline.

\section{Consistency of MIST Predictions with Chemical Intuition}
\label{sec:si:chemical_consistency}

To check that the predictions of fine-tuned \ac{MIST} models align with chemical principles, we analysed trends in predicted properties for hydrocarbons and fatty acids.

\subsection{Trends in Hydrocarbons}
We first consider hydrocarbons of the form \texttt{R-X}, where \texttt{R} is an alkyl chain and \texttt{X} is a functional group (\cref{fig:hc_trends_size}).
Across all functionalized hydrocarbons studied, \ac{MIST} predicts a general decrease in Gibbs free energy \(G^{\degree} = H^{\degree} - TS^{\degree}\) with increasing chain length.
This trend reflects enhanced enthalpic stabilization from successive methylation, which outweighs entropic losses for short chains due to increased conformational freedom \cite{Wade2016}.
At longer chain lengths, where the hydrocarbons are typically solid at room temperature, reduced configurational entropy causes \(G^{\degree}\) to plateau or rise slightly.
A similar trend is observed in polarizability \(\alpha\), which increases with chain length due to larger electron clouds and enhanced dispersion forces, before tapering at longer chains \cite{Reichardt2003,Grimme2004}.
These predictions suggest that \ac{MIST} captures the interplay between enthalpy, entropy, and intermolecular forces.

We also find that predicted melting $T_m$ and boiling points $T_b$ converge as chain length increases, indicating the dominance of alkyl-chain dispersion interactions over functional group effects (\cref{fig:hc_trends_size}).
\ac{MIST} correctly predicts \(T_b > T_m\) for nearly all molecules, except for methane {\color{cat:blue}\smiles{C}} and acetylene {\color{cat:green}\smiles{C\#C}}, which are known outliers.
Acetylene sublimes under standard conditions, while methane’s predicted values (\(T_b = -120 \pm 13\si{\degreeCelsius}\) and \(T_m = -140 \pm 11\si{\degreeCelsius}\)) are within the model's error bounds compared to ground truth.
Fine-tuned models have a validation \ac{MAE} of \(30.1\si{\degreeCelsius}\) and \(20.0\si{\degreeCelsius}\) for melting and boiling points, respectively.

\subsection{Trends in Fatty Acids}
\begin{figure}[h!]
    \centering
    \labelphantom{fig:omega_fatty_acids}
    \labelphantom{fig:unsat_fat_trends}
    \includegraphics[width=\linewidth]{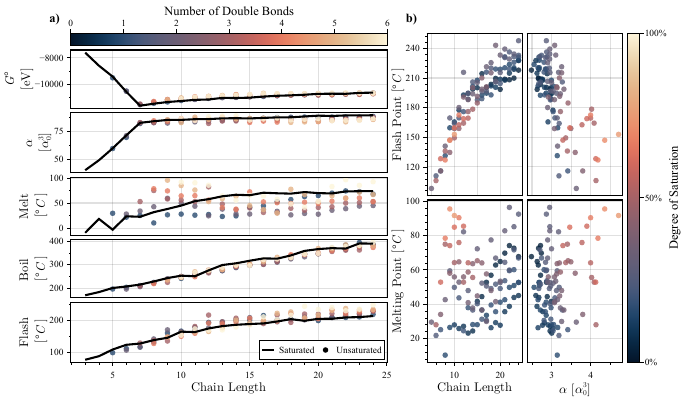}
    \caption{
        \label{fig:fatty_acids}
        (\subref*{fig:omega_fatty_acids})~Trends in quantum, chemical, and thermodynamic properties for Omega-3 unsaturated and saturated fatty acids with chain length.
        (\subref*{fig:unsat_fat_trends})~Covariance plots for \ac{MIST} predictions.
        Overall, \ac{MIST} predictions align with expected thermodynamic trends, namely increasing melting, boiling and flash points as the size of the fatty acids increases.
    }
\end{figure}

\begin{figure}[h!]
    \centering
    \includegraphics[width=\linewidth]{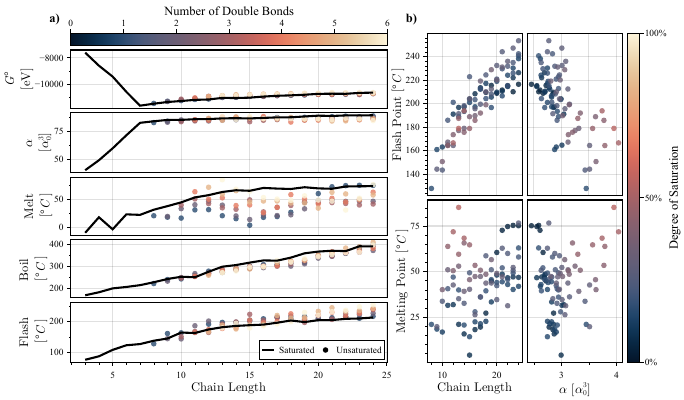}
    \caption{
        \label{fig:si:omega_6_fatty}
        (\subref*{fig:omega_fatty_acids})~Trends in quantum, chemical, and thermodynamic properties for Omega-6 unsaturated and saturated fatty acids with chain length.
        (\subref*{fig:unsat_fat_trends})~Covariance plots for \ac{MIST} predictions.
        Overall, \ac{MIST} predictions align with expected thermodynamic trends, namely increasing melting, boiling and flash points as the size of the fatty acids increases.
    }
\end{figure}

\begin{figure}[h!]
    \centering
    \includegraphics[width=\linewidth]{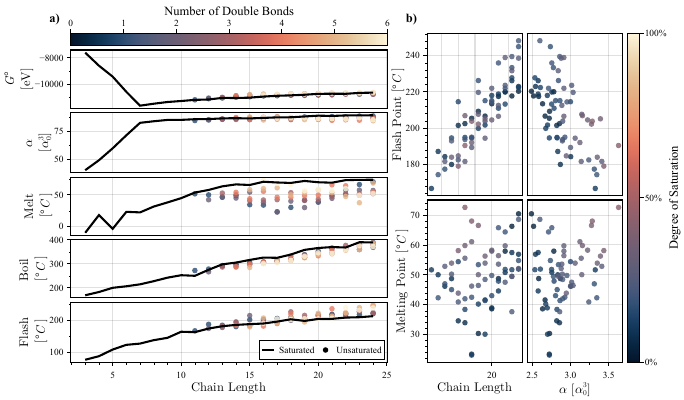}
    \caption{
        \label{fig:si:omega_9_fatty}
        (\subref*{fig:omega_fatty_acids})~Trends in quantum, chemical, and thermodynamic properties for Omega-9 unsaturated and saturated fatty acids with chain length.
        (\subref*{fig:unsat_fat_trends})~Covariance plots for \ac{MIST} predictions.
        Overall, \ac{MIST} predictions align with expected thermodynamic trends, namely increasing melting, boiling and flash points as the size of the fatty acids increases.
    }
\end{figure}

We examined \(n\)-3 (omega-3) polyunsaturated fatty acids of varying chain lengths (\cref{fig:fatty_acids}).
\ac{MIST} predicts that Gibbs free energy \(G^{\degree}\) decreases with chain length, consistent with enhanced van der Waals stabilization \cite{Reichardt2003}.
At fixed chain length, saturated fatty acids exhibit lower \(G^{\degree}\) than their unsaturated counterparts, reflecting their greater thermodynamic stability \cite{Wade2016}.
Similar trends were observed for omega-6 and omega-9 fatty acids (\cref{fig:si:omega_6_fatty,fig:si:omega_9_fatty}).

These trends extend to phase change properties.
Melting, boiling, and flash points increase with chain length and are higher for saturated fatty acids due to better molecular packing and stronger dispersion interactions \cite{Reichardt2003}.
In contrast, unsaturation introduces \(\pi\)-bonds that kink the chain and disrupt packing.
Although conjugation can modestly increase polarizability \cite{Grimme2004}, the net effect is a reduction in intermolecular cohesion, leading to lower phase change temperatures (\cref{fig:unsat_fat_trends}).

At short chain lengths, we find consistent with expectations, that unsaturated fatty acids sometimes exhibit higher melting points due to conjugation-enhanced polarizability (\cref{fig:unsat_fat_trends}).
At longer chains, however, packing disruption dominates and melting points fall below those of saturated analogs\cite{Wade2016}.
Flash points follow similar trends, increasing with chain length and decreasing with unsaturation.
However, highly unsaturated long-chain molecules occasionally exhibit elevated flash points due to increased delocalization\cite{Wade2016}.

Overall, MIST captures the nuanced interplay between chain length, saturation, and intermolecular forces underpinning the thermodynamics of fatty acids.
Notably, our analysis explored trends for carbon chain lengths far beyond what occurs in our dataset, suggesting that \ac{MIST} can reasonably extrapolate chemical trends beyond its training data.

\begin{figure}[h!]
    \centering
    \includegraphics[width=\linewidth]{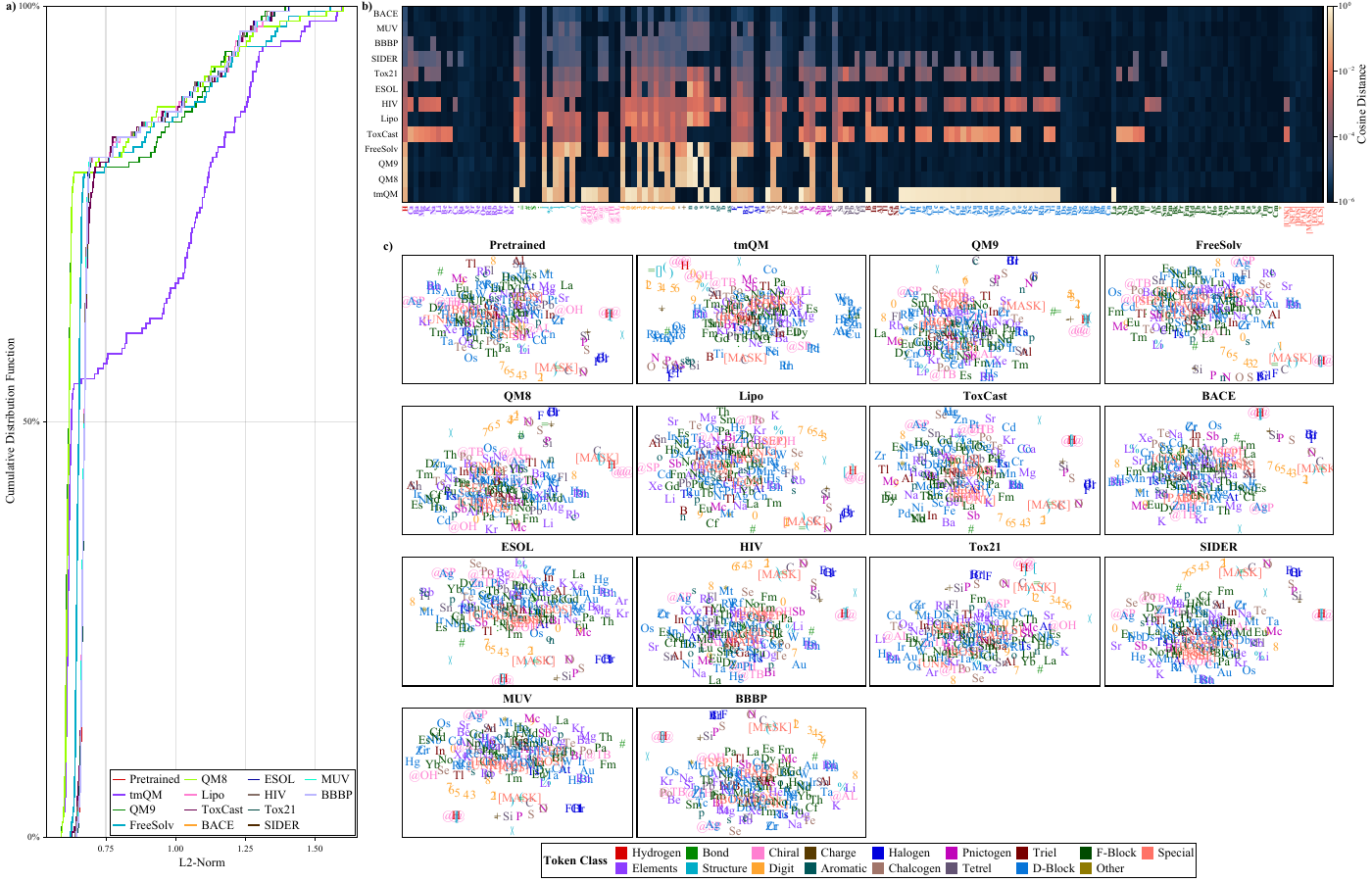}
    \caption{
        Token embedding vector updates and \ac{t-SNE} maps for all fine-tuned variants of MIST-28M.
    }
\end{figure}

\section{Validation of Embedding Cluster Structure with \acsp{SVM}}
\label{sec:si:embeddings}

Low-dimensional projections of the pretrained \ac{MIST} embeddings using \ac{t-SNE} and \ac{UMAP} revealed visually coherent clusters corresponding to chemically meaningful classes (aromaticity and \ac{PAH} structure), as seen in \cref{fig:interp_emb_pah,fig:interp_emb_rings}.
To validate that these clusters reflect discriminative structure in the original high-dimensional embedding space, we employed \acp{SVM} as diagnostic classifiers.
For each binary label set (aromatic/ anti-aromatic and cata/ peri-condensed), raw labels were mapped to $\{ +1, -1\}$ and embeddings were standardized to zero mean and unit variance per dimension.

To assess linear structure consistent with the projected clusters, we trained a \ac{SVM} with a linear kernel and a large regularization constant $C$ on stratified training splits.
The learned decision functional margins are $m_i = y_i(w^Tx_i + b)$, where $w$ and $b$ are the normal vector and bias.
Perfect hard-margin separability requires $m_i \geq 1$ for all molecule embeddings $x_i$.
In our results, both chemical class distinctions have some margin violations (i.e., that no single hyperplane achieved strict separation of all points).
Despite this, the linear decision scores retained strong discriminative power: the \ac{AUROC} was 0.9759 for aromatic/ anti-aromatic molecules and 0.9920 for the cata/ peri-condensed \acp{PAH} classification when evaluated on held-out validation sets.
This demonstrates  that the embedding space still encodes the class distinctions in a mostly ordered (soft-separable) manner even when some individual samples lie ambiguously near the boundary.

\section{Compute-Optimal Neural Scaling}\label{sec:si:compute_optimal_scaling}
In this section, we derive the compute-optimal frontier, following the convention of Hoffmann et al.\cite{HBM+TrainingComputeOptimalLarge2022}, and the resulting compute-optimal scaling of data and model size.
Specifically, Hoffmann et al.\cite{HBM+TrainingComputeOptimalLarge2022} modeled the pretraining loss \(L\) as a function of the number of non-embedding parameters \(N\) and dataset size \(D\):
\begin{equation}\label{eq:pretrainingloss}
    L = \frac{A}{N^{\alpha}} + \frac{B}{D^{\beta}} + E ,
\end{equation}
where \(A, \alpha, B, \beta,\) and \(E\) are coefficients empirically fitted to the measured minimum loss across different model parameterizations.

During the forward pass, each token requires approximately \(2N\) \acp{FLOP}\cite{KMH+ScalingLawsNeural2020,HBM+TrainingComputeOptimalLarge2022}.
The backward pass requires roughly \(4N\) \acp{FLOP} to update both the optimizer state and the model weights\cite{KMH+ScalingLawsNeural2020,HBM+TrainingComputeOptimalLarge2022}.
This yields a total training compute of \(C \approx 6 N D\) \acp{FLOP}.
Taking the compute cost to be \(C = 6 N D\), the optimal parameter $N_{\mathrm{opt}}$ and dataset $D_{\mathrm{opt}}$ sizes for a fixed compute budget are given by:
\begin{equation}
    (N_\mathrm{opt},\, D_\mathrm{opt}) = \underset{\substack{N,\, D \\ \text{s.t. } 6 N D \leq C}}{\arg\min} \frac{A}{N^{\alpha}} + \frac{B}{D^{\beta}} + E .
\end{equation}

\noindent
Solving yields:
\begin{align}
    N_\mathrm{opt} &=
        \left( \frac{C}{6} \right)^{\frac{\beta}{\alpha + \beta}}
        \left( \frac{\alpha A}{\beta B} \right)^{\frac{1}{\alpha + \beta}}
        \label{eq:chinchilla_vals_n} \\
    D_\mathrm{opt} &=
        \left( \frac{C}{6} \right)^{\frac{\alpha}{\alpha + \beta}}
        \left( \frac{\alpha A}{\beta B} \right)^{-\frac{1}{\alpha + \beta}}
        \label{eq:chinchilla_vals_d} \\
    D_\mathrm{opt} &= \left( \frac{\alpha A}{\beta B} \right)^{-\frac{1}{\beta}} \, (N_\mathrm{opt})^{\frac{\alpha}{\beta}} .
\end{align}

\noindent
Substituting \cref{eq:chinchilla_vals_n,eq:chinchilla_vals_d} into \cref{eq:pretrainingloss} gives the compute-optimal loss:
\begin{align}
    L_\mathrm{opt}(C) &= E + A G^{-\alpha}\left( \frac{C}{6} \right)^{-\frac{\alpha\beta}{\alpha + \beta}}
       + B G^{\beta} \left( \frac{C}{6} \right)^{-\frac{\alpha\beta}{\alpha + \beta}} \\
    &= E + \left(A G^{-\alpha} + B G^{\beta} \right) \left( \frac{C}{6} \right)^{-\frac{\alpha\beta}{\alpha + \beta}} ,\label{eq:si:compute_optimal_loss}
\end{align}
where \(G=\left( \frac{\alpha A}{\beta B} \right)^{\frac{1}{\alpha+\beta}}\).
Since \(L_\mathrm{opt}\propto C^{-\frac{\alpha\beta}{\alpha+\beta}}\), the convergence rate to the intrinsic entropy as a function of \(C\) is governed by \(\tfrac{\alpha\beta}{\alpha+\beta}\).

\subsection{Convergence rate as a function of \(\alpha\) and \(\beta\)}\label{sec:si:alpha_beta_optim}
The convergence rate of \(L_\mathrm{opt}\) (\cref{eq:si:compute_optimal_loss}) is maximized when \(\alpha / \beta = 1\) for \(\alpha,\beta>0\) under a well-behaved symmetric constraint, as we now show.
Let \(f(\alpha, \beta) = \frac{\alpha \beta}{\alpha + \beta}\), and let \(g(\alpha, \beta)\) be a continuously differentiable constraint satisfying the symmetry condition \(g(\alpha, \beta) = g(\beta, \alpha)\).
Suppose the constraint surface \(g(\alpha, \beta) = c\) admits an interior optimum \((\alpha^*, \beta^*)\) with \(\alpha^*, \beta^* > 0\) and \(\nabla g(\alpha^*, \beta^*) \neq (0, 0)\).
Then, by symmetry and the Lagrange multiplier conditions, the optimum occurs at \(\alpha^* = \beta^*\).
Without any constraint \(g\) on \(f\), \(f(\alpha,\beta)\) can be made arbitrarily large by increasing one exponent while holding the other fixed;
a symmetric constraint prevents this degeneracy and ensures a well-defined interior optimum.

Explicitly, consider:
\[
\max_{\alpha,\, \beta > 0} f(\alpha, \beta) \quad \text{subject to} \quad g(\alpha, \beta) = c .
\]
The Lagrangian is \(\mathcal{L} = f(\alpha, \beta) - \lambda \, (g(\alpha, \beta) - c)\).
Stationarity gives
\[
\frac{\partial f}{\partial \alpha} = \lambda \, \frac{\partial g}{\partial \alpha}
\quad\text{and}\quad
\frac{\partial f}{\partial \beta} = \lambda \, \frac{\partial g}{\partial \beta}
\ \Rightarrow \
\frac{\partial f/\partial \alpha}{\partial f/\partial \beta}
=
\frac{\partial g/\partial \alpha}{\partial g/\partial \beta} .
\]
At the optimum, we have
\[
\left.\frac{\partial f}{\partial \alpha}\right|_{(\alpha^*, \beta^*)} = \frac{(\beta^*)^2}{(\alpha^* + \beta^*)^2}
\quad
\left.\frac{\partial f}{\partial \beta}\right|_{(\alpha^*, \beta^*)} = \frac{(\alpha^*)^2}{(\alpha^* + \beta^*)^2} .
\]
By the symmetry of \(g\), \(\left.\frac{\partial g}{\partial \alpha}\right|_{(\alpha^*, \beta^*)} = \left.\frac{\partial g}{\partial \beta}\right|_{(\alpha^*, \beta^*)}\), which implies
\[
\frac{(\beta^*)^2}{(\alpha^*)^2} = 1 \ \Rightarrow\ \frac{\alpha^*}{\beta^*} = 1 .
\]

\section{Penalized Bayesian Neural Scaling Laws}
\label{sec:si:bayesian_neural_scaling}
\begin{figure}[ht!]
    \centering
    \includegraphics[width=\linewidth]{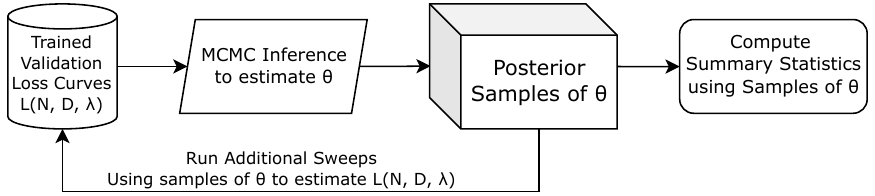}
    \caption{
        \label{fig:scaling_workflow}
        \textbf{Workflow for developing neural scaling laws.}
        Neural scaling law parameters \(\theta = (A, \alpha, B, \beta, E, c_i, \dots)\) were estimated via \ac{MCMC} sampling using \texttt{DynamicHMC.jl}\cite{TDD+TpappDynamicHMCjlV3502025}.
        Posterior draws of \(\theta\) were then used to compute summary statistics (e.g., \(\alpha / (\alpha + \beta)\) in \cref{fig:compute_scaling_parameters}), evaluate parameter covariances (\cref{fig:bayes_covariance}), and predict model loss along the compute-optimal frontier (\cref{fig:compute_optimal_frontier}).
    }
\end{figure}
In this section, we document our methodology for fitting neural scaling laws using Bayesian parameter estimation, including our priors for all coefficients and alternative formulations that were evaluated.
As outlined in \cref{sec:scaling} of the main manuscript, we extended Hoffmann et al.\cite{HBM+TrainingComputeOptimalLarge2022}'s model to include additional penalty terms to capture the impact of hyperparameter selection on the minimum loss.
Our workflow for fitting neural scaling parameters is shown in \cref{fig:scaling_workflow}.
The neural scaling laws presented in the manuscript were fit to validation-loss curves collected during our final sweep (\cref{tab:scaling_campaign}) in December 2024, prior to training our largest model (MIST-1.8B) in January 2025.
A smaller variant (MIST-228M) was trained to validate our neural scaling laws prior to training MIST-1.8B.
Since training MIST-1.8B, we have fit alternative formulations of hyperparameter penalized neural scaling laws to validate our methodology.

\paragraph{Estimating Compute Cost}
Our scaling plots and fitted neural scaling laws (\cref{fig:neural_scaling_laws,sec:si:alt_scaling_laws}) approximate the compute cost in \ac{FLOP}s using \(C \approx 6 N D\), consistent with the literature\cite{HBM+TrainingComputeOptimalLarge2022,KMH+ScalingLawsNeural2020}.
However, here we treat \(D\) as the number of molecules seen by the model and not the number of tokens.
This was primarily for simplicity, tracking the number of tokens masked or seen during distributed training incurs added complexity.
Treating \(D\) as the number of molecules also maps closer to our training workflow: we iterate over molecules not tokens.
If desired, this effect can be accounted for by scaling \(D\) or \(C\) by the average number of tokens per molecule: \(65.2 \pm 9.57\)\cite{WBVTokenizationMolecularFoundation2026}.

\subsection{Neural Scaling Law Development Campaign}
\label{sec:si:scaling_campaign}
To develop our neural scaling laws, we pretrained numerous models (\cref{tab:scaling_campaign}) following \cref{sec:si:pretraining}.
Consistent with prior work\cite{VSP+AttentionAllYou2017,KMH+ScalingLawsNeural2020,HBM+TrainingComputeOptimalLarge2022}, we parameterize the size and shape of the RoBERTa encoder\cite{LOG+RoBERTaRobustlyOptimized2019} using:
\begin{description}
    \item[Hidden size \(d_{\textrm{model}}\):] Dimension of the residual stream (the hidden size of the transformer layer).
    \item[Feed-forward size \(d_{\textrm{ff}}\):] Dimension of the intermediate feed-forward layer.
    \item[Number of attention heads \(n_{\textrm{heads}}\):] Number of attention heads within multi-head attention.
    \item[Number of layers \(n_{\textrm{layer}}\):] Number of transformer layers within the encoder.
\end{description}

\noindent
Following Kaplan et al.\cite{KMH+ScalingLawsNeural2020}, these define several non-dimensional ratios describing encoder shape:
\begin{description}
    \item[Feed-forward ratio \(r_{\textrm{ff}}\):] \(d_{\textrm{ff}} / d_{\textrm{model}}\).
    \item[Aspect ratio \(r_{\textrm{aspect}}\):] \(d_{\textrm{model}} / n_{\textrm{layer}}\).
    \item[KVQ size \(r_{\textrm{kv}}\):] \(d_{\textrm{model}} / n_{\textrm{heads}}\).
\end{description}

\noindent
Consistent with Kaplan et al.\cite{KMH+ScalingLawsNeural2020}, we compute model size \(N\) as the number of non-embedding parameters:
\[
    N = 2 d_{\mathrm{model}} n_{\mathrm{layer}} \left( 2 \, d_{\mathrm{model}} + d_{\mathrm{ff}} \right) .
\]

Model hyperparameters and sizes for all development models are tabulated in \cref{tab:scaling_campaign} using the above definitions.
The remaining columns are:
\begin{description}
    \item[\(N\):] Number of non-embedding parameters.
    \item[\(D\):] Number of molecules seen during pretraining.
    \item[\(\mathcal{B}\):] Total batch size after accounting for \ac{GAS} and global world size.
    \item[\(\eta\):] Maximum learning rate for the schedule.
    \item[\(L\):] Minimum validation loss during pretraining the model.
\end{description}

We used \ac{GSD}\cite{SVH+GeneralizedSubsetDesigns2017} to guide the design of our development campaign (\cref{tab:scaling_campaign}).
Specifically, \ac{GSD} selected hyperparameters (learning rate, number of steps, batch size, feed-forward ratio, KVQ size) for a given hidden size \(d_{\text{model}}\) for a series of model.
We then launched training runs with selected hyperparameters.
This process was repeated for progressively larger hidden sizes.
Rather than directly sweeping learning rates, we enforced \(\eta \propto \sqrt{\mathcal{B}}\) and varied only the prefactor via \ac{GSD}.
This incremental approach enabled us to scale models while verifying training stability before advancing to larger configurations.
We did not employ a fully automated pipeline given the complexity of such a system and the cost of individual runs; manual submission served as an effective sanity check.
Detailed training logs---including loss traces, hyperparameters, system diagnostics, and metadata---are provided as JSON files in our data release.

\begin{center}
\begin{longtable}{rrccccrr}
    \caption{\label{tab:scaling_campaign}Neural Scaling Law Development Models}\\\toprule
    \(N\) & \(D\) & \(\mathcal{B}\) & \(r_{\text{ff}}\) & \(r_{\text{aspect}}\) & \(r_{\text{kv}}\) & \(\eta\) & \(L\) \\\hline
  \endfirsthead
  \multicolumn{8}{c}%
{{\bfseries \tablename\ \thetable{} -- continued from previous page}} \\
\hline
    \(N\) & \(D\) & \(\mathcal{B}\) & \(r_{\text{ff}}\) & \(r_{\text{aspect}}\) & \(r_{\text{kv}}\) & \(\eta\) & \(L\) \\\hline
  \endhead

  \hline \multicolumn{8}{r}{{Continued on next page}} \\
  \endfoot

  \hline
  \endlastfoot
  393,216 & 16,384,000 & 4096 & 1 & 32.00 & 32 & \sn{1}{-3} & \sn{5.97}{-1} \\
  393,216 & 32,768,000 & 4096 & 1 & 32.00 & 32 & \sn{2}{-3} & \sn{2.36}{-1} \\
  393,216 & 32,768,000 & 4096 & 1 & 32.00 & 64 & \sn{1}{-3} & \sn{4.29}{-1} \\
  393,216 & 65,536,000 & 8192 & 1 & 32.00 & 32 & \sn{1.41}{-3} & \sn{2.96}{-1} \\
  393,216 & 65,536,000 & 8192 & 1 & 32.00 & 64 & \sn{2.83}{-4} & \sn{8.93}{-1} \\
  393,216 & 131,072,000 & 16384 & 1 & 32.00 & 32 & \sn{4}{-4} & \sn{5.82}{-1} \\
  589,824 & 32,768,000 & 4096 & 1 & 21.33 & 32 & \sn{1}{-3} & \sn{3.2}{-1} \\
  589,824 & 65,536,000 & 8192 & 1 & 21.33 & 32 & \sn{2.83}{-4} & \sn{7.91}{-1} \\
  786,432 & 32,768,000 & 4096 & 4 & 32.00 & 32 & \sn{1}{-3} & \sn{3.54}{-1} \\
  786,432 & 65,536,000 & 8192 & 4 & 32.00 & 32 & \sn{2.83}{-4} & \sn{8.83}{-1} \\
  2,359,296 & 20,480,000 & 4096 & 1 & 42.67 & 32 & \sn{1}{-3} & \sn{2.18}{-1} \\
  2,359,296 & 32,768,000 & 4096 & 1 & 42.67 & 32 & \sn{1}{-3} & \sn{1.52}{-1} \\
  2,359,296 & 32,768,000 & 4096 & 1 & 42.67 & 64 & \sn{2}{-4} & \sn{5.98}{-1} \\
  2,359,296 & 40,960,000 & 4096 & 1 & 42.67 & 32 & \sn{2}{-3} & \sn{9.79}{-2} \\
  2,359,296 & 40,960,000 & 4096 & 1 & 42.67 & 64 & \sn{1}{-3} & \sn{1.46}{-1} \\
  2,359,296 & 65,536,000 & 8192 & 1 & 42.67 & 32 & \sn{2.83}{-4} & \sn{4.05}{-1} \\
  2,359,296 & 81,920,000 & 4096 & 1 & 42.67 & 32 & \sn{4}{-3} & \sn{6.17}{-2} \\
  2,359,296 & 81,920,000 & 4096 & 1 & 42.67 & 64 & \sn{2}{-3} & \sn{7.34}{-2} \\
  2,359,296 & 81,920,000 & 8192 & 1 & 42.67 & 32 & \sn{1.41}{-3} & \sn{1.1}{-1} \\
  2,359,296 & 163,840,000 & 8192 & 1 & 42.67 & 32 & \sn{2.83}{-3} & \sn{6.8}{-2} \\
  2,359,296 & 163,840,000 & 8192 & 1 & 42.67 & 64 & \sn{1.41}{-3} & \sn{9.17}{-2} \\
  2,359,296 & 327,680,000 & 16384 & 1 & 42.67 & 32 & \sn{2}{-3} & \sn{5.64}{-2} \\
  3,145,728 & 32,768,000 & 4096 & 1 & 32.00 & 32 & \sn{2}{-4} & \sn{5.59}{-1} \\
  3,145,728 & 40,960,000 & 4096 & 1 & 32.00 & 32 & \sn{1}{-3} & \sn{1.39}{-1} \\
  3,145,728 & 81,920,000 & 4096 & 1 & 32.00 & 32 & \sn{2}{-3} & \sn{7.26}{-2} \\
  3,145,728 & 81,920,000 & 4096 & 1 & 32.00 & 64 & \sn{1}{-3} & \sn{1.06}{-1} \\
  3,145,728 & 163,840,000 & 8192 & 1 & 32.00 & 32 & \sn{1.41}{-3} & \sn{8.7}{-2} \\
  3,932,160 & 81,920,000 & 4096 & 1 & 25.60 & 32 & \sn{1}{-3} & \sn{9.48}{-2} \\
  4,718,592 & 32,768,000 & 4096 & 4 & 42.67 & 32 & \sn{2}{-4} & \sn{6.17}{-1} \\
  4,718,592 & 40,960,000 & 4096 & 4 & 42.67 & 32 & \sn{1}{-3} & \sn{1.41}{-1} \\
  4,718,592 & 81,920,000 & 4096 & 4 & 42.67 & 32 & \sn{2}{-3} & \sn{6.95}{-2} \\
  4,718,592 & 81,920,000 & 4096 & 4 & 42.67 & 64 & \sn{1}{-3} & \sn{9.69}{-2} \\
  4,718,592 & 163,840,000 & 8192 & 4 & 42.67 & 32 & \sn{1.41}{-3} & \sn{8.09}{-2} \\
  6,291,456 & 81,920,000 & 4096 & 4 & 32.00 & 32 & \sn{1}{-3} & \sn{8.93}{-2} \\
  7,864,320 & 32,768,000 & 16384 & 4 & 25.60 & 64 & \sn{8}{-3} & \sn{7.51}{-2} \\
  7,864,320 & 81,920,000 & 16384 & 4 & 25.60 & 64 & \sn{8}{-3} & \sn{5}{-2} \\
  14,155,776 & 204,800,000 & 8192 & 1 & 192.00 & 32 & \sn{1.41}{-4} & \sn{1.19}{-1} \\
  14,155,776 & 409,600,000 & 8192 & 1 & 192.00 & 32 & \sn{1.98}{-4} & \sn{6.59}{-2} \\
  14,155,776 & 409,600,000 & 8192 & 1 & 192.00 & 64 & \sn{1.41}{-4} & \sn{8.18}{-2} \\
  14,155,776 & 614,400,000 & 8192 & 1 & 192.00 & 32 & \sn{2.83}{-4} & \sn{4.88}{-2} \\
  14,155,776 & 614,400,000 & 8192 & 1 & 192.00 & 64 & \sn{1.98}{-4} & \sn{6.05}{-2} \\
  14,155,776 & 819,200,000 & 16384 & 1 & 192.00 & 32 & \sn{2}{-4} & \sn{6.01}{-2} \\
  14,155,776 & 1,228,800,000 & 16384 & 1 & 192.00 & 32 & \sn{2.8}{-4} & \sn{4.57}{-2} \\
  14,155,776 & 1,228,800,000 & 16384 & 1 & 192.00 & 64 & \sn{2}{-4} & \sn{5.46}{-2} \\
  14,155,776 & 2,457,600,000 & 32768 & 1 & 192.00 & 32 & \sn{2.83}{-4} & \sn{4}{-2} \\
  15,728,640 & 32,768,000 & 4096 & 1 & 51.20 & 32 & \sn{2}{-4} & \sn{2.18}{-1} \\
  15,728,640 & 40,960,000 & 4096 & 1 & 51.20 & 32 & \sn{1}{-3} & \sn{8.24}{-2} \\
  15,728,640 & 81,920,000 & 4096 & 1 & 51.20 & 32 & \sn{2}{-3} & \sn{4.75}{-2} \\
  15,728,640 & 81,920,000 & 4096 & 1 & 51.20 & 64 & \sn{1}{-3} & \sn{6.1}{-2} \\
  15,728,640 & 163,840,000 & 8192 & 1 & 51.20 & 32 & \sn{1.41}{-3} & \sn{7.07}{-2} \\
  18,874,368 & 81,920,000 & 4096 & 1 & 42.67 & 32 & \sn{1}{-3} & \sn{6.04}{-2} \\
  22,020,096 & 32,768,000 & 16384 & 1 & 36.57 & 64 & \sn{8}{-3} & \sn{5.92}{-2} \\
  22,020,096 & 81,920,000 & 16384 & 1 & 36.57 & 64 & \sn{8}{-3} & \sn{4.78}{-2} \\
  28,311,552 & 409,600,000 & 8192 & 1 & 96.00 & 32 & \sn{1.41}{-4} & \sn{6.18}{-2} \\
  28,311,552 & 409,600,000 & 8192 & 4 & 192.00 & 32 & \sn{1.41}{-4} & \sn{7.48}{-2} \\
  28,311,552 & 614,400,000 & 8192 & 1 & 96.00 & 64 & \sn{1.41}{-4} & \sn{5.09}{-2} \\
  28,311,552 & 614,400,000 & 8192 & 4 & 192.00 & 32 & \sn{1.98}{-4} & \sn{5.48}{-2} \\
  28,311,552 & 614,400,000 & 8192 & 4 & 192.00 & 64 & \sn{1.41}{-4} & \sn{6.3}{-2} \\
  28,311,552 & 1,228,800,000 & 16384 & 1 & 96.00 & 32 & \sn{2}{-4} & \sn{4.18}{-2} \\
  28,311,552 & 1,228,800,000 & 16384 & 4 & 192.00 & 32 & \sn{2}{-4} & \sn{4.77}{-2} \\
  31,457,280 & 81,920,000 & 4096 & 4 & 51.20 & 32 & \sn{1}{-3} & \sn{6.5}{-2} \\
  37,748,736 & 32,768,000 & 16384 & 4 & 42.67 & 64 & \sn{8}{-3} & \sn{7.47}{-2} \\
  37,748,736 & 81,920,000 & 16384 & 4 & 42.67 & 64 & \sn{8}{-3} & \sn{5.52}{-2} \\
  42,467,328 & 614,400,000 & 8192 & 1 & 64.00 & 32 & \sn{1.41}{-4} & \sn{5.04}{-2} \\
  44,040,192 & 16,384,000 & 8192 & 4 & 36.57 & 64 & \sn{5.66}{-3} & \sn{7.13}{-2} \\
  44,040,192 & 32,768,000 & 16384 & 4 & 36.57 & 32 & \sn{8}{-3} & \sn{7.38}{-2} \\
  44,040,192 & 32,768,000 & 16384 & 4 & 36.57 & 64 & \sn{4}{-3} & \sn{6.59}{-2} \\
  44,040,192 & 40,960,000 & 8192 & 4 & 36.57 & 64 & \sn{5.66}{-3} & \sn{5.61}{-2} \\
  44,040,192 & 65,536,000 & 16384 & 4 & 36.57 & 64 & \sn{8}{-3} & \sn{6.17}{-2} \\
  44,040,192 & 81,920,000 & 16384 & 4 & 36.57 & 32 & \sn{8}{-3} & \sn{5.23}{-2} \\
  44,040,192 & 81,920,000 & 16384 & 4 & 36.57 & 64 & \sn{4}{-3} & \sn{4.99}{-2} \\
  44,040,192 & 163,840,000 & 16384 & 4 & 36.57 & 64 & \sn{8}{-3} & \sn{4.95}{-2} \\
  49,545,216 & 81,920,000 & 4096 & 1 & 54.86 & 32 & \sn{1}{-3} & \sn{5.67}{-2} \\
  50,331,648 & 409,600,000 & 8192 & 1 & 128.00 & 32 & \sn{1.41}{-4} & \sn{5.33}{-2} \\
  50,331,648 & 614,400,000 & 8192 & 1 & 128.00 & 32 & \sn{1.98}{-4} & \sn{4.12}{-2} \\
  50,331,648 & 614,400,000 & 8192 & 1 & 128.00 & 64 & \sn{1.41}{-4} & \sn{4.47}{-2} \\
  50,331,648 & 1,228,800,000 & 16384 & 1 & 128.00 & 32 & \sn{2}{-4} & \sn{3.72}{-2} \\
  56,623,104 & 81,920,000 & 16384 & 1 & 48.00 & 64 & \sn{8}{-3} & \sn{1.01}{-1} \\
  56,623,104 & 614,400,000 & 8192 & 4 & 96.00 & 32 & \sn{1.41}{-4} & \sn{4.99}{-2} \\
  63,700,992 & 40,960,000 & 8192 & 1 & 42.67 & 64 & \sn{5.66}{-3} & \sn{6.64}{-2} \\
  63,700,992 & 81,920,000 & 16384 & 1 & 42.67 & 32 & \sn{8}{-3} & \sn{9.84}{-2} \\
  63,700,992 & 81,920,000 & 16384 & 1 & 42.67 & 64 & \sn{4}{-3} & \sn{5.86}{-2} \\
  63,700,992 & 163,840,000 & 16384 & 1 & 42.67 & 64 & \sn{8}{-3} & \sn{1}{-1} \\
  75,497,472 & 614,400,000 & 8192 & 1 & 85.33 & 32 & \sn{1.41}{-4} & \sn{4.41}{-2} \\
  99,090,432 & 81,920,000 & 16384 & 4 & 54.86 & 64 & \sn{8}{-3} & \sn{9.98}{-2} \\
  100,663,296 & 614,400,000 & 8192 & 4 & 128.00 & 32 & \sn{1.41}{-4} & \sn{4.41}{-2} \\
  113,246,208 & 8,192,000 & 2048 & 1 & 56.89 & 32 & \sn{9.9}{-5} & \sn{4.58}{-1} \\
  113,246,208 & 8,192,000 & 2048 & 1 & 56.89 & 64 & \sn{7.07}{-5} & \sn{6.16}{-1} \\
  113,246,208 & 16,384,000 & 2048 & 1 & 56.89 & 32 & \sn{1.41}{-4} & \sn{1.78}{-1} \\
  113,246,208 & 16,384,000 & 2048 & 1 & 56.89 & 64 & \sn{9.9}{-5} & \sn{2.11}{-1} \\
  113,246,208 & 16,384,000 & 4096 & 1 & 56.89 & 32 & \sn{1}{-4} & \sn{3.69}{-1} \\
  113,246,208 & 32,768,000 & 4096 & 1 & 56.89 & 32 & \sn{1.4}{-4} & \sn{1.57}{-1} \\
  113,246,208 & 32,768,000 & 4096 & 1 & 56.89 & 64 & \sn{1}{-4} & \sn{1.81}{-1} \\
  113,246,208 & 40,960,000 & 4096 & 1 & 56.89 & 32 & \sn{2}{-4} & \sn{1.07}{-1} \\
  113,246,208 & 40,960,000 & 8192 & 4 & 48.00 & 64 & \sn{5.66}{-3} & \sn{5.69}{-2} \\
  113,246,208 & 65,536,000 & 8192 & 1 & 56.89 & 32 & \sn{1.41}{-4} & \sn{1.46}{-1} \\
  113,246,208 & 81,920,000 & 4096 & 1 & 56.89 & 32 & \sn{1}{-3} & \sn{6.47}{-2} \\
  113,246,208 & 81,920,000 & 4096 & 1 & 56.89 & 64 & \sn{2}{-4} & \sn{7.75}{-2} \\
  113,246,208 & 81,920,000 & 16384 & 4 & 48.00 & 32 & \sn{8}{-3} & \sn{1.1}{-1} \\
  113,246,208 & 81,920,000 & 16384 & 4 & 48.00 & 64 & \sn{4}{-3} & \sn{5.51}{-2} \\
  113,246,208 & 122,880,000 & 4096 & 1 & 56.89 & 32 & \sn{2}{-3} & \sn{5.37}{-2} \\
  113,246,208 & 122,880,000 & 4096 & 1 & 56.89 & 64 & \sn{1}{-3} & \sn{6.09}{-2} \\
  113,246,208 & 122,880,000 & 4096 & 1 & 56.89 & 128 & \sn{2}{-4} & \sn{6.81}{-2} \\
  113,246,208 & 163,840,000 & 8192 & 1 & 56.89 & 32 & \sn{2.83}{-4} & \sn{6.72}{-2} \\
  113,246,208 & 163,840,000 & 16384 & 4 & 48.00 & 64 & \sn{8}{-3} & \sn{1.13}{-1} \\
  113,246,208 & 245,760,000 & 8192 & 1 & 56.89 & 32 & \sn{1.41}{-3} & \sn{5.8}{-2} \\
  113,246,208 & 245,760,000 & 8192 & 1 & 56.89 & 64 & \sn{2.83}{-4} & \sn{5.59}{-2} \\
  113,246,208 & 491,520,000 & 16384 & 1 & 56.89 & 32 & \sn{4}{-4} & \sn{3.35}{-2} \\
  125,829,120 & 8,192,000 & 2048 & 1 & 51.20 & 32 & \sn{7.07}{-5} & \sn{6.4}{-1} \\
  125,829,120 & 16,384,000 & 2048 & 1 & 51.20 & 32 & \sn{9.9}{-5} & \sn{2.2}{-1} \\
  125,829,120 & 16,384,000 & 2048 & 1 & 51.20 & 64 & \sn{7.07}{-5} & \sn{2.69}{-1} \\
  125,829,120 & 32,768,000 & 4096 & 1 & 51.20 & 32 & \sn{1}{-4} & \sn{1.86}{-1} \\
  125,829,120 & 81,920,000 & 4096 & 1 & 51.20 & 32 & \sn{2}{-4} & \sn{8.25}{-2} \\
  125,829,120 & 122,880,000 & 4096 & 1 & 51.20 & 32 & \sn{1}{-3} & \sn{6.21}{-2} \\
  125,829,120 & 122,880,000 & 4096 & 1 & 51.20 & 64 & \sn{2}{-4} & \sn{6.74}{-2} \\
  125,829,120 & 245,760,000 & 8192 & 1 & 51.20 & 32 & \sn{2.83}{-4} & \sn{4.91}{-2} \\
  127,401,984 & 20,480,000 & 4096 & 4 & 42.67 & 64 & \sn{4}{-3} & \sn{5.78}{-2} \\
  127,401,984 & 40,960,000 & 8192 & 4 & 42.67 & 32 & \sn{5.66}{-3} & \sn{7.61}{-2} \\
  127,401,984 & 40,960,000 & 8192 & 4 & 42.67 & 64 & \sn{2.83}{-3} & \sn{5.06}{-2} \\
  127,401,984 & 81,920,000 & 8192 & 4 & 42.67 & 64 & \sn{5.66}{-3} & \sn{5.35}{-2} \\
  127,401,984 & 81,920,000 & 16384 & 4 & 42.67 & 32 & \sn{4}{-3} & \sn{5.05}{-2} \\
  127,401,984 & 81,920,000 & 16384 & 4 & 42.67 & 64 & \sn{2}{-3} & \sn{4.75}{-2} \\
  127,401,984 & 163,840,000 & 16384 & 4 & 42.67 & 32 & \sn{8}{-3} & \sn{1.25}{-1} \\
  127,401,984 & 163,840,000 & 16384 & 4 & 42.67 & 64 & \sn{4}{-3} & \sn{4.49}{-2} \\
  127,401,984 & 327,680,000 & 16384 & 4 & 42.67 & 64 & \sn{8}{-3} & \sn{1.29}{-1} \\
  138,412,032 & 16,384,000 & 2048 & 1 & 46.55 & 32 & \sn{7.07}{-5} & \sn{3.09}{-1} \\
  138,412,032 & 122,880,000 & 4096 & 1 & 46.55 & 32 & \sn{2}{-4} & \sn{7.07}{-2} \\
  226,492,416 & 8,192,000 & 2048 & 4 & 56.89 & 32 & \sn{7.07}{-5} & \sn{5.18}{-1} \\
  226,492,416 & 16,384,000 & 2048 & 4 & 56.89 & 32 & \sn{9.9}{-5} & \sn{2.21}{-1} \\
  226,492,416 & 16,384,000 & 2048 & 4 & 56.89 & 64 & \sn{7.07}{-5} & \sn{2.61}{-1} \\
  226,492,416 & 32,768,000 & 4096 & 4 & 56.89 & 32 & \sn{1}{-4} & \sn{1.94}{-1} \\
  226,492,416 & 81,920,000 & 4096 & 4 & 56.89 & 32 & \sn{2}{-4} & \sn{8.19}{-2} \\
  226,492,416 & 122,880,000 & 4096 & 4 & 56.89 & 32 & \sn{1}{-3} & \sn{6}{-2} \\
  226,492,416 & 122,880,000 & 4096 & 4 & 56.89 & 64 & \sn{2}{-4} & \sn{6.83}{-2} \\
  226,492,416 & 245,760,000 & 8192 & 4 & 56.89 & 32 & \sn{2.83}{-4} & \sn{4.48}{-2} \\
  251,658,240 & 16,384,000 & 2048 & 4 & 51.20 & 32 & \sn{7.07}{-5} & \sn{2.75}{-1} \\
  251,658,240 & 122,880,000 & 4096 & 4 & 51.20 & 32 & \sn{2}{-4} & \sn{6.82}{-2} \\
  603,979,776 & 614,400,000 & 8192 & 1 & 85.33 & 32 & \sn{2}{-4} & \sn{2.91}{-2} \\
\end{longtable}

\end{center}

\subsubsection{Estimate Compute Cost Reduction.}\label{sec:si:doe_cost_reduction}

Combining \ac{GSD}\cite{SVH+GeneralizedSubsetDesigns2017} with Bayesian parameter estimation to fit our neural scaling laws was key to keeping the development costs of MIST within our computational budget.
Specifically, we estimate this reduced our compute costs by at least an order of magnitude (\cref{tab:scaling_campaign_doe}) relative to the grid-search campaigns used by the literature\cite{li2025misfittingsurveyscalinglaws,KMH+ScalingLawsNeural2020,HBK+ScalingLawsComputeOptimal2024}.
In total, developing our neural scaling laws costs 4,760 GPU-hours compared to the 5,344 GPU-hours used to train MIST 1.8B;
MIST 228M was trained as part of the scaling campaign.

\begin{table}[h]
\centering
\resizebox{\columnwidth}{!}{%
\begin{tabular}{lccc}
\hline
Case & Design Factors & Total Compute (PF-days) & Rel. Compute \\
\hline
Full Factorial & $d_{\textrm{model}}$(6), $n_{\textrm{layer}}$(11), $r_{\textrm{ff}}$(2), $\mathcal{B}$(5), $r_{\textrm{kv}}$(3), $D$(19), $\eta_{0}$(7) & 27100.0 & 3520.0 \\
$N \times D \times \eta_0$ & $N$(32), $D$(19), $\eta_{0}$(7) & 561.0 & 73.0 \\
$N \times D$ & $N$(32), $D$(19) & 80.2 & 10.4 \\
Our Campaign & --- & 7.7 & 1.0 \\
\hline
\end{tabular}
}
\caption{\label{tab:scaling_campaign_doe}Training cost comparison across experimental design strategies.
The number of levels for each design factor is shown in parenthesis.
}
\end{table}

\paragraph{Design Levels for our Campaign}
\begin{description}

\item[$N$ (32)]
$
[\sn{3.93216}{5},\;
\sn{5.89824}{5},\;
\sn{7.86432}{5},\;
\sn{2.3593}{6},\;
\sn{3.14573}{6},\;
\sn{3.93216}{6},\;
\sn{4.71859}{6},\;
\sn{6.29146}{6},\;
\sn{7.86432}{6},\;
\sn{1.41558}{7},\;
\sn{1.57286}{7},\;
\sn{1.88744}{7},\;
\sn{2.20201}{7},\;
\sn{2.83116}{7},\;
\sn{3.14573}{7},\;
\sn{3.77487}{7},\;
\sn{4.24673}{7},\;
\sn{4.40402}{7},\;
\sn{4.95452}{7},\;
\sn{5.03316}{7},\;
\sn{5.66231}{7},\;
\sn{6.3701}{7},\;
\sn{7.54975}{7},\;
\sn{9.90904}{7},\;
\sn{1.00663}{8},\;
\sn{1.13246}{8},\;
\sn{1.25829}{8},\;
\sn{1.27402}{8},\;
\sn{1.38412}{8},\;
\sn{2.26492}{8},\;
\sn{2.51658}{8},\;
\sn{6.0398}{8}]
$

\item[$D$ (19)]
$
[\sn{8.192}{6},\;
\sn{1.6384}{7},\;
\sn{2.048}{7},\;
\sn{3.2768}{7},\;
\sn{4.096}{7},\;
\sn{6.5536}{7},\;
\sn{8.192}{7},\;
\sn{1.2288}{8},\;
\sn{1.31072}{8},\;
\sn{1.6384}{8},\;
\sn{2.048}{8},\;
\sn{2.4576}{8},\;
\sn{3.2768}{8},\;
\sn{4.096}{8},\;
\sn{4.9152}{8},\;
\sn{6.144}{8},\;
\sn{8.192}{8},\;
\sn{1.2288}{9},\;
\sn{2.4576}{9}]
$

\item[$d_{\textrm{model}}$ (6)]
$
[128,\;
256,\;
512,\;
768,\;
1024,\;
2048]
$

\item[$n_{\textrm{layer}}$ (11)]
$
[4,\;
6,\;
8,\;
10,\;
12,\;
14,\;
16,\;
18,\;
20,\;
22,\;
24]
$

\item[$r_{\textrm{ff}}$ (2)]
$
[1,\;
4]
$

\item[$r_{\textrm{kv}}$ (3)]
$
[32,\;
64,\;
128]
$

\item[$\mathcal{B}$ (5)]
$
[2048,\;
4096,\;
8192,\;
\sn{1.6384}{4},\;
\sn{3.2768}{4}]
$

\item[$\eta_{0}$ (7)]
$
[\sn{1.56}{-6},\;
\sn{2.19}{-6},\;
\sn{2.21}{-6},\;
\sn{3.12}{-6},\;
\sn{1.56}{-5},\;
\sn{3.12}{-5},\;
\sn{6.25}{-5}]
$

\end{description}

\subsection{Numerical and Statistical Considerations}
Models were evaluated for convergence, predictive accuracy, and residuals using the metrics in \cref{tab:bayes_reg_quality_metrics}.
Exact definitions and computational details appear in \cref{sec:si:metrics}.
Consistent with Ref.~\cite{GCS+BayesianDataAnalysis2014}, model selection was guided by \ac{WAIC} (\(\downarrow\)).
Metrics for released models are in the \texttt{score} field within \texttt{chains.jld2} in our data release.
See \texttt{opt/BayesianRegression/scripts} in our source code for example code using the chains.

\begin{table}[h]
\resizebox{\columnwidth}{!}{%
\begin{tabular}{ll}
\toprule
\textbf{Metric} & \textbf{Description} \\
\midrule
\acf{MAPE} (\(\downarrow\)) & Primary metric for expected predictive accuracy. \\
\acf{WAIC} (\(\downarrow\)) & Primary metric for model selection and goodness of fit. \\
\acf{AIC} (\(\downarrow\)) & Reported for reference. \\
\acf{BIC} (\(\downarrow\)) & Reported for reference. \\
\acf{DIC} (\(\downarrow\)) & Reported for reference. \\
\acf{ESS} (\(\uparrow\)) & Check that sufficient samples were drawn during \ac{MCMC}. \\
\(\hat{R}\) (\(\downarrow\)) & Convergence check with criterion \(\hat{R} < 1.01\)\cite{VGS+RankNormalizationFoldingLocalization2021,GCS+BayesianDataAnalysis2014}. \\
\bottomrule
\end{tabular}%
}
\caption{
    \label{tab:bayes_reg_quality_metrics}
    Summary of evaluation metrics for assessing the quality of fitted neural scaling laws.
    Definitions are provided in \cref{sec:si:metrics}.
}
\end{table}

\paragraph{Bayesian parameter inference.}
We used \ac{MCMC}, specifically the No-U-Turn Sampler\cite{HGNouturnSamplerAdaptively2014}, to draw samples from the posterior of \(\theta = (A, \alpha, B, \beta, E, c_i, \dots)\).
We distributed chains over multiple threads.
However, this was not required in practice, as sampling completes in minutes on consumer hardware.
After discarding warm-up iterations, draws were treated as approximate samples from the target posterior.
We used \texttt{DynamicHMC.jl}\cite{TDD+TpappDynamicHMCjlV3502025}.
Gradients of the unscaled log posterior, \(\nabla_{\theta} \log p(\theta \mid L, N, D, \lambda)\), were obtained via reverse-mode automatic differentiation with \texttt{Enzyme.jl}\cite{MCP+ReversemodeAutomaticDifferentiation2021,MNP+ScalableAutomaticDifferentiation2022} in Julia\cite{BEKSJuliaFreshApproach2017}.

\paragraph{Model inference.}
For flexibility, downstream quantities were computed directly from posterior samples.
For example, the expected loss of a new configuration is:
\[
    \mathbb{E}\left[L(N, D, \lambda; \theta)\right] = \frac{1}{m} \sum_{j=1}^{m} L\!\left(N, D, \lambda; \theta_j\right) ,
\]
where \(m\) is the number of posterior draws.
Credible intervals were computed using the \(P^2\) approximate quantile algorithm\cite{JCP2AlgorithmDynamic1985}, which enables a single-pass estimate of the interval.

\subsection{Alternative formulations}
\label{sec:si:alt_scaling_laws}

In addition to the penalized formulation in the manuscript, we explored the following variations (\cref{tab:alt_scaling_laws}).
A brief context for each is provided below.
Posterior chains and statistics for each formulation are provided in the \texttt{scaling/chains} folder of our data release.

\begin{table}[h]
\centering
\begin{tabularx}{\linewidth}{lX}
\toprule
\textbf{Setting} & \textbf{Description} \\
\midrule
LR scaling with \(d_{\text{model}}\) & \(\eta_* \propto d_{\text{model}}^{\gamma}\) \\
Exp.\ prior on \(\sigma^2\) & \(\operatorname{Exponential}(0.1)\) prior on \(\sigma^2\) \\
No penalty terms & Fit models without penalty terms \\
Additive penalties & \(L(N, D) + \sum P\) instead of \(L(N, D) \times \prod \exp P\) \\
Harmonic shape penalties & \(P(\lambda_i - \lambda_*)\) for \(r_{ff}, r_{aspect}, r_{kv}\) \\
Smoothed loss & Average validation loss over the final \(n\) samples \\
\bottomrule
\end{tabularx}
\caption{
    \label{tab:alt_scaling_laws}
    Alternative formulations of neural scaling laws evaluated in this work.
}
\end{table}

\paragraph{LR scaling with \(d_{\text{model}}\).}
\begin{figure}[h]
    \centering
    \includegraphics[width=\linewidth]{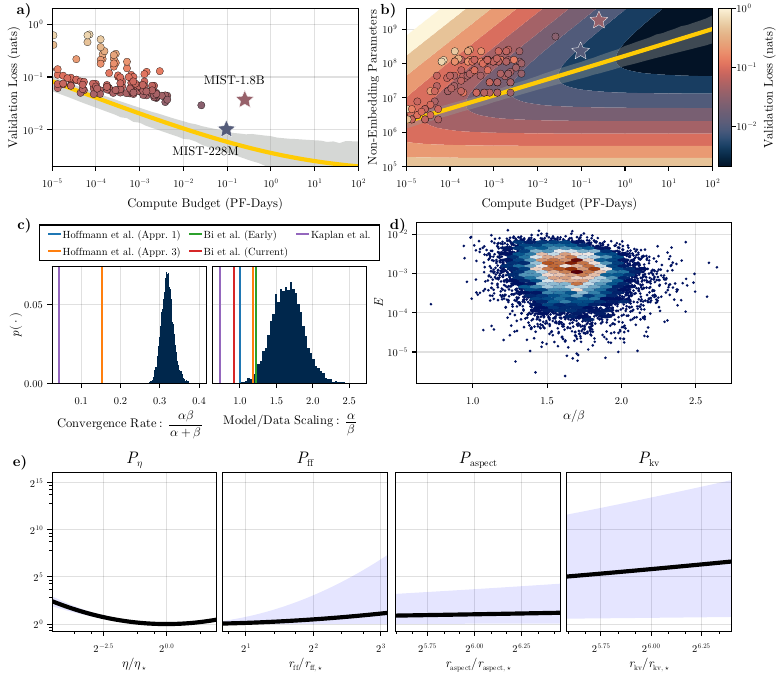}
    \caption{
        \label{fig:bayesian_scale_d_model}
        Penalized Bayesian neural scaling with \(\eta_* = \eta_0 d_{\text{model}}^{\gamma}\mathcal{B}^{\delta}\), where \(\mathcal{B}\) is the effective batch size and \(\eta_0, \gamma, \delta\) are fitted.
        (a)~95\% credible interval for the compute-optimal frontier (\(L_{\mathrm{opt}}\), \cref{eq:compute_optimal_loss}).
        (b)~Posterior mean of \(L(N, D)\) (\cref{eq:hoffmann_scaling}) and 95\% credible interval for \(L_{\mathrm{opt}}\).
        (c)~Histograms of convergence rate and model/data balance, with historical \ac{NLP} values indicated.
        (d)~Hexbin covariance plot of posterior samples for \(E\) and \(\alpha/\beta\) after fitting \cref{eq:penalized_neural_scaling}.
        (e)~Fitted penalty terms for the learning-rate (\(P_{\eta}\)) and encoder shape (\(P_{ff}\), \(P_{aspect}\) \& \(P_{kv}\)).
        Overall, this fit leads to quantitatively similar observations regarding \(\alpha/\beta\) as the baseline (\cref{fig:neural_scaling_laws}).
    }
\end{figure}
After training frontier models using learning rates chosen from the baseline fits, we explored \(\eta_* = \eta_0d_{\text{model}}^{\gamma} \mathcal{B}^{\delta}\), where \(\mathcal{B}\) is the effective batch size and \(\eta_0, \gamma, \delta\) are fitted.
This yielded improved fit statistics over baseline: \ac{MAPE} \(20\%\) and \ac{WAIC} \(-580\) versus \(29.1\%\) and \(-517\).
Both specifications led to qualitatively similar conclusions for \(\alpha/\beta\) and \(\tfrac{\alpha\beta}{\alpha + \beta}\) (\cref{fig:bayesian_scale_d_model}) ---
we consistently recover \(\alpha/\beta > 1\), deviating from the \(\approx 1\) reported for \ac{NLP} models\cite{BCC+DeepSeekLLMScaling2024,HBM+TrainingComputeOptimalLarge2022}.
Because this variant was fit retrospectively, and was not used to select the hyperparameters of MIST-1.8B, we present it here as a Supplement.

\paragraph{Prior on \(\sigma^2\).}
\begin{figure}[h]
    \centering
    \includegraphics[width=\linewidth]{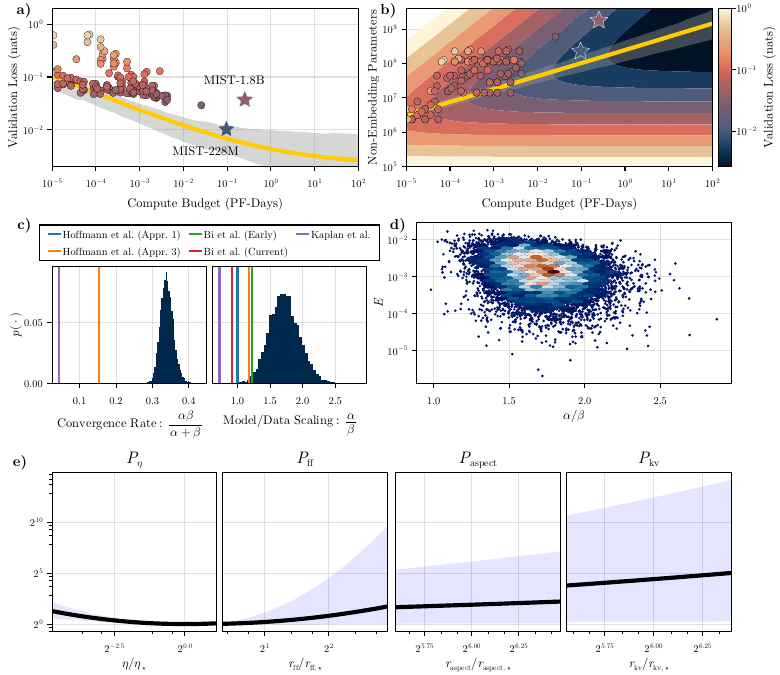}
    \caption{
        \label{fig:bayesian_scale_exp_sigma}
        Penalized Bayesian neural scaling with an Exponential prior on \(\sigma^2\).
        (a)~95\% credible interval for the compute-optimal frontier (\(L_{\mathrm{opt}}\), \cref{eq:compute_optimal_loss}).
        (b)~Posterior mean of \(L(N, D)\) (\cref{eq:hoffmann_scaling}) and 95\% credible interval for \(L_{\mathrm{opt}}\).
        Overlaid scatter plots in (a,b) show minimum validation loss for models used in fitting; frontier base models (MIST-228M, MIST-1.8B) are marked with \(\bigstar\).
        (c)~Histograms of convergence rate and model/data balance, with historical \ac{NLP} values indicated.
        (d)~Hexbin covariance plot of posterior samples for \(E\) and \(\alpha/\beta\).
        (e)~Fitted penalty terms for the learning-rate (\(P_{\eta}\)) and encoder shape (\(P_{ff}\), \(P_{aspect}\) \& \(P_{kv}\)).
        Overall, this fit leads to qualitatively similar observations as \cref{fig:neural_scaling_laws}.
    }
\end{figure}
We fit penalized neural scaling laws using an Exponential prior for \(\sigma^2\) --- rather than a Gamma prior (\cref{tab:neural_scaling_priors}).
Using an Exponential (\cref{fig:bayesian_scale_exp_sigma}) or Gamma (\cref{fig:neural_scaling_laws}) prior gave comparable results.
We elected to adopt the Gamma prior as recommended by Chung et al.\cite{CRD+NondegeneratePenalizedLikelihood2013} and the Stan documentation\cite{AVB+PriorChoiceRecommendations2025,CRD+NondegeneratePenalizedLikelihood2013}.

\begin{figure}
    \centering
    \includegraphics[width=\linewidth]{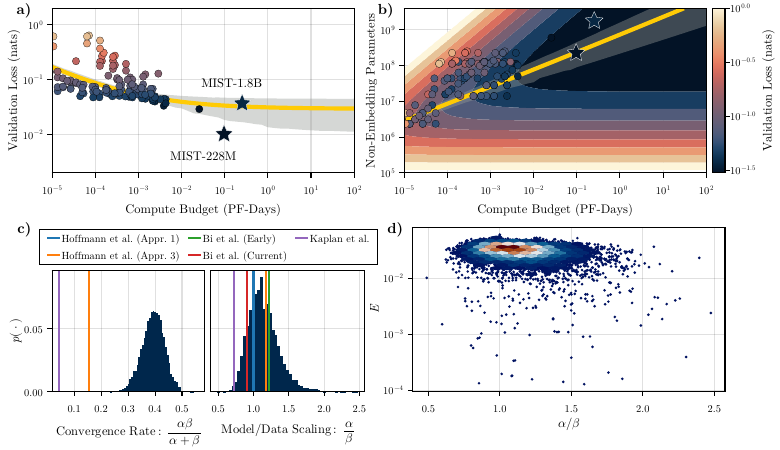}
    \caption{
        \label{fig:no_penalty_all}
        Neural scaling without penalty terms (\cref{eq:hoffmann_scaling}) fitted to all development models.
        (a)~95\% credible interval for the compute-optimal frontier (\(L_{\mathrm{opt}}\), \cref{eq:compute_optimal_loss}).
        (b)~Posterior mean of \(L(N, D)\) and 95\% credible interval for \(L_{\mathrm{opt}}\).
        Overlaid points show minimum validation loss; frontier base models (MIST-228M, MIST-1.8B) marked with \(\bigstar\).
        (c)~Histograms of convergence rate and model/data balance.
        (d)~Hexbin covariance of \(E\) and \(\alpha/\beta\).
        Predictive quality by \acs{WAIC} is worse than baseline (\(-452\) vs. \(-517\)).
    }
\end{figure}

\begin{figure}
    \centering
    \includegraphics[width=\linewidth]{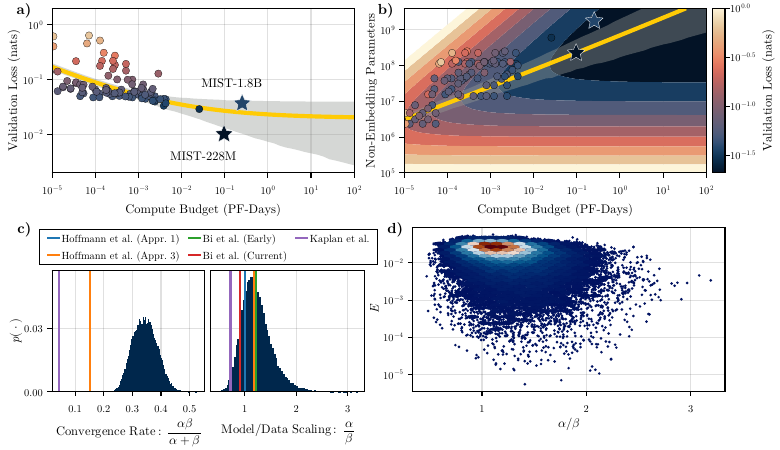}
    \caption{
        \label{fig:no_penalty_tuned}
        Neural scaling without penalty terms (\cref{eq:hoffmann_scaling}) fitted to the lowest loss for each \((N, D)\) in \cref{tab:scaling_campaign}.
        (a)~95\% credible interval for \(L_{\mathrm{opt}}\) (\cref{eq:compute_optimal_loss}).
        (b)~Posterior mean of \(L(N, D)\) and 95\% credible interval for \(L_{\mathrm{opt}}\).
        Overlaid points show minimum validation loss; frontier base models marked with \(\bigstar\).
        (c)~Histograms of convergence rate and model/data balance.
        (d)~Hexbin covariance of \(E\) and \(\alpha/\beta\).
        Predictive quality by \acs{WAIC} is worse than baseline (\(-310\) vs. \(-517\)).
    }
\end{figure}

\paragraph{No penalty terms.}
We also fit neural scaling laws without penalty terms, using \cref{eq:hoffmann_scaling} for \(L(N, D)\).
This matches Hoffmann et al.’s Approach 3\cite{HBM+TrainingComputeOptimalLarge2022}, but with Bayesian estimation instead of non-linear curve fitting.
We fit variants to (i) all development models (No Penalty; \cref{fig:no_penalty_all}) and (ii) the lowest-loss run for each \((N, D)\) (No Penalty—Tuned; \cref{fig:no_penalty_tuned}).
Beyond losing the ability to model off-optimal hyperparameters explicitly, both variants degraded predictive quality: \ac{WAIC} of \(-452\) (No Penalty) and \(-310\) (Tuned) versus \(-517\) for baseline.

\paragraph{Additive penalties.}
\begin{figure}
    \centering
    \includegraphics[width=\linewidth]{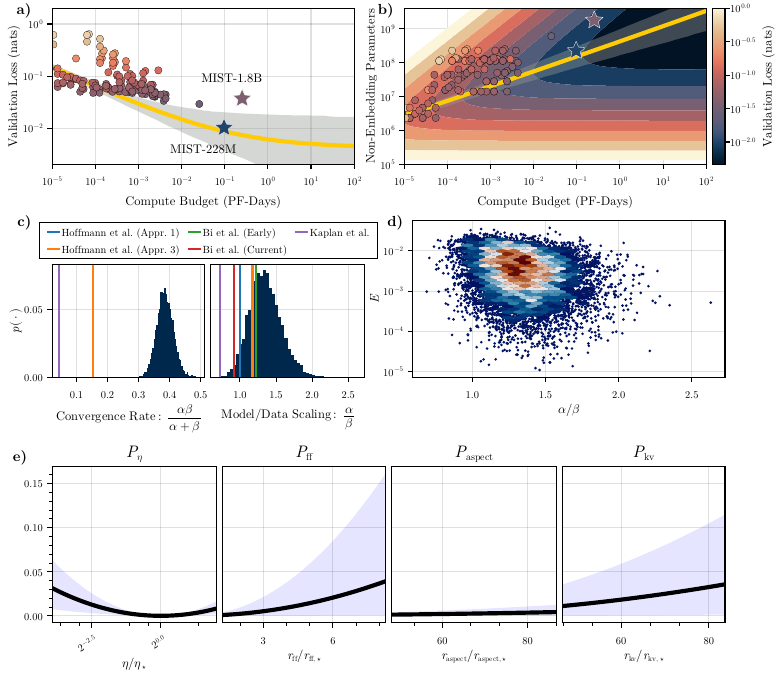}
    \caption{
        \label{fig:additive_penalty}
        Neural scaling with additive penalty terms \(L(N, D, \lambda) = L(N, D) + P(\lambda)\).
        (a)~95\% credible interval for \(L_{\mathrm{opt}}\) (\cref{eq:compute_optimal_loss}).
        (b)~Posterior mean of \(L(N, D)\) and 95\% credible interval for \(L_{\mathrm{opt}}\).
        Overlaid points show minimum validation loss; frontier base models marked with \(\bigstar\).
        (c)~Histograms of convergence rate and model/data balance.
        (d)~Hexbin covariance of \(E\) and \(\alpha/\beta\).
        (e)~Fitted penalty terms for the learning-rate (\(P_{\eta}\)) and encoder shape (\(P_{ff}\), \(P_{aspect}\) \& \(P_{kv}\)).
        Predictive quality by \acs{WAIC} is worse than baseline (\(-464\) vs. \(-517\)).
    }
\end{figure}
Our baseline (\Cref{eq:penalized_neural_scaling}) models off-optimal hyperparameters multiplicatively, \(L(N, D, \lambda) = L(N, D)\,\exp P(\lambda)\).
Additive penalties (\cref{fig:additive_penalty}) instead use \(L(N, D, \lambda) = L(N, D) + P(\lambda)\).
This notably degraded model performance: \ac{WAIC} \(-464\) vs. \(-517\) baseline.

\paragraph{Harmonic shape penalties.}
\begin{figure}
    \centering
    \includegraphics[width=\linewidth]{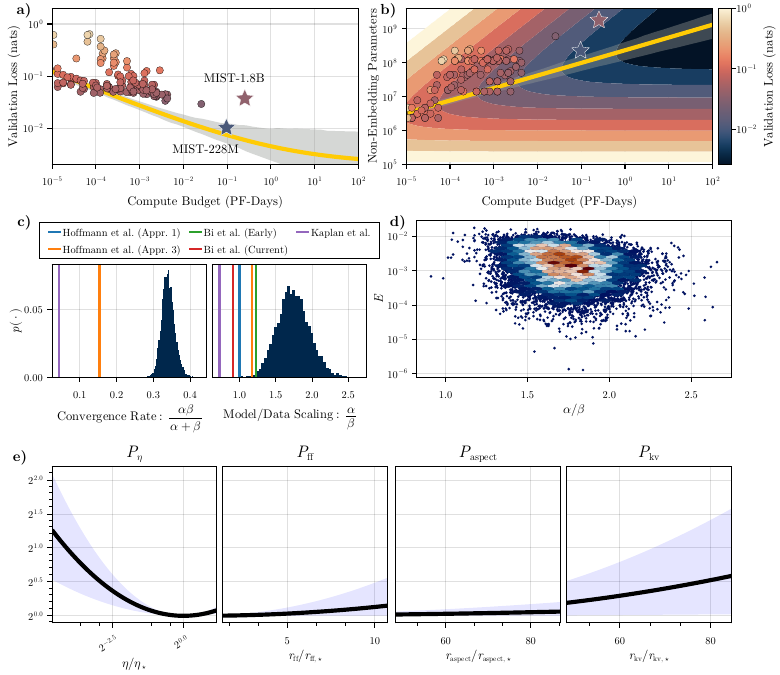}
    \caption{
        \label{fig:harmonic_penalty}
        Neural scaling with harmonic shape penalty terms for \(r_{\text{ff}}\), \(r_{\text{aspect}}\) and \(r_{\text{kv}}\).
        (a)~95\% credible interval for \(L_{\mathrm{opt}}\) (\cref{eq:compute_optimal_loss}).
        (b)~Posterior mean of \(L(N, D)\) and 95\% credible interval for \(L_{\mathrm{opt}}\).
        Overlaid points show minimum validation loss; frontier base models marked with \(\bigstar\).
        (c)~Histograms of convergence rate and model/data balance.
        (d)~Hexbin covariance of \(E\) and \(\alpha/\beta\).
        (e)~Fitted penalty terms for the learning-rate (\(P_{\eta}\)) and encoder shape (\(P_{ff}\), \(P_{aspect}\) \& \(P_{kv}\)).
        Predictive quality by \acs{WAIC} matches baseline performance (\(-517\) vs. \(-517\)).
    }
\end{figure}
Baseline penalties (\cref{eq:penalty_term}) scale with log relative distance, \(\ln \lambda - \ln \lambda_{\star}\).
Harmonic shape penalties instead use a linear distance, \(\lambda - \lambda_{\star}\):
\[
    P_i(\lambda_i) = c_i \bigl( \lambda_i - \lambda_{\star,i} \bigr)^2 .
\]
We retained log relative distance penalties for the learning-rate penalty due to numerical issues when using linear distances for \(P_{\eta}\).
Overall, harmonic penalties had a negligible impact on performance (Extended Data~\cref{tab:scaling_law_parameters}) --- \ac{WAIC} \(-517\) vs. \(-517\) baseline.
However, they caused initialization difficulties during \ac{MCMC} sampling and required truncated priors for \(c_r\) (\cref{tab:neural_scaling_priors}); we therefore use our baseline formulation.

\paragraph{Smoothed validation loss.}
\begin{figure}
    \centering
    \includegraphics[width=\textwidth]{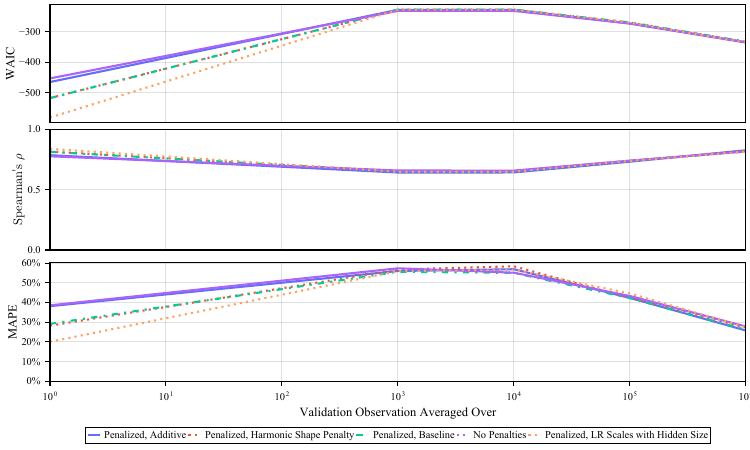}
    \caption{
        \label{fig:bayesian_smoothed_validation}
        Penalized Bayesian neural scaling fit to \(L\) averaged over the final \(n\) validation observations;
        Model quality metrics: \acs{WAIC} (\(\downarrow\)), Spearman's \(\rho\) (\(\uparrow\)) and \acs{MAPE} (\(\downarrow\)).
    }
\end{figure}
In addition to fitting to the minimum validation loss, we fit neural scaling laws to the average validation loss over the last \(10^3, 10^4, 10^5,\) or \(10^6\) validation samples.
This guards against minima that reflect outliers rather than typical performance and normalizes across configurations with differing validation-epoch sizes (\cref{fig:bayesian_smoothed_validation}).
In practice, this smoothing degraded model quality (higher \ac{WAIC}, worse \ac{MAPE}) and reduced usable observations as \(n\) increased (\(n\) sets a floor on \(D\)).

\begin{figure}
    \centering
    \includegraphics[width=\linewidth]{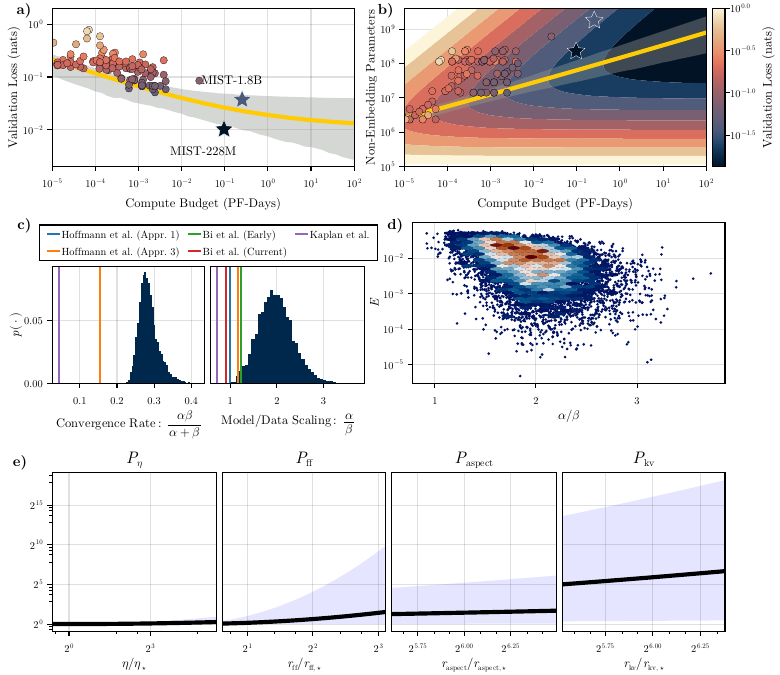}
    \caption{
        \label{fig:bayesian_scale_d_model}
        Baseline Penalized Bayesian neural scaling fit to the validation loss averaged over 1M observations.
        (a)~95\% credible interval for the compute-optimal frontier (\(L_{\mathrm{opt}}\), \cref{eq:compute_optimal_loss}).
        (b)~Posterior mean of \(L(N, D)\) (\cref{eq:hoffmann_scaling}) and 95\% credible interval for \(L_{\mathrm{opt}}\).
        (c)~Histograms of convergence rate and model/data balance, with historical \ac{NLP} values indicated.
        (d)~Hexbin covariance plot of posterior samples for \(E\) and \(\alpha/\beta\) after fitting \cref{eq:penalized_neural_scaling}.
        (e)~Fitted penalty terms for the learning-rate (\(P_{\eta}\)) and encoder shape (\(P_{ff}\), \(P_{aspect}\) \& \(P_{kv}\)).
        Overall, this fit leads to quantitatively similar observations regarding \(\alpha/\beta\) as the baseline (Manuscript \cref{fig:neural_scaling_laws}).
    }
\end{figure}

\subsection{Priors for Bayesian neural scaling laws}
\label{sec:si:bayesian_priors}

Priors are shown in \cref{tab:neural_scaling_priors}.
Priors for \(\alpha\) and \(\beta\) follow Hoffmann et al.\cite{HBM+TrainingComputeOptimalLarge2022}.
For \(A\) and \(B\) we use LogNormal distributions centered at the values reported by Hoffmann et al.\cite{HBM+TrainingComputeOptimalLarge2022} with high variance to be weakly informative.
We use a weakly informative, slightly optimistic prior for \(E\), based on experience stabilizing training.

Learning-rate priors are anchored to MoLFormer’s reported value (\(\sn{1.6}{-4}\)\cite{RBC+LargescaleChemicalLanguage2022}), square-root scaling\cite{KriOneWeirdTrick2014,YLR+LargeBatchOptimization2020}, and observed insensitivity of \(\eta\) to \(N\).
We maintain these priors when switching to \(d_{\text{model}}\) scaling and set \(\eta(\mathcal{B}=1024)=\sn{1.6}{-4}\).
Penalty coefficients \(c\) use weakly informative Exponential priors encouraging \(c\to 0\).
For harmonic-shape models, we additionally truncate \(c \leq \sn{1}{-2}\) to avoid overflow from large \(P\).

\begin{table}
\resizebox{\columnwidth}{!}{%
\begin{tabular}{l|llllll}\toprule
Parameter        & Hoffmann et al.\cite{HBM+TrainingComputeOptimalLarge2022} & Baseline \& Additive                         & Scaling with \(d_{\text{model}}\)            & Harmonic shape penalty                        \\\midrule
A                & \(\mathrm{LogNormal}(\ln 500, 2)\)                 & \(\mathrm{LogNormal}(\ln 500, 2)\)           & \(\mathrm{LogNormal}(\ln 500, 2)\)           & \(\mathrm{LogNormal}(\ln 500, 2)\)            \\
\(\alpha\)       & \(\mathrm{Uniform}(0, 2)\)                         & \(\mathrm{Uniform}(0, 2)\)                   & \(\mathrm{Uniform}(0, 2)\)                   & \(\mathrm{Uniform}(0, 2)\)                    \\
B                & \(\mathrm{LogNormal}(\ln 500, 2)\)                 & \(\mathrm{LogNormal}(\ln 500, 2)\)           & \(\mathrm{LogNormal}(\ln 500, 2)\)           & \(\mathrm{LogNormal}(\ln 500, 2)\)            \\
\(\beta\)        & \(\mathrm{Uniform}(0, 2)\)                         & \(\mathrm{Uniform}(0, 2)\)                   & \(\mathrm{Uniform}(0, 2)\)                   & \(\mathrm{Uniform}(0, 2)\)                    \\
E                & \(\mathrm{LogNormal}(\ln\sn{1}{-2}, 2)\)           & \(\mathrm{LogNormal}(\ln\sn{1}{-2}, 2)\)     & \(\mathrm{LogNormal}(\ln\sn{1}{-2}, 2)\)     & \(\mathrm{LogNormal}(\ln\sn{1}{-2}, 2)\)      \\
\(\eta_0\)       & —                                                & \(\mathrm{LogNormal}(\ln\sn{5}{-6}, 0.5)\)   & \(\mathrm{LogNormal}(\ln\sn{1.4}{-4}, 0.8)\) & \(\mathrm{LogNormal}(\ln\sn{5}{-6}, 0.5)\)    \\
\(\delta\)       & —                                                & \(\mathcal{N}(0.5, 0.05)\)                   & \(\mathcal{N}(0.5, 0.05)\)                   & \(\mathcal{N}(0.5, 0.05)\)                    \\
 \(\gamma\)      & —                                                & \(\mathcal{N}(0, 0.1)\)                      & \(\mathcal{N}(-0.5, 0.2)\)                   & \(\mathcal{N}(0, 0.1)\)                       \\
\(c_{\eta}\)     & —                                                & \(\mathrm{Exponential}(1.0)\)                & \(\mathrm{Exponential}(1.0)\)                & \(\mathrm{Exponential}(1.0)\)                 \\
\(r_{ff,*}\)     & —                                                & \(\mathrm{LogNormal}(\ln 4, 0.5)\)           & \(\mathrm{LogNormal}(\ln 4, 0.5)\)           & \(\mathrm{LogNormal}(\ln 4, 0.5)\)            \\
\(r_{aspect,*}\) & —                                                & \(\mathrm{LogNormal}(\ln 64, 0.1)\)          & \(\mathrm{LogNormal}(\ln 64, 0.1)\)          & \(\mathrm{LogNormal}(\ln 64, 0.1)\)           \\
\(r_{kv,*}\)     & —                                                & \(\mathrm{LogNormal}(\ln 64, 0.1)\)          & \(\mathrm{LogNormal}(\ln 64, 0.1)\)          & \(\mathrm{LogNormal}(\ln 64, 0.1)\)           \\
\(c_r\)          & —                                                & \(\mathrm{Exponential}(1.0)\)                & \(\mathrm{Exponential}(1.0)\)                & \(\mathrm{Trunc}\!\left(\mathrm{Exponential}(1.0); \text{upper}=\sn{1}{-2}\right)\) \\
\(\sigma^2\)     & \(\mathrm{Gamma}(2, 0.1)\)                        & \(\mathrm{Gamma}(2, 0.1)\)                   & \(\mathrm{Gamma}(2, 0.1)\)                   & \(\mathrm{Gamma}(2, 0.1)\)                    \\\bottomrule
\end{tabular}%
}
\caption{
    \label{tab:neural_scaling_priors}
    Priors for Bayesian neural scaling laws fit in this work.
    Penalty terms for shape hyperparameters (\(r_{ff}, r_{kv}, r_{aspect}\)) share a common prior \(c_r\).
}
\end{table}

\clearpage

\section{Impact of Data Quality on Compute-Optimal Scaling} \label{sec:si:data_quality}
In this section, we explain the relatively high ratio of model to dataset scaling exponents ($\frac{\alpha}{\beta}$) computed for \ac{MIST} models, and we extend this to draw insights about the impact of data quality on the scaling efficiency of \acp{FM}.

Scaling laws often assume that training data is uniformly high-quality and fungible. 
In practice, however, real-world datasets contain redundancy, imbalance, and gaps in concept coverage. 
Such imperfections can alter the effective scaling behaviour, shifting the balance between model and data size at compute-optimality. 
Here, we connect theoretical predictions from recent scaling analyses \cite{BDK+ExplainingNeuralScaling2024,NVInformationTheoryComputeOptimal2024,AALTheoryInferenceCompute2025} with our empirical observations for \ac{MIST}. 
We first interpret our measured exponents (\(\alpha/\beta > 1\)) in the variance- and resolution-limited framework based on linear random feature models~\cite{BDK+ExplainingNeuralScaling2024}.
We then model learning as an iterative process linking text pieces to latent concepts, grounded in information-theoretic analyses of compute-optimality~\cite{NVInformationTheoryComputeOptimal2024,AALTheoryInferenceCompute2025}.

\subsection{Variance- and Resolution-Limited Regimes}

In the variance-limited regime with respect to model size, Bahri et al.\cite{BDK+ExplainingNeuralScaling2024}
show that the excess loss scales as:
\begin{equation*}
    \mathcal{L}(N) - \mathcal{L}_{\infty} \;\propto\; N^{-\alpha}
    \qquad \alpha = 1 ,
\end{equation*}
\noindent
provided the model width is large enough for fluctuations around the infinite-width limit to be variance-controlled \cite{BDK+ExplainingNeuralScaling2024}.
Our empirical estimate of $\alpha \approx 1$ (Extended Data~\cref{tab:scaling_coeffs}) therefore places \ac{MIST} models in variance-limited scaling regime, with respect to model size.
This is consistent with the view that the network is already wide enough to behave as a linear random-feature model\cite{BDK+ExplainingNeuralScaling2024}.

For dataset size \(D\), the resolution-limited scaling regime, with respect to data size, predicts \cite{BDK+ExplainingNeuralScaling2024}:
\begin{equation*}
    \mathcal{L}(D) \;\propto\; D^{-\beta}
    \qquad
    \beta = \frac{n}{d} ,
    \label{eq:resolution_limited}
\end{equation*}
where \(d\) is the intrinsic dimension of the data manifold and \(n \ge 4\) is the order of the first non-vanishing term in an expansion of the loss around each training point.
Our fitted value \(\beta \approx 0.54\) implies an effective manifold dimension
\begin{equation*}
    d \;=\; \frac{n}{\beta}
    \;\;\longrightarrow\;\;
    d \gtrsim \frac{4}{0.54} \approx 7.4 ,
\end{equation*}
likely indicative of a low-dimensional structure in the REALSpace\cite{EnaREALSpace2024} corpus that \ac{MIST} models exploit during pretraining.
While the bound $d \geq n/\beta$ is technically an inequality, Bahri et al.\cite{BDK+ExplainingNeuralScaling2024} argue that equality is often achieved in practice\cite{BDK+ExplainingNeuralScaling2024}, lending further support to this interpretation.

Taken together, the pair \((\alpha,\beta) \approx (1,\,0.54)\) suggests that \ac{MIST} appears to be highly efficient with respect to model scaling, shifting the primary bottleneck towards the data side.
Since the inferred manifold dimension is low, the limiting factor is likely the quality or extent of coverage across that manifold.

\subsection{Super-linear Data Scaling under Concept Concentration}

Following the framework introduced by Nayak and Varshney \cite{NVInformationTheoryComputeOptimal2024} and extended in \cite{AALTheoryInferenceCompute2025}, we model the relationship between text and pretraining loss as a bipartite graph between text pieces $\mathfrak{T}$ and concepts $\mathfrak{R}$, with $|\mathfrak{T}| = T$ and $|\mathfrak{R}| = R$.
Each text node connects to one or more concept nodes, reflecting the idea that a given training example embodies several latent ideas.
The number of text pieces $T$ is directly proportional to $D$, the size of the data set.
The number of learnable concepts $R$ is directly proportional to the model size $N$.
Initially, all concepts are unlearned.
Learning proceeds as an iterative process, akin to the peeling form of \ac{LDPC} decoding for erasure channels.
A concept is learned once a neighboring text provides a sufficient signal such that for a concept $r\in\mathfrak R$ there exists a connected $t\in\mathfrak T$.
The peeling process terminates after either failing to or succeeding to learn all $R$ concepts.

Then, by taking an underlying distribution and noting that the training compute $C$ approximately scales with the product of the model and dataset sizes $ND$, this formulation allows us to analyse the resulting scaling laws which maximizes the number of learned concepts given a compute budget\cite{NVInformationTheoryComputeOptimal2024}.
In prior work, this model was used to recover the Chinchilla-style compute-optimal scaling laws, balancing model and data size $(N, D)$ to minimize loss under a fixed compute budget.
While initial formulations assumed a uniform connection pattern between texts and concepts, sufficient to recover the Chinchilla scaling $D \propto N$ \cite{NVInformationTheoryComputeOptimal2024}, we show that such uniformity is not necessary.
In practice, non-uniform exposure is common.
For example, repeating data necessarily concentrates multiple text pieces on a limited subset of concepts \cite{MRB+ScalingDataConstrainedLanguage2023}.
We interpret this kind of concept concentration as a form of lower data quality, which has been empirically\cite{BCC+DeepSeekLLMScaling2024,MRB+ScalingDataConstrainedLanguage2023} linked to deviations from standard scaling behaviour (i.e., \(\alpha/\beta \approx 1\)).

We now show that once the concept-frequency distribution becomes sufficiently heavy-tailed, namely, when its Zipf exponent exceeds the critical value $\alpha/\beta = 1$, the bipartite-graph model predicts super-linear data scaling with model size:
\[
D \propto N^{\alpha/\beta} \qquad \alpha/\beta > 1  .
\]
In this regime, the probability of encountering the rarest concepts decays faster than $1/R$, violating the conditions under which linear data scaling suffices.
As a result, each additional parameter requires disproportionately more data to ensure all learnable concepts remain accessible, leading to super-linear compute-optimal scaling (\(\alpha/\beta > 1\)).
We formalize this next.
Empirically, concept frequencies in many domains exhibit Zipfian tails. Natural-language corpora cluster around an exponent of 1 \cite{PiaZipfsWordFrequency2014}, and infochemical datasets in chemistry show their own Zipfian rank–frequency structure \cite{HTCompressionPrincipleZipfs2022}.
An exponent of 1 in natural language aligns with the linear scaling regime observed in Chinchilla-style laws\cite{HBK+ScalingLawsComputeOptimal2024}, where data and model size grow proportionally.
Since heavier tails reflect lower effective data quality, the ratio $\alpha/\beta$ may capture the degree of concept concentration.

\begin{theorem}[Super-linear Data Scaling under Concept Concentration]
\label{thm:super_linear_data_scaling}
Following Ref.~\cite{NVInformationTheoryComputeOptimal2024}, consider a bipartite graph between \( T \) text nodes and \( R \) concept nodes.
Suppose the probability of a text node connecting to a concept \( r \in \{1, \dots, R\} \) follows a Zipf distribution:
\begin{equation*}
p_r = \frac{r^{-\theta}}{H_R^{\theta}} \quad \theta > 1 ,
\end{equation*}
where \( H_R^{\theta} = \sum_{r=1}^R r^{-\theta} \) normalizes the distribution.
Then, to learn \( R(1-\delta) \) concepts via iterative decoding, up to any \( \delta \in (0,1) \) tolerance, the number of required text nodes \( T \)  satisfies:
\begin{equation*}
T \propto R^\theta \implies D\propto N^{\theta} \quad \theta > 1 ,
\end{equation*}
and in particular must grow faster than \( R \).
\end{theorem}

\subsection{Proof of Theorem~\ref{thm:super_linear_data_scaling}}

\paragraph{Setup and Rate.}
Let $T$ text nodes be the right nodes (check nodes), and let $R$ concept nodes be the left nodes (variable nodes) in a standard \ac{BEC} decoding formulation.

Following Ref.~\cite{NVInformationTheoryComputeOptimal2024}, we introduce a set of auxiliary concept nodes that are deterministically erased by the physical channel.
This expands the set of variable nodes to:
\begin{equation*}
\tilde{R} = R + R \cdot \frac{1 - \varepsilon}{\varepsilon} = \frac{R}{\varepsilon} ,
\end{equation*}
so that exactly an $\varepsilon$ fraction of the enlarged concept set corresponds to physical (non-auxiliary) concepts. In this formulation, the communication rate is defined as the fraction of non-erased symbols:
\begin{equation*}
\mathcal{R} = 1 - \frac{T}{\tilde{R}} = 1 - \frac{\varepsilon T}{R} .
\end{equation*}
This implies an effective erasure probability of:
\begin{equation*}
P = \frac{\varepsilon T}{R} .
\end{equation*}

Since auxiliary nodes are always erased, while physical nodes are erased with some physical probability $P_{\mathrm{phys}}$, we have the convex combination:
\begin{equation*}
P = \varepsilon P_{\mathrm{phys}} + (1 - \varepsilon) .
\end{equation*}
Solving for the physical erasure probability yields: $P_{\mathrm{phys}} = \frac{P - (1 - \varepsilon)}{\varepsilon} = \frac{T}{R} - \frac{1 - \varepsilon}{\varepsilon}.$
In order for this to represent a valid erasure probability ($0 \leq P \leq 1$), we require
\begin{equation}
\label{cond:eps1}
0 < \varepsilon < \frac{R}{T} .
\end{equation}

We now specify the degree distribution on the concept nodes.
Each text node selects $d_t$ concepts independently according to a Zipf distribution:
\[
p_r = \frac{r^{-\theta}}{H_R^{\theta}} \quad \text{where } H_R^{\theta} = \sum_{r=1}^{R} r^{-\theta} .
\]
For large $R$, we have the approximations:
\begin{equation}\label{eq:zetaapprox}
H_R^{(\theta)} \approx
\begin{cases}
    \zeta(\theta) & \theta>1\\
    \frac{R^{1-\theta}}{1-\theta} & 0<\theta<1\\
    R & \theta=0 ,
\end{cases}
\end{equation}
where $\zeta(\cdot)$ is the Riemann zeta function.

\paragraph{Right and Left-Degree Distribution.}

We assume the right-degree distribution is regular: each text node has fixed degree $d_t \geq 2$. This implies the total number of edges in the bipartite graph is:
\[
E = T d_t .
\]
Let $\rho(x)$ denote the generating function of the right-degree distribution. Since every right node has degree $d_t$, we have:
\begin{equation}\label{eq:rightgen}
    \rho(x) = x^{d_t - 1} .
\end{equation}

Now consider the left-degree distribution. Each concept node $r$ receives edges according to the edge-perspective probability $p_r$, so the expected number of edges connected to concept $r$ is:
\begin{equation}\label{eq:meanedges_r}
    k_r = \mathbb{E}[\deg(r)] = T d_t\, p_r = T d_t\, \frac{r^{-\theta}}{H_R^{(\theta)}} .
\end{equation}
Define the generating function for the left-degree distribution as:
\begin{equation}\label{eq:leftgen}
\Lambda(x) = \sum_{r=1}^R p_r \, x^{k_r - 1} = \sum_{r=1}^R \frac{r^{-\theta}}{H_R^{\theta}} \, x^{k_r - 1} .
\end{equation}

These generating functions appear in the standard density evolution and extrinsic information transfer chart (EXIT) chart analyses for \ac{LDPC} decoding.

\paragraph{EXIT Criterion.}

The iterative decoding process can be analysed via density evolution, which tracks the expected fraction of erasures across iterations. A key condition is the absence of additional fixed points in the decoding equation:
\begin{equation}\label{eq:ldpc_constraint}
    z = 1 - \rho\bigl(1 - P \Lambda(z)\bigr) = 1 - \bigl(1 - P \Lambda(z)\bigr)^{d_t - 1} ,
\end{equation}
where $P$ is the effective erasure probability and $z$ denotes the fraction of erased variable nodes in a given iteration.

A fixed point $z > 0$ would indicate that decoding stalls before recovering all concepts. To prevent this, we require that the function
\[
g(z) := 1 - \bigl(1 - P \Lambda(z)\bigr)^{d_t - 1}
\]
has no fixed point on $z \in (0,1)$. This is ensured if its derivative satisfies $g'(z) < 1$ for all $z \in (0,1)$, since the identity function $z$ has slope 1. Differentiating, we obtain
\[
g'(z) = (1 - P \Lambda(z))^{d_t - 2} \cdot P \Lambda'(z) \cdot (d_t - 1) .
\]
Now consider the derivative of $\Lambda(z)$:
\[
\Lambda'(z) = \sum_{r=1}^R p_r z^{k_r - 2}(k_r - 1) .
\]
We now impose a sparsity condition: for all $r \in \{1, \dots, R\}$, the probability that $\deg(r) \le 1$ is bounded by
\begin{equation}\label{eq:delta_accurate}
    \Pr[\deg(r) \le 1] \le \delta ,
\end{equation}
for some small $\delta > 0$ which is the tolerance for the fraction of concepts learned. Thus, with high probability, $k_r \ge 2$ for all $r$, which implies $z^{k_r - 2} < 1$ for $z \in (0,1)$.
Using this, we bound the derivative:
\[
\Lambda'(z) < \sum_{r=1}^R p_r (k_r - 1) = \mu_k - 1 ,
\]
where $\mu_k$ is the expected degree of a concept node:
\[
\mu_k = \sum_{r=1}^R p_r k_r = T d_t \sum_{r=1}^R p_r^2= T d_t \sum_{r=1}^R \left(\frac{r^{-\theta}}{H_R^{(\theta)}}\right)^2<Td_t\;\mbox{.}
\]
Moreover, since $\Lambda(z) \in (0,1)$ and $P > 0$, we have $0 < 1 - P \Lambda(z) < 1$, so:
\[
(1 - P \Lambda(z))^{d_t - 2} < 1  .
\]
Putting this together, we obtain the bound:
\begin{align}
g'(z) &= (1 - P \Lambda(z))^{d_t - 2} \cdot P \Lambda'(z) \cdot (d_t - 1) \\
      &< P (Td_t - 1)(d_t - 1)  .
\end{align}
Thus, a sufficient condition for convergence (i.e., for $g(z)$ to stay below the identity line and avoid a nontrivial fixed point) is:
\begin{equation}
\label{eq:prob_constraint2}
P < \frac{1}{Td_t^2} .
\end{equation}
This inequality ensures that the derivative $g'(z) < 1$ for all $z \in (0,1)$, thereby guaranteeing successful decoding of all but a $\delta$-fraction of the concept nodes.

\paragraph{Relationship Between Text Size and Concept Count for Reliable Peeling.}

To ensure there are sufficient edges to enable peeling-based decoding, we must satisfy \cref{eq:delta_accurate}.
Since the minimal expected degree occurs at $r = R$, it is sufficient to enforce:
\begin{equation}
    \Pr[\deg(R) \le 1] \le \delta .
\end{equation}

The degree distribution for each concept node is binomial: $k_r \sim \mathrm{Binom}(T d_t, p_r)$. For small $p_R$ and large $R$, which is the case in the concentrated regime, we invoke a standard Poisson approximation: $k_R \sim \mathrm{Pois}(\lambda)$, where the mean is
\[
\lambda = T d_t \frac{R^{-\theta}}{H_R^{(\theta)}} .
\]

Then \cref{eq:delta_accurate} is satisfied if:
\begin{equation}
    \Pr[\deg(R) \le 1] = (1 + \lambda)\, e^{-\lambda} \le \delta .
\end{equation}
This expression is monotonically decreasing in $\lambda$, so the tightest bound occurs at equality:
\[
(1 + \lambda_\delta)\, e^{-\lambda_\delta} = \delta .
\]
After algebraic manipulation, the expression $-(1 + \lambda_\delta)\, e^{-(1+\lambda_\delta)} = -\delta/e$ can be rewritten in terms of the Lambert $W$ function, yielding:
\[
\lambda_\delta = -1 - W_{-1}(-\delta/e) ,
\]
where $W_{-1}(x)$ denotes the negative branch of the Lambert $W$ function, defined over $x \in (-1/e, 0)$ for $\delta \in (0,1)$. Note that $\lambda_\delta > 0$.

Imposing the threshold condition $\lambda \ge \lambda_\delta$, we obtain:
\begin{equation}\label{eq:scaling_inequality_full}
    T d_t \frac{R^{-\theta}}{H_R^{(\theta)}} \ge \lambda_\delta .
\end{equation}
Taking the approximation of $H_R^{(\theta)}\approx \zeta(\theta)$ when $\theta>1$, \cref{eq:zetaapprox} gives the lower bound on the number of text pieces $T$:
\begin{equation}\label{eq:scaling_inequality}
    T \ge \frac{\zeta(\theta) \lambda_\delta}{d_t}  R^{\theta} \quad \Longrightarrow \quad T \propto R^{\theta} .
\end{equation}
This relation gives the minimal text budget $T$ required to ensure decoding up to $\delta$-tolerance, with constants $\zeta(\theta), \lambda_\delta$, and $d_t$ for $\theta>1$.

The converse follows by considering the zero-tolerance limit $\delta \to 0^+$. In that case, the final concept $r=R$ must strictly satisfy the EXIT constraint to avoid a crossing in \cref{eq:ldpc_constraint}. Thus, if $T$ fails to scale at least as $R^\theta$, the iterative peeling process---and hence concept learning---must fail.

Finally, the choice of $\varepsilon$ must satisfy both \cref{cond:eps1} and the EXIT constraint. From \cref{cond:eps1} and the scaling in \cref{eq:scaling_inequality}, we require:
\[
\varepsilon < \frac{d_t}{\zeta(\theta) \lambda_\delta} R^{-\theta + 1} .
\]
Moreover, from the EXIT condition on $P$, namely \cref{eq:prob_constraint2}, and substituting \cref{eq:scaling_inequality}, we obtain:
\[
P = \varepsilon \frac{T}{R} < \frac{1}{T d_t^2}
\quad \Longrightarrow \quad
\varepsilon < \frac{R^{-2\theta + 1}}{\zeta(\theta)^2 \lambda_\delta^2} .
\]
To satisfy both conditions, we conservatively choose $\varepsilon$ to be one half the minimum of the two upper bounds:
\[
\varepsilon = \frac{1}{2} \min\left(
    \frac{d_t}{\zeta(\theta) \lambda_\delta} R^{-\theta + 1},\,
    \frac{R^{-2\theta + 1}}{\zeta(\theta)^2 \lambda_\delta^2}
\right) .
\]

Since $D \propto T$ and $N \propto R$, then as $T \propto R^\theta$, it follows that $D\propto N^{\theta}$.
Thus, under a fixed compute budget $C\propto N\,D$, with this Zipf distribution relationship between text pieces and concepts, minimization of pre-training loss then replaces the Chinchilla relation $D\propto N$ with $D\propto N^{\theta}$, completing the proof.

\subsection{Connections to Empirical Scaling Exponents}
The results above complement empirical findings reported in Refs.~\cite{BCC+DeepSeekLLMScaling2024,MRB+ScalingDataConstrainedLanguage2023}, where lower data quality led to sharper scaling in model size relative to dataset size.
In Nayak and Varshney\cite{NVInformationTheoryComputeOptimal2024}'s framework, low data quality leads to overlapping concept coverage, requiring more text samples to reliably expose each additional concept.
This shifts the compute-optimal allocation toward larger dataset sizes as scale increases.

Importantly, our result does not contradict the Chinchilla rule $D \propto N$ established in Ref.~\cite{NVInformationTheoryComputeOptimal2024};
rather, it extends that rule to non-uniform data.
For a strictly uniform concept distribution, $\theta = 0$ and $H_R^{(0)} = R$, so \cref{eq:scaling_inequality_full} gives $T \propto R$. 
Likewise, inserting the large-$R$ approximation $H_R^{(\theta)} \sim R^{1-\theta}/(1-\theta)$ for $0 < \theta < 1$ yields the same leading relation $T \propto R$.
Hence, any Zipf distribution with tail exponent $0 \le \theta < 1$, where the rarest concept appears no less frequently than $1/R$, still enforces the linear data-model scaling $D \propto N$.
Only when the distribution becomes heavier than Zipf-1 ($\theta > 1$), so that the rarest‐concept probability decays faster than $1/R$, does the compute-optimal allocation transition to the super-linear regime $D \propto N^{\theta}$ for \(\theta > 1\).
These are not contradictory regimes: for $\theta \leq 1$ the rarest concepts are still encountered at least $1/R$ of the time, recovering the standard linear law $D \propto N$.
Only when $\theta > 1$ do the tails become too heavy, forcing a super-linear departure.

In the baseline analysis, we assumed each text piece had fixed degree $d_t$, independent of scale and data quality. However, this ignores a fundamental information-theoretic constraint: any individual datum can only support a finite number of learnable connections. Asymptotically sub-linear scaling would thus imply that a single text piece carries increasing conceptual load, eventually exceeding its information capacity. In practice, the degree $d_t$ should increase with the number of concepts $R$, since larger concept sets yield more opportunities for overlap. 
However, this growth must saturate at some ceiling $D_{\max}(\theta)$, which depends on data quality: lower $\theta$ (more uniformly distributed data) supports a higher cap. Moreover, in low-quality regimes ($\theta > 1$), the degree should saturate more quickly as $R$ grows, reflecting the sparsity of overlap. One can capture these effects with the expression: $d_t(R, \theta) = D_{\max}(\theta) [ 1 + ( \frac{R_c(\theta)}{R})^{|1 - \theta|}]^{-1},$ where $R_c(\theta)$ is a crossover scale that decreases with $\theta$.

Inserting this into the scaling relation $T \propto R^\theta / d_t(R, \theta)$, we find that in the super-linear regime ($\theta > 1$), the saturation occurs early ($R \gg R_c(\theta)$), so $d_t \approx D_{\max}(\theta)$ and $T \propto R^\theta$ remains unchanged. In contrast, for $\theta < 1$ and $R \ll R_c(\theta)$, we have $d_t \approx D_{\max}(\theta) (R / R_c(\theta))^{1 - \theta}$, and thus $T \propto R^\theta$, capturing the softer, sub-linear scaling enabled by increased overlap.

We hypothesize that this sub-linear scaling behaviour could also arise from a finitary analysis of a compression problem, analogous to how the super-linear regime emerges from \ac{BEC} behaviour in \ac{LDPC} peeling analyses.
In particular, the dual of the \ac{BEC} --- \acf{BEQ}\cite{MYIterativeQuantizationUsing2004} --- may naturally yield the sub-linear form observed here.
Recent work has shown that compression-based sample selection can improve data quality and enhance \ac{LLM} training \cite{YWW+EntropyLawStory2024}.

\section{Metrics and Statistical Measures}
\label{sec:si:metrics}

In this section, we document all metrics and statistical measures used.
As we consistently follow the widely accepted definitions for all metrics, this section's primary purpose is to provide a readily accessible point of reference.

\paragraph{Coefficient of Variation.}
The coefficient of variation is the ratio of the standard deviation to the mean:
\(CV = \sigma / \mu\),
where \(\sigma\) is the standard deviation and \(\mu\) is the arithmetic mean of the sample/population.

\paragraph{Pearson's Correlation Coefficient.}
Pearson's correlation coefficient measures the linear relationship between two variables \(X\) and \(Y\), and is defined as:
\begin{equation}
r = \frac{\sum_{i=1}^n (x_i - \bar{x})(y_i - \bar{y})}
         {\sqrt{\sum_{i=1}^n (x_i - \bar{x})^2} \sqrt{\sum_{i=1}^n (y_i - \bar{y})^2}} ,
\end{equation}
where \(\bar{x}\) and \(\bar{y}\) are the sample means of \(X\) and \(Y\), respectively.
The value of \(r\) ranges from \(-1\) (perfect negative linear correlation) to \(1\) (perfect positive linear correlation), with \(0\) indicating no linear correlation.
In simple linear regression with an intercept, \(r^2 = R^2\).

\paragraph{Spearman's Correlation Coefficient.}
Spearman's correlation coefficient, denoted \(\rho\), assesses monotonic relationships by computing Pearson's correlation on the ranks of the data. It is defined as:
\begin{equation}
\rho = 1 - \frac{6 \sum_{i=1}^n d_i^2}{n(n^2 - 1)} ,
\end{equation}
where \(d_i\) is the difference between the ranks of \(x_i\) and \(y_i\), and \(n\) is the number of observations.
Like Pearson's \(r\), Spearman's \(\rho\) ranges from \(-1\) (perfect negative rank correlation) to \(1\) (perfect positive rank correlation), with \(0\) indicating no correlation.

\paragraph{\(\chi^2\) Test.}
The chi-square (\(\chi^2\)) test assesses whether observed counts differ significantly from those expected under a null hypothesis. For a goodness-of-fit test, the test statistic is defined as:
\begin{equation}
\chi^2 = \sum_{i=1}^{k} \frac{(O_i - E_i)^2}{E_i},
\end{equation}
where \(O_i\) and \(E_i\) are the observed and expected counts in category \(i\), respectively, and \(k\) is the number of categories.
Under the null hypothesis, \(\chi^2\) approximately follows a chi-square distribution with an appropriate number of degrees of freedom.
Larger values of \(\chi^2\) indicate greater discrepancy between the observed and expected distributions, whereas smaller values indicate better agreement.

\paragraph{Combined \(\chi^2\) Goodness-of-Fit Statistic (Fisher's Method).}
To aggregate evidence across multiple hypothesis tests, we used Fisher's method, which combines \(m\) independent \(p\)-values \(p_1, \dots, p_m\) into a single test statistic:
\begin{equation}
X^2 = -2 \sum_{i=1}^{m} \log p_i .
\end{equation}
Under the null hypothesis that all tests are jointly consistent with the model, and assuming the individual \(p\)-values are independent, \(X^2\) approximately follows a chi-square distribution with \(2m\) degrees of freedom.
Larger values of \(X^2\) indicate stronger joint evidence against the null hypothesis, whereas smaller values indicate better overall agreement between the model and the observed data.
In our setting, this provides a single goodness-of-fit statistic summarizing whether multiple topological summary statistics are jointly consistent with a candidate geometric model.

\paragraph{\acf{WAIC}.}
The Watanabe-Akaike or Widely Applicable Information Criterion is ``a more fully Bayesian approach for estimating the out-of-sample expectation'' for a fitted model\cite{GCS+BayesianDataAnalysis2014}, defined as:
\begin{equation}
    \mbox{WAIC} = -2 \mbox{lpdd} + 2 p_{\mbox{WAIC}}  ,
\end{equation}
where \(\mbox{lppd}\) is the log-likelihood pointwise predictive density (\cref{eq:lppd}) and \(p_{\mbox{WAIC}}\) is the WAIC bias correction (\cref{eq:p_waic}), defined as:

\begin{align}
    \mbox{lppd} &=& \sum_i^n \ln \left ( \frac{1}{S} \sum_i^S p(y_i | \theta^s) \right) \label{eq:lppd} \\
    p_{\mbox{WAIC}} &=& 2 \sum_i^n \left(
        \ln \left ( \frac{1}{S} \sum_s^S p(y_i | \theta^s ) \right)
        - \frac{1}{S}\sum_s^S \ln p(y_i | \theta^s)
    \right) .
    \label{eq:p_waic}
\end{align}

Note that \(\theta^s\) referees to the \(s\)-th sample of the model parameters \(\theta\), and not \(\theta\) raised to the \(s\).
Here, \(S\) is the total number of samples drawn using Monte Carlo sampling, specifically \ac{NUTS}, and \(n\) is the total number of observations of \(y\).

Similar to the \acf{AIC}, \ac{WAIC} gives the negative of the expected log pointwise predictive density; and, in limit of \(n \to \infty\), it approaches the Bayesian leave-one-out cross validation lppd\cite{GCS+BayesianDataAnalysis2014}.
Summations over \(\ln p(y_i| \theta^2)\) and the outer summations were computed in a single-pass using Kahan summation\cite{KahPracniquesFurtherRemarks1965}.
For summations over \(p(y_i|\theta^s)\), we used the single pass compensated log-sum-exp implemented in \texttt{LogExpFunctions.jl}\cite{SebStreamingLogsumexpComputation2016,TSJuliaStatsLogExpFunctionsjl2024}.

\paragraph{\acf{AIC}.}
The Akaike Information Criteria is defined as follows\cite{GCS+BayesianDataAnalysis2014}:
\begin{equation}
    \mbox{AIC} = -2\ln p(y | \theta_{MLE}) + 2k  ,
\end{equation}
where \(k\) is the number of fitted parameters and \(\theta_{MLE}\) is the \ac{MLE} estimate for \(\theta\), computed here as \(\theta_{MLE} = \underset{\theta^s \in \{\theta^1, \dots \theta^S\}}{\mbox{argmax}} p(y | \theta^s) \).
That is, we take the draw \(\theta^s\) with the maximum likelihood as our \ac{MLE} estimate, instead of refitting the model entirely.

\paragraph{\acf{BIC}.}
The Bayesian Information Criteria is defined as follows\cite{GCS+BayesianDataAnalysis2014}:
\begin{equation}
    \mbox{BIC} = -2\ln p(y | \theta_{MLE}) + k \ln n  ,
\end{equation}
where \(k\) is the number of fitted parameters,
\(n\) is the number of observations of \(y\), and
\(\theta_{MLE}\) is the \ac{MLE} estimate for \(\theta\).

\paragraph{\acf{DIC}.}
The Deviance Information Criteria is defined as follows\cite{GCS+BayesianDataAnalysis2014}:
\begin{align}
    \mbox{BIC} &= -\ln p(y | \hat{\theta}_{bayes}) - p_{DIC} \\
    \hat{\theta}_{bayes} &= \frac{1}{S} \sum_{s=1}^S \theta^s \\
    p_{DIC} &= 2 \left( \ln p(y | \hat{\theta}_{bayes}) - \frac{1}{S}\sum_{s=1}^S \ln p(y|\theta^s) \right)  ,
\end{align}
where \(s \in 1 \dots S\) indexes draws \(\theta^s\) from the posterior distribution \(p(y | \theta^s)\).

\paragraph{Potential Scale Reduction Factor.}
We used the bulk-\(\hat{R}\) metric with the criteria that \(\hat{R} < 1.01\) for all fitted parameters\cite{VGS+RankNormalizationFoldingLocalization2021,GCS+BayesianDataAnalysis2014}.
Computation of \(\hat{R}\) was handled using \texttt{MCMCDiagnosticTools.jl}\cite{DSTuringLangMCMCDiagnosticToolsjl2025}, and it is defined as
\begin{align*}
    \hat{R} &= \sqrt{\frac{\hat{\mbox{var}}(\psi|y)}{W}}
        & \hat{\mbox{var}}(\psi|y) &=  \frac{n-1}{n} W + \frac{1}{n} B
    \\
    B &= \frac{n}{m-1} \sum_{j=1}^m \left( \overline{\psi_{.j}} - \overline{\psi_{..}}\right)^2
        & \overline{\psi_{.j}} &= \frac{1}{n} \sum_{i=1}^n \psi_{ij}
        & \overline{\psi_{..}} &= \frac{1}{m} \sum_{j=1}^m \overline{\psi_{.j}}
    \\
    W &= \frac{1}{m} \sum_{j=1}^m s_j^2 &
        s_j^2 &= \frac{1}{n-1} \left( \psi_{ij} - \overline{\psi_{.j}} \right)^2  ,
\end{align*}
where \(\psi_{ij}\) is the \(i\)-th of \(n\) samples of a scalar parameter \(\psi\) drawn from the \(j\)-th of \(m\) chains.
The overall parameter vector \(\theta\) is a tuple of multiple scalar parameters \(\psi\).
For example \(\theta = [A, \alpha, B, \beta, E, \sigma^2]\) for the Chinchilla\cite{HBM+TrainingComputeOptimalLarge2022} neural scaling laws.
The fitted models provided in our \datadrop[data release] report the maximum \(\hat{R}\) over all model parameters.

\paragraph{\acf{ESS}.}
We estimated the number of independent simulation draws using \ac{ESS}, as implemented by \texttt{MCMCDiagnosticTools.jl}\cite{DSTuringLangMCMCDiagnosticToolsjl2025}.
For all models, we used the criterion that the \ac{ESS} for each parameter \(\psi\) should exceed 100 times the number of chains, adjusting our simulation parameters as needed to achieve this\cite{VGS+RankNormalizationFoldingLocalization2021}.
We used eight (8) chains for our Bayesian regression models, setting the target \ac{ESS} at 800.
In practice, we achieved an \ac{ESS} around 2,000 -- 250x the number of chains.
The fitted models provided in our \datadrop[data release] report the maximum \ac{ESS} over all model parameters.
\ac{ESS} is defined as:
\begin{equation}
    n_{eff} = \frac{mn}{1 + 2 \sum_{t=1}^T \hat{\rho_t}} ,
\end{equation}
where \(m\) is the number of chains, \(n\) is the number of post-warmup samples per chain, \(\hat{\rho}_t\) is the autocorrelation between samples of \(\psi\) at lag \(t\), and \(T\) is the first lag at which the sum of two successive \(\hat{\rho}_t\) values becomes negative~\cite{GCS+BayesianDataAnalysis2014}.

\paragraph{Regression Error Metrics.}
Various regression metrics used throughout the paper are defined as follows.
\begin{align*}
    \mbox{MAE} &= \frac{1}{N} \sum_{i=1}^N \left | y_i - \hat{y}_i \right | \\
    \mbox{MAPE} &= \frac{1}{N} \sum_{i=1}^N \frac{ \left | y_i - \hat{y}_i \right | }{y_i} \\
    \mbox{MSE} &= \frac{1}{N} \sum_{i=1}^N \left ( y_i - \hat{y}_i \right )^2 \\
    \mbox{RMSE} &= \sqrt{\frac{1}{N} \sum_{i=1}^N \left| y_i - \hat{y}_i \right|^2} .
\end{align*}
In all cases, \(y\) is the true value and \(\hat{y}\) is the predicted value, with subscripts \(i\) is used to index the observation, and
the total number of observation is given by \(N\).

\paragraph{R\textsuperscript{2}.}
The coefficient of determination, or \(R^2\), is the proportion of the data's variance explained by the model.
It is defined as \(R^2 = 1 - SS_{res} / S_{tot}\) where the definitions for \(S_{tot}\) and \(S_{res}\) are:
\begin{align*}
  SS_{tot} &= \sum_i \left( y_i - \bar{y} \right)^2 \\
  SS_{res} &= \sum_i \left( y_i - \hat{y}_i \right)^2 ,
\end{align*}
where \(y\) is the true value, \(\hat{y}\) is the predicted value and \(\bar{y}\) is the mean value over all observation.

\paragraph{Multi-Target Regression Metrics.}
Consistent with MoleculeNet\cite{WRF+MoleculeNetBenchmarkMolecular2018}, we report the Average-MAE for benchmarks returning with multiple targets.
Taking \(y_i = [y_{i,1}, y_{i,2}, \ldots, y_{i,C}]\) to be a vector in
\( \mathbb{R}^{C} \), the Average-MAE is given by:
\begin{equation}
    \frac{1}{C} \sum_c \mbox{MAE}( y_{i,c}, \hat{y}_{i,c} )  ,
\end{equation}
where \(y_{i,c}\) is the \(i\)-th observation of the \(c\)-th channel.
We compute the Avg-\(R^2\) in an identical manner.

\paragraph{Cross-Entropy Loss.}
Cross-entropy loss is the primary training objective for language models~\cite{VSP+AttentionAllYou2017,DCLTBERTPretrainingDeep2019}.
Given a model that assigns a logit score \(Q(x)\) to each token \(x \in V\), the cross-entropy loss over a sequence \(x_0, x_1, \dots, x_N\) is defined as:
\begin{equation}
\label{eq:crossentropy}
  H(x_0, x_1, \dots, x_N) = - \sum_{i=0}^N \ln \frac{\exp Q(x_i)}{\sum_{x \in V} \exp Q(x)} .
\end{equation}
\noindent
This formulation corresponds to the negative log-likelihood of the observed sequence under the model distribution, and it can be interpreted as an online estimate of the cross-entropy \(H(P, Q) = -\sum_x P(x) \ln Q(x)\) between the empirical data distribution \(P(x)\) and the model’s predicted distribution \(Q(x)\).
Cross-entropy is reported in units of \emph{nats}—the natural unit of information per token—since it uses base-\(e\) logarithms.
Alternative units include the \emph{shannon} (base-2) and \emph{hartley} (base-10)\cite{BS80000132008}, which are related to the choice of logarithmic base.
This formulation is identical to our proposed definition for molecular surprise, presented in \cref{sec:screening}.

\paragraph{KL-Divergence.}
The Kullback-Leibler divergence, \(D_{KL}(P || Q)\), is a measure of the difference between two probability distributions \(P\) and \(Q\).
Similarly to cross-entropy, \(D_{KL}\) is reported in units of nats:
\begin{equation}
    D_{KL}(P || Q) = \sum_{x \in V} P(x) \ln \frac{P(x)}{Q(x)} ,
\end{equation}
where \(x\) is a token in the vocabulary \(V\).

\paragraph{\acs{AUROC}.}
The \acf{AUROC} is a metric for the performance of a binary classifier.
It is defined as the area under the curve when plotting the true positive rate (TPR) against the false positive rate (FPR).
The True Positive Rate (TPR) is the proportion of actual positive cases that are correctly identified by the model, while the False Positive Rate (FPR) is the proportion of actual negative cases that were incorrectly identified as positive.

\paragraph{Maximum Eigenvalue ($\lambda_{max}$).}
The $\lambda_{max}$ metric used to diagnose spiking is based on prior \ac{HT-SR} work \cite{MTPMPredictingTrendsQuality2021}.
It is the maximum eigenvalue of $\mathbf{X}$, where $\mathbf{X}$ is the normalized correlation matrix for layer weight matrix $\mathbf{W}$ ($\mathbf{X} = \frac{1}{N}\mathbf{W}^T\mathbf{W}$).

\FloatBarrier

\clearpage

\printbibliography[
    title={Supplementary References},
    segment=\therefsegment,
    heading=bibintoc,
]
\end{refsegment}

\end{document}